\documentclass[useAMS,usenatbib]{mn2e}

\usepackage{xspace}
\usepackage{graphicx}
\usepackage{rotate} 
\usepackage{subfigure}
\usepackage{placeins}
\usepackage{float}
\usepackage{bm}
\renewcommand\deg {\hbox{$^\circ$}}
\graphicspath{{./figs/}}
\newcommand{\Msun}{\ensuremath{\rm M_\odot}\xspace}

\newcommand{\etal}{et al.\xspace}

\newcommand{\fluxunits}{erg\,cm$^{-2}$\,s$^{-1}$\xspace}
\newcommand\mic   {\hbox{$\mu{\rm m}$}}
\newcommand\C     {\textsc{Clumpy}}        
\newcommand\Ni    {\hbox{${\cal N}_{i}$}}  
\newcommand\Nc    {\hbox{${\cal N}_{C}$}}  
\newcommand\No    {\hbox{${\cal N}_0$}}    
\newcommand\sig   {\hbox{$\sigma$}}
\newcommand\tv    {\hbox{$\tau_{\rm V}$}}  
\newcommand\pobsc {\hbox{$P_{\rm obsc}$}}

\title[X-ray Constraints for Clumpy Tori]{First X-ray-Based Statistical Tests for Clumpy-Torus Models: Eclipse Events from 230 Years of Monitoring of Seyfert AGN}

\author[Markowitz, Krumpe, Nikutta]
{A.\ G.\ Markowitz$^{1,2,3}$\thanks{E-mail: almarkowitz@ucsd.edu},
M.\ Krumpe$^{4,1}$, 
R.\ Nikutta$^{5}$ \\
$^{1}$University of California, San Diego, Center for Astrophysics 
  and Space Sciences, 9500 Gilman Dr., La Jolla, CA 92093-0424, USA\\
$^{2}$Dr.\ Karl Remeis Sternwarte, Sternwartstrasse 7, D-96049 Bamberg, Germany\\
$^{3}$Alexander van Humboldt Fellow\\
$^{4}$European Southern Observatory, Karl-Schwarzschild-Strasse 2, 
     D-85748 Garching bei M\"{u}nchen, Germany\\
$^{5}$Departamento de Ciencias Fisicas,
  Universidad Andr\'{e}s Bello, Av.\ Rep\'{u}blica 252, Santiago, Chile   }

\begin{document}

\date{Accepted 2013 December 24. Received 2013 November 29; in
  original form 2013 September 3}

\pagerange{\pageref{firstpage}--\pageref{lastpage}} \pubyear{2013}

\maketitle

\label{firstpage}

\begin{abstract}
  We present an analysis of multi-timescale variability in
  line-of-sight X-ray absorbing gas as a function of optical
  classification in a large sample of Seyfert active galactic nuclei
  (AGN) to derive the first
  X-ray statistical constraints for clumpy-torus models. We
  systematically search for discrete absorption events in the vast
  archive of \textit{Rossi X-ray timing Explorer} 
  monitoring of dozens of nearby type I and
  Compton-thin type II AGN. We are sensitive to discrete absorption
  events due to clouds of full-covering, neutral or mildly ionized gas
  with columns $\ga 10^{22-25}$~cm$^{-2}$ transiting the line of
  sight.

  We detect 12 eclipse events in 8 objects, roughly tripling the
  number previously published from this archive. Peak column densities
  span $\sim4-26 \times 10^{22}$~cm$^{-2}$, i.e., there are no
  full-covering Compton-thick events in our sample. Event durations
  span hours to months. The column density profile for an eclipsing
  cloud in NGC~3783 is doubly spiked, possibly indicating a cloud that
  is being tidally sheared.

  We infer the clouds' distances from the black hole to span $\sim 0.3
  - 140 \times 10^{4} R_{\rm g}$. In seven objects, the clouds'
  distances are commensurate with the outer portions of Broad Line
  Regions (BLR), or outside the BLR by factors up to $\sim10$ (the
  inner regions of infrared-emitting dusty tori). We discuss implications
  for cloud distributions in the context of clumpy-torus models. Eight
  monitored type II AGN show X-ray absorption that is consistent with
  being constant over timescales from 0.6 to 8.4~yr. This can either
  be explained by a homogeneous medium, or by X-ray-absorbing clouds
  that each have $N_{\rm H} \ll 10^{22}$~cm$^{-2}$. The probability of
  observing a source undergoing an absorption event, independent of
  constant absorption due to non-clumpy material, is
  $0.006^{+0.160}_{-0.003}$ for type Is and $0.110^{+0.461}_{-0.071}$
  for type IIs.

\end{abstract}

\begin{keywords}
galaxies: active -- X-rays: galaxies -- galaxies: Seyfert
\end{keywords}
\section{INTRODUCTION}

While it is generally accepted that active galactic nuclei (AGN) are
powered by accretion of gas on to supermassive ($10^{6-9}\Msun$) black
holes, the exact geometry and mechanisms by which material gets
funneled from radii of kpc/hundreds of pc down to the accretion disk
at sub-pc scales remains unclear. This unsolved question is, however,
linked to the overall efficiency of how supermassive black holes are
fed and to AGN duty cycles.

Clues to the morphology of the circumnuclear gas come from different
studies across the electromagnetic spectrum. For instance, Seyfert AGN
are generally classified into two broad categories based on the detection
or lack of broad optical/UV emission lines (type I or II,
respectively), with intermediate subclasses (1.2, 1.5, 1.8, 1.9)
depending on broad lines' strengths (e.g., Osterbrock 1981).
Unification theory posits that all Seyferts host the same central
engine, but observed properties depend on orientation due to the
presence of dusty circumnuclear gas blocking the line of sight to the
central illuminating source in type II Seyferts.  The classical model
holds that a primary component of this obscuring gas is a pc-scale,
equatorial, dusty torus (Antonucci \& Miller 1985; Urry \& Padovani
1995), where ``torus'' generally denotes a donut-shaped morphology,
supplying the accretion disk lying co-aligned inside it. However, some
recent papers (e.g., Pott et al.\ 2010) use the term ``torus'' to
denote simply the region where circumnuclear gas can exist, with the
precise morphology of that region still to be determined; we adopt
that notation in this paper.

A donut morphology can explain the observed bi-conical morphology of
Narrow Line Regions (NLR), e.g., via collimation of ionizing radiation
(e.g., Evans \etal\ 1994), and also why most type II Seyferts show evidence for
X-ray obscuration along the line of sight.  However, the relation
between optical/UV-reddening dust and X-ray absorbing gas is not
straightforward.  Very roughly 5 per cent of all AGN have differing X-ray
and optical obscuration indicators (e.g., X-ray-absorbed type Is,
Perola \etal\ 2004; Garcet \etal\ 2007); a complication is that
optical obscuration probes only dusty gas, while X-ray absorption
probes both dusty and dust-free gas, and X-ray columns can exceed
inferred optical-reddening columns by factors of 3--100 (Maiolino
\etal\ 2001).  Lutz et al.\ (2004) and Horst et al.\ (2006)
demonstrated that the infrared (IR) emission by dust, which 
thermally re-radiates higher energy radiation, is isotropic, with
type I and II Seyferts having similar ratios of X-ray to mid-IR
luminosity. This is not expected from classical unification models,
since the classical donut-shaped torus both absorbs and re-emits
anisotropically.  Furthermore, if the torus is comprised of primarily
Compton-thick gas, i.e., with X-ray-absorption column densities
$N_{\rm H}$ above $1.5 \times 10^{24}$~cm$^{-2}$, then it is not clear
how Compton-thin columns can be attained unless lines of sight happen
to ``graze'' the outer edges.  A unification scheme based solely on
inclination angle and the presence of a donut-shaped absorbing
morphology may thus be an oversimplification (Elvis 2012).  It is also
not clear how geometrically thick structures (scale heights $H/R \sim
1$) comprised of cold gas (100 K) are supported vertically over long
timescales (Krolik \& Begelman 1988).

Meanwhile, the community has accumulated observations of variations in
the X-ray absorbing column $N_{\rm H}$ in both (optically classified)
type Is and IIs, with timescales of variability ranging from hours to
years. For instance, Risaliti, Elvis, \& Nicastro (2002; hereafter
REN02) claim that variations in $N_{\rm H}$ in a sample of
X-ray-bright, Compton-thin and moderately Compton-thick type IIs are
ubiquitous, with typical variations up to factors of $\sim1.5-3$.  
Current X-ray missions such as \textit{XMM-Newton}, \textit{Chandra}
and \textit{Suzaku} have enabled high precision studies of variations
in $N_{\rm H}$ in numerous AGN; major findings include the following:

$\bullet$ Numerous moderately Compton-thick variations
($\Delta$$N_{\rm H} \sim 10^{23-24}$~cm$^{-2}$) on timescales $\la
1-2$ d in NGC~1365 (Risaliti \etal\ 2005, 2007, 2009a) and ESO
323--G77 (Miniutti \etal\ 2014); moderately Compton-thick variations
on timescales from days to months in NGC~7582 (Bianchi \etal\ 2009).

$\bullet$ Changes in covering fraction of partial-covering absorbers
on timescales from $<$1 d (NGC~4151, Puccetti \etal\ 2007; NGC~3516,
Turner \etal\ 2008; Mkn~766, Risaliti \etal\ 2011; SWIFT J2127.4+5654,
Sanfrutos \etal\ 2013) to months--years in NGC~4151 (de Rosa \etal\
2007 and Markowitz \etal\ in preparation, using \textit{BeppoSAX} and
\textit{Rossi X-ray Timing Explorer} (\textit{RXTE}) data, respectively).

$\bullet$ Time-resolved spectroscopy of full eclipse events (ingress
and egress), yielding constraints on clouds' density profiles for
long-duration (3--6 months) eclipses in NGC~3227 (Lamer \etal\ 2003)
and Cen~A (Rivers \etal\ 2011b) and for eclipses $\la$1 d in NGC~1365
(Maiolino \etal\ 2010) and SWIFT J2127.4+5654 (Sanfrutos \etal\ 2013).

These results suggest that the circumnuclear absorbing gas is
\textit{clumpy}, with non-homogeneous or clumpy absorbers being
invoked to explain time-variable X-ray absorption as far back as
\textit{Ariel V} and \textit{Einstein} observations (Barr \etal\ 1977;
Holt \etal\ 1980).  With the concept of a homogeneous, axisymmetric
absorber thus under scrutiny, the community has been developing torus
models incorporating clumpy gas, e.g., Elitzur \& Shlosman (2006) and
Nenkova \etal\ (2008a, 2008b; see H\"{o}nig 2013 for a review),
although suggestions that the torus should consist of clouds go as far
back as, e.g., Krolik \& Begelman (1986, 1988).  In the most recent
models, total line-of-sight absorption for a given source is
quantified as a viewing dependent \textit{probability} based on the
size and locations of clouds, although typically, clouds are
preferentially distributed towards the equatorial plane.  The fraction
of obscured sources depends on the average values and distributions of
such parameters as the average number of clouds lying along a radial
path, and the thickness of the cloud distribution.  Clouds are
possibly supported vertically by, e.g., radiation pressure (Krolik
2007) or disk winds (Elitzur \& Shlosman 2006).

Observations so far suggest that clouds are typically on the order of
$10^{13-15}$~cm in diameter, with number densities $\sim 10^{8-11}$
cm$^{-3}$. Inferred distances from the black hole, usually based on
constraints from X-ray ionization levels and assumption of Keplerian
motion, range from light-days and commensurate with clouds in the
Broad Line Region (BLR; e.g., for NGC~1365 and SWIFT J2127.4+5654) to
several light-months (Rivers \etal\ 2011b for Cen~A) and commensurate
with the IR-emitting torus in that object. We emphasize
``commensurate,'' as X-ray absorbers lie along the line of sight, but
BLR clouds likely contain components out of the line of sight.

Recent IR interferometric observations have spatially resolved
distributions of dust down to radii of tenths of pc (e.g., Kishimoto
\etal\ 2009, 2011; Tristram \etal\ 2009; Pott \etal\ 2010).  In
addition, there are suggestions from reverberation mapping of the
thermal continuum emission in four Seyferts that warm dust gas extends
down to $\sim10-80$ light-days, likely just outside the outer BLR
(Suganuma \etal\ 2006).  As suggested by Netzer \& Laor (1993),
Elitzur (2007), and Gaskell et al.\ (2008), the BLR and dusty torus
may be part of a common radially extended structure, spanning radii
inside and outside the dust sublimation radius $R_{\rm d}$,
respectively, since dust embedded in the gas outside $R_{\rm d}$
suppresses optical/UV line emission.  Optical obscuration is due to
dusty gas, while X-ray obscuration can come from dusty or dust-free
gas. X-ray obscuration is thus the only way to probe obscuration
inside $R_{\rm d}$.

In order to gain a more complete picture of the geometry of
circumnuclear gas in AGN, however, the community needs to gauge the
relevance of clumpy-absorber models over a wide range of length
scales, including both inside and outside $R_{\rm d}$, and to clarify
the links between the distributions of dusty gas, X-ray-absorbing gas,
and the BLR.  To date, however, observational constraints to limit
parameter space in clumpy-torus models has been lacking because there
has been no statistical survey so far.  One of our goals for this
paper is to derive constraints on clumpy-torus models via variable
X-ray-absorbing gas, including estimates of the probability that the
line of sight to the AGN is intercepted by a cloud.

We use the vast archive of \textit{RXTE} 
multi-timescale light curve monitoring of AGN, as
described in $\S$2.  We search for changes in full- or
partial-covering $N_{\rm H}$ in Seyferts.  We use a combination of
light curve hardness ratios and time-resolved spectroscopy to identify
and confirm eclipses, which are summarized in $\S$3. We use the
observed eclipse durations and the observation sampling patterns to
estimate the probability to observe a source undergoing an eclipse in
$\S$4.  In $\S$5, we infer the eclipsing
clouds' radial locations and physical properties, relate the
X-ray-absorbing clouds to other AGN emitting components, and describe
the resulting observational constraints for key parameters of
clumpy-torus models.  The results are summarized in $\S$6.  In a
separate paper (Nikutta et al., in preparation), we will extend our analysis
of the cloud properties based on the X-ray data presented in this
paper.

\section{OBSERVATIONS, DATA REDUCTION, AND ECLIPSE IDENTIFICATION}

Our strategy to identify eclipse events in general follows that used
in successful detections by, e.g., Risaliti \etal\ (2009b, 2011),
Rivers \etal\ (2011b), etc.  We first extract sub-band light curves
for all objects with sufficient X-ray monitoring, and examine hardness
ratios to identify possible eclipse events, which manifest themselves
via sudden increases in hardness ratio. We then perform time-resolved
spectroscopy, binning individual adjacent observations as necessary to
achieve adequate signal to noise ratio, to attempt to confirm such events as
being due to an increase in $N_{\rm H}$ as opposed to a flattening of
the continuum photon index.

\textit{RXTE} has already revealed long-term (months in duration)
eclipse events with $\Delta$$N_{\rm H} \sim 10^{23}$~cm$^{-2}$ for
four objects: as mentioned above, complete eclipse events were
confirmed via time-resolved spectroscopy for NGC~3227 in 2000--1
(Lamer \etal\ 2003) and Cen~A in 2010--2011 (Rivers \etal\ 2011b).  As
discussed by Rivers \etal\ (2011b) and Rothschild \etal\ (2011), there
is evidence for an absorption event in Cen~A in 2003--2004 with a similar
value of $\Delta$$N_{\rm H}$ to the 2010--2011 event.  Smith \etal\
(2001) and Akylas \etal\ (2002) presented evidence for a decrease in 
$N_{\rm H}$ during
1996--1997 in the Compton-thin Sy~2 Mkn~348, suggesting that
\textit{RXTE} witnessed the tail end of an absorption event with
$\Delta$$N_{\rm H} \ga 14 \times 10^{22}$~cm$^{-2}$.

\subsection{Target selection from the \textit{RXTE} database}

\textit{RXTE} operated from 1995 December until 2012 January.  We consider
data from its Proportional Counter Array (PCA; Jahoda \etal\ 2006),
sensitive over 2--60~keV. \textit{RXTE}'s unique attributes -- large
collecting area for the PCA, rapid slewing and flexible scheduling --
made it the first X-ray mission to permit sustained monitoring
campaigns, with regularly spaced visits, usually 1--2~ks each, over
durations of weeks to years.

\textit{RXTE} visited 153 AGN 
during the mission, with sustained monitoring (multiple individual
observations) spanning durations of $\sim4$ d or longer occurring for
118 of them.  \textit{RXTE} monitored AGN for a variety of science
pursuits, including, for example, interband correlations to probe
accretion disk structure and jet-disk links (e.g., Ar\'{e}valo
\etal\ 2008; Breedt \etal\ 2009; Chatterjee \etal\ 2011), X-ray timing
analysis to constrain variability mechanisms in Seyferts (e.g.,
Markowitz \etal\ 2003; McHardy \etal\ 2006), and coordinated
multi-wavelength campaigns on blazars during giant outbursts to
constrain spectral energy distributions (SEDs) 
and thus models for particle injection/acceleration in
jets (e.g., Krawczynski \etal\ 2002; Collmar \etal\ 2010).  The
archive thus features a wide range of sampling frequencies and
durations from object to object. Typical long-term campaigns consisted
of one observation every 2--4 d for durations of months to years
(15.4~yr in the longest case, NGC~4051).  A few tens of objects were
subject to more intensive monitoring consisting of e.g., 1--4 visits
per day for durations of weeks.

For this paper, we do not consider sources visited less than four times
during the mission; many sources were in fact visited hundreds to
more than a thousand times during \textit{RXTE}'s lifetime.  We
rejected sources whose mean 2--10~keV flux is $\la 8 \times
10^{-12}$ \fluxunits; such sources had very large error bars in the
sub-band light curves and hardness ratios, and poor constraints from
spectral modeling.  

We also searched for eclipse events in the 29 blazars that were
monitored with \textit{RXTE} and have average 2--10~keV fluxes
$>8\times 10^{-12}$~erg cm$^{-2}$ s$^{-1}$.  These are sources
considered under unification schemes to have jets aligned along the
line of sight, with jet emission drowning out emission from the
accretion disk and corona. We found no evidence for eclipses from the
hardness ratio light curves.\footnote{This is not surprising, given
  blazars' orientation and that lines of sight along the poles have
  the lowest likelihood to have obscuring clouds in $\C$ models
  ($\S$5.5), and additionally given that jets might destroy clouds or
  push them aside.}  We do however include 3C 273 in our final sample,
since its X-ray spectrum is likely a composite of typical Sy 1.0 and
blazar spectra (e.g., Soldi \etal\ 2008).

As per $\S2.3$ below, we can detect if a source changes from
Compton thin to Compton thick. However, we cannot accurately quantify
changes in $N_{\rm H}$ where there is already a steady full-covering
Compton-thick absorber present.  This is due to the energy
resolution and bandpass.  The presence of Compton reflection in some
of the Compton-thick sources observed with \textit{RXTE} (Rivers
\etal\ 2011a, 2013) combined with the fact that Compton-thick
absorption causes a rollover around 10~keV means there is not
sufficient ``leverage'' in the spectrum to fully deconvolve the power
law, Compton reflection, and absorption with short exposure times.  In
such cases, we also cannot detect addition absorption by $10^{22-23}$
cm$^{-2}$.  We checked the hardness ratio light curves (see below) of
the seven Compton-thick sources monitored with \textit{RXTE}, but
there was no significant evidence for such sources becoming
Compton thin or unabsorbed; the only variations in hardness ratio were
small and consistent with modest variations in the photon index of the
coronal power law.  We exclude Compton-thick-absorbed sources from our
final sample and do not discuss them further.

The final target list consists of 37 type I and 18 Compton-thin type
II AGN, listed in Tables~\ref{tab:monit1} and \ref{tab:monit2},
respectively. We use the optically classified subtypes from the
NASA/IPAC Extragalactic Database as follows: we group subtypes 1.0,
1.2, and 1.5 together into the type I category. We have no Sy 1.8s in
our sample.  We group subtypes 1.9 (objects where H$\alpha$ is the
only broad line detected in non-polarized optical light) and 2
(objects with no broad lines detected in non-polarized optical light)
into the type II category.  15 of the 18 type IIs are Sy 2s. In
general, it is not entirely clear whether Sy 2s intrinsically lack
BLRs or if the BLR in those objects is present but
``optically hidden,'' manifesting itself only in scattered/polarized
optical emission or via Paschen lines in the IR band.  However, 9/15
Sy 2s and 2/3 Sy 1.9s show evidence for ``optically hidden'' BLRs,
with references listed in Table~\ref{tab:monit2}.  We adhere to the
assumption that BLRs exist in all our objects.  Our distinction
between type Is and type IIs is thus based on the assumption of
relatively increasing levels of obscuration in the optical band,
independent of assumptions about system orientation.
The average redshift of the type Is is $\langle z \rangle = 0.045$,
with all but five objects (3C~273, MR~2251--178, PDS~456, PG~0052+251,
and PG~0804+761) having $z<0.100$. For the type IIs, the average
redshift is $\langle z \rangle = 0.011$. 

The majority of these 55 objects are well studied with most major
X-ray missions.  Previous publications (e.g., REN02, Patrick et
al.\ 2012) usually find $N_{\rm H} > \sim 5\times 10^{21}$~cm$^{-2}$
for each of the 18 type IIs, with NGC~6251 a possible exception
($N_{\rm H} \la 5 \times 10^{20}$~cm$^{-2}$, Dadina 2007;
Gonz\'{a}lez-Mart\'{i}n et al.\ 2009).  X-ray obscuration in type Is
is generally not as common, and with generally lower columns.  For
example, 28 of the 37 type Is in our monitored sample overlap with the
\textit{Suzaku} sample of Patrick et al.\ (2012). Those authors model
full- or partial-covering neutral or warm absorbers in 21 of the 28
sources, but only 12 have at least one component with $N_{\rm H} > 1
\times 10^{22}$~cm$^{-2}$, with only two of those having neutral
absorbers (NGC~4151 and PDS~456).

We define an ``object-year'' as one target being monitoring for a
total of one year including smaller gaps in monitoring due to, e.g.,
satellite sun-angle constraints or missing individual observations,
but excluding lengthy gaps $>75$ per cent\footnote{Other values, e.g.,
  between 50 and 90 per cent yield virtually identical results for
  this calculation.} of the full duration, e.g., yearly observing cycles when no
observations were scheduled.  We estimate totals of 189 and 41
object-years for the type Is and the Compton-thin type IIs,
respectively.  (These values fall to 169 and 26 object-years when
sun-angle gaps are removed.)  Consequently, with a total of roughly
230 years of monitoring 55 AGN, and with a wide dynamic range in
timescales sampled, the present data set is by far the largest ever
available for statistical studies of cloud events in AGN to timescales
from days to years.

\begin{table*}
\begin{minipage}{180mm}
\caption{Sample of type I AGN monitored with \textit{RXTE}}
\label{tab:monit1}
\begin{tabular}{llccccl}
\hline
Source        & Opt.\               & Redshift  & log($L_{\rm 2-10~keV}$) & $D_{\rm min}$ &  $D_{\rm max}$   &   \\ 
name          & class.\             & $z$    & (erg s$^{-1}$)   & (d) &  (Tot.\ gap frac.)   &   Comments  \\ \hline
3C~111        & BLRG/Sy1            & 0.0485 & 43.9            & 0.41        & 14.77 yr  ($54.3\%$) & \\
3C~120        & BLRG/Sy1            & 0.0330 & 43.4            & 0.21        & 11.14 yr  ($32.5\%$) & \\
3C~273        & QSO/FSRQ/Sy1        & 0.1583 & 45.7            & 0.37        & 15.91 yr  ($28.2\%$) &   \\ 
3C~382        & BLRG/Sy1            & 0.0579 & 44.0            & 0.38        & 7.59 yr   ($99.7\%$) &  \\
3C~390.3      & BLRG/Sy1            & 0.0561 & 43.8            & 0.66        & 8.66 yr   ($73.6\%$) & \\ 
Ark 120       & Sy1                 & 0.0327 & 43.4            & 0.33        & 5.50 yr   ($72.4\%$) & \\
Ark 564       & NLSy1               & 0.0247 & 42.9            & 0.21        & 6.19 yr   ($32.6\%$) & \\ 
Fairall 9     & Sy1                 & 0.0470 & 43.4            & 0.24        & 6.32 yr   ($16.1\%$) &  Candidate event, 2001.3 \\
Mkn 335       & NLSy1               & 0.0258 & 42.7            & 5.2         & 0.99 yr   ($15.0\%$) &  $^{\dagger}$ \\
Mkn 766       & NLSy1               & 0.0129 & 42.5            & 0.47        & 10.65 yr  ($27.2\%$) &  \\
NGC 3783      & Sy1                 & 0.0097 & 42.6            & 0.37        & 15.92 yr  ($30.1\%$) & Double eclipse, 2008.3. \\
              &                     &        &                 &             &                  & Candidate event, 2008.7 \\
              &                     &        &                 &             &                  & Candidate event, 2011.1 \\
NGC 3998      & Sy1/LINER           & 0.0035 & 40.8            & 0.60        &  0.99 yr  ($0\%$) & $^{\dagger}$ \\
NGC 4593      & Sy1                 & 0.0090 & 42.3            & 0.34        & 10.51 yr  ($39.2\%$) & \\
PDS 456       & Sy1/QSO             & 0.1840 & 44.2            & 0.20        & 11.32 yr  ($88.9\%$) &  $^{\dagger}$  \\
PG 0804+761   & Sy1                 & 0.1000 & 43.9            & 0.91        &  5.91 yr  ($80.2\%$) & \\
PG 1121+343   & NLSy1               & 0.0809 & 43.4            & 1.8         &  0.89 yr  ($40.7\%$) &   $^{\dagger}$  \\
Pic A         & BLRG/Sy1/LINER      & 0.0351 & 43.2            & 3.3         & 3.3 d     ($0\%$) &   \\
PKS 0558--504 & RL NLSy1            & 0.1370 & 44.3            & 1.3         & 14.21 yr ($52.8\%$) &   \\
PKS 0921--213 & FSRQ/Sy1            & 0.0520 & 43.2            & 0.27        & 4.6 d    ($0\%$) &  $^{\dagger}$ \\
4U~0241+622   & Sy1.2               & 0.0440 & 43.6            & 2.2         & 31.8 d   ($0\%$) & \\
IC 4329a      & Sy1.2               & 0.0161 & 43.2            & 0.25        & 11.01 yr ($64.2\%$) & \\
MCG--6-30-15  & NLSy1.2             & 0.0077 & 42.2            & 0.20        & 14.78 yr ($26.1\%$) & \\
Mkn 79        & Sy1.2               & 0.0222 & 42.8            & 0.39        & 11.81 yr ($20.0\%$) & Eclipses, 2003.5, 2003.6, \& 2009.9  \\
Mkn 509       & Sy1.2               & 0.0344 & 43.4            & 0.21        & 10.24 yr ($71.4\%$) &  Eclipse, 2005.9  \\
Mkn 590       & Sy1.2               & 0.0264 & 43.2            & 5.9         & 0.80 yr  ($0\%$)       &  \\
NGC 7469      & Sy1.2               & 0.0163 & 42.7            & 0.20        & 13.71 yr  ($56.9\%$)  &  \\
PG 0052+251   & Sy1.2               & 0.1545 & 44.1            & 3.0         & 7.51 yr   ($89.2\%$) &  $^{\dagger}$ \\
MCG--2-58-22  & Sy1.5               & 0.0469 & 43.6            & 0.32        & 160.5 d   ($88.7\%$) & \\
Mkn 110       & NLSy1.5             & 0.0353 & 43.4            & 0.41        & 11.81 yr  ($35.5\%$) &  \\
Mkn 279       & Sy1.5               & 0.0305 & 43.1            & 0.26        & 6.00 yr   ($99.2\%$) & \\ 
MR 2251--178  & QSO/Sy1.5           & 0.0640 & 44.0            & 1.7         & 15.05 yr  ($58.\%$) & 1996 obsn.\ during eclipse. \\
NGC 3227      & Sy1.5               & 0.0039 & 41.5            & 0.53        & 6.92 yr   ($36.4\%$) & Eclipses 2000--1 (Lamer \etal\ 2003) \\
              &                     &        &                 &             &                  &    and 2002.8 \\
NGC 3516      & Sy1.5               & 0.0088 & 42.3            & 0.64        & 14.79 yr  ($62.4\%$) & Candidate event, 2011.7 \\
NGC 4051      & NLSy1.5             & 0.0023 & 40.8            & 0.20        & 15.69 yr  ($0.7\%$) &  \\ 
NGC 4151      & Sy1.5               & 0.0033 & 42.2            & 0.21        & 8.36 yr   ($70.8\%$) & Var.\ partial covering $N_{\rm H}\sim$ \\
              &                     &        &                 &             &                  & $10^{23.5}$~cm$^{-2}$ (De Rosa \etal\ 2007; \\ 
              &                     &        &                 &             &                  & Markowitz in prep.) \\
NGC 5548      & Sy1.5               & 0.0172 & 42.9            & 0.26        & 15.65     ($26.8\%$) &  \\
NGC 7213      & RLSy1.5             & 0.0058 & 41.7            & 1.1         & 3.83 yr   ($0\%$) &  \\ \hline
\end{tabular}\\
2--10~keV luminosities refer to the hard X-ray
  power-law component, which are corrected for absorption, and are
  taken from Rivers \etal\ (2013).  Optical classifications are taken
  from NED.  We collectively refer to type Is as including all Seyfert
  1.0s, 1.2s and 1.5s.   
  $^{\dagger}$ denotes that the 10--18~keV band had poor signal-to-noise ratio.
  $D_{\rm min}$ and $D_{\rm max}$ denote the lengths of the minimum
  and maximum campaign durations, respectively, where one campaign is
  defined as a minimum of four observations, with no single gap $>75$ per cent
  of the duration (see $\S$4), i.e., it does not necessarily mean
  sustained, regular monitoring for the entire duration.  The values
  in parentheses denote the accumulated fraction of missing time of
  the single longest campaign due to gaps in monitoring (e.g.,
  sun-angle constraints, or campaigns consisting of only a few
  observations concentrated into $\sim$days--weeks separated by
  $\sim$years).  
\end{minipage}
\end{table*}

\begin{table*}
\begin{minipage}{175mm}
\caption{Sample of Compton-thin type II AGN monitored with \textit{RXTE}}
\label{tab:monit2}
\begin{tabular}{llccccl}
\hline
Source        & Opt.\               & Redshift  & log($L_{\rm 2-10~keV}$) & $D_{\rm min}$ &  $D_{\rm max}$   &   \\ 
name          & class.\             & $z$    & (erg s$^{-1}$)   & (d)       &  (Tot.\ gap frac.)   &   Comments  \\ \hline
NGC 526a      & NELG/Sy1.9          & 0.0192 & 43.1            & 0.20        & 5.9 d     ($99.0\%$) &   \\
NGC 5506      & Sy1.9               & 0.0062 & 42.4            & 0.20        & 8.39 yr   ($38.6\%$) & Eclipse, 2000.2; BLR:Pa.(N02)\\
NGC 7314      & Sy1.9               & 0.0048 & 41.7            & 0.30        & 3.55 yr   ($65.0\%$) & BLR:Pol.(L04)  \\
Cen A         & NLRG/Sy2            & 0.0018 & 41.6            & 0.24        & 15.37 yr  ($85.5\%$) &  Eclipses in 2010--1 and $\sim$2003--4: \\
              &                     &        &                 &             &                      & Rivers \etal\ (2011b), \\
              &                     &        &                 &             &                      & Rothschild \etal\ (2011) \\ 
NGC 4258      & Sy2/LLAGN           & 0.0015 & 40.3            & 0.34        & 15.07 yr  ($37.2\%$) &   $^{\dagger}$           \\
ESO 103--G35  & Sy2                 & 0.0133 & 42.4            & 0.66        & 0.59 yr   ($95.3\%$) &  \\
IC~5063       & Sy2                 & 0.0113 & 42.1            & 150.1       & 0.81 yr   ($99.8\%$) &  BLR:Pol.(L04) \\
IRAS~04575--7537 & Sy2              & 0.0181 & 42.7            & 5.8         & 0.61 yr   ($81.3\%$) &         \\
IRAS~18325--5926 & Sy2              & 0.0202 & 42.8            & 0.67        & 2.07 yr   ($98.5\%$) &  BLR:Pol.(L04)  \\
MCG--5-23-16   & Sy2/NELG           & 0.0085 & 42.7            & 0.46        & 9.63 yr   ($99.9\%$) &   BLR:Pa.(G94)\\ 
Mkn 348        & Sy2                & 0.0150 & 42.5            & 0.36        & 1.13 yr   ($70.2\%$) &  Eclipse 1996-7: Akylas \etal\ (2002); \\
               &                    &        &                 &             &                      &  BLR:Pol.(M90) \\
NGC 1052       & RLSy2              & 0.0050 & 41.2            & 18.1        & 4.56 yr   ($13.8\%$) &  $^{\dagger}$; BLR:Pol.(B99)     \\
NGC 2110       & Sy2                & 0.0076 & 42.2            & 0.65        & 2.0 d     ($0\%$) &    BLR:Pa.(R03) \\  
NGC 2992       & Sy2                & 0.0077 & 42.0            & 27.9        & 0.90 yr   ($10.3\%$) &  BLR:Pa.(R03) \\
NGC 4507       & Sy2                & 0.0118 & 42.8            & 1.1         & 15.2 d    ($0\%$) &   BLR:Pol.(M00) \\          
NGC 6251       & LERG/Sy2           & 0.0247 & 42.1            & 7.7         & 0.99 yr   ($0\%$) & $^{\dagger}$  \\
NGC 7172       & Sy2                & 0.0087 & 42.1            & 5.0         & 12.2 d    ($0\%$) &    \\
NGC 7582       & Sy2                & 0.0053 & 41.3            & 0.61        & 1.24 yr   ($97.3\%$) & BLR:Pa.(R03) \\ \hline
\end{tabular}\\
Same as Table~\ref{tab:monit1}, but for our sample of type IIs, which includes all Seyfert 1.9s and 2s.
``BLR:Pa.'' indicates a ``hidden'' BLR with broad
  Paschen lines; references are: G94 = Goodrich et al.\ (1994), N02 =
  Nagar et al.\ (2002), R03 = Reunanen et al.\ (2003).  ``BLR:Pol.''
  indicates a ``hidden'' BLR with broad Balmer lines detected in
  polarized emission; references are : B99 = Barth et al.\ (1999), L04
  = Lumsden et al.\ (2004), M90 = Miller \& Goodrich (1990), M00 =
  Moran et al.\ (2000).
\end{minipage}
\end{table*}

\subsection{Summary of data reduction and light curve extraction}  

We extract light curves and spectra for each observation for each
target following well-established data reduction pipelines and
standard screening criteria.  We use \textsc{heasoft} version 6.11 software.
We use PCA background models pca\_bkgd\_cmbrightvle\_e5v20020201.mdl
or pca\_bkgd\_cmfaintl7\_eMv20051128.mdl for source fluxes brighter or
fainter than $\sim18$ mCrb, respectively.  We extract PCA STANDARD-2
events from the top Xenon layer to maximize signal-to-noise.  We use
data from Proportional Counter Units (PCUs) 0, 1 and 2 prior to 1998
December 23; PCUs 0 and 2 from 1998 December 23 until 2000 May 12; and
PCU 2 only after 2000 May 12\footnote{PCUs 3 and 4 (and also PCU 1
  starting late 1998/early 1999) suffered from repeated discharge
  problems.  PCU 0 lost its propane veto layer following a suspected
  micrometeroid hit on 2000 May 12.}.  We ignore data taken within 20
min of the spacecraft's passing through the South Atlantic
Anomaly, during periods of high particle flux as measured by the
ELECTRON2 parameter, when the spacecraft was pointed within 10$\degr$
of the Earth, or when the source was $>$0.02$\degr$ from the optical
axis.  Good exposure times per observation ID after screening were
usually near 1~ks: 80.2 per cent of the 22,934 observations had good exposure
times between 0.5 and 1.5~ks, although a few tens of observations had
good exposure times 10--20~ks.

The wide range of sampling patterns (durations, presence of gaps)
means that our sensitivity to eclipses of a given duration can vary
strongly from one object to the next. We do not split up individual
observations. The shortest timescale we consider is 0.20~d, which
corresponds to four points separated by three satellite orbits (3
$\times$ 5760~s). Tables~\ref{tab:monit1} and \ref{tab:monit2} list
the minimum and maximum durations to which we are sensitive. For
example, NGC 4051 was subjected to sustained monitoring with one
pointing every $\sim$2--14 d regularly for 15.69 yr. There were
also six intensive monitoring campaigns, with the frequencies of
individual observations spanning 0.26 to a few days, and with
durations spanning 9.6--147~d. The full light curve thus affords us
sensitivity to full eclipse events on a virtual continuum of
timescales from $D_{\rm min}$ = 0.26~d to $D_{\rm max}$ = 15.69~yr.

\begin{figure*}
\includegraphics[angle=-90,width=1.00\textwidth]{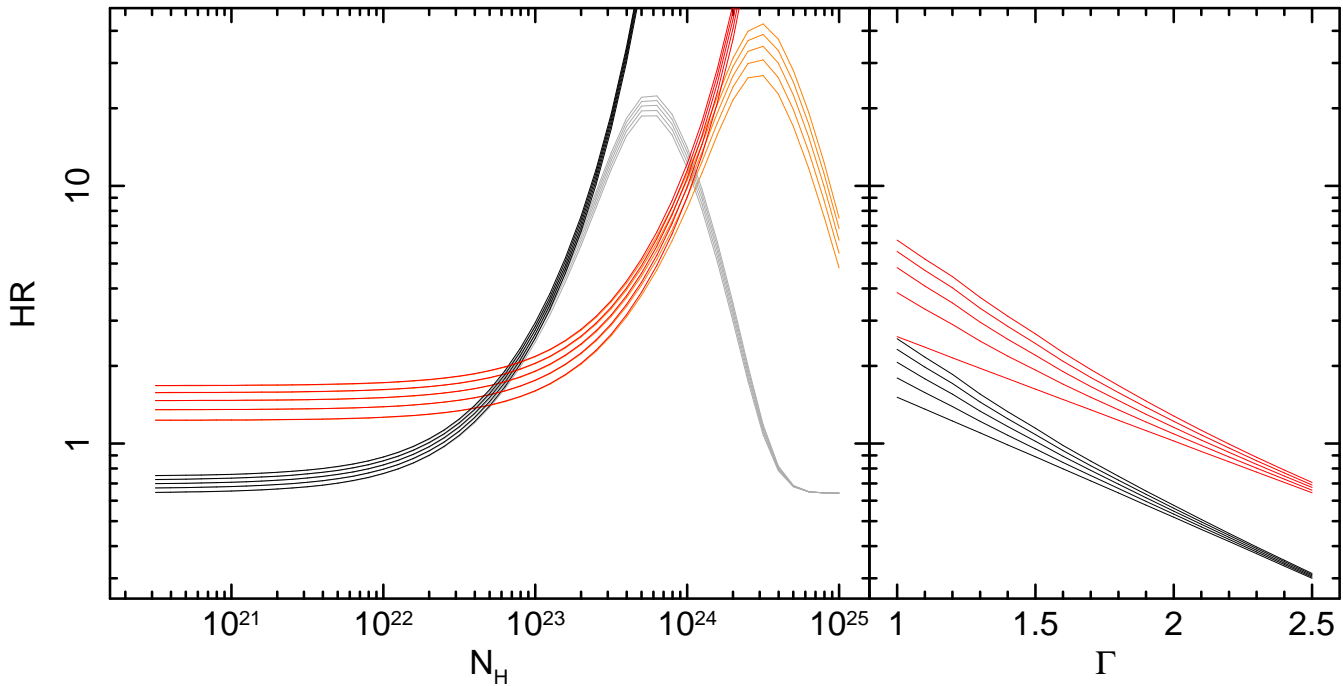}   
\caption{The left-hand panel illustrates the dependence of $HR1 \equiv
  F_{7-10}/F_{2-4}$ (black) and $HR2 \equiv F_{10-18}/F_{4-7}$ (red)
  on $N_{\rm H}$, the total column of gas obscuring an X-ray spectrum
  with $\Gamma$=1.8 and an Fe K$\alpha$ emission line with an
  equivalent width of 100~eV.  The orange and gray 
  lines respectively denote HR1
  and HR2 when there exists an additional power-law component, with
  $\Gamma=1.8$ and normalization set to 0.01 times that of the
  primary power law, and absorbed only by $N_{\rm H,Gal}$ = $3 \times
  10^{20}$~cm$^{-2}$.  
  The right-hand panel illustrates the dependence of
  $HR1$ and $HR2$ on $\Gamma$ when fitting only a simple power law.
  In both panels, a Compton reflection
  component modeled with \textsc{pexrav} is included in all cases; the
  five sets of lines denote spectra with $R$=0.0, 0.5, 1.0, 1.5 and
  2.0.  Assuming no absorption other than $N_{\rm H,Gal}$ and assuming
  $R\leq2.0$, values of $HR1$ greater than $\sim2$ are only attainable
  when $\Gamma \la 1.2$.}
\label{fig:HRplotNH}
\end{figure*}

We extract continuum light curves, one point averaged over each
observation, in the 2--10~keV band and in the sub-bands 2--4, 4--7,
and 7--10~keV. We also extract 10--18~keV light curves for the 37 objects
with average 10--18~keV flux $\ga 6 \times 10^{-12}$ \fluxunits
(lower fluxes yield large statistical uncertainties and/or large
uncertainties due to background systematics).  Errors for each light
curve point are obtained by dividing the standard deviation of the $N$
16 s binned count rate light curve points in that observation by
$\sqrt{N}$. We define the hardness ratio $HR1$ as $F_{7-10}/F_{2-4}$.
Assuming full-covering absorption, $HR1$ peaks at column densities
roughly $1-3 \times 10^{23}$~cm$^{-2}$ while giving us sensitivity
down to $\sim$ a few $10^{22}$~cm$^{-2}$, as illustrated in
Fig.~\ref{fig:HRplotNH}.  
We define $HR2$ as $F_{10-18}/F_{4-7}$, which
peaks at full-covering columns of roughly $0.8 - 3 \times 10^{24}$
cm$^{-2}$.\footnote{Values of $F_{10-18}/F_{2-4}$ peak at column densities
$\sim 0.8-2 \times 10^{23}$~cm$^{-2}$, quite similar to $HR1$, so we
use 4--7~keV as our lower energy band for $HR2$.}

The model $HR$ values in Fig.~\ref{fig:HRplotNH} are calculated in
\textsc{xspec} assuming neutral gas fully covering a power-law
component with $\Gamma$=1.8, an Fe K$\alpha$ emission line with an
equivalent width of 100 eV and a Compton reflection component modeled
with \textsc{pexrav}. We include $N_{\rm H,Gal}$, assumed to be
$3\times 10^{20}$~cm$^{-2}$. 
Models were calculated at values of $N_{\rm H}$ every 0.1 in the log.
Black and red lines denote $HR1$ and
$HR2$, respectively. The orange and gray lines denote $HR1$ and $HR2$
when there exists an additional power-law component to represent
nuclear power-law emission scattered off diffuse, extended gas; such
emission is frequently observed in the $<$5--10~keV spectra of
Compton-thick absorbed Seyferts (e.g., Lira et al.\ 2002).  This
component is modeled by a power law that is unabsorbed (except by
$N_{\rm H,Gal}$). It has the same photon index but a normalization
$f = 0.01$ times that of the primary power law.  
The value of $f$ will vary from one object to the next, but 
values $\la1 - 5$ per cent are typical for type II Seyferts, e.g., 
Bianchi \& Guainazzi (2007), Awaki \etal\ (2008), and Yang \etal\ (2009). Modest changes 
in $f$ do not significantly impact our analysis. 
For values of $N_{\rm H}$ below
$\sim 3 \times10^{23}$~cm$^{-2}$, this component does not strongly
affect $HR1$, and $HR2$ is similarly not strongly affected below $\sim
2 \times10^{24}$~cm$^{-2}$.  For Compton-thick absorption, $HR1$
returns to values below 2, but $HR2$ remains very high, above $\sim$5,
thus breaking the degeneracy in $HR1$.

\subsection{Limitations and caveats}  
 
The PCA's moderate energy resolution ($E$/$\Delta$$E$ $\sim6$ at 6
keV) means the archive is one of \textit{spectral} monitoring as
opposed to simply X-ray photometry.  A typical 1--2~ks exposure for a
typical flux (1 to a few mCrb) AGN yielded PCA energy spectra covering
3 up to $\sim20$~keV, with 2--10~keV continuum flux constrained to
within $\sim2$ per cent, and photon index of the coronal power-law continuum
$\Gamma$ constrained to $\pm\sim$0.01--0.02, or $\sim 1$ per cent.

The energy resolution and bandpass of the PCA means that we are sensitive
only to full-covering or near-full-covering events.  
Specifically, for 2--10~keV fluxes less than roughly 3 $\times
10^{-10}$ \fluxunits (almost all our sources) and/or for accumulated
spectra with less than $\sim10^6$ 2--10~keV counts, we are able to
detect an absorption event and rule out spectral pivoting of the power
law only if the covering fraction is greater than approximately
80--90 per cent. An exception
is NGC~4151 (average $F_{2-10}$ from \textit{RXTE} observations =
$1.2\times 10^{-10}$ \fluxunits); its X-ray spectrum is frequently
modeled with complex absorption using neutral or moderately-ionized
partial coverers (e.g., Schurch \& Warwick 2002).  Time-resolved
spectroscopy using both \textit{BeppoSAX} (de~Rosa \etal\ 2007) and
\textit{RXTE} (Markowitz \etal\ in preparation), respectively, provides
evidence for variable absorption by columns of gas near $1-3 \times
10^{23}$~cm$^{-2}$ and with covering fractions ranging from
approximately 30--70 per cent over timescales of months to years. This
implies numerous eclipses moving in and out of the line of sight over
these timescales.  However, NGC~4151 is likely the only type I Seyfert
in the sample whose 2--10~keV spectrum is strongly affected by neutral
partial covering.  To avoid biasing our determination of the number of
eclipses for the type I class, we do not count eclipses implied by
NGC~4151's variable partial covering and we limit ourselves to full-
or near-full-covering absorption for this study, henceforth assuming
the origin of the X-ray continuum to be a point source.

We are not highly sensitive to absorption by highly ionized gas. We
estimate that our PCA spectra are sensitive to ionization levels up to
log ($\xi$, erg~cm~s$^{-1}$) $\ga 1-2$ [$\xi \equiv L_{\rm
  ion}/(nr^2)$, where $L_{\rm ion}$ is the luminosity of the ionizing
continuum, $n$ is the number density of the gas, and $r$ is the
distance from the source of the ionizing continuum to the gas cloud].

In summary, our study is probing restricted regions of the full
parameter space that can be used to quantify circumnuclear absorbing
gas in Seyferts. Our findings are complementary to those derived using
the archival databases of other X-ray missions. \textit{XMM-Newton},
\textit{Chandra}, and \textit{Suzaku} have higher energy resolution
and bandpasses extending to lower energies than the PCA, and can probe
events with durations $\la1$~d, (as \textit{XMM-Newton} and
\textit{Chandra} observations do not suffer from Earth occultation),
lower column densities, partial covering, and/or high-ionization
absorption (e.g., Turner \etal\ 2008; Risaliti \etal\ 2011). However,
unlike \textit{RXTE}, these missions generally do not perform
sustained monitoring and are not able to detect absorption events
longer than $\sim$a few days. The other uniqueness of the
\textit{RXTE} archive compared to \textit{XMM-Newton} or
\textit{Chandra} is the sensitivity to relatively higher column
eclipses thanks to energy coverage $>$10~keV.


\subsection{Identification of candidate eclipse events}   

In this section, we introduce criteria that we use to identify and
confirm eclipse events.  In brief, ``candidate'' events are identified
via the light curves of hardness ratios and $\Gamma_{\rm app}$,
defined below, but confirmation of increased $N_{\rm H}$ via
time-resolved spectroscopy is required to move an event into the
``secure'' category.

Ideally, a full eclipse event, wherein we can constrain ingress and
egress, must consist of (at least) two adjacent observations with
elevated values of hardness ratio and $N_{\rm H}$ (see $\S$2.4), plus
two observations on either side to define the ``baseline'' values of
hardness ratio and $N_{\rm H}$.  In cases where there exist large gaps
in monitoring, we can still identify eclipses based on elevated values
of hardness ratio and $N_{\rm H}$, but we may lack information on when
ingress/egress occurred.

Sudden increases in light curve hardness ratio and/or decreases in
overall spectral slope could indicate increased absorption, but such
spectral variability can also potentially be caused by variations in
the photon index of the coronal power-law component that typically
dominates Seyfert X-ray spectra. The overall energy spectra of
Seyferts lacking substantial X-ray absorption are frequently observed
to flatten as the total 2--10~keV flux lowers (e.g., Papadakis \etal\
2002; Sobolewska \& Papadakis 2009). Some objects' long-term 2--10~keV
flux behavior is consistent with pivoting of the continuum power law
at some energy $>$10~keV (e.g., Taylor \etal\ 2003). Other objects are
consistent with the ``two component'' model across the range of fluxes
observed; in this model, the power law remains constant in $\Gamma$
but its normalization varies, and the presence of a Compton reflection
component with constant absolute normalization causes a flattening in
the overall observed spectral slope as power-law flux lowers (e.g.,
Shih \etal\ 2002). In either case, $\Gamma$ is observed to usually be
higher than $\sim1.5-1.6$, after accounting for absorption, if
present. For example, from a large sample of X-ray-selected type~I
Seyferts, Mateos \etal\ (2010) find a mean photon index of $1.96$ with
a standard deviation of 0.27. In black hole X-ray binary systems,
thought to also possess a disk and corona surrounding the black hole,
$\Gamma$ is also usually observed to not be lower than about 1.5
(e.g., McClintock \& Remillard 2003; Done \& Gierli\'{n}ski 2005).

We fit the spectrum of each individual observation with a simple power
law using \textsc{xspec} version 12.7.1, accounting for absorption
\textit{only} by the Galactic column, $N_{\rm H,Gal}$ (Kalberla
\etal\ 2005).  For each spectrum of NGC~4151, Cen~A, and NGC~5506, we
add systematics of 5, 5, and 3 per cent, respectively.  Because we are
neglecting the absorption $N_{\rm H}$ in excess of the Galactic column
and other commonly observed spectral features such as, Compton
reflection and Fe K emission, the power-law index we measure in this
way is not a direct measure of the photon index of the power law, but
is instead only a general indicator of hard X-ray spectral shape,
which we label $\Gamma_{\rm app}$ (``apparent'' photon index).  A
value of $\Gamma_{\rm app}$ lower than 1.5 for objects normally
lacking absorption $>10^{22-23}$~cm$^{-2}$, particularly if they do
not occur near times of low power-law continuum flux as probed by
2--10 or 10--18~keV flux, very likely indicates an increase in $N_{\rm H}$.
Following Mateos \etal\ (2010), $\Gamma=1.5$ (1.4) indicates a
1.7$\sigma$ (2.1$\sigma$) deviation from $\langle\Gamma\rangle=1.96$.
These values correspond to $HR1 = 0.9-1.2$ (1.0--1.3) and $HR2 =
1.9-3.6$ (2.2--4.7), assuming values of the Compton reflection
strength $R$ spanning 0.0--2.0\footnote{Here, $R$ is defined following
  the convention of the \textsc{xspec} model \textsc{pexrav}, with a
  normalization defined relative to that of the illuminating power law
  and with $R=1$ corresponding to a sky-covering fraction of 2$\pi$ sr
  as seen from the illuminating source.}.
 
Hardness ratios alone cannot confirm an increase in absorption.  
Time-resolved spectra, in many cases, also lack the statistics to rule
out such scenarios, and as discussed below, we sometimes must freeze
$\Gamma$ to avoid degeneracy between $\Gamma$ and $N_{\rm H}$.  For
the purposes of identifying \textit{candidate} eclipse events, we
adhere to the assumption that $\Gamma$ intrinsically does not vary to
values below 1.4--1.5, and so values of $\Gamma_{\rm app}$ below
roughly 1.3 and consequently values of $HR1$ above 1.7 are very likely
only due to the presence of significant line-of-sight absorption. We
assume that the strength of the Compton reflection hump relative to
the power law, also a source of spectral hardening, remains constant
over time.  For sources that are perpetually X-ray obscured (e.g., the
Compton-thin type IIs), we rely on the deviations of $HR1$, $HR2$, and
$\Gamma_{\rm app}$ from their mean values to identify potential
eclipse events.


Our criteria for a ``candidate'' eclipse event are thus as follows:

$\bullet$ Criterion 1: trends in $HR1$ (and/or $HR2$ depending on
$\Delta$$N_{\rm H}$) that indicate a statistically significant
deviation compared to the average spectral state,
with a minimum of 2$\sigma$ (standard deviations), and at
least a 50 per cent increase above the mean value of $HR1$.

$\bullet$ Criterion 2: at least two consecutive points in a row in the
hardness ratio light curve have elevated values. This removes
single-point outliers due to statistical fluctuations.

$\bullet$ Criterion 3: trends in $\Gamma_{\rm app}$ that deviate at
the $\geq$2$\sigma$ level.  ($\Gamma_{\rm app}$ tends to be more
statistically noisy than $HR1$, and in most of our candidate events,
the deviation in $HR1$ is commonly 1--2$\sigma$ more than the
corresponding deviation in $\Gamma_{\rm app}$.  Consequently, $HR1$ is
our primary selector.)  For X-ray unobscured sources, 
$\Gamma_{\rm app}$ must be lower than 1.3. \smallskip 

To classify an eclipse candidate as ``secure,'' we impose a fourth criterion:

$\bullet$ Criterion 4: the increase in $N_{\rm H}$ must be confirmed
via follow-up time-resolved X-ray spectroscopy in binned spectra. 

``Secure'' events are divided into ``secure A'' events, wherein
$N_{\rm H}$ and $\Gamma$ are deconvolved in the time-resolved spectra,
or ``secure B'' events, wherein $N_{\rm H}$ has to be determined with
an assumed frozen value of $\Gamma$.  We define ``candidate'' events as
those satisfying criteria 1--3, but not 4.  Such events can be
attributed to low signal-to-noise ratio, in terms of either low continuum
fluxes or due to the inferred weakness of the event, e.g., values of
$N_{\rm H}$ only barely above $1 \times 10^{22}$~cm$^{-2}$.
Qualitatively speaking, the candidate events 
had large uncertainties on $N_{\rm H}$ even when $\Gamma$ is frozen
(e.g., errors spanning factors of more than a few) and/or had 
large uncertainties on $\Gamma$ (e.g., several tenths or more)
when modeling no excess absorption.

 
\section{OVERVIEW OF ECLIPSE RESULTS}  

We defer all details on individual objects to Appendix A.  There, we
provide long-term light curves for sub-band continuum fluxes, hardness
ratios $HR1$ and $HR2$, and $\Gamma_{\rm app}$.  For brevity, we
include light curve plots only for the 10 objects with candidate or
secure eclipse events; in the other 45 objects, we find no significant
sustained deviations in $HR1$, $HR2$ or $\Gamma_{\rm app}$ at the
2$\sigma$ level or greater.  However, we also include plots of
selected type II objects with evidence for constant, non-zero
absorption.  Appendix~A contains all the details on the observed
deviations in the $HR1$, $HR2$, and/or $\Gamma_{\rm app}$ light curves
and the subsequent time-resolved spectroscopy.  The reader is referred
to Lamer \etal\ (2003), Rivers \etal\ (2011a), and Akylas
\etal\ (2002) for NGC~3227/2000--1, Cen~A, and Mkn~348, respectively;
we present new results of time-resolved spectroscopy for all other
events. 

In this paper, we confirm a total of 12 ``secure (A+B)'' X-ray
absorption eclipses in eight objects (confirmed with spectral fitting)
plus four ``candidate'' eclipses in three objects, all summarized in
Table~\ref{tab:ecl1summ}. For our 16 secure/candidate events, each
with $N_{\rm H} \sim 10^{22-23}$ cm$^{-2}$ (see below), each object is
bright enough for us to use the 10--18 keV band to probe the
uneclipsed continuum. In Appendix A10, we present flux-flux plots that
demonstrate that the 2--4~keV flux is affected independently of the
behavior of the 10--18~keV continuum during our secure events and that
one can distinguish the spectral variability from a discrete eclipse
event from that due to variability in the power law. For candidate
events, though, the flux-flux plots are unable to fully separate the
two types of spectral variability, and this is tied to ambiguity in
modeling the time-resolved spectra. Six of these 12 are complete
absorption events, with \textit{RXTE} witnessing both ingress and
egress. The events span a range of quality, depending on source
brightness, the sampling of the observations, the duration, and
$N_{\rm H}$: four events' column density profiles are well-resolved in
time even after binning the observations for time-resolved
spectroscopy, allowing constraints on the density profile:
Cen~A/2010--1, Mkn~348/1996--7, NGC~3227/2000--1, and NGC~3783/2008.3. In
contrast, other, more rapid events subtend only two points in the
$HR1$ light curve, with only one binned energy spectrum demonstrating
increased absorption. Inferred durations range from $<$1~d to a few
years, with six secure events' durations in the $\sim$tens of days
range, and three of them $\geq$5~months.

We are sensitive to nearly full-covering absorption by columns in a
range of $\sim10^{22-25}$~cm$^{-2}$, but the detected absorption
events span only peak column densities of $\sim 4 - 26 \times
10^{22}$~cm$^{-2}$, i.e., we see no evidence from our sample for
full-covering absorption by Compton-thick clouds.  This is not to say
that full-covering Compton-thick eclipse events do not occur in
Seyferts in general.  In other words, although our survey is sensitive
to such events, we do not detect any for our monitored sample.

The average values of $N_{\rm H}$ in log space for the two classes
(considering only secure events) are nearly identical: 12 and 10
$\times10^{22}$~cm$^{-2}$.  In the cases where the column density
profiles can be well accessed by several data points after binning for
time-resolved spectroscopy, we see no strong evidence for
non-symmetric events (along the direction of motion across the line of
sight) such as the ``comet'' shaped clouds inferred by Maiolino
\etal\ (2010) in NGC~1365.  The $N_{\rm H}$($t$) and $HR1$($t$)
profiles of NGC~3783/2008.3, however, are particularly intriguing:
they suggest two peaks separated by $\sim$11~d, but with the values
in between the peaks not returning to post/pre-eclipse levels. The
possibility of a double-cloud absorption event is discussed further in
$\S$\ref{sec:3783double}.  In Table~\ref{tab:ecl1summ}, we list
parameters both for the full duration and for each clump separately.

Surprisingly, only three type II objects show cloud events.  That is,
based on their $HR$ light curves, the majority of all other monitored
type IIs display no strong evidence (sustained trends in $HR$ more
than $50$ per cent above the mean value) for variations in $N_{\rm
  H}$, although several were individually monitored for several years
or more in some cases.  In $\S$5.6, we will focus on the nine type II
objects that have sustained monitoring for $\geq$0.6~yr. We will show
that seven of them lack long-term ($>$1~d) variations in $N_{\rm H}$
down to $\Delta$$N_{\rm H} \sim 1-9 \times 10^{22}$~cm$^{-2}$
depending on the signal-to-noise ratio and discuss the applicability
of clumpy-absorber models to these objects.

In the case of Cen~A, in addition to the two eclipses identified, we
present in Appendix~A evidence (at $\sim$2.2$\sigma$ confidence) that
the baseline level of $N_{\rm H}$ dipped by $\sim14$ per cent and then
recovered during the first three months of 2010.  The reader should
bear in mind, however, that Cen~A is a radio galaxy and no BLR has
been detected, so it may not be representative of all Seyferts, most
of which are radio quiet.  This observation and its implications are
discussed in $\S$5.6.1.

We caution that ``peak $N_{\rm H}$''refers only to what we
have measured with time-resolved spectroscopy.  Since clouds are not
point sources, we can only probe that two-dimensional slice of the
cloud that transits the line of sight, and there may exist other lines
of sight outside that slice which intersect parts of the cloud with
higher columns. Consequently, the intrinsic maximum column density of
the cloud could be greater than what we measure if \textit{RXTE} was
not monitoring the source when that part of the cloud transited the
line of sight.

The durations we measure also refer only to that slice of the cloud
intersected by the line of sight. In the case of a
spherically-symmetric cloud, observed eclipse durations will be, on
average, $\pi/4$=0.79 times the maximum durations that could have been
observed. (Consequently, average inferred cloud diameters, estimated in
$\S$\ref{sec:diams}, may be underestimated by this modest factor.)
In the case of spherical clouds with maximum column density at the
center and radial density profiles similar to those for Cen~A/2010--1
and NGC~3227/2000--1, then the corresponding effect on peak $N_{\rm H}$
will likely be less than $\sim20$ per cent.

\begin{table*}
\begin{minipage}{180mm}
\caption{Summary of eclipse events}   
\label{tab:ecl1summ}  
\begin{tabular}{lllllll}   \hline
Source       &           &                  &           & Duration                  & Peak  $N_{\rm H}$     & \\  
name         & Type      & Event            & Category  & (d)                       & ($10^{22}$~cm$^{-2}$)              &   Comments   \\ \hline 
\multicolumn{7}{c}{\textbf{Type I Secure Events}} \\  \hline
NGC 3783     & Sy1       & 2008.3           & Secure B  & 14.4--15.4                & $11.2^{+1.7}_{-1.5}$                &   $N_{\rm H}$($t$) resolved. \\
             &           &                  &           & $\sim$9.2  \& $\sim$4.6   & $11.2^{+1.7}_{-1.5}$ \& $8.6^{+1.5}_{-1.3}$  &        \medskip   \\ 
Mkn 79       & Sy1.2     & 2003.5           & Secure B  & 12.0--39.4                & $14.4^{+4.8}_{-4.2}$$^{\ddagger}$     &  Ingress only. \\  
             &           & 2003.6           & Secure B  & 34.5--37.9                & $11.5^{+3.2}_{-2.8}$                &  \\ 
             &           & 2009.9           & Secure B  & 19.6--40.0                & $7.6\pm2.2$                       &   \medskip \\       
Mkn 509      & Sy1.2     & 2005.9           & Secure B  & $26-91$                   & $8.8\pm1.7$$^{\ddagger}$           & Ingress only. $N_{\rm H}$($t$) resolved.   \medskip \\ 
MR 2251--178 & Sy1.5/QSO & 1996             & Secure A  & 3  -- 1641                & $6.6^{+0.8}_{-1.4}$$^{\ddagger}$      & Egress before Jun.\ 1998   \medskip \\            
NGC 3227     & Sy1.5     & 2000--1          & Secure A  & 77--94                    & 19--26                           & $N_{\rm H}$($t$) resolved. \\
             &           & 2002.8           & Secure B  & 2.1--6.6                  & $13.3^{+2.6}_{-2.2}$                 & \\   \hline  
\multicolumn{7}{c}{\textbf{Type II Secure Events}} \\  \hline
Cen A        & NLRG/Sy2  & $\sim$2003--4    & Secure A  & $356-2036$                &  8$\pm$1$^{\ddagger}$              & \\     
             &           & 2010--1          & Secure A  & $170.2-184.5$             &  8$\pm$1                         & $N_{\rm H}$($t$) resolved. \medskip  \\ 
NGC 5506     & Sy1.9     & 2000.2           & Secure A  & 0.20--0.80                &  $4.0\pm1.4$                     & \medskip  \\     
Mkn 348      & Sy2       & 1996--7          & Secure A  & $399-693$                 & $18\pm3$$^{\ddagger}$              &  Egress only. $N_{\rm H}$($t$) resolved. \\ \hline 
\multicolumn{7}{c}{\textbf{Candidate Events}} \\  \hline
Fairall~9    & Sy1       & 2001.3           & Candidate & 5.4--15.0                 & $<21$                            &  \medskip  \\  
NGC 3783     & Sy1       & 2008.7           & Candidate & 17--28                    & $<4$                             &  \\ 
             &           & 2011.2           & Candidate & 4.1--15.8                 & $10.9^{+6.0}_{-5.3}$                &  \medskip \\ 
NGC~3516     & Sy1.5     & 2011.7           & Candidate & $\sim57$                  & 4.7                              & Possible variation in covering fraction \\ 
             &           &                  &           &
             &                                  & of partial covering, moderately \\ 
             &           &                  &           &
             &                                  & highly ionized absorber       \\ \hline
\end{tabular}\\
Summary of the 12 secure and 4 candidate eclipse events detected.  We
list events in the order of: secure events in type Is, secure events
in type IIs, and then candidate events (which are all for type I
objects).  ``$N_{\rm H}$($t$) resolved'' means that at least several
consecutive binned spectra confirm increased $N_{\rm H}$ levels.  The
double-dagger ($^{\ddagger}$) indicates that intrinsic peak value of
$N_{\rm H}$($t$) may be higher than that observed if it occurred
during a gap in monitoring. 
\end{minipage}
\end{table*}

\section{PROBABILITIES FOR OBSERVING AN X-RAY ECLIPSE EVENT}  

Ideally, we would like to derive the instantaneous probability
$\overline{P_{\rm ecl}}$ of catching a given source while it is
undergoing an X-ray eclipse event, and then relate that probability to
the source's optical classification and/or the constant
presence/absence of X-ray-absorbing gas along the line of sight.
Eclipse events detected in the \textit{RXTE} archive are, however,
evidently rare, with only 10 different objects showing events and
with only 1--3 events in each of those objects. We will thus focus in
this section on the average instantaneous probability to detect
absorption due to an eclipse event for a given \textit{class} of
objects as opposed to individual objects.  However, deriving such a
probability is not straightforward, because we must factor in biases
resulting from our observation sampling, which yields a sensitivity to
eclipse durations that is very heterogeneous as a function of
timescale, both for object classes and for individual objects.  For 
sustained monitoring, for instance, we are sensitive
to a relatively larger number of short-duration eclipses than to
long-duration eclipses.  For example, given a hypothetical campaign
consisting of one observation daily for 64~d, we can detect a
maximum of 16 full eclipses of duration 3~d (4 points), 8 full
eclipses of duration 7~d (8 points), etc.

We thus quantify an instantaneous probability \textit{density} 
$p_{\rm ecl}$ as a function of eclipse duration as follows: we first define
for each individual object a ``selection function'' $SF_{\rm ind}$ to
quantify our sensitivity, as a function of eclipse duration, to the
maximum total number of eclipses which \textit{could} have been
potentially observed, given that object's sampling with \textit{RXTE};
this procedure is described in $\S4.1$.  We then produce summed
selection functions $SF_{\rm sum I,II}$ to quantify the average
sensitivity to the total number of eclipses for each object class
(type I and II), as well as to identify potential biases affecting a
given object class as a whole.  In $\S4.2$, we quantify the total
number of eclipse events $N_{\rm ecl}$ actually observed within each
object class as a function of eclipse duration. Then, in $\S4.3$, as
we have binned our observed events and the selection functions
(maximum possible event number) on to the same grid as a function of
eclipse duration, we divide $N_{\rm ecl}$ by $SF_{\rm sum I,II}$ to
obtain the instantaneous probability density $p_{\rm ecl}$($t_{\rm i}$). 
This quantifies the likelihood to witness an eclipse as a
function of that eclipse's duration, defined over a discrete set of
timescale bins $t_{\rm i}$.  Finally, we integrate 
$p_{\rm ecl}$($t_{\rm i}$) over all timescale bins to obtain
$\overline{P_{\rm ecl}}$, the instantaneous probability of witnessing
a source in eclipse for \textit{any} eclipse duration.

As a reminder, $\overline{P_{\rm ecl}}$ and $p_{\rm ecl}$($t_{\rm i}$)
do not necessarily denote the likelihood to simply observe a source
with non-zero X-ray obscuration; this holds true \textit{only} if the
source normally devoid of X-ray-absorbing gas along the line of
sight. $\overline{P_{\rm ecl}}$ and $p_{\rm ecl}$($t_{\rm i}$) refer
to the likelihood of catching the source in state with a
higher-than-usual value of $N_{\rm H}$ at any instant due specifically
to a discrete (localized in time) eclipse event by a cloud of gas, one
that transits the line of sight with an observed duration between
$t_{\rm i}$ and $t_{\rm i+1}$ in the case of $p_{\rm ecl}$($t_{\rm
  i}$).  In $\S$5.5, we will compare our estimates of
$\overline{P_{\rm ecl}}$ for each class to the predictions for a
clumpy torus to cause a given source to be observed in eclipse.

\subsection{Selection functions} 

This section describes how we generate ``selection functions'' for
each individual object $SF_{\rm ind}$($t_{\rm i}$) and sum selection
functions for each object class $SF_{\rm sumI,II}$($t_{\rm i}$) to
quantify the number of light curve observation segments
(``campaigns'') of a given duration as a function of event duration.
Here we define a ``campaign'' as consisting of a minimum of four
observations, with no single gap between adjacent observations being
greater than 75 per cent of the duration.\footnote{Other values, e.g.,
  between 50 and 90 per cent yield virtually identical results for
  this calculation as well; the effects on the summed selection
  functions are always negligible compared with the uncertainties
  stemming from the varying contributions of individual source
  selection functions ($\S$4.1).}

We define 19 time bins $t_{\rm i}$ spanning from 0.20 to 5850 d,
with time bins equally spaced by 0.235 in log space; in linear space,
the $i$th time bin is defined by $0.20\times$($1.718^{i}$) d $\leq t <
0.20\times$($1.718^{i+1}$) d, with $i = 0 ... 18$.  $SF_{\rm
  ind}$($t_{\rm i}$) effectively tells us the potential maximum number
of eclipse events of duration $t_{\rm D}$ satisfying $t_{\rm i} \leq
t_{\rm D} < t_{\rm i+1}$ that \textit{RXTE} was capable of potentially
catching, given the observed sampling for that object.

The $SF_{\rm ind}$($t_{\rm i}$) are constructed as follows: consider a
light curve with $J$ total observations (data points), and observation
times denoted by $X_{j=1}$ ... $X_{j=J}$.  For each time bin
$t_{\rm i}$, we start at $j=1$ and we identify the first light curve
segment $X_{1} .. X_{q} $ whose duration $t_{\rm D}$ satisfies
$t_{\rm i} \leq t_{\rm D} < t_{\rm i+1}$.  We require a minimum of
four data points, since we defined above an eclipse event relying on a
minimum of two consecutive points to form a significant peak in the
$HR$ light curves, plus one point before and one point after the
putative eclipse to denote the ``baseline'' levels of $HR$.  The goal
is to have consistency between detecting real eclipse events and how
we determine the maximum possible number of potential eclipse events.
Consequently, we disqualify a segment if any gap between two adjacent
points is $>75$ per cent of the segment's duration.  If the segment 
$X_{1} ... X_{q}$ satisfies these criteria, then $SF_{\rm
  ind}$($t_{\rm i}$) = $SF_{\rm ind}$($t_{\rm i}$) +1, and we restart
counting from point $X_{q+1}$.  Otherwise, we restart from $j=2$.

\begin{figure}
\includegraphics[angle=-90,width=0.48\textwidth]{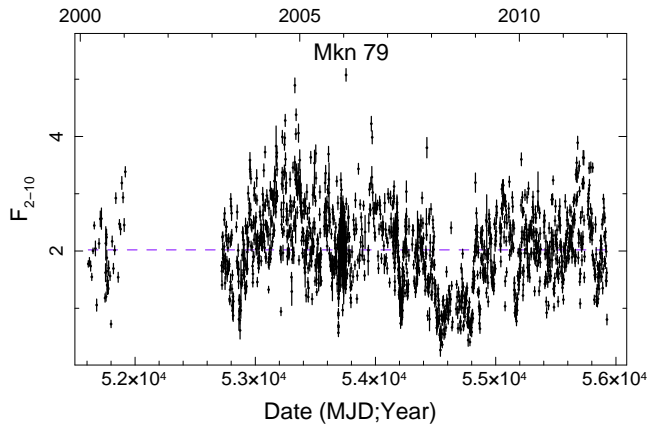}  
\caption{2--10~keV observed light curve for Mkn~79.  
  This plot only shows the general trend of the hard flux and does not
  serve to identify any candidate eclipse events.}
\label{fig:mkn79flux}
\end{figure}

\begin{figure}
\includegraphics[angle=-90,width=0.48\textwidth]{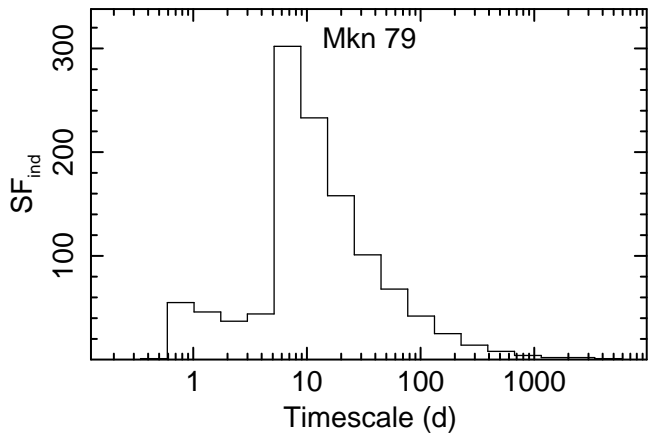} 
\caption{Selection function $SF_{\rm ind}$($t_{\rm i}$) 
for Mkn~79. The histogram quantifies the number of ``campaigns'' per timescale bin 
and is computed on the base of the light curve in Fig.~\ref{fig:mkn79flux}.}
\label{fig:mkn79selfxn}
\end{figure}

When there existed sustained monitoring with a sampling interval
$\Delta$$t$ for a duration $t_{\rm D}$, we are sensitive to eclipse
events on a continuous range of timescales from $4\Delta$$t$ up to
$t_{\rm D}$, with $SF_{\rm ind}$($t_{\rm i}$) increasing towards
shorter timescales.  \textit{RXTE} monitored 42 sources
continuously for $\sim$1~yr and longer (sun-angle gaps
notwithstanding), and one of the most common sampling times for these
programmes was 3--4 d.  Many objects' selection functions thus
feature peaks in the 8.8--15.2 d time bin, with a sharp drop-off in
$SF_{\rm ind}$($t_{\rm i}$) below 8.8 d.  In addition, there were
multiple intensive-monitoring programmes featuring observations several
times daily for durations of days to weeks.  These programmes lead to
large contributions in $SF_{\rm ind}$($t_{\rm i}$) from $\sim$1 to
several days.  As an example, we consider the case of Mkn~79, which
had monitoring observations
every $\sim$10 d for a duration of 309 d (MJD 51610--51919), 
every $\sim$2 d for 15.6 d (MJD 51754.5--51770.1),
every $\sim$2 d for 3205 d (MJD 52720--55925) with 
$\sim$26 d gaps once a year due to sun-angle constraints,
and every $\sim6$ hr for 64.6 d (MJD 53691.4--53756.0).
The $F_{2-10}$ light curve and the corresponding selection function are plotted
in Figs.~\ref{fig:mkn79flux} and \ref{fig:mkn79selfxn}, respectively.

\begin{figure*}
\includegraphics[angle=-90,width=0.999\textwidth]{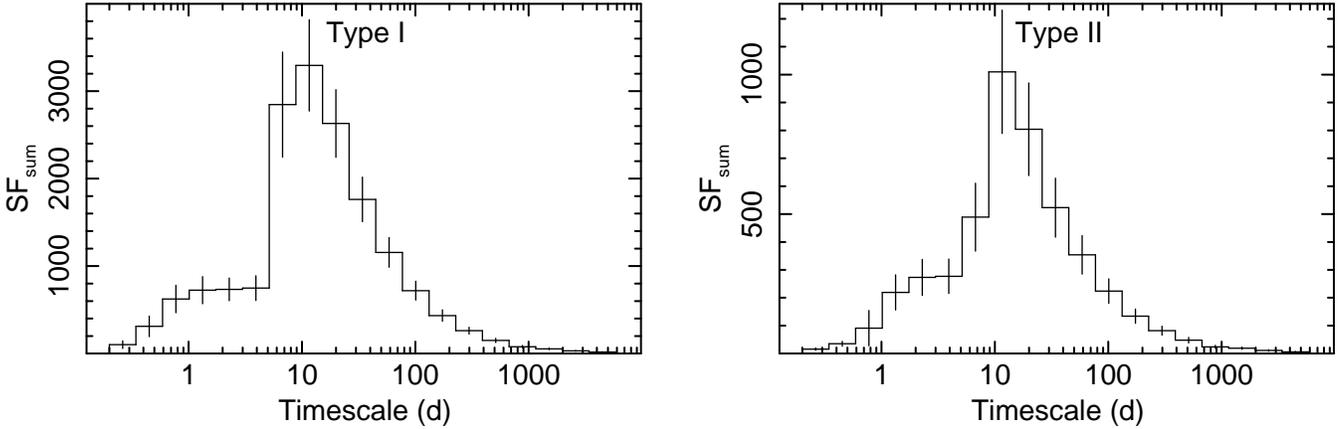}  
\caption{Summed selection functions $SF_{\rm sumI}$($t_{\rm i}$) and $SF_{\rm sumII}$($t_{\rm i}$) for
  the 37 type I and 18 type II AGN, respectively.
  These figures illustrate the total number of observation
  ``campaigns'' (maximum number of potential detected eclipses)
   per timescale bin $t_{\rm i}$. The 1$\sigma$ uncertainties are
  determined by a Monte Carlo bootstrap method.}
\label{fig:selfxnALL}
\end{figure*}

We create summed selection functions for each object class as a function of
timescale bin ($SF_{\rm sumI,II}$($t_{\rm i}$)) by summing up the
individual selection functions for the 37 type I and 18 type II AGN
separately (left- and right-hand panels of Fig.~\ref{fig:selfxnALL}).  We
see that the number of campaigns per given timescale is on average 
3.9 times greater for the type Is than for the type IIs.  For both
classes, the most common campaign timescales are $\sim$tens of days.

Due to the heterogeneous sampling in the archive, some objects can
contribute as many as 200--300 campaigns to an individual $t_{\rm i}$
bin (usually the 8.8--15.2~d bin) while other objects were observed
much more sparsely and yield selection functions containing only a few
campaigns to a few timescale bins.  To estimate the uncertainty on
each binned $SF_{\rm sumI,II}$($t_{\rm i}$) point stemming from the
varying contributions of individual source selection functions, we
employ a Monte Carlo bootstrap procedure.  For each object class (type
I or type II) of $N_{\rm S}$ objects, we do the following $m=1000$
times: we select at random one individual selection function from the
pool of observed $SF_{\rm ind}$ functions, until a total of $N_{\rm
  S}$ individual selection functions was accumulated.  We explicitly
allow for individual objects to be selected multiple times.  We sum
those to create a simulated summed selection function $SF^{i}_{\rm
  sum}$, where $i \in {1 ... m}$.  Once we have $m$ $SF^{i}_{\rm sum}$
functions, we calculate the standard deviation within each bin to
yield the relative uncertainties plotted in Fig.~\ref{fig:selfxnALL}.
This procedure yields typical bin uncertainties of $\sim$12 and
25 per cent for type Is and IIs, respectively.


\begin{figure*}
\includegraphics[angle=-90,width=0.999\textwidth]{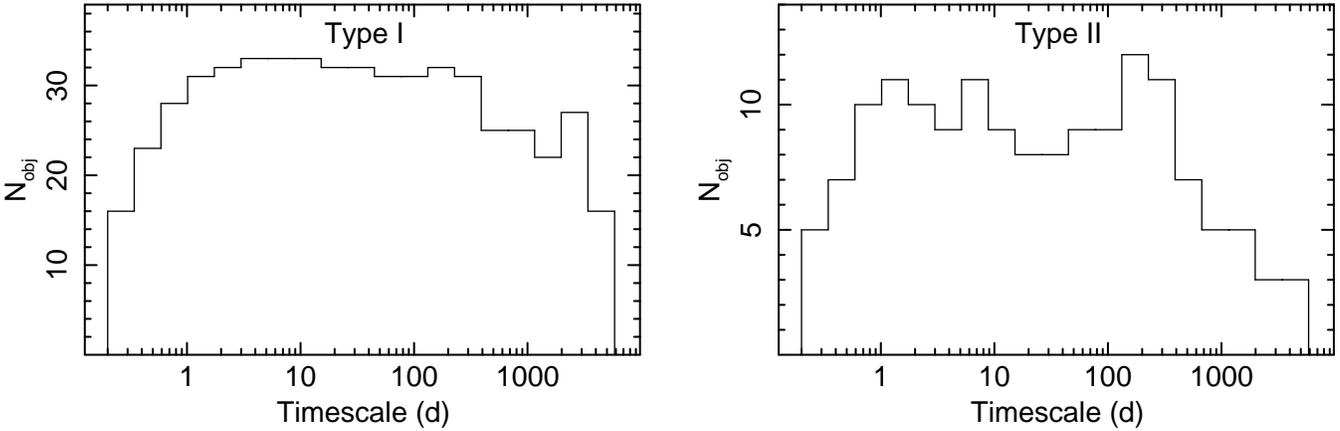}  
\caption{Number of different objects that contribute to a given timescale 
bin for type I (left) and type II (right) AGN.}
\label{fig:numobjpertimescaleALL}
\end{figure*}

We also plot for each object class the number of different objects
that contribute to a given timescale bin
(Fig.~\ref{fig:numobjpertimescaleALL}).  These plots can identify
those timescales over which biases arise because only a small number
of different objects contribute to the relevant eclipse duration bin.
For both classes of objects, the number of objects contributing to a
given timescale is roughly constant from $\sim$1 to $\sim$300 d, but
drops off rapidly above $\sim$300 d in the type IIs.  In fact, for the type IIs,
timescales above 400 d are probed by only 3--5 objects.  Especially in
the case of type IIs, we are thus relying on the assumption that these
few objects (Cen~A, NGC~1052, NGC~4258, NGC~5506, and NGC~7314) are
truly representative of the whole class in terms of their variable
X-ray absorption properties, since they strongly influence our
inferences about variable absorption in type IIs on these long
timescales.

\subsection{Number of observed events per timescale bin} 

\begin{figure}
\includegraphics[angle=-90,width=0.48\textwidth]{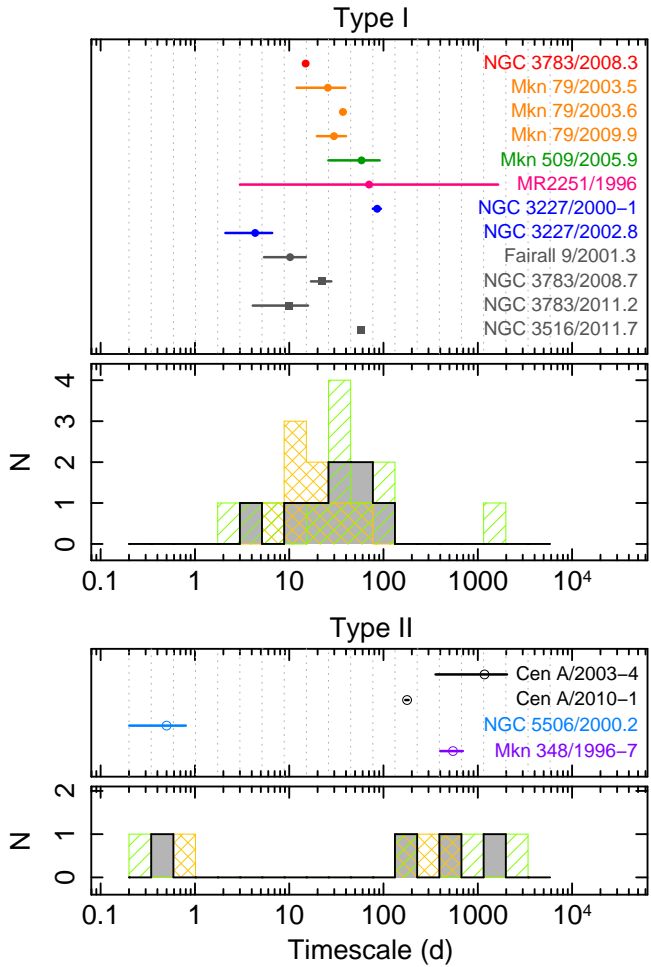}           
\caption{Number of observed eclipse events $N_{\rm ecl}$ as a function
  of timescale bin $t_{\rm i}$ for the type I AGN (top pair of panels)
  and type II (bottom pair). Within each object class, we plot the
  observed durations for each eclipse in the top of each pair.  
  In the case of MR~2251--178,
  the star denotes the average of the minimum/maximum eclipse
  durations in log space, as constraints for the best estimate of the
  actual duration are poor.  Candidate eclipses are shown in gray.  
  In each lower panel, the
  black line indicates the ``best estimate'' of $N_{\rm ecl}$($t_{\rm
    i}$), using only the best-estimate values of the durations and
  excluding candidate eclipses.  The orange line is the minimum of
  $N_{\rm ecl}$($t_{\rm i}$) (i.e., the minimum possible number of
  events with duration $t_{\rm i} \leq D < t_{\rm i+1}$).  The green
  line is the maximum of $N_{\rm ecl}$($t_{\rm i}$) (i.e., the maximum
  possible number of events with duration $t_{\rm i} \leq D < t_{\rm
    i+1}$), including candidate eclipses.}
\label{fig:durbars}
\end{figure}

In Fig.~\ref{fig:durbars}, we plot the number of eclipse events as a
function of each of their durations, using the same 19 timescale bins
as above.  In the upper panel for each class, each eclipse event is
depicted by a separate symbol/bar.  In these panels we plot all
eclipse events independent of the data quality of the eclipse (13
secure A+B and 3 candidate events).  In the lower panel for each class
we plot histograms denoting the number of observed eclipses as a
function of timescale bin $N_{\rm ecl}$($t_{\rm i}$).  The black solid
line is the ``best estimate'' of $N_{\rm ecl}$($t_{\rm i}$), using
only the best-estimate values of the durations and ignoring the
candidate events.

We also present estimates of $N_{\rm ecl}$($t_{\rm i}$) that
correspond to the maximum and minimum likelihoods of witnessing
sources in eclipse, given that an event's contributions to $N_{\rm
  ecl}$($t_{\rm i}$) may shift in $t_{\rm i}$ given the measured
uncertainties in the observed duration. For each eclipse event, we
thus considered those timescale bins $t_{\rm i}$ consistent with the
limits of the observed duration (as identified in
Fig.~\ref{fig:durbars}), and identified which one of those timescale
bins $t_{\rm i}$ had the highest corresponding value of $SF^{i}_{\rm
  sumI,II}$ (lowest probability associated with one single eclipse
event). The orange line denotes the final $N_{\rm ecl}$($t_{\rm i}$)
histogram corresponding to the minimum probability densities,
accumulated over all secure events within each class. The green solid
line in each lower panel, meanwhile, denotes $N_{\rm ecl}$($t_{\rm
  i}$) corresponding to the \textit{highest} probability densities,
computed in a similar fashion as the orange solid line, but this time
including the candidate events as well.

For type I AGN, the "typical" observed duration timescale is tens of
days, although this may not be surprising given the strong peak of the
selection function at tens of days.  The case of
MR~2251--178 is very exceptional, as we detect clear signs for a
secure eclipse event but have only poor constraints on its duration
($3$ d $ < t < $ 4.5~yr).  If we ignore this event, we do not have any
type I eclipses that are consistent with an event duration longer than
100 d despite the numerous campaigns lasting for several hundreds
of days to several years.  Even though the shapes of the selection
functions for type I and IIs are extremely similar, for type II
objects, we do not detect any eclipse events with durations in the
range 1--100~d. The three type II eclipses with durations of $>$100~d
have more than five times fewer observing campaigns than between
10 and 30~d where the maximum of the summed selection function for
type II objects occurs. In other words, the histogram of type I
eclipses agrees well with the expected distribution based on the
summed selection function of type I objects, while type II eclipses do
not appear at the expected duration if only their summed selection
function is considered.


\subsection{Calculating the instantaneous eclipse probability densities $p_{\rm ecl}$($t_{\rm i}$) and
the instantaneous eclipse probability $\overline{P_{\rm ecl}}$}  
\label{sec:sect43}

\begin{figure}
\includegraphics[angle=-90,width=0.48\textwidth]{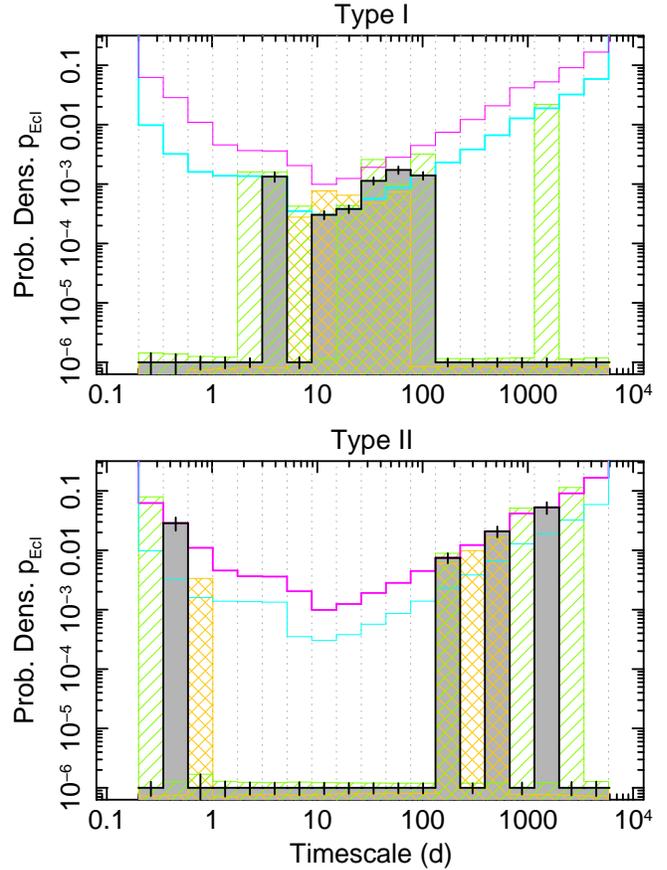}    
\caption{The probability density functions $p_{\rm ecl}$($t_{\rm i}$)
  for each class denote the instantaneous likelihood of witnessing a
  source undergoing an eclipse event with eclipse duration between
  $t_{\rm i}$ and $t_{\rm i+1}$. They are calculated by dividing the
  $N_{\rm ecl}$($t_{\rm i}$) functions in Fig.~\ref{fig:durbars} by
  the summed selection functions $SF_{\rm sumI,II}$($t_{\rm i}$) in
  Fig.~\ref{fig:selfxnALL}.  The black solid line
  corresponds to the best-estimate durations; the vertical black bars
  denote the relative uncertainties derived from uncertainties in the
  summed selection function. The orange and green solid lines denote
  the functions $p_{\rm ecl}$($t_{\rm i}$) that minimize or maximize,
  respectively, the likelihood of occurrence given the limits on each
  observed event duration (see text for details).  Because $p_{\rm
    ecl}$($t_{\rm i}$) is in log space, zero probabilities are plotted
  as $10^{-6}$ for clarity.  The cyan and magenta lines denote
  1/$SF_{\rm sumI}$($t_{\rm i}$) and 1/$SF_{\rm sumII}$($t_{\rm i}$),
  respectively.  The thin magenta line in the top panel and the
  thin cyan line in the lower panel denote 1/$SF$ of the other
  object class to allow visual comparison between both object
  classes.}
\label{fig:durprobs}
\end{figure}

Given the values of $N_{\rm ecl}$($t_{\rm i}$) corresponding to the
best estimates of duration, we can now estimate the corresponding
function of $p_{\rm ecl}$($t_{\rm i}$) for each class by dividing each bin
of $N_{\rm ecl}$($t_{\rm i}$) by $SF_{\rm sumI,II}$($t_{\rm i}$).  
The resulting function of $p_{\rm ecl}$($t_{\rm i}$) for each
class is plotted as the black solid lines in Fig.~\ref{fig:durprobs}.
As a reminder, $p_{\rm ecl}$($t_{\rm i}$) describes the likelihood to
catch a source undergoing an eclipse due to one specific kind of
cloud: one that results in an eclipse with a duration $t_{\rm i} \leq
D < t_{\rm i+1}$. The errors on individual points of 
$p_{\rm ecl}$($t_{\rm i}$) take into account the error in the 
selection function obtained from the bootstrapping method only.

The minimum- and maximum-likelihood $p_{\rm ecl}$($t_{\rm i}$)
functions are calculated in a similar way, although we divide by
$SF_{\rm sumI,II}$($t_{\rm i}$) $\pm$ $\sigma_{\rm SF}$($t_{\rm
  i}$), where $\sigma_{\rm SF}$($t_{\rm i}$) denotes the
uncertainties in $SF_{\rm sumI,II}$($t_{\rm i}$) as determined by the
Monte Carlo bootstrap method; the minimum and maximum functions are,
respectively, the orange and green solid lines in
Fig.~\ref{fig:durprobs}.  Because $p_{\rm ecl}$($t_{\rm i}$) is in
log space, any bin with zero probability density is plotted as
$10^{-6}$.  ``Typical'' values of $p_{\rm ecl}$($t_{\rm i}$) when
eclipses are detected are $\sim10^{-(2.8-3.4)}$ for type Is and
$\sim10^{-(1.3-2.2)}$ for type IIs. We must caution the reader that
most probability density values are on the order of 1/$SF_{\rm
  sum}$($t_{\rm i}$). We display 1/$SF_{\rm sum}$($t_{\rm i}$) in
Fig.~\ref{fig:durprobs} via cyan and magenta lines for types I and
II, respectively.  Consequently, the detection of one eclipse within
these duration bins would strongly influence the inferred $p_{\rm
  ecl}$($t_{\rm i}$) profile.  Nonetheless, the data suggest that
eclipses of any duration generally occur more frequently in type IIs.
 
In type IIs, we do not detect events of durations $\sim$tens of days,
for which one event for a given timescale bin would have yielded a
probability density of $\sim$ 1/200 -- 1/1000.  If it were the case
that eclipses with durations of $\sim$tens of days occurred in type
IIs with the same probability density as in type Is,
$\sim10^{-(2.8-3.4)}$, our observations would likely not have detected
them.  This is because the values of those $p_{\rm ecl}$($t_{\rm i}$) points
would lie below the detection threshold for
eclipses in type IIs, represented by 1/$SF_{\rm sumII}$($t_{\rm i}$),
the magenta line in the lower panel of Fig.~\ref{fig:durprobs}; recall
that there were $\sim$3-4 times fewer campaigns for type IIs as there
were for type Is.

In type Is, we did not detect events of durations $\sim$hundreds of
days (although such a duration cannot be ruled out in the case of
MR~2251--178), for which one event for a given timescale bin would
have yielded a probability of $\sim$ 1/200 -- 1/1000.  If it were the
case that eclipses with durations of $\sim$hundreds of days occurred
in type Is with the same probability density as in type IIs, then the
number of monitoring campaigns for type Is would have clearly allowed
us to detect them.  If this were the case, then the resulting $p_{\rm
  ecl}$($t_{\rm i}$) values for type Is would appear in the upper
panel of Fig.~\ref{fig:durprobs} with values higher than the detection
threshold, represented by 1/$SF_{\rm sumI}$($t_{\rm i}$), the cyan
lines.

Finally, we sum up the values of $p_{\rm ecl}$($t_{\rm i}$) over all
timescale bins to obtain estimates of $\overline{P_{\rm ecl}}$.  More
specifically, best-estimate, minimum and maximum values of
$\overline{P_{\rm ecl}}$ can be obtained by summing the functions of
$p_{\rm ecl}$($t_{\rm i}$) denoted by the black, orange, and green
histogram lines, respectively, in Fig.~\ref{fig:durprobs}.  However, we must bear in
mind that we have a very small number of total events, and many
timescale bins contain only $\sim0-1$ events.  When summing to obtain
the maximum value of $\overline{P_{\rm ecl}}$, we opt to be
conservative and take into account the selection function, the
reciprocal of which denotes the likelihood of witnessing just one
event with duration $t_{\rm D}$ satisfying $t_{\rm i} \leq D < t_{\rm
  i+1}$. For each $t_{\rm i}$, we use the maximum of either $1/SF_{\rm
  sumI,II}$($t_{\rm i}$) (cyan/magenta lines in
Fig.~\ref{fig:durprobs}) or $p_{\rm ecl}$($t_{\rm i}$) (green line).
To reiterate, our uncertainties on $\overline{P_{\rm ecl}}$ are
conservative estimates in that they take into account the effect of
uncertainties in duration on $N_{\rm ecl}$($t_{\rm i}$), the
uncertainties in $SF_{\rm sumI,II}$($t_{\rm i}$) as determined by the
Monte Carlo bootstrap method, the effects of the selection function,
and the presence of candidate events on the maximum probability.  Our
final estimate for $\overline{P_{\rm ecl}}$ for type Is is 
0.006 with
a range of 0.003--0.166 (minimum -- maximum probability).  For type
IIs, $\overline{P_{\rm ecl}}$=0.110, with a range of 0.039--0.571. 
That is, based on our best estimates, type II AGN have a $\sim18$ times
higher chance of showing an eclipse (of any duration).  These values
are summarized in Table~\ref{tab:finalprobs}, and we
return to these values later in $\S$5.

We caution that these probabilities refer only to eclipses by
full-covering, neutral or lowly-ionized clouds with columns densities
$\ga 10^{22}$ up to $\sim10^{25}$~cm$^{-2}$; when one considers the
full range of clouds (larger range of $N_{\rm H}$, partial-covering
clouds, wider range of ionization) the resulting probabilities will
almost certainly be higher.

\begin{table}
\caption{Inferred probabilities $\overline{P_{\rm ecl}}$ to witness a source undergoing an eclipse event}   
\label{tab:finalprobs} 
\begin{tabular}{lccc} \hline
 Object       & $\overline{P_{\rm ecl}}$  &   $\overline{P_{\rm ecl}}$   & Limits on   \\
 class        & (Best                   & (Min.--max.\               & $\overline{P_{\rm ecl}}$ for Compton-  \\
              & estimate)               & range)                     & thick events  \\ \hline
Type Is       & 0.006                   & 0.003--0.166               & $<0.158$   \\
Type IIs      & 0.110                   & 0.039--0.571               & $<0.520$   \\ \hline
\end{tabular}
\end{table}

Ramos~Almeida et al.\ (2011) noted that the IR SEDs of Sy 1.8--1.9 AGN
had spectral slopes intermediate between those of Sy 1.0--1.5s and Sy
2s, and noted similarities in SED fitting parameter results between
the 1.8--1.9s and the 1.0--1.5s.  However, we simply do not have
enough events to break up the probability estimates for type IIs into
further subclasses, and cannot address this point.  If we break up the
type Is into Sy 1.0s/1.2/1.5s, tantalizingly, the best-estimate values
for $\overline{P_{\rm ecl}}$ increase with subclass: 0.0003, 0.0024,
and 0.0036, consistent with the notion that relatively higher levels
of obscuration exist in higher numbered subclasses. However, the
(conservatively-determined) minimum and maximum values indicate nearly
identical ranges ([0.0003-0.1586], [0.0014-0.1597], and
[0.0013-0.1619]) because with only $\sim0-1$ events per timescale bin,
the maximum values tended towards the integral of $1/SF_{\rm
  sum1.0/1.2/1.5}$($t_{\rm i}$).  Furthermore, any conclusions would
be based on a mere 1--4 secure eclipse events and 1--3 objects per
subclass, and so we do not address the subclasses further.

In $\S$3, we noted that we did not detect any Compton-thick
obscuration events, but this is not to say with full certainty that
such events cannot exist. A conservative upper limit on the
probability to observe a Compton-thick event of any duration between
0.2~d and 16~yr is obtained by summing $1/SF_{\rm sumI,II}$($t_{\rm
  i}$): $\overline{P_{\rm ecl}} < 0.158$ (type Is) or $< 0.520$ (type
IIs).

\section{DISCUSSION}  

There has been much work so far into variable X-ray absorption in AGN.
The accumulated research so far, including this work, has revealed
eclipse events spanning a wide range of observed durations, ionization
levels, and including full or partial covering.

Most of the short-term events (durations $\la 1-3$ d) have been
observed in the prominent case of the Sy 1.8 NGC~1365, where rapid and
strong variations in $N_{\rm H}$ are observed in a very large fraction
of observations.  The clouds are inferred to exist in the
BLR and are frequently modeled via variations in covering fraction of
partial-covering material (e.g., De Rosa et al.\ 2007; Turner et al.\
2008; Risaliti et al.\ 2011).  In the context of clumpy-absorber
models, there can exist a variable number of clouds in the line of
sight at any given time, with a relatively extended X-ray continuum
source located behind them.  

\textit{RXTE}, in contrast, has detected full-covering, cold or at
most modestly ionized clouds.  An emerging picture is that clouds in
the BLR and clouds in the torus may be two manifestations of the same
radially extended cloud distribution, existing inside and outside,
respectively, the dust sublimation region (Elitzur 2007).  That is, in
the simplest picture, the full-covering clouds detected with
\textit{RXTE} are part of the same population of clouds as those
causing the short-duration events, just occurring at larger radii, as
we will demonstrate in $\S$5.2.

It is potentially interesting that 4 of the 10 sources in our sample
with secure or candidate eclipse events have multiple such events
(certain objects thus seem to be prone to a higher frequency of events
than others).  The high frequency of events observed in NGC~1365 may
potentially be explained by a favorable geometry/viewing angle.
However, NGC~1365 may be a special object; a statistical sample on a
large number of AGN may be better suited to quantify eclipse events
across all AGN.

\subsection{Dependence of cloud events on AGN physical parameters}   

Having identified secure and candidate eclipses in 10 of the 55
objects monitored with \textit{RXTE}, we can attempt to explore if
certain system parameters govern the presence or lack of eclipsing
clouds.  For example, the 10 sources' black hole masses (see $\S$5.2)
span log($M_{\rm BH}$) = 6.9--8.8, and these values are not extreme
compared to black hole masses for Seyferts/quasi-stellar objects
(QSOs) in general or our whole sample.

There is no evidence that having an extreme value of luminosity
governs the presence/lack of eclipses: the range of log($L_{2-10}$)
and log($L_{\rm Bol}$) where eclipses are confirmed with \textit{RXTE}
to occur are 41.93--44.73 and 42.9--45.6,
respectively.\footnote{Values for $L_{\rm Bol}$ for most sources are
  taken from Vasudevan et al.\ (2010), except for Fairall~9 (Vasudevan
  \& Fabian 2009), and MR~2251--178 and Cen~A ($L_{\rm 2-10}$ from 
  Rivers et al.\ 2013 and 
  $L_{\rm Bol}/L_{2-10}$ corrections from Marconi et al.\ 2004 for both 
  objects).} Again, these values are not extreme.

In the context of clouds being produced in a disk wind, Elitzur \& Ho
(2009) estimate that such winds cannot exist when the bolometric
luminosity drops below $5 \times 10^{39}$ ($\frac{M_{\rm
    BH}}{10^7\Msun}$)$^{2/3}$ erg s$^{-1}$. However, we cannot test
this with our sample, as the lowest-luminosity objects monitored with
\textit{RXTE} have 2--10~keV luminosities of $2-6\times 10^{40}$ erg
s$^{-1}$ (NGC~4258, NGC~3998, and NGC~4051).

Accretion rate relative to Eddington, $L_{\rm
  Bol}/L_{\rm Edd}$, is also not an obvious factor; values of
$\dot{m}$ for the 10 objects span $\sim0.1$ per cent for Cen~A up to
$\sim16$ per cent for MR~2251--178, with most other objects' values in the
range of 3--10 per cent. These values are typical for Seyferts in
general as well as for our sample.


Can radio loudness be a factor? One of our eight sources with secure
eclipses is radio loud, compared to 13/55 (24 per cent) sources in the
original sample. If we start with 55 objects, 13 of which are
radio loud, and select at random eight of the 55 for a sub-sample, the
probability $P(X=k)$ that exactly $k$ out of eight sources in the
subsample will be radio loud is given by the hypergeometric
distribution   
\begin{equation}
  \label{eq:hgd}
  P(X=k) = {K \choose k} {N-K \choose n-k} \Big/ {N \choose n}\ ,
\end{equation}
where $N=55$ is the population size, $K=13$ is the number of successes
in the population (radio-loud sources), and $n=8$ and $k$ are the
sample size and the number of successes in the sample,
respectively. ${\cdot \choose \cdot}$ is a binomial coefficient. The
chance of having exactly $k=0, 1, 2, 3$ or $4$ radio-loud sources in a random
sample of eight is $P(X=0, 1, 2, 3,$ or $ 4) =$ (0.10, 0.29, 0.34, 0.20, or
0.07). In other words, it is almost as likely to have exactly 1/8
radio-loud sources in the sub-sample as it is to have 2/8 radio-loud
sources. There is thus no significant statistical evidence that our
sub-sample of sources with secure eclipses
is unusual compared to the original sample in terms of
fraction of radio-loud sources.

In summary, we find no evidence for a strong dependence of the
presence of eclipse events as a function of the most common AGN
parameters. One has to keep in mind that only strong correlations
could have been detected with our relatively small number of observed
events.  That being said, the reader is reminded that with a high
dynamic range of sampling for 55 objects totaling 230 object-years of
monitoring, this study is by far the largest available data set for
testing the environment close to supermassive black holes via eclipse
events over timescales longer than a few days.

\subsection{Locations of X-ray obscuring clouds}    
\label{sec:locations}

Constraints on the distance from the X-ray continuum source to each
cloud $r_{\rm cl}$ can be derived from the cloud's ionization level,
column density, and eclipse duration.  Following, e.g., $\S$4 of Lamer
\etal\ (2003), we assume for simplicity that each cloud has a uniform
density and ionization parameter, and that clouds are in Keplerian
orbits. The cloud diameter $D_{\rm cl}$ = $N_{\rm}/n_{\rm H} =  v_{\rm cl}t_{\rm D}$, 
where $t_{\rm D}$ is the cloud's crossing time across the line of sight and
$v_{\rm cl}$ is the transverse velocity, equal to $\sqrt{GM_{\rm BH}/r_{\rm cl}}$.
Solving for $r_{\rm cl}$ and using the definition of $\xi$,
one obtains (see, e.g., Eqn.\ 3 of Lamer \etal\ 2003):
$r_{\rm cl} = 4\times10^{16} M_7^{1/5} L_{42}^{2/5} t_{\rm D}^{2/5} N_{\rm H,22}^{-2/5} \xi^{-2/5}$~cm,
where $M_7 = M_{\rm BH}/10^6 \Msun$,   $L_{42}=L_{\rm ion}/$($10^{42}$ erg s$^{-1}$), 
$N_{\rm H,22}=N_{\rm H}/$($10^{22}$~cm$^{-2}$), and $t_{\rm D}$ is in units of days.

Table~\ref{tab:ecl1dist} lists the inferred values of $r_{\rm cl}$,
obtained as follows: we use estimates of $M_{\rm BH}$ from
reverberation mapping where available; otherwise, estimates are from
stellar kinematics or gas dynamics, based on empirical relations
between optical luminosity and optical line widths, or from modeling
reprocessing in accretion disks. The references are listed in column~4
of Table~\ref{tab:ecl1dist}.

For all objects except Cen A, we estimate $L_{\rm ion}$ as follows: we
use the best-fitting unabsorbed power law from Rivers et al.\ (2013), and
extrapolate to the 0.1--13.6 keV range to estimate the luminosity
$L_{\rm 0.1-13.6 keV}$. The luminosity below $\sim$0.1 keV, however,
is expected to contain significant contributions from the thermal
accretion disk continuum emission. Vasudevan \& Fabian (2009) estimate
that the ratio of the 0.0136--0.1 keV luminosity to the bolometric
luminosity, $L_{\rm 0.0136-0.1 keV}/L_{\rm Bol}$, ranges from 0.2 to
0.6 for values of $L_{\rm Bol}/L_{\rm Edd}$ ranging from 0.01 to 0.6,
respectively. We take values of $L_{\rm Bol}$ and $L_{\rm Bol}/L_{\rm
  Edd}$ from Vasudevan et al.\ (2010) when available. However, for
Fairall~9 the value of $L_{\rm Bol}$ is taken from Vasudevan \& Fabian
(2009), and for MR~2251--178, we use the 2--10 keV flux from Rivers et
al.\ (2013) and we estimate $L_{\rm Bol}/L_{\rm Edd} = 40$ from
Marconi et al.\ (2004). We then assume for simplicity a linear
dependence of log10($L_{\rm 0.0136-0.1 keV}/L_{\rm Bol}$) on
log10($L_{\rm Bol}/L_{\rm Edd}$) to estimate $L_{\rm 0.0136-0.1 keV}$,
and add that to $L_{\rm 0.1-13.6 keV}$ to obtain the values of $L_{\rm
  ion}$ listed in Table~\ref{tab:ecl1dist}. The broadband SED of Cen~A
is more like that of blazars than radio-quiet Seyferts, with a broad
inverse-Compton hump dominating the emission from the optical band to
gamma-rays. We assume $\Gamma=1.84$ (Rivers et al.\ 2013) over
0.4--13.6 keV, breaking to $\Gamma=1.0$ over 0.0136--0.4 keV (Steinle
2010; Roustazadeh \& B\"{o}ttcher 2011); this yields $L_{\rm ion}$ =
$10^{42.3}$ erg s$^{-1}$.

Our constraints on the ionization parameters for our clouds are poor.
Given the column densities of our clouds and the energy resolution of
the PCA, we can safely rule out values of log($\xi$) above $\sim2$. In
that case, the absorber would manifest itself above 2~keV only via
discrete lines and edges, as opposed to a broad rollover towards lower
energies. We therefore calculate distances assuming log($\xi$) = --1,
0, or +1.  Uncertainty on $r_{\rm cl}$ is thus dominated by
uncertainty on $\xi$.  For brevity, we list in
Table~\ref{tab:ecl1dist} only values assuming log($\xi$)=0.  Because
$r_{\rm cl} \propto \xi^{-2/5}$, values of $r_{\rm cl}$ assuming e.g.,
log($\xi$) = --1 or +1 are only factors of 2.5 larger or smaller,
respectively. However, for Cen~A, we do not consider log($\xi$) = +1.
We fit the spectrum of the most absorbed spectral bin used in Rivers
\etal\ (2011b), model the absorption with an \textsc{xstar} table, and
obtain log($\xi$) $\la -0.1$.  Similarly, for NGC~3227/2000--1,
Lamer \etal\ (2003) found log($\xi$) = --0.3--0 from a contemporaneous
\textit{XMM-Newton} observation.  These two events have among the
highest quality energy spectra in the group, and one might speculate
that such an ionization level may be representative of the rest of the
sample.  For the candidate event in NGC~3516, we use the best-fitting
ionization from Turner et al.\ (2008), log($\xi$)=2.19, as explained
in Appendix A.  Finally, we note that although for a few events, it is
difficult to estimate reliable uncertainties on $t_{\rm d}$, a
$\pm10$ per cent uncertainty on $t_{\rm d}$ translates into only $\pm4$ per cent
uncertainty in $r_{\rm cl}$.


We obtain best-estimate values of $r_{\rm cl}$ that are typically tens
to hundreds of light-days from the central engine, and lie in the
range $1 - 50 \times 10^{4} R_{\rm g}$ ($0.3 - 140 \times 10^{4}
R_{\rm g}$ accounting for uncertainties), where $R_{\rm g} \equiv
GM_{\rm BH}c^{-1}$. We now compare these inferred radii to the locations
of typical emitting components in AGN. In Table~\ref{tab:BLRdust} we
list inferred locations for the origins of optical and near-IR broad
emission lines and ``hidden'' broad lines observed in polarized
emission for type IIs, locations of IR-emitting dust determined via
either interband correlations or IR interferometry, and locations of
Fe K$\alpha$ line-emitting gas determined via either X-ray
spectroscopy or variability. For reverberation mapping of emission
lines, we use time-lag results when available, otherwise we use full
width at half-maximum (FWHM)
velocities $v_{\rm FWHM}$, assumed for simplicity Keplerian motion,
and estimate radial locations via $r = GM_{\rm BH}/( \frac{3}{4}
v_{\rm FWHM}^2)$.

\begin{figure*}
\includegraphics[angle=-90,width=1.00\textwidth]{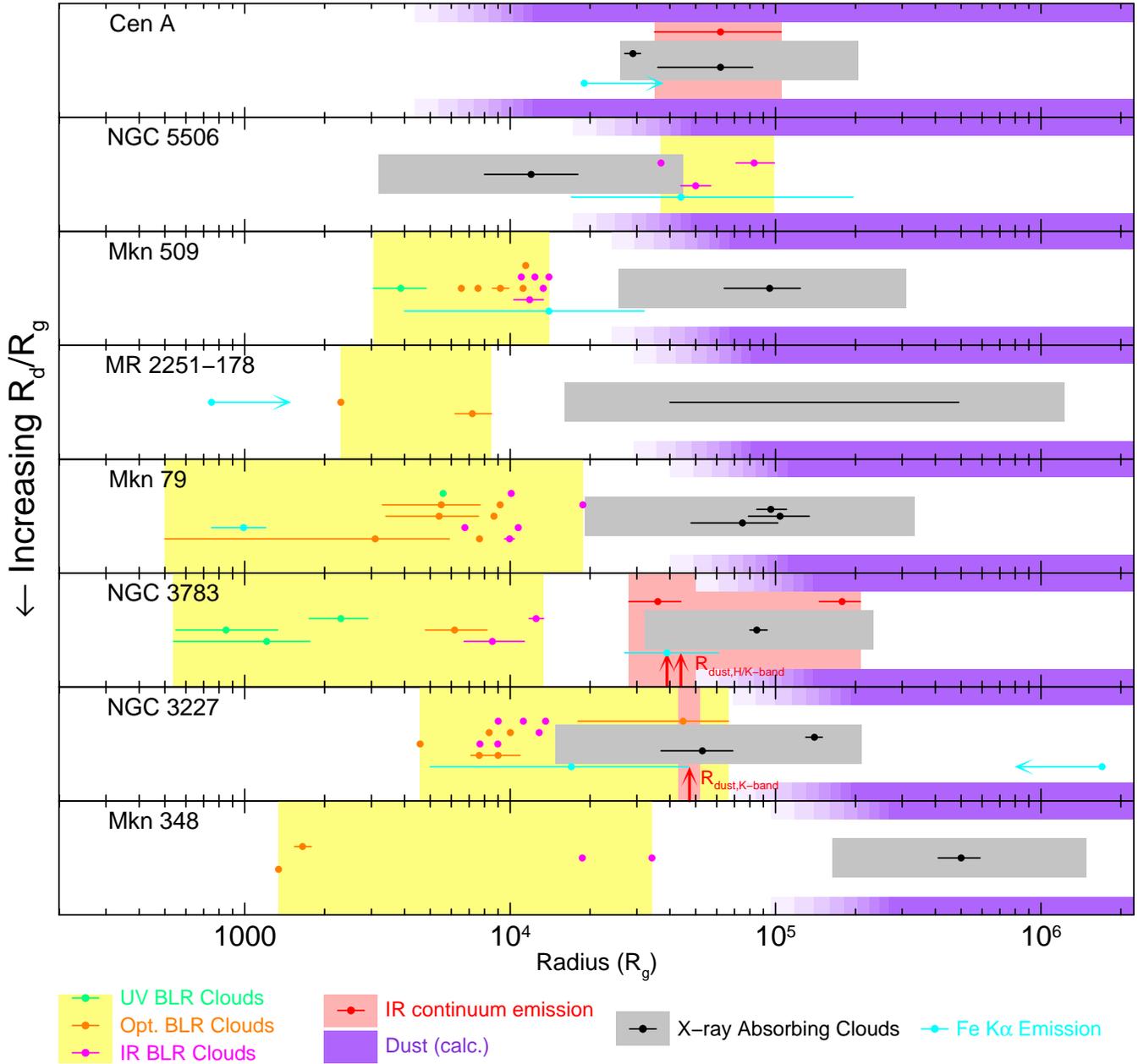}   
\caption{Plot of inferred radial locations of the secure X-ray-absorbing
  clouds $r_{\rm cl}$, plus inferred radial locations of IR
  continuum-emitting dust and BLR clouds, in units of $R_{\rm g}$.
  See Table~\ref{tab:BLRdust} for details.  Sources are ranked in
  order of increasing $R_{\rm d}$/$R_{\rm g}$.  Locations of each
  of the X-ray absorbing clouds for an assumed ionization value of
  log($\xi$)=0 are denoted by black points, with the full range
  obtained with log($\xi$) spanning values from --1 to +1 (with --1 to
  0 for Cen~A and --0.3 to 0 for NGC~3227/2000--1) indicated by the
  gray shaded areas. The data point for MR~2251--178 is omitted due to
  the poor constraints on $t_{\rm D}$ and thus $r_{\rm cl}$.  The
  purple shaded areas denote radial distances expected to contain
  dust, with the fading purple areas (1.0 down to 0.4$R_{\rm d}$)
  denoting the ``dust sublimation zone.''  Known locations of IR
  continuum emitting dust are plotted in red points, with ranges
  indicated by pink shaded areas.  BLR ranges are indicated by the
  yellow areas.  Green points indicate origins of UV broad emission
  lines (e.g., He~\textsc{ii}, C~\textsc{iv}, S~\textsc{iv}). Orange points show
  origins of optical broad emission lines (e.g., Balmer series). 
  Magenta points show origins of IR broad emission lines (e.g.,
  Bracken and Paschen series, He~\textsc{i}$\lambda$2.058$\mu$).
  Locations of Fe K$\alpha$-emitting gas are plotted in cyan.  As a
  reminder, radial locations of these structures may be commensurate
  but structures may not physically overlap, e.g., the X-ray-absorbing
  clouds lie on the direct line of sight to the central engine, but
  emission components may lie off the line of sight.}
\label{fig:plotrs}
\end{figure*}

The results are also plotted in Fig.~\ref{fig:plotrs} in units of
$R_{\rm g}$ for each object. All structures for a given object in
Fig.~\ref{fig:plotrs} are plotted on one dimension for clarity, but
they do not necessarily overlap in space (e.g., X-ray clouds must lie
along the line of sight, but other structures may lie out of it). The
objects are listed in order of increasing $R_{\rm
d}$/$R_{\rm g}$. The black points denote the estimates for
$r_{\rm cl}$ assuming log($\xi$)=0, but we denote the minimum/maximum
range assuming log($\xi$)= --1 to +1 by the gray areas (again, with
--1 to 0 for Cen~A and --0.3 to 0 for NGC~3227/2000--1).


The radial distance from the black hole where dust sublimates can be
estimated via $R_{\rm d} \sim 0.4 (L_{\rm bol}/10^{45} {\rm erg
  s}^{-1})^{1/2} (T_{\rm d}/T_{\rm 1500})^{-2.6}~{\rm pc}$, where
$T_{\rm d}$ is the dust sublimation radius, here assumed to be 1500 K
(Eqn.\ 2.1 of Nenkova et al.\ 2008b). However, the boundary between
dusty and dust-free zones is likely highly blurred, because relatively
larger grains can survive at higher temperatures. In addition,
individual components of dust can sublimate at different radii, e.g,
graphite grains sublimate at a slightly higher temperature than
silicate grains (Schartmann et al.\ 2005). We use the above equation
as an approximation for the outer boundary of the "dust sublimation
zone" (DSZ), i.e., we assume that no dust sublimation occurs
outside $R_{\rm d}$. We take the inner boundary of the DSZ to be a
factor of $\sim$2--3 smaller than $R_{\rm d}$. For example, Nenkova et
al.\ (2008b) point out that the $V$--$K$ band reverberation lags measured
by Minezaki et al.\ (2004) and Suganuma et al.\ (2006) are $\sim2-3$
times shorter than the light travel times predicted by the above
equation. Those experiments thus may be tracking the innermost, larger
grains. For simplicity, we assume that the central engine emits
isotropically.\footnote{This may not be true for Cen~A, whose
  continuum emission is likely mildly beamed. A typical value of
  Doppler $\delta$ for Cen~A is 1.2 (Chiaberge et al.\ 2001). Using
  0.66 as the spectral index (Abdo et al.\ 2010), the flux will be
  boosted by $\delta^{2+\alpha} \sim 1.6$, which translates into only
  a $\sim26$ per cent effect on our estimate of $R_{\rm d}$.} The
estimated values of $R_{\rm d}$ are listed in Table~B1, with dusty
regions represented by the purple areas in Fig.~\ref{fig:plotrs}, and
the inferred DSZs represented by the fading purple areas.

We list the radial locations of each cloud in units of $R_{\rm d}$ in
Table~\ref{tab:ecl1dist}.  The clouds in 7/8 objects are consistent
with residing in the DSZ considering uncertainties on $r_{\rm cl}$.  
However, this ``clustering'' may be in part associated with our
observational bias to select eclipses with events of $\sim$tens of
days, as per our selection function.  In Cen~A, the clouds are
inferred to be consistent with residing entirely in the dusty zone.
In contrast, NGC~5506 has the lowest value of $r_{\rm cl}$/$R_{\rm
  d}$; those clouds are likely the least dusty of the secure events in
our sample.

The IR-emitting structures for Cen~A, NGC~3783, and NGC~3227 as mapped
by either interferometric observations or optical-to-near-IR
reverberation mapping are inferred to exist at radii $r_{\rm IR} \sim
3-20 \times 10^4 R_{\rm g}$. Although we can only make a statement
based on three objects, values of $r_{\rm cl}$ are generally
consistent with these structures, again supporting the notion that the
X-ray-absorbing clouds detected with \textit{RXTE} are thus likely
dusty.

For six of the seven objects with known BLRs (MR~2251--178, Mkn~509,
Mkn~79, NGC~3783, NGC~3227, and Mkn~348), the BLR clouds are inferred
to exist at $r_{\rm BLR} \sim 0.1 - 2 \times 10^4 R_{\rm g}$ from the
black hole. We thus discuss our results in the context of the notion
put forth by Netzer \& Laor (1993) that the outer radius of the BLR
may correspond to $R_{\rm d}$ and the inner radius of the dusty torus,
supported by the results of Suganuma et al.\ (2006). In these six
objects, one can see from Fig.~\ref{fig:plotrs} that the radial ranges
of $r_{\rm BLR}$ are generally smaller than those for both $r_{\rm
cl}$ and $r_{\rm IR}$. That is, best-estimate values of $r_{\rm cl}$
are generally commensurate with the outer portions of the BLR or exist
at radii up to $\sim$15 times that of the outer boundary of the BLR,
although we must caution that we are dealing with a very small sample
of only six objects.

We can conclude that, at least for these six objects, the
X-ray-absorbing clouds are likely more consistent with the dusty torus
than with the BLR. Note that this conclusion is relatively robust to
our assumed values of log($\xi$); even if log($\xi$)=+2, the
X-ray-absorbing clouds would be closer to the black hole by only a
factor of 2.5. Our results thus demonstrate that radii commensurate
with the outer BLR and with the inner dusty torus are radii where
X-ray-absorbing clouds may exist.

The seventh object with measured BLR parameters is NGC~5506, where
$r_{\rm BLR} \sim 4-10 \times 10^4 R_{\rm g}$ based on IR emission
lines. The inferred radial location of the X-ray cloud is
$0.3-4.5\times 10^4 R_{\rm g}$ -- roughly commensurate with the
innermost BLR or distances slightly smaller than the BLR as traced by
IR emission lines, and thus likely non-dusty. The eclipse event in
NGC~5506 may thus be analogous to short-term ($t_{\rm D} \la$ 1 d)
eclipse events observed in other objects, namely NGC~1365 and Mkn~766
(Risaliti et al.\ 2005, 2007, 2009b, 2011; Maiolino et al.\ 2010). The
variable absorbers detected in those observations have column
densities typically $N_{\rm H} \sim 1-5 \times 10^{23}$~cm$^{-2}$ and
can be partial covering, as for Mkn~766. They are inferred to exist
$\sim10^{2.5} - 10^4 R_{\rm g}$ from the black hole, and are
frequently identified as BLR clouds.

 
We return to a point regarding the type Is from $\S$4.3, regarding
eclipse events longer than $\sim100$~d in duration.  Our selection
function analysis shows that if such events occurred in type Is with
the same frequency as in type IIs, we should have detected them.  It
is thus possible that type Is are devoid of clouds at these radial
distances along the line of sight.  Another possibility is that these
clouds do exist at these radii in type Is, but intersect our line of
sight so rarely [probabilities $<$ 1/(10--300)] that 16 years of
monitoring 37 type I AGN was insufficient to catch even one event
(excluding the possibility that the event in MR~2251--178 falls into
this category).  In either case, we can estimate the corresponding
radial distances.  We assume a black hole mass of 10$^{7.7} \Msun$,
$N_{\rm H} = 12 \times 10^{22}$~cm$^{-2}$, log($\xi$)=0, and
log($L_{\rm ion}$) = 44.3, the non-weighted average of the type Is
with secure events.  Arbitrarily-chosen durations of 100, 316, 1000,
and 3162 d correspond to inferred radial distances of 14, 23, 36, 
and 57 $\times 10^4$ $R_{\rm g}$. 

\begin{figure*}
 \includegraphics[angle=-90,width=1.00\textwidth]{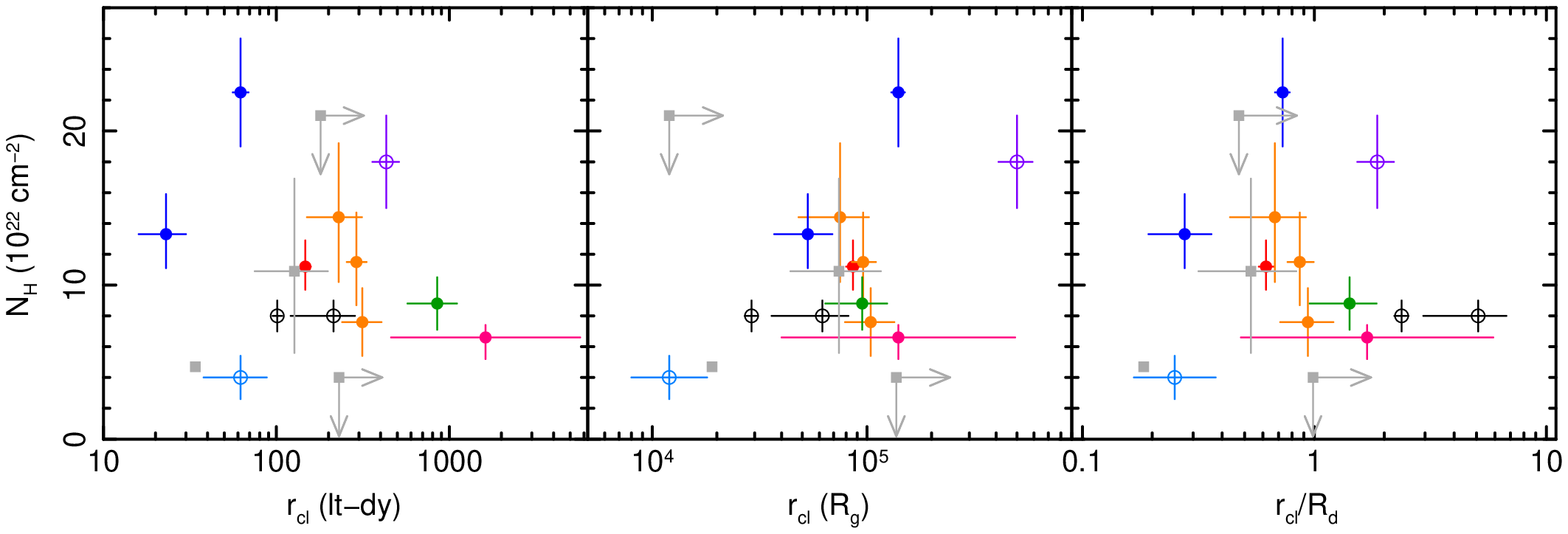} 
 \caption{Best-fitting values of $N_{\rm H}$ plotted against inferred
   radial location $r_{\rm cl}$ in units of light-days (left), in units of
   gravitational radii $R_{\rm g}$ (center; $R_{\rm g} \equiv GM_{\rm BH}/c^2$), and relative to dust sublimation radius
   $R_{\rm d}$. Filled and open symbols correspond to type I and II
   objects, respectively. Colors are the same as in
   Fig.~\ref{fig:durbars}. }
 \label{fig:NHvsrad}
 \end{figure*}

Finally, we investigate any possible trend between the best-fitting values
of peak $N_{\rm H}$ and the inferred values of $r_{\rm cl}$ listed in
Table~\ref{tab:ecl1dist}, in units of either light-days, $R_{\rm g}$, or
$r_{\rm cl}$/$R_{\rm d}$ (Fig.~\ref{fig:NHvsrad}).
Given the small number of eclipse events, and given that
values of $N_{\rm H}$ span only a factor of $\sim$7, we find no robust
correlations; inclusion/exclusion of only 1--2 secure or candidate
points significantly changes the resulting 
coefficients and null hypothesis probabilities.  Note that the
overdensity of eclipse events around 200 light-days (left-hand panel),
$8\times10^4 R_{\rm g}$ (middle panel), and $\la 1 R_{\rm d}$
(right-hand panel) is likely not intrinsic to the sources but mainly caused
by our selection function. This figure should thus not be used to
derive constraints on the radial distribution of clouds.

\begin{table*}
\begin{minipage}{170mm}
\caption{Inferred cloud distances from the black hole}   
\label{tab:ecl1dist}  
\begin{tabular}{llcccccccc} \hline
Source   &           & log($M_{\rm BH}$) &       & $L_{2-10}$, $L_{\rm ion}$, $L_{\rm Bol}$, & log($\xi$, erg         &  $r_{\rm cl}$      &  $r_{\rm cl}$    &                       &  \\
name     & Event     &  ($\Msun$)      & Ref.\ & (erg s$^{-1}$)          & cm s$^{-1}$)  & (light-days)          & ($10^4$ $R_{\rm g}$) & $r_{\rm cl}$/$R_{\rm d}$ &  Notes \\ \hline 
NGC 3783 & 2008.3        & 7.47        & VP06 &  43.2, 44.1, 44.4       &  0                & $147^{+11}_{-10}$  & $8.6^{+0.7}_{-0.6}$   & $0.62\pm0.04$         &     \\
         & (Full event)  &             &      &                         &                   &                  &                    &                        &   \medskip    \\
Mkn 79   & 2003.5        &  7.72       & Pe04 &  43.3, 44.3, 44.7       &  0                & $229^{+83}_{-79}$  & $7.5\pm2.7$        & $0.68\pm0.24$           &   \\
         & 2003.6        &             &      &                         &  0                & $290^{+42}_{-35}$  & $9.6^{+1.4}_{-1.1}$   & $0.86^{+0.13}_{-0.09}$     &   \\
         & 2009.9        &             &      &                         &  0                & $314^{+91}_{-74}$  & $10.4^{+3.0}_{-2.5}$   & $0.93^{+0.27}_{-0.22}$     &   \medskip \\   
Mkn 509  & 2005.9        &  8.19       & VP06 &  44.3, 44.9, 45.2       &  0                & $851^{+255}_{-278}$ & $9.5^{+2.9}_{-3.1}$  & $1.42^{+0.42}_{-0.46}$     &  $^{\dagger}$ \medskip \\ 
MR 2251--178& 1996  & 8.3 & W09  &  44.7, 45.3, 45.6          &  0                & 460--5700        & 4--49                   &  0.5--6.0              &   $^{\dagger}$$^{\ddagger}$ \medskip \\  
NGC 3227 & 2000--1       &  6.88       & D10  &  42.5, 43.0, 43.5       &  --0.3            & $82^{+9}_{-8}$     & $19\pm2$           &  $0.97\pm0.11$         &    \\
         &               &             &      &                         &  0                & $62^{+7}_{-6}$     & $14\pm1$           &  $0.74\pm0.05$         &  \\ 
         & 2002.8        &             &      &                         &  0                & $23\pm7$         & $5.3\pm1.6$        &  $0.28^{+0.08}_{-0.09}$   &  \medskip \\  
Cen~A    & $\sim$2003--4 &  7.78       & R11  &  41.9, 42.3, 42.9       &  0                & $214^{+70}_{-93}$   & $6.2^{+2.0}_{-2.6}$  & $5.3^{+1.6}_{-2.1}$      &   $^{\dagger}$ \\
         & 2010--1       &             &      &                         &  0                & $101^{+7}_{-6}$    & $2.9\pm0.2$        & $2.4\pm0.2$           &   \medskip \\     
NGC 5506 & 2000.2        &  7.94       & Pa04 &  43.0, 43.9, 44.3       &  0                & $62^{+26}_{-24}$    & $1.2^{+0.6}_{-0.4}$  & $0.26^{+0.11}_{-0.08}$    &  \medskip \\    
Mkn 348  & 1996--7       &  7.18       & WZ07 &  43.0, 44.1, 44.4       &  0                & $432^{+79}_{-73}$    & $50\pm9$          & $1.82\pm0.37$       &   $^{\dagger}$  \medskip \\ 
Fairall 9 & 2001.3 (cand.) & 8.41      & VP06 &  43.4, 44.6, 44.8       &  0                & $>$180            & $>$1.2            & $>0.5$              &   $^{*}$ \medskip  \\  
NGC 3783 & 2008.7 (cand.)& 7.47        & VP06 &  43.2, 44.1, 44.4       &  0                & $>$230            & $>$13.7            & $>1.0$              &   $^{*}$   \\   
         & 2011.2 (cand.)&             &      &                         &  0                & $127^{+71}_{-52}$   & $7.4^{+4.2}_{-3.0}$   & $0.54^{+0.30}_{-0.22}$    &   \medskip   \\    
NGC 3516 & 2011.7 (cand.) & 7.50       & D10  & 43.0, 43.7, 44.2        & +2.19             & 34                & 1.9               & $\sim0.6$           & $^{**}$ \\      \hline

\multicolumn{2}{c}{Average for 12 Secure Events} & 7.63  &    &    43.1, 43.8, 44.2 &               & 200             & 8.0               & 1.0                 &   \\ \hline                     
\end{tabular}\\
Estimates of the distance from the central black hole to each
eclipsing cloud $r_{\rm cl}$ following $\S$5.2.  Uncertainties listed
above do not include uncertainty on the ionization parameter $\xi$.
Reasonable ranges for log($\xi$) are --1 to +1 for most objects, which
translates into factors of 2.5 larger/smaller (for Cen~A, values of
log($\xi$) ranging from --1 to 0 are plausible; for NGC~3227/2000--1,
Lamer et al.\ (2003) measured log($\xi$) = --0.3--0).  
$L_{2-10}$ is the 2--10~keV of the hard X-ray power-law component from
the long-term time-averaged \textit{RXTE} spectrum (Rivers
\etal\ 2013).  
$L_{\rm ion}$ is the 1--1000 Ryd ionizing luminosity; please see $\S5.2$ for details.
$L_{\rm Bol}$ is the bolometric luminosity; please see $\S5.1$ for details.
Luminosities are corrected for all (intrinsic and Galactic) absorption.  
$R_{\rm d}$ denotes the outer boundary of the ``dust sublimation zone''
(Nenkova et al.\ 2008b), i.e., dust residing at distances greater than
$R_{\rm d}$ likely does not sublimate, while distances smaller than
$\sim\frac{1}{2}-\frac{1}{3}R_{\rm d}$ are expected to be dust-free.
Average values (last row) were determined in log space.
References for $\Msun$ are: D10 = Denney \etal\ (2010), Pa04 = Papadakis (2004), Pe04 = Peterson
\etal\ (2004), R11 = References in Rothschild \etal\ (2011), VP06 =
Vestergaard \& Peterson (2006), W09 = Wang \etal\ (2009),
and WZ07 = Wang \& Zhang (2007).   \\
$^{\dagger}$Estimates of $r$ may be lower if peak $N_{\rm H}$ was higher than that measured                          
because either the beginning or end of the eclipse was not covered by monitoring data. \\
$^{\ddagger}$For MR~2251--178, we have only limits to the duration, but no reliable ``best estimate.''\\
$^{*}$Upper limits on $N_{\rm H}$ were used to derive lower limits on $r_{\rm cl}$. \\                          
$^{**}$Assumed ionization following Turner et al.\ (2008); see Appendix A.
\end{minipage} 
\end{table*}

\subsection{A possible double-eclipse event in NGC~3783}   
\label{sec:3783double}  
The $N_{\rm H}$($t$) and $HR1$ profiles for the NGC~3783/2008.3 event
are highly intriguing, as the two spikes suggest two absorption events
separated by only 11~d.  The fact that, assuming that $\Gamma$
remains constant during this time, $N_{\rm H}$($t$) remains near
$4\times 10^{22}$~cm$^{-2}$ in between the spikes is also intriguing.
No such complex $N_{\rm H}$($t$) profile has been observed for any
other AGN to date.

\begin{figure}
\includegraphics[angle=-90,width=0.48\textwidth]{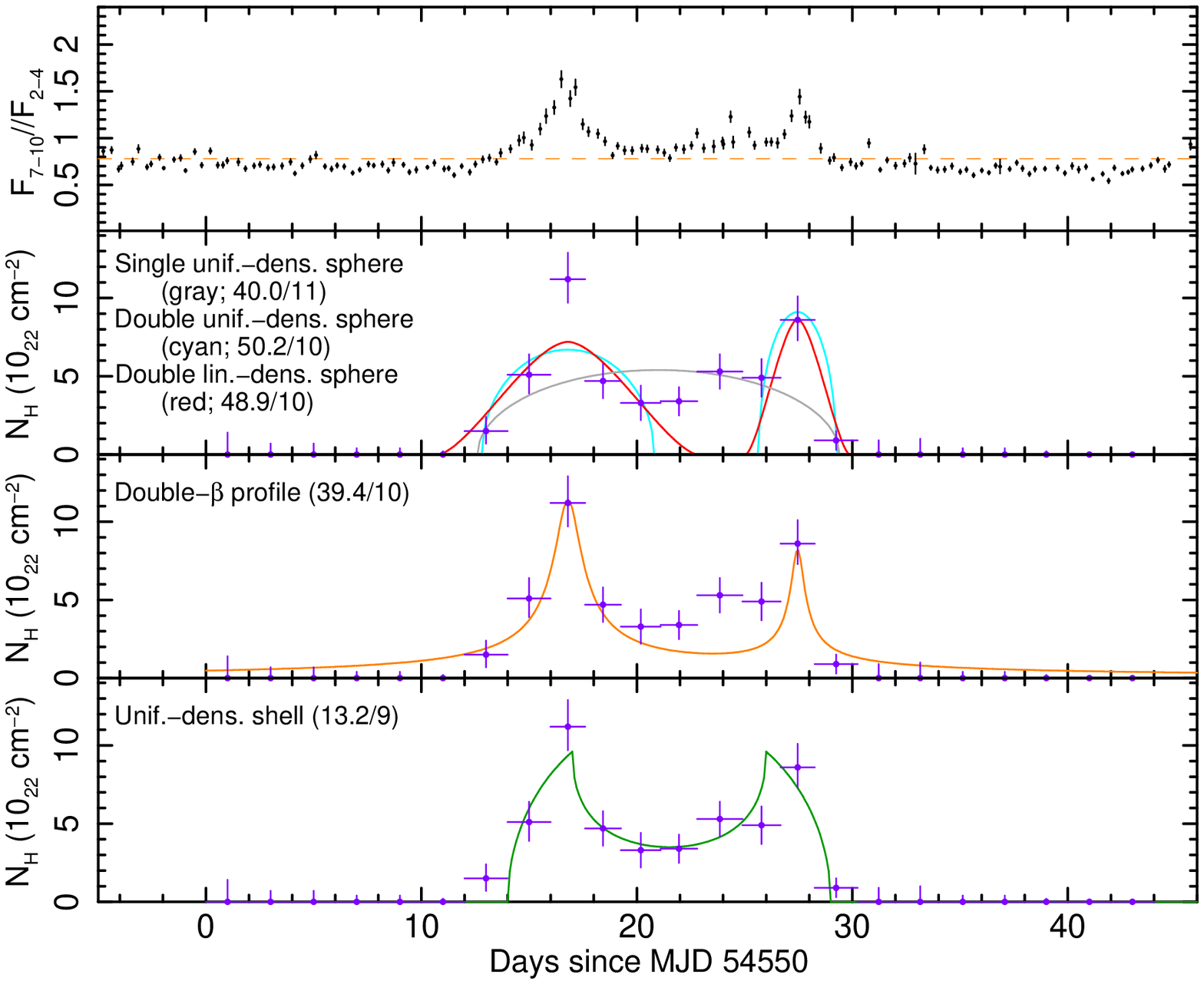}      
\caption{$HR1$ and $N_{\rm H}$($t$) light curves for the
  NGC~3783/2008.3 event, with its suspected ``dual-clump'' profile.
  The solid lines indicate the best-fitting models to the $N_{\rm H}$($t$)
  profile; we fit only bins \#5--14 (those bins with non-zero values
  of $N_{\rm H}$ plus two bins with upper limits before and after;
  see Table~\ref{tab:3783TRtable} for bin definition).
  We test a single uniform-density sphere (gray), a dual
  uniform-density sphere model (cyan), a dual linear-density-sphere
  model, a dual $\beta$-profile (orange), and a uniform-density shell
  (green).  Values of $\chi^2/dof$ are in parentheses.}
\label{fig:3783doubleprofile}
\end{figure}

As illustrated in Fig.~\ref{fig:3783doubleprofile}, we model the
$N_{\rm H}$($t$) profile in several ways, fitting only bins \#5--14
(see Table~\ref{tab:3783TRtable} for bin definition; refers to 
those bins with non-zero values of $N_{\rm H}$ plus two bins with
upper limits before and after).  We
first test a single uniform-density sphere (gray line), which yields
$\chi^2/dof$ = 40.1/11=3.65 and underestimates the spikes in $N_{\rm H}$ in bins
\#9 and \#15.

We also fit the profile assuming two independent eclipsing events.  We
test a profile from two uniform-density spheres (cyan line), and one from 
two linear-density spheres profile (red line). For the latter, the
density profile is $n$($r$) = $n_{\rm max}$(1--$r/R$), where $n_{\rm
  max}$ is the number density at the center and $R$ is the outer
radius of the cloud along the transverse axis (red line).  Each
cloud's center point in time space is held frozen at the midpoints of
bins \#9 or 15.  The best-fitting models yield $\chi^2/dof$ = 50.2/10=5.02 and
48.9/10=4.89, respectively, and did not correctly fit bin \#9 nor the
$\sim$2--4 bins between the spikes.  We also fit each spike with a
phenomenological $\beta$-profile of the form $N_{\rm H}$($t$) = $N_{\rm
  H, max} \sqrt{1 - r/R_{\rm c}}$ (panel c), where $N_{\rm H, max}$ is
the column density along a line of sight through the cloud's center
and $R_{\rm c}$ is the core radius (Dapp \& Basu 2009).  The best fit
yields $\chi^2/dof$ = 39.5/10=3.95 but still underestimates the middle part of
the profile.


Furthermore, the probability that two separate eclipse events could
independently occur so close together in time is low: NGC~3783 was
monitored for 3.27~yr, had no observations for 1.88~yr, and then was
monitored for an additional 7.84~yr.  We perform Monte Carlo
simulations wherein we assume a 1.0-d grid and we randomly place
three eclipses in the 3.27 or 7.84-yr campaigns, and empirically
estimate the probability that any two of their peaks can occur $<$15 d
apart: only 0.24 per cent.  Combined with the fact that $N_{\rm H}$($t$)
does not return to zero between the two spikes, the likelihood that
the profile is comprised of two independent eclipse events seems low.
The addition of a \textit{third} independent cloud with $N_{\rm H}
\sim 4 \times 10^{22}$~cm$^{-2}$ transiting the line of sight roughly
halfway between the maximum transits of the other two would of course
yield a good fit to the observed $N_{\rm H}$($t$) profile, but this is
statistically highly unlikely ($\sim5\times 10^{-6}$).


Finally, we model a uniform-density shell (green line in
Fig.~\ref{fig:3783doubleprofile}; subtracting a smaller
uniform-density sphere from another) to successfully model fit the
middle part and the two spikes.  The best fit has $\chi^2/dof$ =
13.2/9=1.47, with the outer and inner shell boundaries taking
$7.5^{+0.5}_{-1.0}$ and $4.5\pm0.5$~d, respectively, to transit the
line of sight.  However, we stress that this fit is purely
phenomenological; it is not clear how such a structure could be
created or survive for long periods in an AGN environment.


We speculate that the observed profile could be caused by two dense
clumps with some sort of connecting structure (a dumbbell shape), with
parts aligned along the direction of travel across the line of sight.
We discuss a possible physical origin for this behavior in $\S$\ref{sec:profiles}.

\subsection{Physical properties of obscuring clouds}     
\label{sec:diams}

%

We estimate diameters, number densities and masses for each eclipsing
cloud using the equations in $\S$\ref{sec:locations} and listed them
in Table~\ref{tab:clumpsizemass}. The average diameter across the
sample of secure events is $\sim0.25$ light-days, with an average
number density of $\sim3\times10^8$~cm$^{-3}$. Nenkova \etal\ (2008b)
estimate theoretically that a total of $\sim10^4 - 10^5$ clouds
comprises the dusty torus. The mass estimates for individual clouds in
Table~\ref{tab:clumpsizemass} span $\sim10^{-8}-10^{-2} \Msun$ with a
mean value in log space of $10^{-5.2} \Msun$, 
thus suggesting a total mass for the clumpy component of the torus
(i.e., excluding any thin accretion disk or intercloud diffuse medium)
of $\sim 10^{-4} - 10^{3} \Msun$.  
If the torus clouds are produced in a disk wind, then this estimate
indicates the maximum mass associated with the outflow.  Since the
value of $\sim10^4 - 10^5$ clouds quoted above likely refers only to
IR-emitting clouds located outside $R_{\rm d}$, this mass refers to
clouds lying outside $r_{\rm d}$ only, and also ignores contributions
from non-clumpy material such as any intercloud medium (Stalevski et
al.\ 2012).

\begin{table*}
\begin{minipage}{145mm}
\caption{ Inferred physical properties of each cloud; inferred size of X-ray-emitting region }   
\label{tab:clumpsizemass}  
\begin{tabular}{llccccc}  \hline
Source    &              &  Diam.            & Diam.                          & Num.\ dens., $n$   & log(mass)             & $D_{\rm X-src}$   \\
name      & Event        & ($10^{14}$~cm)     & (light-days)                       & ($10^8$~cm$^{-3}$)  & ($\Msun$)         & ($R_{\rm g}$)       \\   \hline             
NGC~3783  & 2008.3       & $1.3^{+0.7}_{-0.5}$  & $0.048^{+0.029}_{-0.018}$         & $8.6^{+7.0}_{-3.9}$   & $-6.0^{+0.4}_{-0.5}$     &  $<47$      \\
Mkn~79    & 2003.5       & $2.4^{+2.6}_{-1.5}$  & $0.089^{+0.097}_{-0.057}$          & $6.0^{+15.9}_{-4.0}$  & $-5.4^{+0.5}_{-0.8}$    &  $<64$  \\ 
Mkn~79    & 2003.6       & $3.1^{+1.8}_{-1.2}$  & $0.11^{+0.07}_{-0.04}$            & $3.7^{+3.9}_{-1.9}$   & $-5.3\pm0.5$          & $<65$    \\
Mkn~79    & 2009.9       & $2.4^{+2.1}_{-1.3}$  & $0.088^{+0.078}_{-0.046}$            & $3.2^{+5.5}_{-2.0}$   & $-5.7^{+0.4}_{-0.5}$     & $<57$    \\
Mkn~509   & 2005.9       & $4.9^{+5.7}_{-3.3}$  & $0.18^{+0.21}_{-0.12}$            & $1.8^{+4.5}_{-1.1}$   & $-5.1^{+0.6}_{-0.9}$     & $<46$    \\
MR~2251--178 & 1996       & $0.25-91$          &  $0.009-3.4$                  & $0.06-30$           & $-7.7$ to $-2.7$     & $<310$    \\  
NGC~3227  & 2000--1       & $5.8^{+0.3}_{-1.0}$   &  $0.22^{+0.01}_{-0.04}$          & $3.9^{+1.4}_{-0.8}$    &  $-4.5^{+0.1}_{-0.2}$    & $<540$    \\
NGC~3227  & 2002.8        & $0.49^{+0.56}_{-0.31}$ & $0.018^{+0.020}_{-0.012}$         & $27^{+62}_{-17}$      &  $-6.9^{+0.6}_{-0.8}$    & $<92$    \\
Cen A     & $\sim$2003--4 & $120^{+60}_{-90}$     & $4.5^{+2.3}_{-3.4}$              & $0.07^{+0.23}_{-0.03}$ & $-2.3^{+0.3}_{-0.2}$    & $<2000$    \\
Cen A     & 2010--1       & $26.9^{+0.1}_{-10.1}$  & $1.00^{+0.01}_{-0.38}$           & $0.30^{+0.24}_{-0.04}$   & $-3.6^{+0.1}_{-0.5}$    & $<300$    \\
NGC 5506  & 2000.2        & $0.12^{+0.13}_{-0.08}$ & $4.3^{+5.8}_{-2.9}\times10^{-3}$ & $34^{+111}_{-23}$     & $-8.6^{+0.5}_{-0.9}$    & $<1.9$  \\
Mkn~348   & 1996--7       & $20^{+17}_{-10}$      & $0.74^{+0.62}_{-0.36}$           & $0.90^{+1.17}_{-0.49}$  & $-3.5\pm0.5$        &  $<1600$ \\ 
Fairall~9 & 2001.3        & $<4.5$              & $<0.17$                       & $\sim 5-27$       & $\sim-6.2$ to  $-4.7$  & $<12$\\
NGC~3783  & 2008.7        & $<2.8$              & $<0.10$                       & $\sim 1-5$        & $\sim-7.0$ to $-5.9$   & $<63$ \\
NGC~3783  & 2011.2       & $0.9^{+1.0}_{-0.6}$  & $0.035^{+0.035}_{-0.023}$         & $12^{+41}_{-8}$       & $-6.4^{+0.4}_{-0.8}$      & $<43$   \\
NGC~3516  & 2011.7        & $\sim11$            & $\sim0.4$                     & $\sim 0.4$        & $\sim-4.6$              & $<230$\\  \hline
\multicolumn{2}{c}{Average for 12 Secure Events} &   3.9  & 0.25                & 2.6               & --5.2                    &  $<120$  \\ \hline

\end{tabular}\\
Best-estimate values listed here correspond to best-fitting values of
$N_{\rm H}$ and values of duration listed in Table~\ref{tab:ecl1summ}
and an assumed ionization state of log($\xi$)=0. Uncertainties listed
here are based on the uncertainties on observed duration, $N_{\rm H}$,
and the ranges of $r_{\rm cl}$ and log($\xi$) listed in
Table~\ref{tab:ecl1dist}. $D_{\rm X-src}$ denotes inferred upper
limits on the size of the X-ray continuum source, assuming that the cloud
completely covers the X-ray continuum source. Average values were
determined in log space.
\end{minipage}
\end{table*}

Finally, assuming that each eclipsing cloud fully covers the X-ray
continuum source behind it, the upper limits on the diameter of the
cloud yield inferred upper limits on the size of the X-ray continuum
source, $D_{\rm X-src}$. These limits are listed in
Table~\ref{tab:clumpsizemass} in units of $R_{\rm g}$. Among the
secure events, the limits are as low as a few tens of $R_{\rm g}$ for
NGC~3783, Mkn~79, and Mkn~509, and even as low as $2R_{\rm g}$ in the
case of NGC~5506.

The X-ray continuum in Seyferts is generally thought to originate via
inverse Compton scattering of the soft UV photons from the accretion
disk by hot electrons in a compact corona, possibly located in the
innermost accretion disk or at the base of a jet (e.g., Haardt \etal\
1997; Markoff \etal\ 2005). X-ray microlensing analysis of distant
quasars typically yields upper limits on the size of the X-ray corona
of $\sim10R_{\rm g}$ (e.g., Chartas \etal\ 2012, Chen \etal\ 2012, and
references therein). Recent results using X-ray reverberation lags
imply sizes of a few to $\sim10$ $R_{\rm g}$ (Reis \& Miller 2013).
Geometrical limits from absorption variability in NGC~1365 typically
yield $D_{\rm X-src} \la 30-60 R_{\rm g}$ (Risaliti \etal\ 2007, 
2009b). Our estimates of $D_{\rm X-src}$ are
consistent with these studies.

\subsubsection{Exploring the range in observed density profiles}
\label{sec:profiles}

We speculated in $\S$\ref{sec:3783double} that the unusual density
profile of the NGC~3783/2008.3 event could be caused by the cloud
having a dumbbell shape. Even more speculatively, perhaps this
structure is a cloud that is in the process of getting torn in half.
As pointed out by Krolik \& Begelman (1988), the self-gravity of a
single cloud in the vicinity of a supermassive black hole is not
highly effective against tidal shearing, and clouds can get
significantly stretched out within an orbit.
Resistance to tidal shearing requires that the size of the cloud is
limited to $\la D_{\rm shear} = 10^{16} N_{\rm H,23} r_{\rm pc}^3
M_7^{-1}$~cm, where $N_{\rm H,23} = N_{\rm H}/10^{23} {\rm cm}$,
$r_{\rm pc}$ is the distance from the black hole to the cloud, and
$M_7 = M_{\rm BH}/(10^7 \Msun)$ (Elitzur \& Shlosman 2006). For
NGC~3783/2008.3, $D_{\rm shear} \sim 7\times10^{12}$~cm, a factor of
$\sim1/18$ times the inferred cloud diameter $D_{\rm cl}$, so the
possibility of tidal shearing seems plausible. In fact, for all secure
eclipse events, we obtain values of $D_{\rm cl}/D_{\rm shear}$
spanning 2.3--350, with an average value (in log-space) of 26,
suggesting that many of these clouds' sizes put them at risk of being
tidally disrupted or sheared.

The unusual density profile could also be consistent with models
wherein torus clouds may be in the form of clumpy winds, as opposed to
compact, discrete clouds, originating at the accretion disk.  These
winds may be magnetohydrodynamic (MHD) driven (e.g., Konigl \& Kartje 1994; Fukumura et
al.\ 2010), with local column densities that can be consistent with
values measured in this paper.  Other models feature IR
radiation-driven winds (Dorodnitsyn \&
Kallman 2012; Dorodnitsyn et al.\ 2012): 
UV/X-ray photons are reprocessed into IR thermal
emission in the torus, and IR radiation pressure vertically drives
dust grains. Moderately Compton-thick flows are possible at distances
of $\ga$1 pc from the black hole (if our assumption about
Keplerian motion is not valid, then distance estimates for our clouds
may be in error).  The models generally assume that clouds'
self-gravity is negligible.  Similarly, Czerny \& Hryniewicz (2011)
proposed a turbulent, dusty, disk wind as the origin for the
low-ionization region of the BLR: clouds rise from the disk, intense
external radiation heats the gas and dust sublimates.  In all these
models, the torus is highly dynamic and can feature cloud motions
perpendicular to the disk (including failed winds), and thus not
strictly Keplerian.  In any case, the $N_{\rm H}$($t$) profile of the
NGC~3783/2008.3 event can thus also be explained via a non-homogeneous
mass outflow, featuring two overdensities that crossed the line of
sight 11~d apart.

However, the profile of NGC~3783/2008.3 contrasts with three of the
other eclipse events that have $N_{\rm H}$($t$) profiles that are also
well resolved in time but are symmetric and centrally peaked:
NGC~3227/2000--1 and Cen~A/2010--1, observed with \textit{RXTE}, and
SWIFT J2127.4+5654 (Sanfrutos \etal\ 2013). Such profiles pose a
challenge to MHD/IR-driven wind models. For example, the IR-driven
winds of Dorodnitsyn et al.\ (2012) suggest that $\Delta$$N_{\rm
  H}$/$N_{\rm H}$ is typically $\sim$ a few (less than the factors we
observe). In addition, such winds are inferred to originate $\sim1-3$
pc from the black hole, farther than the distances inferred for our
observed eclipses.

In summary, the self-gravity of these clouds likely dominates over any
tidal shearing/internal turbulence, presenting a challenge to the
expectation that clouds are easily tidally sheared. In these cases,
one or more of the assumptions that go into the calculations of
$D_{\rm cl}$ or $D_{\rm shear}$ may be wrong.  Alternately, some other
physical process may prevent clouds from shearing, e.g., confinement
of these clouds by external forces (e.g., gas pressure from the
ambient medium, external magnetic fields; Krolik \& Begelman 1988) may
be important.


\subsection{Implications for clumpy-absorber models}  

\C\ torus models have found a lot of observational support,
particularly from recent IR observations. For example, smooth-density
torus models always predict that the silicate features at 9.7 and
18\mic\ will be in emission for pole-on viewing and in absorption for
edge-on viewing. However, samples of mid-IR spectra of type I Seyferts can exhibit
silicate features spanning a range of emission \textit{and}
absorption (see Hao et al.\ 2007 for a sample obtained with
\textit{Spitzer}); type IIs generally show only weak absorption but a
few can show emission (Sturm et al.\ 2005; Hao et al.\ 2007). Nenkova
et al.\ (2008b) and Nikutta et al.\ (2009) demonstrate that the
clumpy-torus models are consistent with these observations and can
explain mismatches between the optical classification and the expected
behavior of the silicate features. Clumpy-torus models predict nearly
isotropic mid-IR continuum emission and anisotropic obscuration, as
generally observed (e.g., Lutz et al.\ 2004; Horst et
al.\ 2006). Finally, clumpy tori have been predicted to host both hot
and cooler dust in close proximity to each other (Krolik \& Begelman
1988), in contrast to smooth-density tori. Recent IR interferometric
observations indeed do affirm such a co-existence of hot (800~K) and
cooler ($\sim 200-300$~K) dust components in nearby AGN (Jaffe et
al.\ 2004; Poncelet et al.\ 2006; Tristram et al.\ 2007; Raban et
al.\ 2009).

In this subsection, we derive the first \textit{X-ray} constraints on the
\C\ model parameters, thus obtaining constraints in a manner
independent from IR SED fitting.  We caution, however, that there may not be an exact
correspondence between constraints derived from the two methods, since
IR emission arises only from dusty clouds outside the DSZ while our
X-ray clouds can be dust free.  The clumpy-torus models are generally
defined using the following free parameters: \tv\ is the $V$-band
optical depth of single clouds; usually, all clouds are assumed to
have identical values of \tv. The radial extent of the dusty torus is
characterized by $Y$, the ratio of the outer radial extent of clouds
to the dust sublimation radius $R_{\rm d}$. $\Ni(\theta)$ is the
average number of clouds along a radial line of sight that is an angle
$\theta$ from the equatorial plane. Assuming a Gaussian distribution,
$\Ni(\theta,\sigma,\No) = \No \exp(-(\theta/\sigma)^2)$, where \No\ is
the average number of clouds along a radial line of sight in the
equatorial plane (typically 5--15), and $\sigma$ parametrizes the
angular width of the cloud distribution. The index $q$ describes the
radial dependence of the average number of clouds per unit length,
$\Nc(r,q) \propto r^{-q}$.  The inclination angle of the system is $i
\equiv 90\degr-\theta$, defined such that $i=0$ ($\theta=90\degr$)
denotes a system with its equatorial plane in the plane of the sky.

The predicted shape of the IR SED is sensitive to all of these
parameters. Ramos Almeida et al.\ (2009, 2011; hereafter RA09 and
RA11) have successfully fit the mid-IR spectra of $\la20$ AGN
with \C\ torus models and the \textsc{BayesClumpy} tool (Asensio Ramos \& Ramos
Almeida 2009). There are only three sources in our eclipse sample that
overlap with the samples of RA09 and RA11 (NGC~3227, Cen~A, and
NGC~5506). In the following we attempt to focus on comparing our
derived parameter constraints for the type I/II classes as opposed to
for individual objects.


$\bullet$ \tv: we can translate our inferred values of $N_{\rm H}$ for
individual clouds into values of $\tv$, which we can then compare to
those values typically used in the $\C$ model fitting.  From the
``secure'' clumps we observe, $\Delta$$N_{\rm H} = 4-26 \times
10^{22}~{\rm cm}^{-2}$, with values spanning very similar ranges for
both type Is/IIs. Using the Galactic dust/gas conversion ratio,
$N_{\rm H} = A_{\rm V} {\rm (mag)}  \times 1.8 \times 10^{21} {\rm
  cm}^{-2}$ (Predehl \& Schmidt 1995), these values translate to
$A_{\rm V} = 22-144$ mag, or $\tv = 20-132$.  The \C\ models use one
value of $\tau_{\rm V}$ per object/spectrum to represent the weighted
average value across all clouds in that object.  In contrast, we probe
1--3 individual clouds per object.  Subject to small-number
statistics, we can therefore test both the inter-object scatter of
\tv\ values, as well as the scatter from cloud to cloud in one given
source. From general theoretical considerations, Nenkova et
al.\ (2008b) suggest that \tv\ be typically 30--100. RA11, who
use the entire \C\ database of models in their IR SED fitting,
estimate values of 40--140 for type Is and 5--95 for type
IIs. Alonso-Herrero et al.\ (2011), with the same approach but
different sources, find \tv\ = 105--146 (type Is) and 49--130 (type
IIs). These IR-derived ranges are fully consistent with our X-ray
based values. In contrast to the above studies, we find no
evidence for different cloud optical depths between type I and II sources.

Three of our sources are also present in the IR samples of RA11 and
Alonso-Herrero et al.\ (2011). In two of these (Cen~A and NGC~3227)
our X-ray constraints agree very well with the IR ones. However, our
value for NGC~5506 is smaller than that in RA11 by a factor of $\sim2$
and smaller than that of Alonso-Herrero et al.\ (2011) by $\sim5$. For
our sources with multiple eclipsing events, the measured scatter in
$N_{\rm H}$ or \tv\ is very low, always $<$2. It seems then that assuming a single
average value of \tv\ for all clouds in a given \C\ model is
justified.


$\bullet$ $Y$: the IR SED modeling and interferometry indicate the
presence of dusty clouds out to $Y \sim 20-25$ (e.g., RA11), although
emission from beyond this radius may be difficult to separate from the
surrounding host galaxy emission, especially for emission longer than
$\sim 10\mic$. Our inferred locations of the X-ray-absorbing clouds
cover out to $Y \sim 20$, in agreement with the IR observations. It
should be noted though that clouds at much larger distances from the
black hole might exist, but would only be detectable with even longer
sustained X-ray monitoring campaigns of a larger number of targets.


$\bullet$ $\sigma$, $i$ ($\theta$) and \No: we observe eclipse events
in type I objects, and derive non-zero values of $P_{\rm ecl}$. If it
is the case that all type Is are oriented relatively face-on, then
clumpy-torus models featuring sharp-edged tori are strongly
disfavored, while models featuring a gradual decline of cloud number when
moving away from the equatorial plane are favored (see left- and right-hand
panels of Nenkova et al.\ 2008b, Fig.~1, respectively). IR SED fitting
likewise disfavors sharp-edged tori, which would produce a dichotomy
in the SED shapes that is not observed (Nenkova et al.\ 2008b).

We can derive our constraints on $\sigma$ and \No\ assuming that an
observed baseline level of absorption $N_{\rm H,base}$ corresponds to
an average of \Ni\ clouds along the line of sight ($\Ni = 0$ in the
case of sources normally lacking X-ray obscuration). We also assume
that an eclipse denotes $\Ni \rightarrow \Ni+1$. In the case of Cen~A,
to repeat the exercise from Rothschild et al.\ (2011) and Rivers et
al.\ (2011b), $\Delta$$N_{\rm H}$/$N_{\rm H,base} \sim 3$, suggesting
$\Ni \sim 2$. From IR interferometry, Burtscher et al.\ (2010) model a
best-fitting value for the torus inclination of $i=63\degr$, or $\theta =
27\degr$.  The median value for $\sigma$ modeled by RA11 was 
$\sigma = 20\degr$; in order to agree with this value of $\sigma$,
one needs $\No =12$. 

In both cases of NGC~5506 and Mkn~348, $\Delta$$N_{\rm H}$/$N_{\rm
  H,base} \sim 2$, suggesting $\Ni \sim 1$. Due to the absence of
external information on $i$, we can only derive constraints on the
ratio $\theta/\sigma$. If we assume that $\No = 10$, $\theta/\sigma$ =
1.52. That is, if such a system is inclined relatively edge-on
($\theta\rightarrow 0$), as is commonly suspected for type~IIs, then
$\sigma$ must be rather small, indicating a tightly flattened
distribution. Otherwise, the inclination of the system would have to
deviate more from an equatorial view. This exercise assumes that all
obscuring material along the line of sight is in the form of clouds.
If the baseline absorption $N_{\rm H}$ is not due to clouds (see
$\S$5.6), then $\Ni \sim 0$, suggesting that the inclination of the
system may not be close to edge-on, and/or that $\sigma$ is relatively
small in these systems. A similar condition, $\Ni \sim 0$, applies to
sources normally lacking X-ray absorption, e.g., the type~I objects in
our sample.

\begin{figure*}
  \center
  \includegraphics[width=0.8\textwidth]{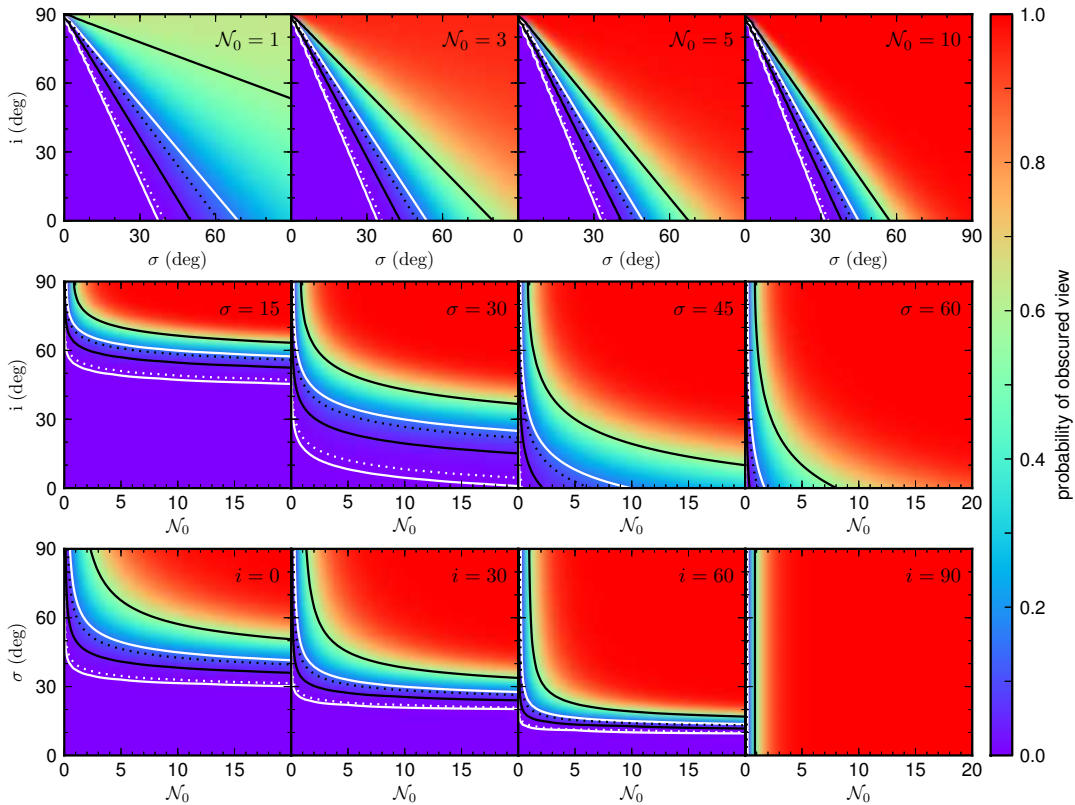} 
  \caption{\C\ parameter constraints based on the theoretical probability \pobsc\
    to see the AGN obscured by torus clouds. The obscuration
    probability is rendered as a color scale in each panel (see color
    bar). The top panels show \pobsc\ as a function of torus angular
    width $\sigma$ and viewing angle $i$ for four different values of
    $\No$. The other two combinations of the three parameters that
    enter \pobsc\ are depicted in the middle and bottom
    rows. Overplotted as contour lines in each panel are the inferred ranges of
    $\overline{P_{\rm ecl}}$ that are compatible with our
    X-ray monitoring data (see Table~\ref{tab:finalprobs}). White and black solid lines 
    outline the compatible ranges for type I and II
    objects, respectively. The dotted lines indicate best estimates in both
    cases.}
  \label{fig:pobscured}
\end{figure*}

For a \C\ torus the probability of the AGN being obscured -- a single
dust cloud along the line of sight is sufficient -- is given by
\begin{equation}
  \label{eq:pobsc}
  \pobsc(\No, \sig, i) = 1 - \exp\left\{-\No \cdot \exp\left\{ -\left(\frac{90-i}{\sigma}\right)^{\!\!\!2}\right\}\right\}
\end{equation}
(e.g., Nenkova et al.\ 2008b; Nikutta et al.\ 2009).  The argument to
the outer exponential is the (negated) number of clouds along the line
of sight \Ni, thus $\pobsc = 1 - \exp\{-\Ni\}$.  We can attempt to
constrain the \C\ parameters \No, \sig, and $i$ by identifying ranges
of \pobsc\ consistent with our X-ray monitoring data. In
Fig.~\ref{fig:pobscured} we plot \pobsc\ as a function of 2 of the 3
parameters, with each row of panels sampling the third parameter. The
scale is linear (see color bar).  In each panel we overplot as contour
lines $\overline{P_{\rm ecl}}$, the inferred values of the probability
to catch an AGN while being obscured by a torus cloud, as computed in
$\S$\ref{sec:sect43} (Table~\ref{tab:finalprobs}).  The solid lines
correspond to the minimum and maximum values of 
$\overline{P_{\rm ecl}}$, and dotted lines 
are best-estimate values. The white lines
thus outline, given the data and the selection function, the range of
permitted values of \pobsc\ for type I objects, and the black lines
trace the allowed range of probabilities for type II objects. The
dotted lines are the best-estimate values of \pobsc. 

%
%

These inferred minimal and maximal probabilities outline the parameter
ranges permissible by our extensive monitoring programmes. Given the
observed (and not observed!) eclipsing events, any torus parameters
which lie outside the enclosed ranges are unlikely to be found in real
sources. We remind the reader, however, that the probability values
$\overline{P_{\rm ecl}}$ we derived in $\S$4 refer only to
lowly-ionized/neutral, full-covering clouds with $N_{\rm H} \ga
10^{22}$ up to $\sim10^{25}$~cm$^{-2}$.  When eclipses are due to clouds
with higher ionizations and/or lower column densities, and
partial-covering events, the values will be higher.  Since three
parameters (\No, \sig, $i$) affect \pobsc\ with some mutual dependence on
each other, it is not straightforward to give one-dimensional ranges
of excluded parameter values. However, for physically reasonable
values some constraints can be given ad hoc. For instance, if \No\ is
between 5 and 15 (see, e.g., Nenkova et al.\ 2008b), it is clear from
Fig.~\ref{fig:pobscured} (bottom row) that \sig\ must be smaller than
$\approx 45\deg$ (even for type II sources); only at the most pole-on
viewing angles, \sig\ can be slightly higher. Similarly, if, for
instance, \sig\ is secured by external information, for the reasonable
range of \No\ the permitted viewing angles $i$ are very
well-constrained (see middle row of Fig.~\ref{fig:pobscured}). Not
surprisingly, the permissible parameter values in type~II sources are
on average higher than in type~I sources, because
\pobsc\ \emph{increases} when any of the three (\No, \sig, or $i$)
grows. There is some overlap, however, and is most likely indicative
of "intermediate" types of AGN.

Overall, Fig.~\ref{fig:pobscured} shows that whenever one of the three
parameters can be secured by other means, the other two allow only
relatively narrow ranges. In practice, external information is
sometimes available, for instance on the orientation of the AGN system
from radio jets or from narrow-line ionization cones, and can be used
to infer the two other parameters. We are not attempting a full
Bayesian parameter inference on the torus parameters here.
Suffice it to say here that our results as
depicted in Fig.~\ref{fig:pobscured} translate into information on the
\emph{priors} $P(\bm\theta)$ of the Bayesian inference problem
$P(\bm\theta|D) \propto P(\bm\theta) \cdot P(D|\bm\theta)$.
Here $\bm\theta = (\No, \sig, i, \ldots)$ is the vector of model
parameters to be estimated, $P(D|\bm\theta)$ is the \emph{likelihood}
that a set of parameter values generates a model compatible with the
data, and $P(\bm\theta|D)$ is the full parameter \emph{posterior}
(also called inverse probability). Faced with perfect ignorance of the
distribution of parameter values before modeling, uniform priors are
often assumed. If there is independent knowledge available on some
parameters it is imperative to incorporate this information into the
Bayesian priors. For example, if the orientation of an AGN was known
to be pole-on, one could assume a narrow Gaussian shape on the prior
of $i$. We presume that then the Bayesian inference process would yield
parameter intervals for \No\ and \sig\ commensurate with the 
lower-left panel of Fig.~\ref{fig:pobscured}. 

The \pobsc\ function transitions quite steeply from very low to very
high probabilities. Our derived probability ranges (see Table~\ref{tab:finalprobs})
typically bracket most of the
transition region, and also exclude a large fraction of the parameter
space that corresponds to extremely low or extremely high \pobsc. We
presume that a source such as NGC~1365, aligned so that \Ni\
(average number of clouds along the line of sight) is $\approx 1$, would
be just at the high end, or slightly above, of the type II region
encompassed by the black lines in Fig.~\ref{fig:pobscured}.

\subsection{Can clumpy-torus models be applied to type II objects?}  

As detailed in Appendix~A and summarized in $\S$3, \textit{RXTE}
provided sustained long-term monitoring for durations $\geq$0.6~yr for
nine type II objects. In eight of them, $HR1$ is consistent with
remaining constant for durations $\geq$0.6~yr (we include the 2011
monitoring of Mkn~348 here).  
This constancy comes despite variations of typical
factors $\sim2-4$ in 2--10~keV continuum flux.
For example, excluding the
$\la$1-d event in NGC~5506, there are no strong variations in
$HR1$ (sustained trends above 2$\sigma$ and/or 
$\sim1.5 \times \langle HR1 \rangle$) lasting longer than 1~d for a period of 
8.39~yr.  Each of these sources routinely show evidence for the presence of
absorbing gas with $N_{\rm H} \sim 10^{21-23}$~cm$^{-2}$ in their
X-ray spectra, suggesting a baseline level of absorption 
$N_{\rm H,base}$ that is consistent with being constant.  Using $HR1$ and
its distribution, and assuming for simplicity that $\Gamma$ remains
constant for each object throughout the monitoring, we estimate the
maximum value of $\Delta$$N_{\rm H}$ possible without significantly
varying $HR1$. Those values are listed in Table~\ref{tab:boringsy2s},
and for the brightest and best-monitored objects is typically
$\sim1-4\times10^{22}$~cm$^{-2}$.

REN02 claim typically 20--80 per cent variations in $N_{\rm H}$ in most of
their sample of type II objects, using individual observations from
multiple X-ray missions spanning $\sim$25~yr.  In this paper, we
rely on measurements from one instrument only, and so we do not
face systematic uncertainties associated with cross-calibration
between various X-ray missions or with fitting spectra data over
different bandpasses.  For our eight type IIs, we can rule out
variations above $\sim20-30$ per cent for IRAS~04575--7537 and Mkn~348 (2011
monitoring only), and variations above $\sim50-60$ per cent for NGC~1052,
NGC~4258, and NGC~5506.


In the context of the clumpy-torus paradigm, having a non-zero value
$N_{\rm H,base}$ requires a non-zero number of clouds along the line
of sight at all times, with the observed value of $N_{\rm H,base}$
being the sum of the values of $N_{\rm H}$ for the individual clouds.
Assuming that all clouds have the same physical characteristics ($n$, $N_{\rm H}$, 
etc.), the measured long-term constancy in $N_{\rm H,base}$ would
require that the number of clouds along the line of sight $\Ni$ 
either remain constant for many years or is sufficiently large such that the
addition/subtraction of one cloud intersecting or transiting along the
line of sight does not change the total measured value of $N_{\rm H}$
by a detectable amount.  However, with most clouds modeled
with $\C$ having $\tau_{\rm V} \sim 30-100$, which corresponds to
$N_{\rm H} \sim 6-20 \times 10^{22}$~cm$^{-2}$ assuming the Galactic
dust/gas ratio, values of $N_{\rm H,base}$ of $10^{22}-10^{23}$
cm$^{-2}$ imply that $\Ni$ cannot be more than a few.

We can explore if it is feasible to have the observed duration for
each source correspond to a period where $\Ni$ remains constant.  That
is, we assume that all cloud contributing to the total observed value
of $N_{\rm H}$ remain in the line of sight for the duration, and we
assume for simplicity that every cloud has identical $N_{\rm H}$.  We
also assume that every such cloud is $D_{\rm cl}$ = 1 light-day
in diameter and on Keplerian orbits.  Furthermore, we assume that the
fastest velocity (closest to the black hole) cloud has just started
to transit the line of sight when the long-term monitoring began, and
that the cloud begins to leave the line of sight just when the
monitoring ends.  In the case of NGC~5506 ($M_{\rm BH}=10^{7.94}
\Msun$), an 8.39-year-long transit by such a cloud would require it to
be located at least $r_{\rm cl} = G M_{\rm BH} (t_{\rm d}/D_{\rm
  cl})^2$ = 39 pc from the black hole.  For NGC~1052 ($M_{\rm
  BH}=10^{8.19} \Msun$; Woo \& Urry 2002), a 4.56-year transit would
place it $\geq20$ pc from the black hole.  For NGC~4258, ($t_{\rm
  d}$=6.81~yr, 2005 March -- 2011 December; $M_{\rm BH} = 3.9\times 10^7
\Msun$, Miyoshi et al.\ 1995; Herrnstein et al.\ 1999), $r_{\rm cl}
\geq$ 11 pc.\footnote{Resulting values of $r_{\rm cl }$ for NGC~2992
  ($t_{\rm d}$=0.90~yr; $M_{\rm BH} = 5.2\times10^7~\Msun$, Woo \&
  Urry 2002), Mkn~348 ($t_{\rm d}$=0.98~yr, 2011 only), and NGC~7314
  ($t_{\rm d}$=1.56~yr, 1999--2000; $M_{\rm BH} = 1.4 \times 10^6$~$\Msun$,
  Vasudevan et al.\ 2010) are 0.26, 0.09, and 0.02 pc, respectively.}
A cloud diameter of 0.1 light-days, also a feasible value given the range
listed in Table~\ref{tab:clumpsizemass}, would yield values of $r_{\rm
  cl}$ an order of magnitude higher.

The clumpy-torus models typically assume that clouds cannot exist in
large numbers out to several tens of $r_{\rm d}$, or very roughly
10, 5 and 0.5 pc for NGC~5506, NGC~1052, and NGC~4258,
respectively\footnote{Bolometric luminosities for NGC~5506, NGC~1052
and NGC~4258 from Vasudevan et al.\ (2010), Woo \& Urry (2002), and
Lasota et al.\ (1996), respectively.}.  In these objects, it is thus
not very likely that the same clouds used in the $\C$ models can
also be responsible for the non-variable baseline level of absorption.

Furthermore, Lamer et al.\ (2003) and Rivers et al.\ (2011b)
demonstrated that the transiting clouds responsible for the
NGC~3227/2000--1 and Cen~A/2010--1 events were non-uniform, with a
number density increasing towards the center.  If it were the case
that these two clouds were representative of all AGN clouds in terms
of their density profiles, then any single cloud moving across the
line of sight should produce a secular variation in observed $N_{\rm
  H}$ and $HR1$. However, we find no evidence for any strong secular
trend in any type II object, to the limits indicated by
Table~\ref{tab:boringsy2s}.  In addition, a transiting spherical cloud
with a constant \textit{column} density profile would require $n$ to
decrease as one goes from the cloud edge to the center.

We conclude that it is difficult for clumpy-torus models to
satisfactorily explain the constancy of $N_{\rm H,base}$, especially
in the cases of the few longest-monitored and X-ray brightest type II
objects observed with \textit{RXTE}.

It is thus more likely that $N_{\rm H,base}$ arises in a non-clumpy,
highly-homogeneous (to within the above $\Delta$$N_{\rm H}$ limits)
medium; such a medium could be located at any radial location along
the line of sight, subject to restrictions from the medium's
ionization parameter.  That is, the cloud responsible for the
$\la1$-d eclipse in NGC~5506 and the material responsible for
$N_{\rm H,base} \sim 2\times10^{22}$~cm$^{-2}$, for instance, are
likely physically separate entities.  One possibility to explain
$N_{\rm H,base}$ is a smooth, relatively low-density, dusty intercloud
medium in which the relatively higher density clouds are embedded
(Stalevski et al.\ 2012).  To be consistent with the approximate range
of baseline values of $N_{\rm H}$ typically measured in type IIs'
X-ray spectra ($0.3 - 30 \times 10^{22}$~cm$^{-2}$), the intercloud
medium could have values of $\tau_{\rm V} \sim 1.6 - 160$ or
$\tau_{10\micron} \sim 0.07 - 7$.\footnote{We employ the optical
  properties of the composite silicate/graphite grains used in the
  \C\ models, with $\tau_{\rm V} = 23.6\tau_{10\micron}$.}  However,
we do not see evidence for such a medium in type~I AGN.

Another possibility we explore is X-ray-absorbing gas distant from the
black hole and associated with the host galaxy, as has been suggested
e.g., by Bianchi \etal\ (2009) for NGC~7582. For example, dust
lanes/patches associated with the host galaxy are considered a
candidate for the obscuration of BLR lines in some systems (Malkan et
al.\ 1998).

With the exception of Cen~A (see $\S$\ref{sec:cenasmalldip}), we rely
on Shao et al.\ (2007) and Driver et al.\ (2007), who derive the
inclination dependence of dust extinction for disk-dominated galaxies.
In the optical bands, the extinction is typically $\la1-2$ mag,
derived over a range of inclination angles from face-on to nearly
edge-on ($i \sim 80-85 \degr$), corresponding to a maximum $N_{\rm H}
\sim 4 \times 10^{21}$~cm$^{-2}$ for a nearly edge-on disk. If the
obscuring dust is distributed uniformly throughout the disk of the
host galaxy, then such a component may explain relatively low observed
values of $N_{\rm H,base}$ in, for instance, NGC~2992 ($N_{\rm H,base}
\sim 4 \times 10^{21}$~cm$^{-2}$) with the disk component of its host
galaxy oriented close to edge-on (Jarrett et al.\ 2003). However, for
objects with host disk inclinations far from edge-on (e.g., NGC~4258
and Mkn~348: Hunt et al.\ 1999; Jarrett et al.\ 2003) and/or for
objects with $N_{\rm H,base} \sim$ a few $\times 10^{22}$~cm$^{-2}$
and higher, it is difficult for such a distribution of dust to be
associated with $N_{\rm H,base}$. Of course, we cannot rule out the
possibility of having more concentrated regions of gas lying along the
line of sight not associated with the large-scale disk structure (such
as a Giant Molecular Cloud). Dust-\textit{free} X-ray-absorbing gas in
any of these configurations is also a candidate.


A final possible explanation for constancy in $N_{\rm H}$ can be, as
mentioned above, the presence of a large number of low-density clouds,
with $\Ni$ remaining nearly constant.  However, this would require
each cloud to have $N_{\rm H} \ll 10^{22}$~cm$^{-2}$.  Such values
contrast with those used by Nenkova et al.\ (2008b), RA09, RA11, and
the values measured in this paper, but such clouds are not physically
implausible.  Contributions from such clouds, if they are dusty, to
the total IR emission are negligible. However, the sum of many clouds
along the line of sight that each have $N_{\rm H} \ll
10^{22}$~cm$^{-2}$, be they dusty or dust-free, could yield
appreciable X-ray absorption.


\subsubsection{A small dip in $N_{\rm H,base}$ in Cen~A} 
\label{sec:cenasmalldip}

Up to this point, our discussion of the baseline level of X-ray
obscuration in type IIs has centered on constant (to within our sensitivity
limits) values of $N_{\rm H, base}$. However, $N_{\rm H, base}$ for
Cen~A warrants more attention in light of the small dip in early 2010.

The famous dust lane crossing the host elliptical of Cen~A supplies 3--6
mag.\ of optical extinction (Ebneter \& Balick 1983), corresponding
to $\sim5-10 \times 10^{21}$~cm$^{-2}$, far short of observed values
of $N_{\rm H,base} \sim 1-2 \times 10^{23}$~cm$^{-2}$.  It is
generally accepted that a more compact and higher density distribution
is responsible for observed values of $N_{\rm H}$ in X-ray spectra of
Cen~A.  

One possibility is that we have witnessed one cloud leaving the line
of sight followed by another cloud entering the line of sight two
months later. Best-fitting values of $N_{\rm H}$($t$) indicate a drop from
$21.7\pm0.9$ to $18.6^{+0.9}_{-0.8}$ and returning to $N_{\rm H}$($t$)
$ = 21.9 \pm 0.7 \times 10^{22}$~cm$^{-2}$. These values suggest a
scenario in which $\Ni$ dropped temporarily from $\sim$ 7 to 6, with
each cloud contributing $\Delta$$N_{\rm H} \sim 3 \pm 1 \times
10^{22}$ cm$^{-2}$, or less than half the column inferred to exist for
the clouds causing the 2003--2004 and 2010--2011 \textit{increases} in total
$N_{\rm H}$. If all these clouds are part of the same clumpy structure
and have a common origin, then the assumption of each cloud in a given
AGN having exactly uniform values of $N_{\rm H}$ is an
oversimplification. $N_{\rm H} = 3 \pm 1 \times 10^{22}$~cm$^{-2}$ is
not an implausible value for a cloud, and only a factor of 2--3 less
than that found for other clouds detected in our sample.


$N_{\rm H,base}$ in Cen~A could also be attributed to non-clumpy,
spatially-extended material that is not entirely homogeneous; perhaps
this component can be associated with an intercloud medium. 
Here, $N_{\rm H,base}$ is usually $\sim 21.8 \times
10^{22}$~cm$^{-2}$, but there exists an \textit{underdense} region
with $\Delta$$N_{\rm H} = -3 \times 10^{22}$~cm$^{-2}$ that transited
the line of sight in 2010. Assuming an arbitrary distance to the black
hole of 100--200 light-days (the location
of the absorbing clouds as per $\S$5.2), 
the $\sim$80-d duration of the dip implies
that the underdense region is on the order of half a light-day across.


Longer-term variability in $N_{\rm H,base}$ may also exist: Rothschild
et al.\ (2011) measured values of $14-19\times10^{22}$~cm$^{-2}$
during the 1996--2009 \textit{RXTE} observations, excluding the
2003--2004 points. Hopefully, future X-ray monitoring will further
resolve the variability in $N_{\rm H, base}$ and allow us to discriminate
among the above scenarios.

\section{SUMMARY}    

The AGN community is in the process of shifting away from quantifying
emission and absorption processes in Seyferts by modeling
circumnuclear gas via a homogeneous, Compton-thick ``donut''
morphology.  Instead, a new generation of models describe the torus
via distributions of numerous individual clouds, usually
preferentially distributed towards the equatorial plane.  Often, the
clouds are embedded in some outflowing wind from the cold, thin
accretion disk that feeds the black hole.  Observational support for
these models so far has come mainly from fitting IR SEDs in small
samples of Seyferts.  However, a statistical survey of the environment
around supermassive black holes has been needed to properly constrain
parameter space in the clumpy-absorber models.

We present the first such survey, the longest AGN X-ray monitoring
study to date. Our survey quantifies line-of-sight X-ray absorption by
clouds that transit the line of sight to the central engine.  Our
goals are to assess the relevance of clumpy-torus models as a
function of optical classification by exploring absorption over a wide
range of length scales (both inside and outside the dust sublimation
zone), and to explore links between X-ray absorbers, IR-emitting dusty
clouds, and the BLR.

We use the vast public archive of \textit{RXTE}
observations of AGN.  The archive features a wide array of sustained
monitoring campaigns that make us sensitive to variability in line of
sight absorption over a high dynamic range of timescales spanning from
0.2~d to 16~yr. Our final sample consists of 37 type I and 18
Compton-thin-obscured type II Seyferts and totals 230
``object-years,'' the largest ever available for statistical studies
of cloud events in AGN on timescales from days to years.

We use hardness ratio light curves to identify potential eclipse
events, and attempt to confirm the events with follow-up time-resolved
spectroscopy.  We are sensitive to full-covering, neutral or
lowly-ionized clouds with columns densities $\ga 10^{22}$ up to
$\sim10^{25}$~cm$^{-2}$.  Our results are thus complementary to those
derived with $\sim$1~d long-look observations using missions with
\textit{Chandra}, \textit{XMM-Newton}, and \textit{Suzaku}.

Our primary results are as follows:

$\bullet$ We find 12 ``secure'' X-ray absorption events in eight
Seyferts (confirmed with spectral fitting) plus four ``candidate''
eclipses in three Seyferts. As four eclipse events were published
previously (NGC~3227/2000--1, Cen~A/2003--4, Cen~A/2010--1, and
Mkn~348/1996--7), we triple the number of events detected in the
\textit{RXTE} archive. The events span a wide range in duration, from
$\la1$ d to over a year. We model the eclipsing clouds to have column
densities spanning $4-26 \times 10^{22}$ cm$^{-2}$. Importantly, we do
not detect any full-covering clouds that are Compton thick, although
if such clouds are partial-covering clouds then our experiment would
not be highly sensitive to them.


$\bullet$ We derived the probability to catch a type I/II source
undergoing an eclipse event that has any duration between 0.2~d and
16~yr, taking into account the inhomogeneous sampling in our X-ray
observations.  For type Is, it is 0.006 (conservative range:
0.003--0.166); for type IIs, 0.110 (0.039--0.571). Our uncertainties
are conservative, as they take into account our selection function,
candidate eclipse events in addition to the secure ones, uncertainties
in the observed durations, and uncertainties in the contributions of
individual objects' sampling patterns to our total sensitivity
function.  As a reminder, these values indicate the probability to
observe a source undergoing an eclipse event (of any duration $t_{\rm
  D}$ between 0.2 d and 16~yr) due only to a cloud passage through the
line of sight, and are independent of long-term constant absorption,
e.g., associated with gas in the host galaxy.  We conservatively
estimate the upper limit to observe a Compton-thick eclipse event
(with 0.2~d $< t_{\rm D} <$ 16~yr) to be $<$0.158 or $<$0.520 in type
I and II objects, respectively.  To repeat the caveat from $\S4$,
these probabilities refer only to eclipses by full-covering, neutral
or lowly-ionized clouds with column densities $\ga 10^{22}$ up to
$\sim10^{25}$~cm$^{-2}$; when one considers the full range of clouds
(larger range of $N_{\rm H}$, partial-covering clouds, wider range of
ionization) the resulting probabilities will almost certainly be
higher.

Although subject to low number statistics, our observations indicate
differences in the distributions of observed eclipse event durations
and probabilities for type I and II objects.  The type I objects have
$\sim4$ times as many campaigns and twice the number of targets as the
type IIs, but despite this, we do not detect eclipse events with
durations longer than 100~d (although we cannot rule this out for
the poorly-constrained event in MR~2251--178).  If it were the case
that eclipses with durations of $\sim$hundreds of days occurred in
type Is with the same frequency density as in type IIs, then the
monitoring campaigns on the type Is should have detected them.  This
implies that we are "missing" clouds in type Is along the line of
sight at radial distances of $\ga$ a few $\times 10^5 R_{\rm g}$.
This does not necessarily imply \textit{intrinsic} differences in the
cloud distributions of between type Is and IIs, just potential
differences along the line of sight.


Despite the generally low probabilities of observing an eclipse, 4/10
objects show secure or candidate eclipse events multiple
times. Perhaps the system is close to edge-on, and/or the total number
of clouds is very high in these objects.  If this is the case, then an
object for which an eclipse has been detected has a higher chance of
showing an additional eclipse compared to sources with no eclipses so
far. In addition, perhaps there exists some dispersion between the
averaged derived probabilities of cloud eclipses for a class of
objects and the probabilities for individual objects.

$\bullet$ We see no obvious dependence of the likelihood to observe
eclipses in a given object on the usual AGN parameters ($M_{\rm BH}$,
$L_{\rm Bol}$, radio loudness).

$\bullet$ We estimate the locations of the clouds from the central
black hole $r_{\rm cl}$ based on ionization parameter. Best-estimate
values of $r_{\rm cl}$ are typically tens to hundreds of light-days,
or $1-50 \times 10^4 R_{\rm g}$ ($0.3 - 140 \times 10^{4} R_{\rm g}$
accounting for uncertainties). In 7/8 objects, the clouds are
consistent with residing at the estimated location of the dust
sublimation zone (see Fig.~\ref{fig:plotrs}). The eighth object is
Cen~A, where the clouds are estimated to exist just outside the dust
sublimation zone. For the three objects in our sample whose dusty tori
have been mapped via either IR interferometry or optical-to-near-IR
reverberation mapping (Cen~A, NGC~3783, and NGC~3227), the clouds'
radial distances are commensurate with those of the IR-emitting tori.
For six of the seven objects with estimates of the locations of
optical/UV/IR BLR clouds, the X-ray-absorbing clumps are commensurate
with the outer portions of the BLR or radial distances outside the
known BLR within factors of a few to 10. Only in one object
(NGC~5506) is the eclipsing cloud inferred to be commensurate with the
inner BLR. This cloud thus may be akin to those clouds inferred to
exist in the BLRs of other objects (namely NGC~1365).

Our results thus confirm the existence of X-ray-absorbing,
neutral/low-ionization clouds with $N_{\rm H} \sim10^{22-23}$
cm$^{-2}$ at these indicate radial distances from the black hole. This
is not to say that X-ray-absorbing clouds cannot exist at other
distances; our selection function analysis indicates that we are
biased towards detecting eclipses with durations of $\sim$tens of
days. The 12 ``secure'' clouds have an average diameter of 0.25
light-days, an average number density of $3\times 10^{8}$~cm$^{-3}$,
and an average mass of $10^{-5.2} \Msun$. We find no statistically
significant difference between the individual cloud properties of type
Is and IIs. Assuming that the clouds are 100 per cent full covering, their
inferred diameters imply upper limits to the size of the X-ray
continuum-emitting region. These limits are as low as $2 R_{\rm g}$
for NGC~5506 and $45-65 R_{\rm g}$ for NGC~3783, Mkn~79, and Mkn~509.

Given their inferred locations and diameters, the X-ray-absorbing
clouds are likely too small to obscure the view of the entire BLR in the
type Is studied here. However, such clouds could potentially obscure and
redden large parts of the BLR and temporarily turn a type I into a type
II AGN over timescales of months to years. The clouds would need to be
sufficiently dusty, be at least several light-days in diameter, and be
located many light-months away from the black hole. Such events have been
observed as variations in optical broad line strengths (Goodrich 1989;
Tran et al.\ 1992; Aretxaga et al.\ 1999).
The probability to observe an X-ray eclipse in a type I by either a
dusty or dust-free cloud, derived in $\S$4, thus does not serve as a
prediction for the probability to observe a type classification change
in the optical band. The latter can also occur via changes in geometry
or illumination of the BLR (Cohen et al.\ 1986).


$\bullet$ Three eclipse events (NGC~3783/2008.3, NGC~3227/2000--1, and
Cen~A/2010--1) have column density profiles $N_{\rm H}$($t$) that are
well resolved in time.  The event in NGC~3783 at 2008.3 is the best
time-resolved $N_{\rm H}$($t$) profile not yet published, and seems to
show evidence for a double-peaked profile with peaks separated by 
11~d, with $N_{\rm H}$($t$) not returning to baseline levels in the
period between the peaks. The profile could be explained by a
dumbbell-shaped (along the direction of transit) cloud.  One
possibility is that we are witnessing an eclipse by a cloud in the
process of being tidally disrupted, or may be filamentary in
structure.  Models incorporating clumpy disk winds (e.g., MHD-driven
or IR radiation-driven) as opposed to discrete, compact structures,
may also be relevant for explaining the observed profile.  Such models
characterize the torus as a highly dynamic structure, with uplifts and
failed winds in addition to Keplerian motion. In contrast, the density
profiles for the clouds in Cen~A and NGC~3227 have been modeled by
Rivers et al.\ (2011b) and Lamer et al.\ (2003), respectively, to be
centrally peaked, suggesting that self-gravity likely dominates over
tidal shearing or internal turbulence.


$\bullet$ We provide constraints for the parameters used to quantify
the cloud distributions in the $\C$ models ($\tau_{\rm V}$, $Y$,
$\No$, $i$, $\sigma$), which so far have been constrained
observationally only via IR SED fitting. As some of our clouds are in
the dust sublimation zone, the extent to which our constraints for
models that are based on the IR-emitting region is not fully
understood. However, our study provides the first ever statistical
constraints from \textit{X-ray} observations and constitutes a
completely independent way of studying the structure and geometry of
the environment around supermassive black holes. The X-ray constraints
on these parameters derived in this paper can be incorporated into the
priors of future Bayesian IR SED fitting analyses. Our X-ray column
densities are equivalent to values of $\tau_{\rm V}$ of 20--132,
consistent with typical values used by clumpy-torus theorists (Nenkova
et al.\ 2008b) and constrained so far by IR SED modeling (e.g.,
RA11). We find clouds out to $Y=r_{\rm
  cl}/R_{\rm d} \sim 20$, also consistent with the IR observations.
Constraints on $q$ based on our X-ray data will be provided in a
separate paper (Nikutta et al., in preparation).

The fact that we observe eclipses in both type Is and type IIs
disfavors sharp-edged cloud distributions and supports torus models
with as soft, e.g., Gaussian, distribution above/below the equatorial
plane.  Furthermore, we compare our constraints on the integrated
probability to witness a type I or type II object undergoing an
eclipse (of any duration) to the predictions $P_{\rm obsc}$ from $\C$
to obtain constraints in ($\No$, $\sigma$, $i$) parameter space.
When there is external information on any one of these
parameters (e.g., the inclination of the system), then constraints on
the other two parameters can be obtained following
Fig.~\ref{fig:pobscured}.


$\bullet$ We find evidence in eight type II Seyferts for a baseline
level of X-ray absorption that remains constant (down to $\sim 0.6-9
\times 10^{22}$~cm$^{-2}$) over timescales from 0.6 to 8.4~yr.  The
clouds we detect in this paper are not able to explain the constant
baseline absorbers, since we would have expected many more
``negative'' cloud events than observed.  The constant total amount of
X-ray absorption in type IIs can be explained in the context of
clumpy-absorber models only if there exists a large number of very
low-density ($\ll 10^{22}$~cm$^{-2}$) clouds along the line of sight,
with $\No$ remaining roughly constant within each object.
Alternatively, there can exist in type IIs nearly homogeneous 
X-ray-absorbing gas, whose location along the line of sight in most cases is
unconstrained. One possibility is that this gas lies on the order of
tens to thousands of parsecs away from the black hole, and is
associated with X-ray-absorbing matter in the host galaxy.  Another
possibility is a medium of non-clumpy, relatively homogeneous gas
located further in, such as a relatively low-density intercloud medium
in which higher density clumps are embedded (Stalevski et al.\ 2012).
In the case of Cen~A, in addition to the two eclipses observed, we
find evidence that the baseline level of X-ray absorption dipped by
$\sim14$ per cent and then recovered in early 2010.  This is consistent with
the notion that the material describing $N_{\rm H,base}$ in this
object is relatively close to the black hole and indeed not perfectly
homogeneous, and that an underdense region transited the line of
sight.

\textit{In summary,} our findings are consistent with the notion that
both type I and II objects contain clouds with roughly similar
properties ($N_{\rm H}$, mass, location from the black hole).  The
probability to observe an eclipse in type IIs is higher than in type
Is. However, in addition in type IIs there is frequently a baseline
level of X-ray absorption that is not likely due to clouds, but instead
due to an additional, almost homogeneous absorber of unconstrained location
(intercloud medium or gas associated with the host galaxy).  
An exception is Cen~A, where the observation of a small dip in the
baseline level is consistent with an intercloud medium commensurate  
with the dusty and X-ray tori.


\textit{Future observational and theoretical work:} Much more
observational work needs to be performed to further support and
properly test the characteristics of the clumpy-torus model and its
key parameters.  To date, \textit{RXTE} has been unique in its ability
to provide multi-timescale X-ray monitoring for such a large number of
AGN. The only future mission potentially capable of creating such an
X-ray monitoring archive will be extended R\"ontgen Survey
with an Imaging Telescope (eROSITA) onboard the Russian spacecraft
\textit{Spektrum R\"ontgen/Gamma} (Predehl et al.\ 2011).  eROSITA
will scan the entire sky several times during its mission.
 
Ideally, a repeat of the experiment should aim to uncover X-ray
eclipse events spanning as wide a dynamic range in duration and inferred
radial distance as possible.  In particular, we need better sustained
monitoring (with minimal gaps in coverage) on timescales $\ga$ a
few years, especially in the type II Seyferts.  At the other end,
\textit{RXTE} conducted a number of ``intensive monitoring''
campaigns, featuring, e.g., observations four times daily for 1--2
months.  These intensive campaigns allowed us to detect the
shortest-duration eclipse event in our sample (the $\la$1~d event
in NGC~5506) and enabled us to extract a high-quality $N_{\rm H}$($t$)
profile for the NGC~3783/2008.3 eclipse event, and thus
illustrate the necessity of intensive sampling.  

We need more sustained monitoring of Seyferts spanning a wide range of
luminosities, including in the low-luminosity regime, to test the
suggestion that the disk wind/torus cloud outflow
disappears at low luminosities (Elitzur \& Ho 2009).

Finally, additional knowledge about the inclination angle of these
accreting black hole systems would help us better constrain the
$\sigma$ parameter.  Such information can come from, e.g., X-ray
spectroscopic modeling of relativistically-broadened reflected
emission (Fe K$\alpha$ emission lines and soft excess) from the inner
accretion disk (e.g., Patrick et al.\ 2012).


On the theoretical side, the community can benefit from
\textit{dynamical} (time-dependent) models that yield the expected
durations and frequency of observed eclipse events (depending on
clouds' diameters and distribution, source viewing angle, source
luminosity, etc.).  Advances in understanding can also come from models that
further explore the interaction between the intercloud medium and the
clouds embedded in it.

\section*{acknowledgements}
The authors are very grateful to M.\ Elitzur for helpful comments.
The authors also thank the referee for helpful comments.
This research has made use of data obtained from the \textsl{RXTE}
satellite, a NASA space mission.  This work has made use of HEASARC
online services, supported by NASA/GSFC, and the NASA/IPAC
Extragalactic Database, operated by JPL/California Institute of
Technology under contract with NASA.  The research was supported by
NASA Grant NNX11AD07G and funding from the European Community's
Seventh Framework Programme (/FP7/2007-2013/) under grant agreement
no.\ 229517.  RN acknowledges support by the ALMA-CONICYT fund,
project no.\ 31110001.  The authors thank Wolfgang Steffen for
providing images and video in support of the press release associated
with this paper; 
material can be found at http://cass.ucsd.edu/$\sim$rxteagn/clumpytorus/.


\appendix

\setcounter{figure}{0} \renewcommand{\thefigure}{A.\arabic{figure}} 

\section{Details of eclipse identification in individual objects}

Flux, hardness ratio, and $\Gamma_{\rm app}$ light curves are shown in
Figs.~\ref{fig:mega3783} -- \ref{fig:mega348} for NGC~3783, 
MR~2251--178, Mkn~79, Mkn~509, NGC~3227, Cen~A, NGC~5506, and Mkn~348,
respectively, the objects with ``secure'' events, plus one
``candidate'' event for NGC~3783.  Also included are plots of two
sources with only ``candidate'' events, Fairall~9
(Fig.~\ref{fig:megafrl9}) and NGC~3516 (Fig.~\ref{fig:mega3516}).
Errors on each light curve point are 68 per cent confidence.
All other sources' light curves do not yield any strong statistically significant deviations and we do not
include them in this paper for brevity. 

Yellow shaded areas indicate times of events for which we conducted
follow-up time-resolved spectroscopy to confirm events as ``secure.''
We bin up consecutive individual spectra around the period of each
candidate event to achieve sufficient variability/noise in binned
spectra. Bin sizes are determined as a trade-off between the need
to achieve small errors in $\Gamma$ or $N_{\rm H}$ in binned spectra
versus the need to trace out the $N_{\rm H}$ profile if possible.
The number of 2--10~keV counts needed to deconvolve $N_{\rm H}$ and
$\Gamma$ is typically $\sim$ 80000, with typically $\sim$ 15000
counts needed to achieve reasonable constraints on $N_{\rm H}$
assuming a fixed value of $\Gamma$.  All spectra are grouped to a
minimum of 25 counts bin$^{-1}$.  Due to the gradual evolution of the
PCA response over time, response files are generated for each
observation separately, using \textsc{pcarsp} version 11.7.1.  All
spectral fitting is done with \textsc{isis} version 1.6.2-16. Uncertainties on
all parameters derived from spectral fits, including $N_{\rm H}$ and
$\Gamma_{\rm app}$, are 90 per cent confidence ($\Delta\chi^2 = 2.71$ for
one interesting parameter) unless otherwise noted.

We apply the best-fit model from the time-averaged \textit{RXTE}
spectrum derived by Rivers \etal\ (2011a, 2013).  In all fits, we keep
the energy centroid and width of the Fe K$\alpha$ line and all
parameters associated with the Compton reflection hump (except for
normalization, which was tied to that of the incident power law)
frozen at their time-averaged values.  Fe K$\alpha$ emission line
intensities are kept frozen unless there is significant improvement
in the fit.  In all spectral fits to sources that are normally X-ray
unobscured (all are type I objects),
we consider two models. Model 1 has model components identical to
the time-averaged model: power law with $\Gamma$ and 1~keV
normalization $A_1$ kept free, Fe K$\alpha$ emission line, Compton
reflection hump modeled with \textsc{pexrav}, and $N_{\rm Gal}$
modeled with \textsc{phabs}.  In Model 2, we model $N_{\rm H}$
(absorption by neutral gas only at the source, i.e., in excess of
$N_{\rm Gal}$) with \textsc{zphabs}. We keep $\Gamma$ free only in the
high signal-to-noise cases where $\Gamma$ and $N_{\rm H}$ can be
deconvolved with a minimum of degeneracy (``secure A'' events).
Otherwise $\Gamma$ is kept frozen at either the time-averaged value
or the value derived from unabsorbed spectra surrounding the putative
obscuration event (``secure B'' events).  For sources normally X-ray
obscured (all are type II objects), $N_{\rm H}$ as modeled with 
\textsc{zphabs} is included in all fits.

All analysis in this paper uses the cosmic abundances of Wilms et
al.\ (2000) and the cross sections of Verner \etal\ (1996) unless otherwise
stated. Many previous publications on the objects discussed herein
used cosmic abundances similar to those of Anders \& Grevesse (1989),
which have C, N, O, Ne, Si, S, and Fe elemental abundances relative to
H spanning factors of 1.3--1.9 lower than those in Wilms et
al.\ (2000).  For a model consisting of a power law with photon index
$\Gamma=1.8$ this difference yields values of $N_{\rm H}$ using the
\textsc{angr} abundances a factor of $\sim30-40$ per cent greater than
corresponding values of using the \textsc{wilm} abundances for the
column densities encountered in our sample.

There is a known issue with the estimation of background errors by the
\textit{RXTE} data reduction software: the software models the
background counts spectrum based on long blank-sky observations, and
then assumes Poissonian errors for the background counts spectrum
appropriate for the exposure time of the observation of the target.
However, the unmodeled residual variance in the background is on the
order of 1--2 per cent (Jahoda et al.\ 2006), and for spectra with exposure
times higher than $\sim10-30$~ks, the Poissonian errors may be overestimates.
The errors on the net (background-subtracted) spectrum are thus likely
overestimates (Nandra et al.\ 2000), frequently yielding best fits
with values of $\chi^2_{\rm r}$ $\sim$ 0.6--0.8 for many of the
time-resolved spectral fits below.  Because these errors are
overestimates, uncertainties on best-fitting model parameters reported in
the tables are conservative.


\subsection{NGC 3783 (Sy 1)}

\begin{figure*}
\includegraphics[angle=-90,width=1.00\textwidth]{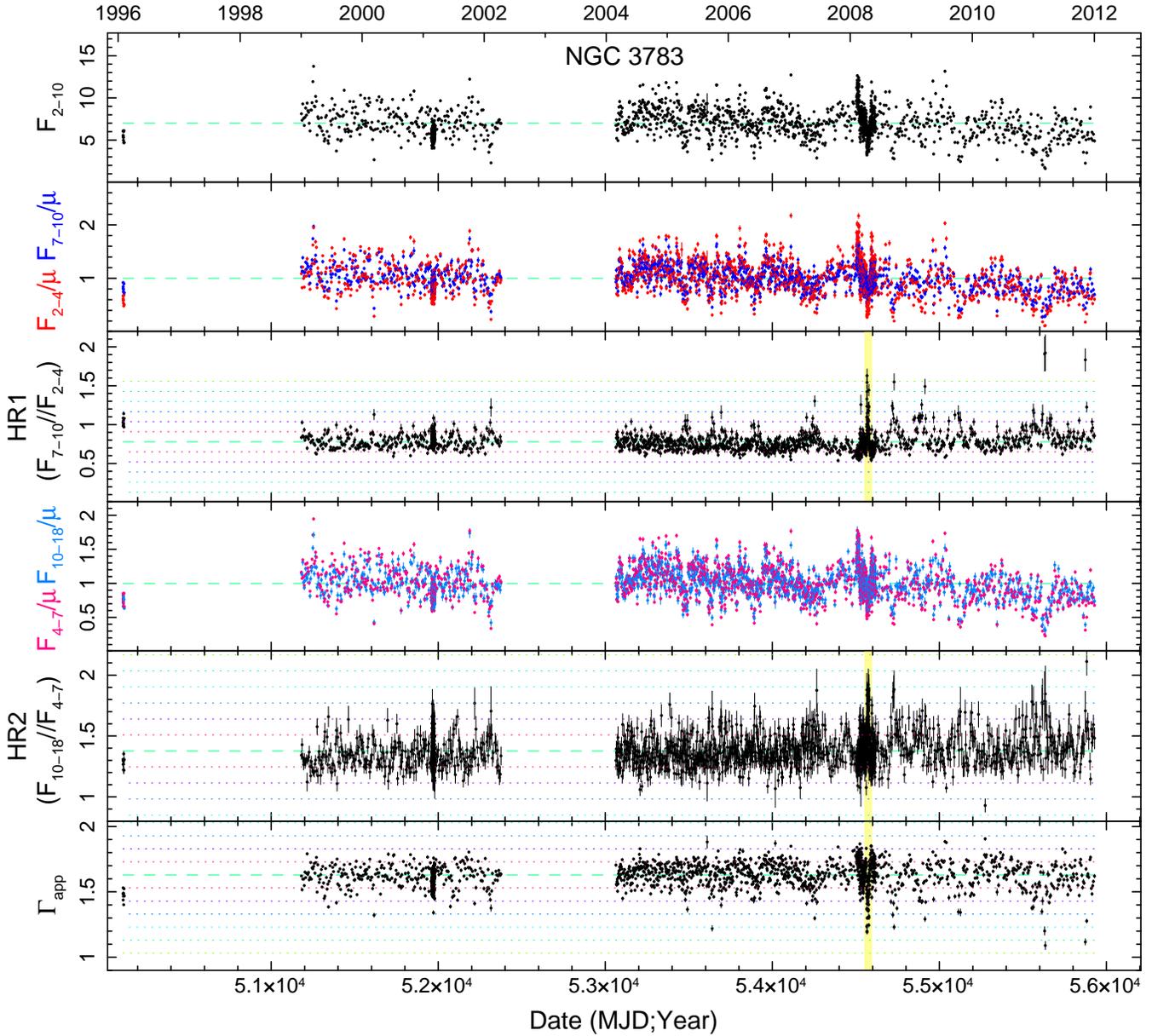}       
\caption{Long-term continuum flux, hardness ratio, and $\Gamma_{\rm
    app}$ light curves for the Sy~1 NGC~3783.  The top panel shows
  observed 2--10~keV flux, $F_{\rm 2-10}$, in units of $10^{-11}$~erg
  cm$^{-2}$ s$^{-1}$, with one point per observation. Error bars are
  frequently smaller than the data points.  The second panel shows
  $F_{\rm 2-4}$ (red) and $F_{\rm 7-10}$ (blue), each normalized by
  their means.  $HR1\equiv F_{\rm 7-10}/F_{\rm 2-4}$ is plotted in the
  third panel.  The fourth panel shows $F_{\rm 4-7}$ (red) and 
  $F_{\rm 10-18}$ (blue), each normalized by their means.  
  $HR2\equiv F_{\rm 10-18}/F_{\rm 4-7}$ is plotted in the fifth panel.  
  The bottom panel shows $\Gamma_{\rm app}$, the photon index obtained from
  fitting a simple power law plus $N_{\rm H,Gal}$ (ignoring all other
  components) to each observed spectrum.  The colored dotted lines in
  panels 3, 5, and 6 indicate the 1--6$\sigma$ (standard deviation)
  levels.  In all panels, the green dashed lines indicate the mean.
  The yellow shaded areas indicate the secure occultation event,
  shown in more detail in Fig.~\ref{fig:mega3783zoom_TR}.}
\label{fig:mega3783}  
\end{figure*}

\begin{figure}
\includegraphics[angle=-90,width=0.48\textwidth]{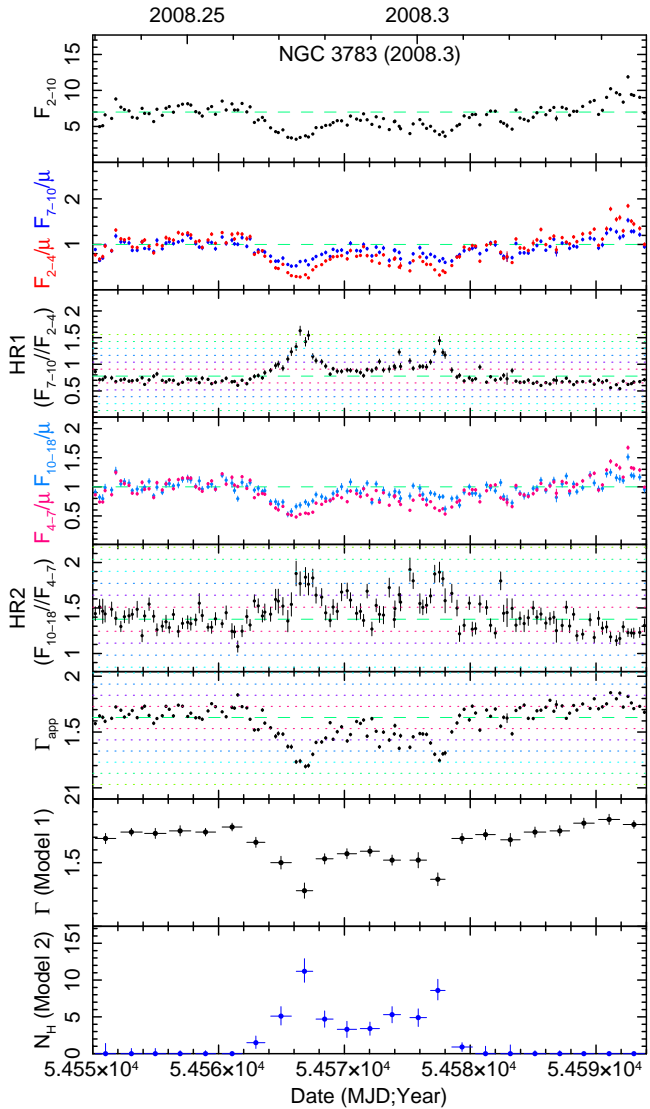}    
\caption{Same as Fig.~\ref{fig:mega3783}, but zoomed in on the
  period of the secure occultation event at 2008.3. The bottom two
  panels show $\Gamma$ for Model~1 (wherein $N_{\rm H}$ is assumed to
  be zero) and $N_{\rm H}$ for Model~2, where $\Gamma$ is assumed to
  be frozen at 1.73. $N_{\rm H}$ is in units
  of $10^{22}$~cm$^{-2}$.  Note the double-peaked behavior
  of $N_{\rm H}$($t$) for this event.}
\label{fig:mega3783zoom_TR}  
\end{figure}

The long-term light curves for NGC~3783 are shown in
Fig.~\ref{fig:mega3783}. The $HR1$ light curve shows several sharp
deviations upward from the mean of 0.9:

$\bullet$ \textit{2008.3: (``Secure B'')} During a period of intensive
monitoring one point every $\sim$6 hrs in early 2008, there are two
``spikes'' to values of $HR1$ = 1.5--1.6 ($4-5\sigma$ deviations),
separated by only 11 d and peaking at MJD $\sim$ 54567.0 and 54577.7.
The light curves are plotted in Fig.~\ref{fig:mega3783zoom_TR}.
$\Gamma_{\rm app}$ falls to 1.2--1.3 during these times (3--4$\sigma$
deviation), although $HR2$ shows only mild deviations, at the
$1-2\sigma$ level.  Interestingly, $HR1$ and $\Gamma_{\rm app}$ do not
return to their mean values in the period between the spikes.

We group data into bins approximately 2 d wide and fit the 3--23
keV PCA spectra, with results plotted in Table~\ref{tab:3783TRtable}.
We first apply Model 1, keeping $R$ frozen at 0.41 and $I_{\rm Fe}$
frozen at $1.5\times 10^{-4}$~ph~cm$^{-2}$ s$^{-1}$. This yields
acceptable fits ($\chi_{\rm red} \sim 0.6-1.3$), but with $\Gamma$
reaching values as low as 1.3 near the positions of the $HR1$ spikes,
as illustrated in Fig.~\ref{fig:mega3783zoom_TR}.

We then apply Model 2, with $\Gamma$ frozen at 1.73; this value is
flatter than that from the time-averaged spectrum, 1.86, but equal to
the average of the values before and after the putative eclipse event.
Values of $\chi_{\rm red}$ are overall similar to the first model.
$N_{\rm H}$($t$) is plotted in the lower panel of
Fig.~\ref{fig:mega3783zoom_TR}.  Choosing slightly different bin
sizes does not change the overall $N_{\rm H}$($t$) profile.  Best-fitting
values of $N_{\rm H}$ to spectra \#9 and \#15 are $11.2^{+1.7}_{-1.5}$ and
$8.6^{+1.5}_{-1.3}$ $\times 10^{22}$~cm$^{-2}$, respectively, with
values $\sim 3-5 \times 10^{22}$~cm$^{-2}$ for the period in between.
Upper limits to $N_{\rm H}$ are obtained for spectra \#1--8 and \#16--23.

We can treat this event as a single event by adopting the maximum
values of $N_{\rm H}$ (during spectrum $\#9$) as the peak value of
$N_{\rm H}$. We use the $HR1$ light curve to assign a duration to the
total event of 14.4--15.4 d.  Alternatively, assuming two independent
occultation events, we derive full durations of 9.2 and 4.6 d based
on a dual linear-density sphere model fit to the column density
profile, discussed in $\S$\ref{sec:3783double}.

We do not include the multiple warm absorbers modeled in previous
X-ray spectra obtained with \textit{XMM-Newton} or \textit{Chandra}
(e.g., Reeves et al.\ 2004 and references therein).  Considering the
three zones of warm absorption modeled by Reeves et al.\ (2004), the
lowly-ionized absorber with log($\xi$)=--0.1 and $N_{\rm H} = 1.1
\times 10^{21}$~cm$^{-2}$ has a negligible effect on the PCA spectrum.
The two other warm absorbers, one with log($\xi$)=2.1 and $N_{\rm H} =
1.2 \times 10^{22}$~cm$^{-2}$, the other with log($\xi$)=3 and $N_{\rm
  H} = $~cm$^{-2}$, have a modest effect on continuum components
above 3~keV.  When we include these two zones in the model, $\Gamma$, 
steepens by $\ga0.05$, but the effect on any values of neutral
$N_{\rm H}$ we measure is only at the $10$ per cent level or
less. Modeling a
simple power-law suffices to quantify the continuum, given the energy
resolution and photon noise of the time-resolved spectra, and we
assume that these absorbers' parameters are constant.

\begin{figure*}
\includegraphics[angle=-90,width=1.00\textwidth]{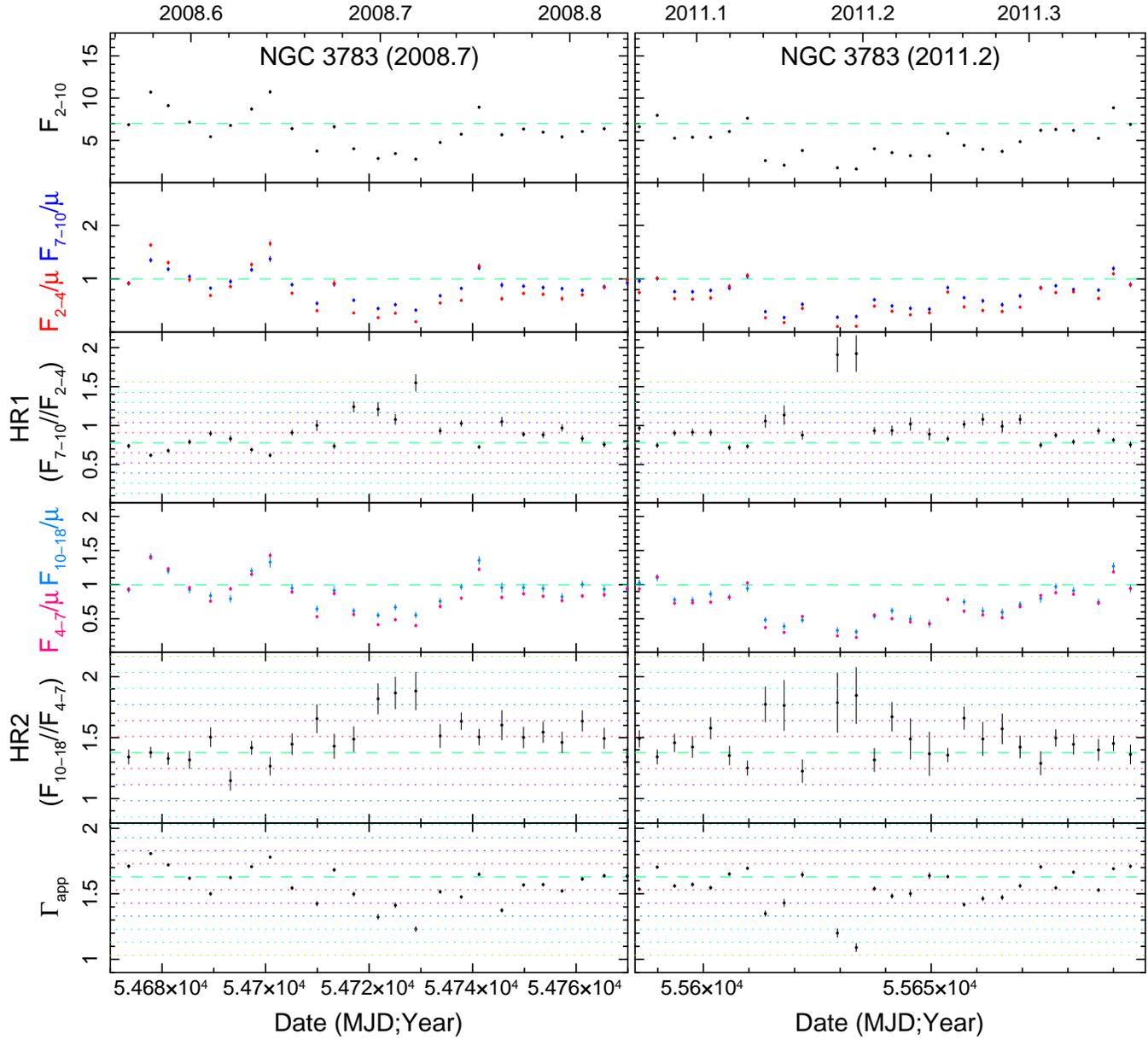}  
\caption{Same as Fig.~\ref{fig:mega3783}, but zoomed in on two ``candidate'' events at 2008.7 and 2011.2.}  
\label{fig:mega3783zoom2008rej}  
\end{figure*}

$\bullet$ At 2008.7 (``Candidate''), $\Gamma_{\rm app}$ drops below a 2$\sigma$
deviation, reaching 1.2, and $HR1$ and $HR2$ increase to $>$3$\sigma$
deviation for 3/4 consecutive observations; see
Fig.~\ref{fig:mega3783zoom2008rej}. However, this also occurs near a
period of low continuum flux in all bands, and time-resolved
spectroscopy cannot confirm absorption in excess of the Galactic
column, with $N_{\rm H} < 4 \times 10^{22}$~cm$^{-2}$.  We tentatively
classify this as a ``candidate'' event of duration 12--28 d.

$\bullet$ \textit{2011.2 (``Candidate'')}: In the $HR1$ light curve (see
Fig.~\ref{fig:mega3783zoom2008rej}), there are two consecutive points at
$HR1\sim1.9$, a $\sim7\sigma$ deviation.  They occur in early 2011 at
MJD 55629 and 55633.  $\Gamma_{\rm app}$ falls from an average of 1.6
in the surrounding $\pm$5 observations to $1.20\pm0.03$ and
$1.09\pm0.03$.

The source was monitored once every 4 d during this time.  We
perform time-resolved spectroscopy on data from MJD
55589.85 to 55693.73, grouping every three observations (exposure times
2--3 ks), except for MJD 55629.43--55633.56 (1.8 ks).  However, this
event also occurs near a period of low continuum flux in all bands,
including the 10--18 keV band, yielding large statistical
uncertainties.  Results from spectral fits for Model 1 (again assuming
$R$ frozen at 0.41 and $I_{\rm Fe}$ frozen at $1.5\times
10^{-4}$~ph~cm$^{-2}$ s$^{-1}$) yields $\Gamma=1.22\pm0.18$ during the
putative eclipse.  For Model 2, $N_{\rm H} = 11^{+6}_{-7} \times
10^{22}$~cm$^{-2}$ for that spectrum, assuming $\Gamma$ is frozen at
1.62, the average of the values during the other spectra.

Given the large uncertainty in $\Gamma$ and $N_{\rm H}$, the
near-identical and low values of $\chi^2_{\rm red}$ for both models
($\sim0.5$), and the low continuum flux in all wavebands, including
the 10--18 keV band (see Appendix A10 and Fig.~\ref{fig:fluxflux}), 
this event does not clearly pass criterion~4. We hence choose
to be conservative and classify this event as a ``candidate'' event,
rather than a ``secure B'' event. Given the start/stop times of the
monitoring observations, we constrain the duration of the obscuration
event to be between 4.1 and 15.8~d.

$\bullet$ There are additional deviations in the long-term $HR1$ and
$\Gamma_{\rm app}$ light curves which tend to catch the eye, but these
deviations did not fulfill our selection criteria.  For example, there
are single-point increases in $HR1$ at the $2.5-4\sigma$ deviation
level near MJD 51610, 52310 (large uncertainty), 53695 (large
uncertainty), 54205, 54253, 54530; $\Gamma_{\rm app}$ was 1.3 or
higher for each of these cases.  In addition, $\Gamma_{\rm app}$ drops to
values $<$1.25 at MJD 53639 and 55873, but again, these 
are single-point decreases, usually occurring during periods
of relatively low 2--10 and/or 10--18~keV flux, and time-resolved
spectroscopy cannot confirm excess absorption.

\begin{table*}
\begin{minipage}{170mm}
\caption{Results of time-resolved spectral fits to NGC~3783, 2008.3}
\label{tab:3783TRtable} 
\begin{tabular}{lccccccccc} \hline
Start--stop        & Expo  & \multicolumn{4}{c}{Model 1: $N_{\rm H}$ frozen at 0}        & \multicolumn{4}{c}{Model 2: $N_{\rm H}$ free; $\Gamma$ frozen} \\
   (MJD)           & (ks)  & $\chi^2_{\rm red}$ & $\Gamma$      & $A_1$        & $F_{2-10}$ & $\chi^2_{\rm red}$ &  $N_{\rm H}$ ($10^{22}$~cm$^{-2}$) & $A_1$  & $F_{2-10}$ \\ \hline
54550.20--54551.83 (\#1) & 5.1   & 0.94 & $1.69\pm0.04$   & $15.0^{+1.1}_{-1.0}$   &   6.07        & 0.86 &  $< 1.4$              & $15.7^{+0.3}_{-0.2}$ &    6.10 \\  
54552.23--54553.93 (\#2) & 5.1   & 0.75 & $1.74\pm0.03$   & $18.1^{+1.2}_{-1.1}$   &   6.82        & 0.76 &  $< 0.7$              & $17.6^{+0.2}_{-0.3}$ &    6.80\\
54554.12--54555.83 (\#3) & 5.2   & 1.02 & $1.73\pm0.04$   & $17.3\pm1.1$        &   6.64        & 1.02 &  $< 0.7$              & $17.2^{+0.2}_{-0.3}$ &    6.63\\
54556.09--54557.79 (\#4) & 4.9   & 0.86 & $1.75\pm0.04$   & $20.8^{+0.3}_{-0.8}$   &   7.61        & 0.92 &  $< 0.4$              & $19.6\pm0.2$         &    7.56\\  
54558.18--54559.76 (\#5) & 5.1   & 0.61 & $1.74\pm0.03$   & $18.6^{+1.3}_{-1.1}$   &   6.93        & 0.64 &  $< 0.4$              & $17.9^{+0.2}_{-0.3}$ &    6.90\\
54560.28--54561.87 (\#6) & 5.4   & 0.97 & $1.78\pm0.03$   & $22.0^{+1.3}_{-1.2}$   &   7.70        & 1.21 &  $< 0.2$             & $19.8\pm0.2$         &    7.61\\
54562.25--54563.68 (\#7) & 5.0   & 1.05 & $1.66\pm0.04$   & $14.4^{+1.1}_{-1.0}$   &   6.17        & 0.95 &  $1.5^{+0.9}_{-0.8}$ & $16.4\pm0.4$         &    6.06\\   
54564.14--54565.79 (\#8) & 5.1   & 0.88 & $1.50\pm0.05$   & $ 7.2\pm0.7$        &   4.04        & 0.78 &  $5.1^{+1.3}_{-1.2}$ & $11.9^{+0.4}_{-0.3}$ &    3.87\\
54566.17--54567.49 (\#9) & 4.9   & 0.75 & $1.28\pm0.06$   & $ 4.3^{+0.5}_{-0.4}$   &   3.48        & 0.70 &  $11.2^{+1.7}_{-1.5}$ & $12.0\pm0.5$         &   3.27\\   
54567.75--54569.11 (\#10)& 5.0   & 1.02 & $1.53\pm0.04$   & $ 9.4^{+0.8}_{-0.7}$   &   5.00        & 0.80 &  $4.7\pm1.1$      & $14.6\pm0.4$         &    4.69\\
54569.43--54570.97 (\#11)& 4.2   & 0.74 & $1.57\pm0.04$   & $11.4^{+1.0}_{-0.8}$   &   5.70        & 0.66 &  $3.3\pm1.1$      & $16.1\pm0.4$         &    5.53 \\
54571.27--54572.79 (\#12)& 5.5   & 1.00 & $1.59\pm0.04$   & $11.9^{+0.9}_{-0.8}$   &   5.70        & 0.68 &  $3.4\pm0.9$      & $16.1\pm0.4$         &    5.51\\       
54573.10--54574.47 (\#13)& 4.5   & 0.94 & $1.52\pm0.04$   & $ 9.5^{+0.8}_{-0.7}$   &   5.13        & 0.58 &  $5.3\pm1.1$       & $15.2^{+0.5}_{-0.4}$ &    4.90   \\
54575.21--54576.53 (\#14)& 4.2   & 1.15 & $1.52\pm0.06$   & $ 9.1\pm0.8$        &   4.93        & 1.06 &  $4.9\pm1.2$       & $14.5^{+0.5}_{-0.4}$ &    4.91  \\   
54576.85--54578.00 (\#15)& 4.7   & 0.96 & $1.37\pm0.05$   & $ 6.0\pm0.6$        &   4.15        & 0.81 &  $8.6^{+1.5}_{-1.3}$  & $13.5\pm0.5$ &    3.92  \\
54578.53--54580.18 (\#16)& 5.0   & 0.81 & $1.69\pm0.04$   & $13.5^{+1.1}_{-1.0}$   &   5.52        & 0.76 &  $0.9\pm0.6$       & $14.3\pm0.4$ &    5.45\\
54580.43--54582.00 (\#17)& 4.4   & 1.06 & $1.72\pm0.04$   & $16.7^{+1.3}_{-1.1}$   &   6.45        & 1.03 &  $< 0.9$            & $16.5\pm0.3$ &    6.40 \\
54582.42--54583.96 (\#18)& 4.1   & 0.78 & $1.68\pm0.05$   & $13.9^{+1.2}_{-1.1}$   &   5.72        & 0.78 &  $< 1.0$            & $14.8\pm0.4$         &    5.69 \\
54584.28--54586.01 (\#19)& 5.3   & 0.74 & $1.74\pm0.04$   & $17.2^{+1.2}_{-1.1}$   &   6.41        & 0.76 &  $< 0.4$             & $16.5\pm0.2$         &    6.39 \\    
54586.34--54587.90 (\#20)& 5.0   & 1.17 & $1.75\pm0.04$   & $19.3^{+1.3}_{-1.2}$   &   7.11        & 1.21 &  $< 0.4$             & $18.3\pm0.3$         &    7.07 \\ 
54588.21--54589.86 (\#21)& 4.5   & 0.97 & $1.81\pm0.04$   & $22.7^{+1.5}_{-1.4}$   &   7.60        & 1.36 &  $< 0.2$             & $19.4\pm0.3$         &    7.48 \\
54590.20--54591.88 (\#22)& 4.4   & 1.34 & $1.84\pm0.04$   & $26.8^{+1.7}_{-1.3}$   &   8.43        & 2.23 &  $< 0.1$             & $21.4\pm0.3$         &    8.24 \\
54592.16--54593.85 (\#23)& 5.1   & 0.97 & $1.80\pm0.03$   & $26.2^{+1.5}_{-1.3}$   &   8.83        & 1.45 &  $< 0.1$             & $22.6^{+0.3}_{-0.2}$ &    8.69  \\ \hline    
\end{tabular}\\
$A_1$ is the 1~keV normalization of the power law in units of $10^{-3}$~ph~cm$^{-2}$ s$^{-1}$ keV$^{-1}$. $F_{2-10}$ is the
observed/absorbed model flux in units of $10^{-11}$~erg cm$^{-2}$ s$^{-1}$.
$I_{\rm Fe}$ is kept frozen at $1.5 \times 10^{-4}$~ph cm$^{-2}$ s$^{-1}$.
For Model 2, $\Gamma$ is frozen at 1.73.  
Each spectral fit is performed over the 3--23~keV bandpass and has 45~$dof$.
\end{minipage}
\end{table*}

 \FloatBarrier

 \FloatBarrier  \clearpage
\setlength{\parskip}{0pt}

\subsection{Mkn~79 (Sy 1.2)}

\begin{figure*}
\includegraphics[angle=-90,width=1.0\textwidth]{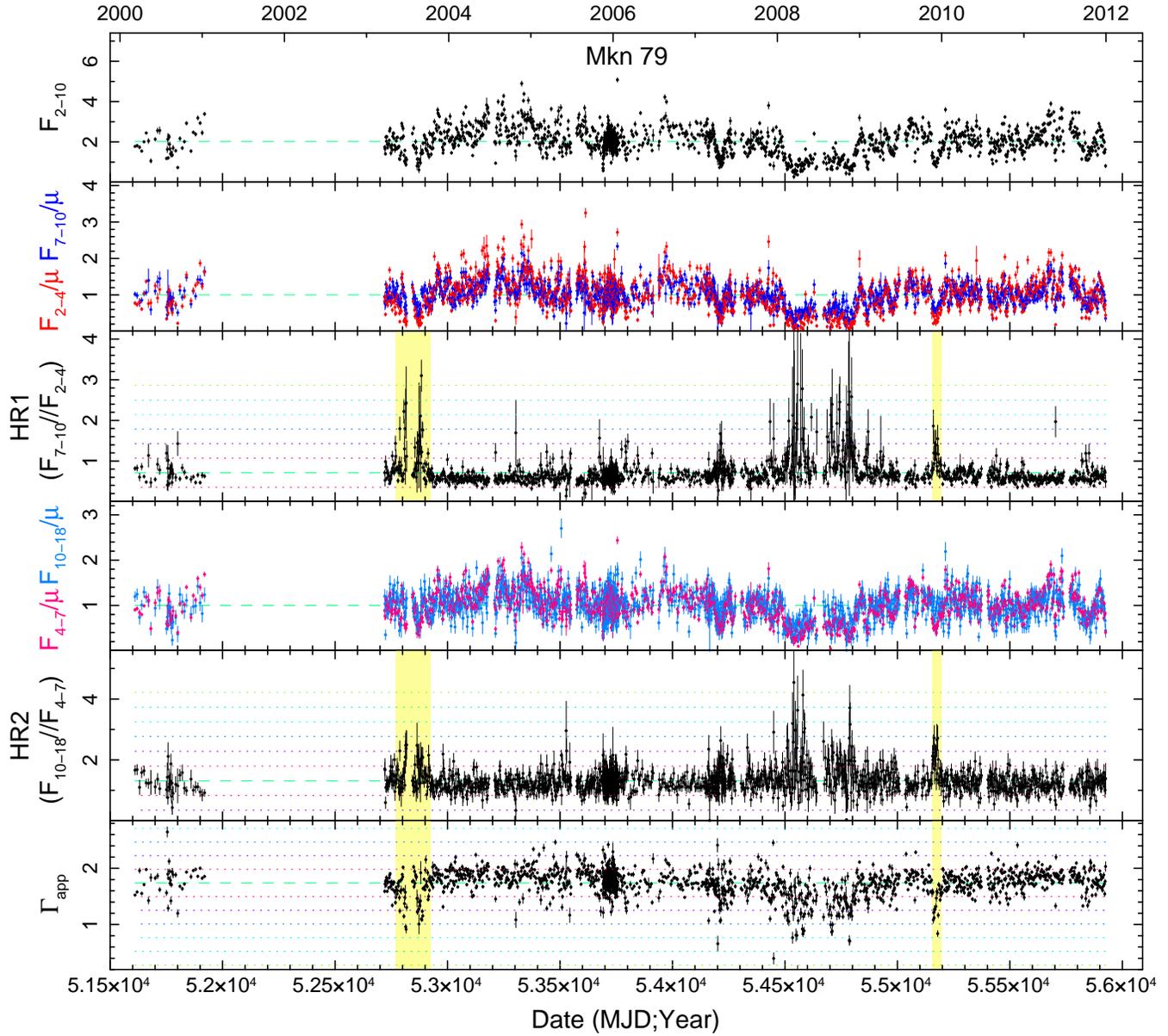}  
\caption{Same as Fig.~\ref{fig:mega3783}, but for the Sy~1.2
  Mkn~79. The yellow shaded areas in mid 2003 and in late 2009
  indicate ``secure'' eclipse events; see
  Figs.~\ref{fig:mega79zoom03_TR} and \ref{fig:mega79zoom99_TR} for
  zoom-ins on these periods.}
\label{fig:mega79}  
\end{figure*}

\begin{figure}
\includegraphics[angle=-90,width=0.48\textwidth]{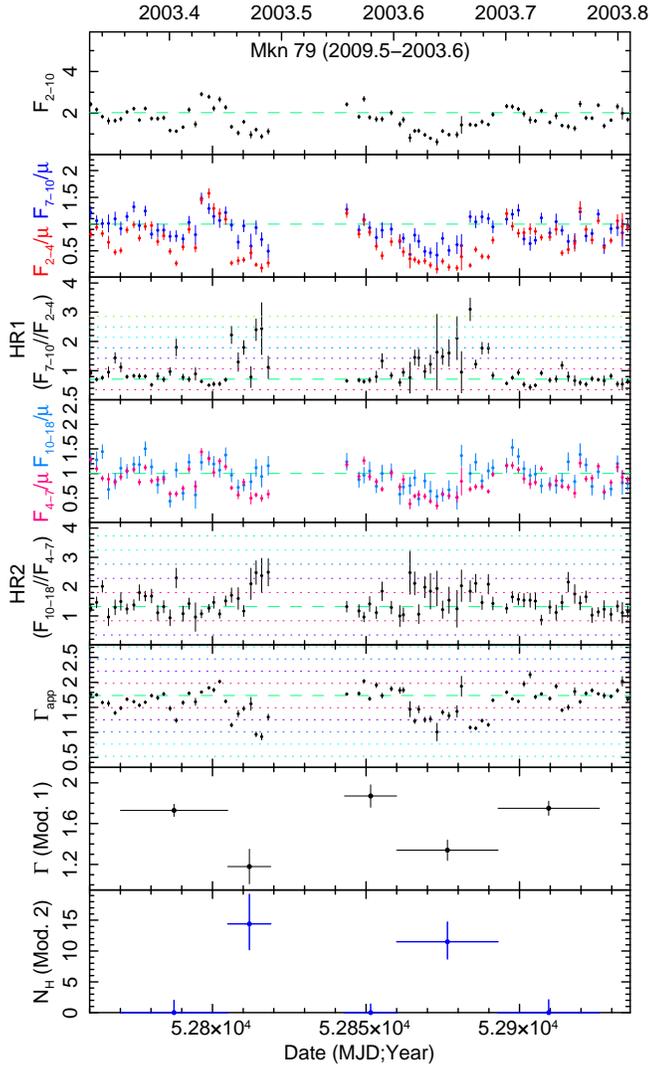}  
\caption{Same as Fig.~\ref{fig:mega79}, but a zoom-in on two
  ``secure B'' occultation events in mid-2003. The bottom two panels
  show $\Gamma$ from Model 1 ($N_{\rm H}$ set to 0) and $N_{\rm H}$ in
  units of $10^{22}$~cm$^{-2}$ from Model 2 ($\Gamma$ frozen at
  1.78).}
\label{fig:mega79zoom03_TR}  
\end{figure}

\begin{figure}
\includegraphics[angle=-90,width=0.48\textwidth]{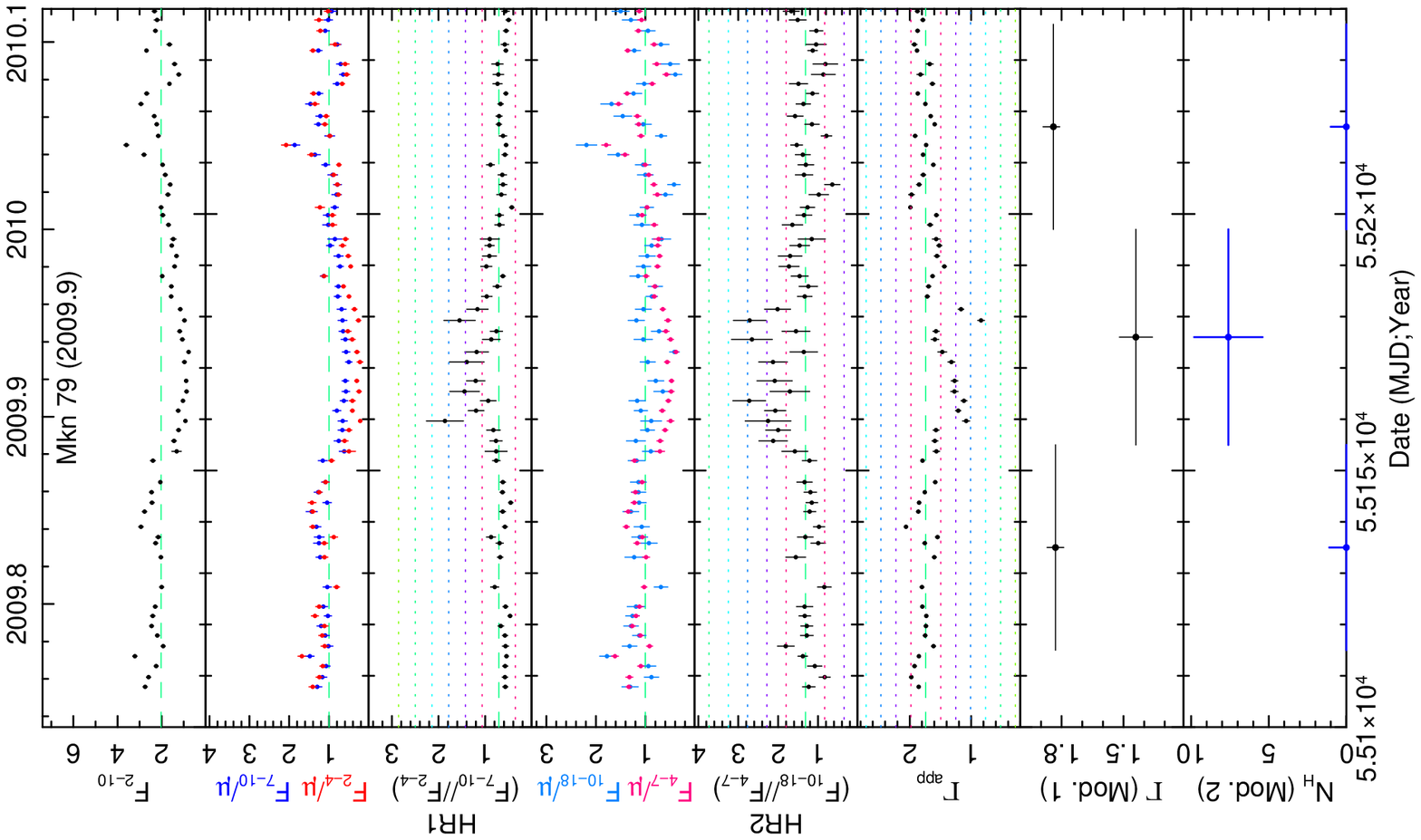}  
\caption{Same as Fig.~\ref{fig:mega79}, but a zoom-in on the
  ``secure B'' occultation event in late 2009. The bottom two panels
  show $\Gamma$ from Model 1 ($N_{\rm H}$ set to 0) and $N_{\rm H}$ in
  units of $10^{22}$~cm$^{-2}$ from Model 2 ($\Gamma$ frozen at
  1.84).}
\label{fig:mega79zoom99_TR}  
\end{figure}

$\bullet$ \textit{2003.5 (``Secure B'') and 2003.6 (``Secure B''):}
The long-term light curves are plotted in Fig.~\ref{fig:mega79}.  As
seen there and in Fig.~\ref{fig:mega79zoom03_TR}, during MJD $\sim$
52805--52818 (2003.5), $HR1$ increases to values ranging from 0.90
(just above $\langle HR1 \rangle$) to 2.4, with large scatter.  The
average and peak values of $HR1$ during this period correspond to 2.8$\sigma$
and 5$\sigma$ deviation, respectively.  $\langle\Gamma_{\rm
  app}\rangle$ during this time is 1.25, with a minimum value of 0.92
(3.2$\sigma$ deviation) at MJD 52816.0.  The 10--18~keV continuum does
not change as rapidly as the $<$7~keV continuum during this time;
$HR2$ roughly doubled, to a $2\sigma$ deviation.  No data were taken during
MJD 50821--50839 due to a sun-angle gap. By MJD 52840, $HR1$ has
returned to its mean value of 0.85.

During MJD $\sim$ 52861--52891 (2003.6), $HR1$ increases again,
showing deviations as high as $3-6\sigma$, with an average deviation
of 2.2$\sigma$ during this time.  During this time, $\Gamma_{\rm app}$
reaches values as low as 1.0--1.1 (2.0--2.8$\sigma$), with large
scatter; $\langle\Gamma_{\rm app}\rangle=1.39$.

We bin up data from MJD 52770--52925 into five bins of
approximately 15--30 d as listed in Table~\ref{tab:79TRtable}. This
yields only one time bin for each of the two candidate events
(spectra 2 and 4), but smaller bin sizes would have yielded poor
parameter constraints.  We fit 3--23~keV data, keeping $R$ frozen at
0.7 (Rivers \etal\ 2013).  We first test Model 1, with $N_{\rm H}$
frozen at 0, $\Gamma$ free, and $I_{\rm Fe}$ frozen at
$5\times10^{-5}$ ph cm$^{-2}$ s$^{-1}$.  Best-fitting values are listed in
Table~\ref{tab:79TRtable}; $\Gamma$ reaches values of 1.18$\pm$0.17
and 1.34$\pm$0.10 in spectra 2 and 4, respectively.  We then add a
neutral column of gas with \textsc{zphabs} for Model 2, keeping
$\Gamma$ frozen at 1.78, the average from spectra 1, 3, and 5.
$N_{\rm H}$ is $14.4^{+4.8}_{-4.2}$ and $11.5^{+3.2}_{-2.8} \times
10^{22}$~cm$^{-2}$ for spectra 2 and 4, respectively, with upper
limits $\sim 2 \times 10^{22}$~cm$^{-2}$ for spectra 1, 3, and
5. Best-fitting values of $\Gamma$ for Model 1 and $N_{\rm H}$ for Model 2
are plotted in Fig.~\ref{fig:mega79zoom03_TR}.

Values of $\chi^2_{\rm red}$ are virtually identical between Models
1 and 2, and span 0.46--0.81.  That is, the observed spectral
flattening \textit{could} potentially be due to a combination of
$\Gamma$ reaching extremely low values and spectral pivoting of the
power law in or near the 10--18~keV band.  However, adhering to our
assumption that $\Gamma$ should not dip below 1.5, the occultation
interpretation is then preferred in each of these cases.

We assign durations based on the $HR1$ light curve, but these
durations are approximate due to large scatter in $HR1$.  For the
2003.5 event, we observed ingress, but egress likely occurred during
the monitoring gap; limits to the duration are 12.0--39.4 d.  We
cannot rule out that the possibility that the peak of the event
occurred during the gap, in which case peak $N_{\rm H}$ might be
higher than that observed.  For the (complete) 2003.6 event, we
estimate a duration of 34.5--37.9 d.

$\bullet$ \textit{2009.9 (``Secure B'')}: At MJD$\sim55155-55197$, 
some values of $HR1$ deviate as high as the 1--3$\sigma$ level,
but $\langle HR1 \rangle$ corresponds to the $\sim1\sigma$ level.
$F_{2-10}$ is about half the average value, but $F_{10-18}$ stays
roughly constant, yielding an $HR2$ peak at the $\sim2-2.5\sigma$
level albeit with large errors.  We sum all spectra taken from MJD
55155 -- 55197, plus all data in the $\sim$40 d periods before and
after this period (spectra 6, 7, and 8 in Table~\ref{tab:79TRtable}).
$I_{\rm Fe}$ is kept frozen at $4\times10^{-5}$~ph~cm$^{-2}$
s$^{-1}$.  For spectrum \#7, we apply Model 2 by freezing $\Gamma$
at 1.84, the average of $\Gamma$ from spectra 6 and 8; this yields
$N_{\rm H} = 7.6\pm2.2 \times 10^{22}$~cm$^{-2}$.  We estimate a
duration of $19.6-40.0$ d from the $HR1$ light curve.

$\bullet$ There are additional deviations in the long-term $HR1$
and/or $\Gamma_{\rm app}$ light curves which tend to catch the eye,
but these deviations do not fulfill our selection criteria and are
rejected:

At 2007.3, at MJD $54208-54219$, 
$HR1$ increases to $\sim$1.2, but these are only $\sim1-2\sigma$
deviations. There are only a few points at $1\sigma$ deviation in the
$HR2$ light curve as well; the 10--18 and 2--10~keV continua are both
$\sim40$ per cent below average during this period.
$\langle\Gamma_{\rm app}\rangle$ during this period is 1.37, just
under a 1$\sigma$ deviation, and with large scatter. Time-resolved
spectroscopy to data summed between MJD $54208$ and $54219$, assuming
that $\Gamma$ is frozen at 1.78, yields $N_{\rm H} < 9.2 \times
10^{22}$~cm$^{-2}$.

At 2008.2, MJD$\sim 54535-54590$,   
and several times near the end of 2008, MJD $\sim54725-54830$, 
there are deviations in $HR1$ up to 2--5$\sigma$ levels, but with very
large scatter and frequently large uncertainties.  Roughly a third of
the values of $\Gamma_{\rm app}$ during these times are $<$1.3, but
there is very large scatter, and average values of $\Gamma_{\rm app}$
are at the $\sim1\sigma$ level of deviation.  Both the 10--18 and
2--10~keV continuua were $\sim30-60$ per cent of their average values, and
follow-up time-resolved spectroscopy cannot confirm significant
increases in $N_{\rm H}$ (upper limits are in the range $3-5 \times
10^{22}$~cm$^{-2}$).

Near 2009.1, at MJD $\sim54856-54876$, 
a couple of $HR1$ points reach the 2$\sigma$ level, but errors are
large, and $\langle HR1 \rangle$ during this time is a
$\ga1\sigma$ deviation.  All continuum flux light curves
experience dips during this time to relatively low
levels. Time-resolved spectroscopy cannot confirm any significant
increase in $N_{\rm H}$.

\begin{figure}
\includegraphics[angle=-90,width=0.48\textwidth]{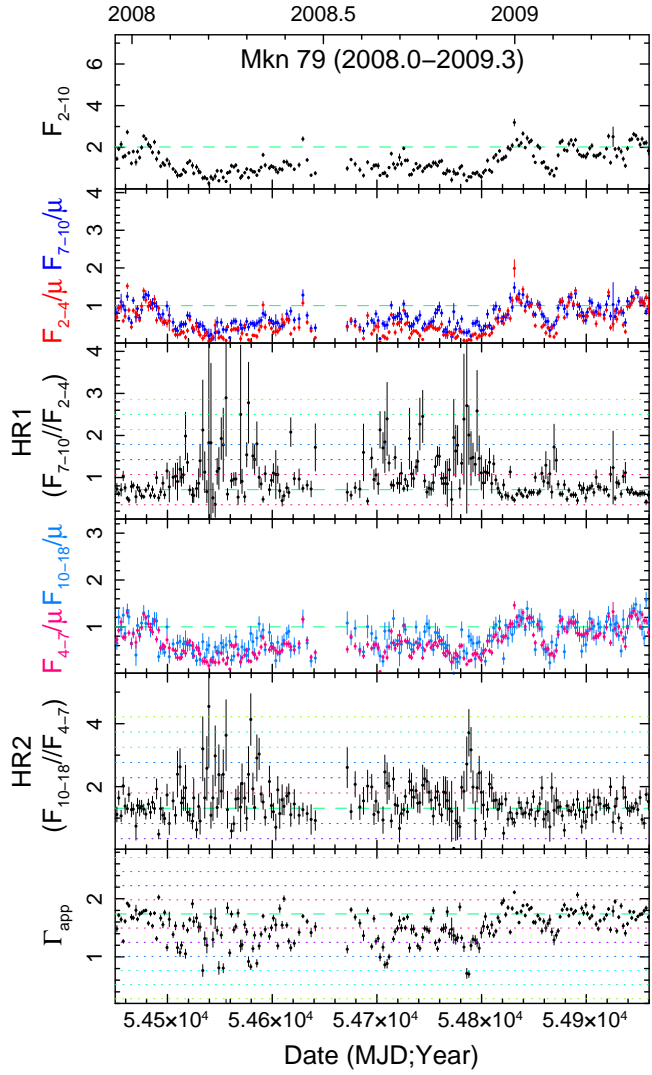}  
\caption{Same as Fig.~\ref{fig:mega79}, but a zoom-in on four
  rejected events in 2008 and early 2009.}
\label{fig:mega79zoom89rej}  
\end{figure}

Fig.~\ref{fig:mega79zoom89rej} shows a zoom-in on the 2008.0--2009.3 data with rejected events.

\begin{table*}
\begin{minipage}{160mm}
\caption{Results of time-resolved spectral fits to Mkn~79, mid-2003 and late 2009}
\label{tab:79TRtable} 
\begin{tabular}{lccccccccc} \hline
Start--stop            & Expo & \multicolumn{4}{c}{Model 1: $N_{\rm H}$ frozen at 0}        & \multicolumn{4}{c}{Model 2: $N_{\rm H}$ free; $\Gamma$ frozen} \\
   (MJD)               & (ks) & $\chi^2_{\rm red}$ & $\Gamma$      & $A_1$        & $F_{2-10}$ & $\chi^2_{\rm red}$ &  $N_{\rm H}$ ($10^{22}$~cm$^{-2}$) & $A_1$  & $F_{2-10}$ \\ \hline
52770.2--52804.3 (\#1) & 14.8 & 0.67             & $1.73\pm0.06$ & $4.7\pm0.5$  & 1.98     & 0.70             & $<$2.0            & $5.1\pm0.1$        & 1.97 \\   
52806.2--52818.2 (\#2) &  3.4 & 0.50             & $1.18\pm0.17$ & $1.2\pm0.3$  & 1.22     & 0.50             & $14.4^{+4.8}_{-4.2}$ & $4.9^{+0.6}_{-0.5}$  & 1.14 \\  
52843.7--52858.3 (\#3) &  4.7 & 0.64             & $1.87\pm0.11$ & $6.0\pm1.2$  & 1.89     & 0.69             & $<$1.4            & $5.1\pm0.2$         & 1.85 \\  
52861.0--52891.3 (\#4) &  9.7 & 0.73             & $1.34\pm0.10$ & $1.7\pm0.3$  & 1.26     & 0.80             & $11.5^{+3.2}_{-2.8}$ & $4.7\pm0.3$        & 1.18 \\ 
52895.8--52925.6 (\#5) & 11.6 & 0.72             & $1.75\pm0.07$ & $4.6\pm0.7$  & 1.81     & 0.74             & $<$2.1            & $5.0^{+0.2}_{-0.1}$   & 1.80 \\ \hline  
55115.8--55153.8 (\#6) & 14.6 & 0.92             & $1.83\pm0.04$ & $6.5\pm0.7$  & 2.12     & 0.94             & $<1.1$            & $6.4\pm0.1$        & 2.08 \\ 
55155.8--55195.1 (\#7) & 18.6 & 0.79             & $1.44\pm0.08$ & $1.8\pm0.3$  & 1.17     & 1.15             & $7.6\pm2.2$       & $4.4\pm0.3$        & 1.17 \\ 
55197.9--55235.7 (\#8) & 18.0 & 0.69             & $1.84\pm0.05$ & $6.2\pm0.4$  & 1.99     & 0.94             & $<1.0$            & $5.9\pm0.1$        & 1.94  \\ \hline 
\end{tabular}\\
$A_1$ is the 1~keV normalization of the power law in
  units of $10^{-3}$~ph~cm$^{-2}$ s$^{-1}$ keV$^{-1}$. $F_{2-10}$ is
  the observed/absorbed model flux in units of $10^{-11}$ erg
  cm$^{-2}$ s$^{-1}$.  $I_{\rm Fe}$ is kept frozen at 5 and $4 \times
  10^{-5}$~ph~cm$^{-2}$ s$^{-1}$ for the 2003 and 2009 spectra,
  respectively.  For Model 2, $\Gamma$ is frozen at 1.78 and 1.84 for
  the 2003 and 2009 spectra, respectively.  Each spectral fit is
  performed over the 3--23~keV bandpass and has 45 $dof$.
\end{minipage}
\end{table*}

\clearpage

 \FloatBarrier

\clearpage

\subsection{Mkn~509 (Sy 1.2)} 

\begin{figure*}
\includegraphics[angle=-90,width=1.0\textwidth]{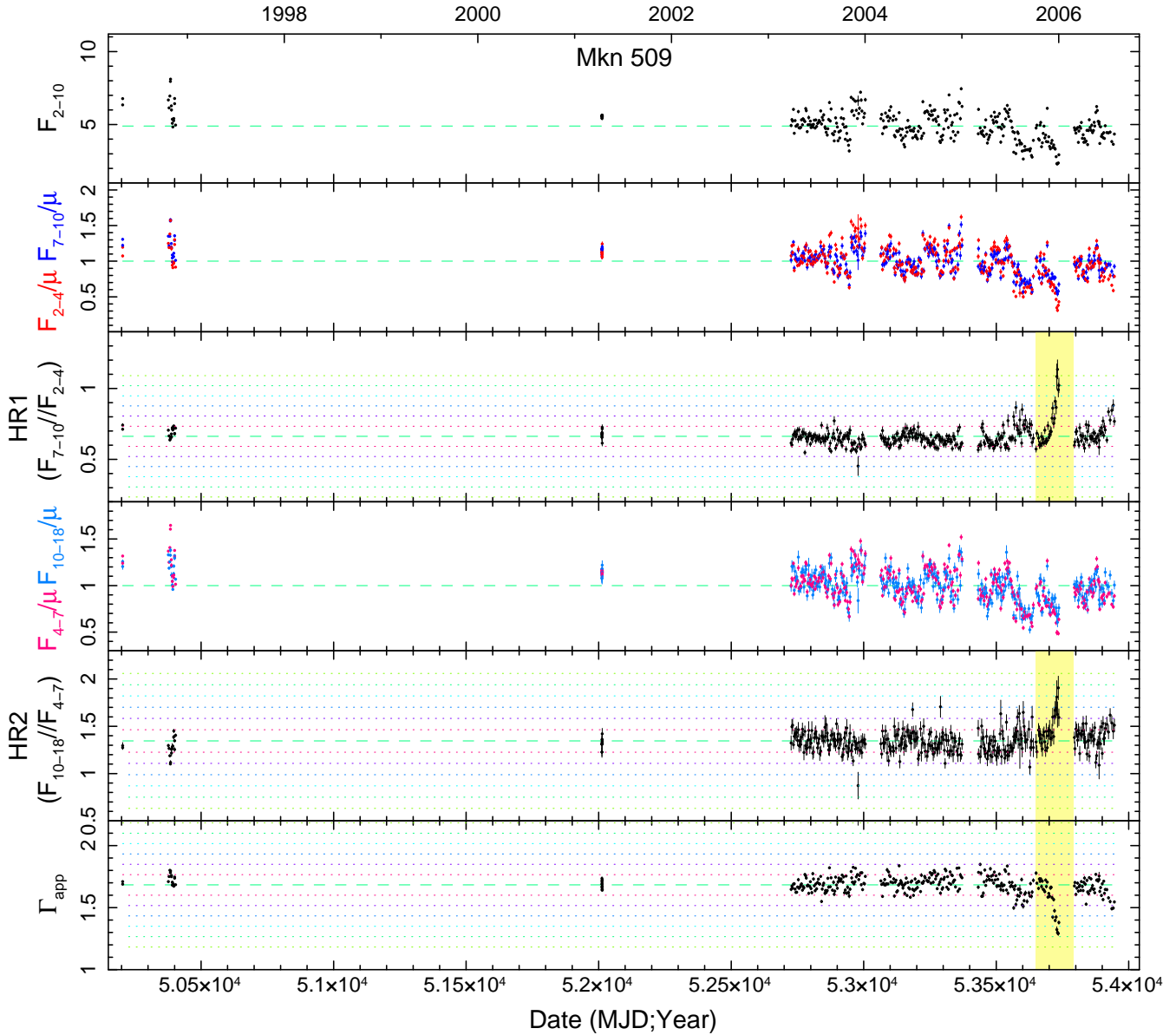}               
\caption{Same as Fig.~\ref{fig:mega3783}, but for the Sy~1.2
  Mkn~509. The yellow shaded area in late 2005 indicates a candidate
  eclipse event; see Fig.~\ref{fig:mega509zoom_TR} for a zoom-in on
  this period. }
\label{fig:mega509}  
\end{figure*}

\begin{figure}
\includegraphics[angle=-90,width=0.48\textwidth]{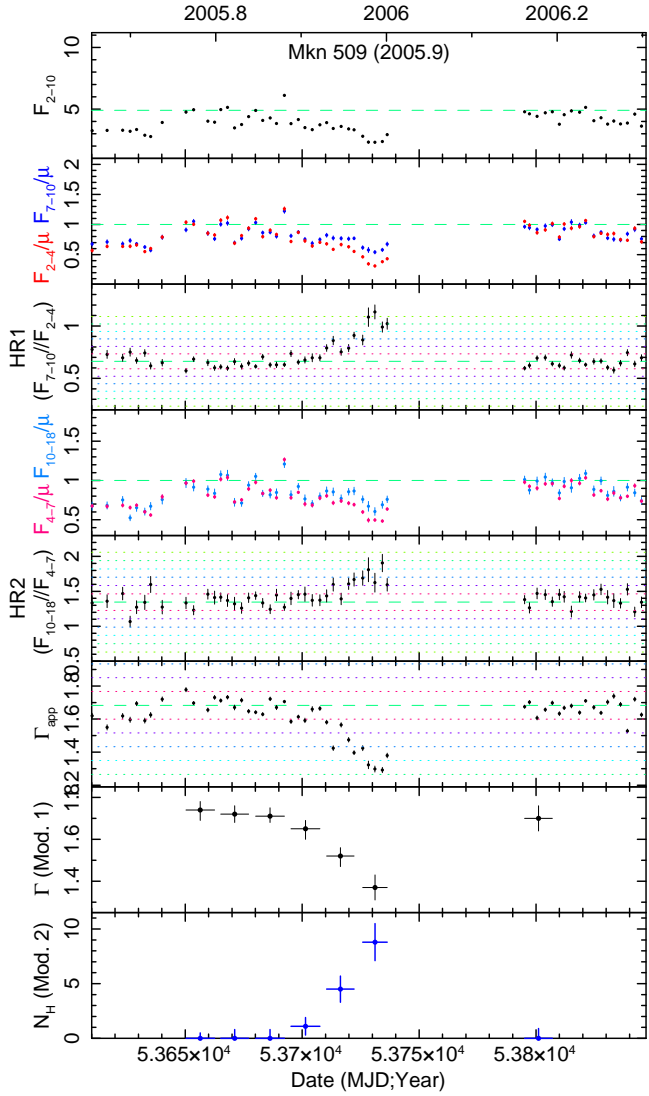}       
\caption{Same as Fig.~\ref{fig:mega509}, but a zoom-in on the candidate
  occultation event in late 2005.  The bottom two panels show $\Gamma$
  from Model 1 ($N_{\rm H}$ set to 0) and $N_{\rm H}$ in units of
  $10^{22}$~cm$^{-2}$ from Model 2 ($\Gamma$ frozen at 1.72).}
\label{fig:mega509zoom_TR}  
\end{figure}

The long-term light curves are plotted in Fig.~\ref{fig:mega509},
with a zoom-in on late 2005 plotted in Fig.~\ref{fig:mega509zoom_TR}.

\textit{2005.9 (``Secure B''):} The $HR1$ light curve in
Fig.~\ref{fig:mega509zoom_TR} shows an upturn from the baseline
value of $\sim$0.6 to 1.2 over a 30 d period in late 2009, with a
$\sim5\sigma$ peak deviation by MJD 53736, after which monitoring was
interrupted for 59 d due to sun-angle constraints.  $\Gamma_{\rm app}$
drops from about 1.7 to 1.3 during this time, a $\sim4\sigma$
deviation.

During this time, \textit{RXTE} monitored the source once every 3 d.
For time-resolved spectroscopy, we grouped the individual spectra into
bins of 15 d each.  We fit the 3--23~keV spectra keeping $R$ and
$I_{\rm Fe}$ frozen at their respective time-averaged values of 0.15
and $7\times 10^{-5}$~ph cm$^{-2}$ s$^{-1}$; results are listed in
Table~\ref{tab:509TRtable}.  With Model 1, acceptable fits
($\chi^2_{\rm red}$ = 0.52--0.68) are obtained, although $\Gamma$
flattens to the very low value of $1.37\pm0.06$ during spectrum \#6.
We then test Model 2, freezing $\Gamma$ to 1.72, the average of the
values in spectra \#1--3 and \#7, and close to the time-averaged value
found by Rivers \etal\ (2011), 1.75.  Acceptable fits spanning a
nearly identical range of $\chi^2_{\rm red}$ are obtained, with
$N_{\rm H}$ peaking at $8.8\pm1.7 \times 10^{22}$~cm$^{-2}$ during
spectrum \#6, and with upper limits during spectra \#1--3 and
7. Best-fitting values of $\Gamma$ for Model 1 and $N_{\rm H}$ for Model 2
are plotted in Fig.~\ref{fig:mega509zoom_TR}.

Under the assumption that an eclipse occurred, it is difficult to pin
down the exact duration because the event ended sometime during the 60
d sun-angle gap, between MJD 53736 and 53795; peak $N_{\rm H}$ could
thus be higher than that measured during spectrum $\#6$, $8.8\pm1.7
\times 10^{22}$~cm$^{-2}$.  From the $HR1$ light curve, we assign a
duration of 26--94 d.  If the occulting cloud is symmetric in density
along the transverse direction, then the total duration can be
$\sim55-90$~d, with peak $N_{\rm H}$ occurring immediately before
the sun-angle gap in the former case (i.e., during the times
constrained by spectrum $\#5$), or with peak $N_{\rm H}$ slightly
higher and occurring $\sim$15 d into the gap in the latter case.

\begin{table*}
\begin{minipage}{160mm}
\caption{Results of time-resolved spectral fits to Mkn~509, 2005.9}
\label{tab:509TRtable} 
\begin{tabular}{lccccccccc}\hline
Start--stop            & Expo & \multicolumn{4}{c}{Model 1: $N_{\rm H}$ frozen at 0}        & \multicolumn{4}{c}{Model 2: $N_{\rm H}$ free; $\Gamma$ frozen at 1.72} \\
   (MJD)               & (ks) & $\chi^2_{\rm red}$ & $\Gamma$      & $A_1$        & $F_{2-10}$ & $\chi^2_{\rm red}$ &  $N_{\rm H}$ ($10^{22}$~cm$^{-2}$) & $A_1$  & $F_{2-10}$ \\ \hline
53650.31--53662.56 ($\#1$) & 6.4 & 0.67 & $1.74^{+0.04}_{-0.05}$  &  $11.7\pm1.0$      & 4.32   & 0.68 &   $<0.5$   & $11.4\pm0.2$  & 4.31 \\
53665.18--53677.10 ($\#2$) & 8.0 & 0.63 & $1.72\pm0.04$        &  $11.1\pm0.9$      & 4.22   & 0.63 &   $<0.8$   & $11.1\pm0.2$  & 4.22 \\
53680.11--53692.37 ($\#3$) & 9.7 & 0.52 & $1.71^{+0.04}_{-0.03}$  &  $11.8^{+0.9}_{-0.7}$ & 4.55  & 0.53 &   $<0.8$   & $12.0^{+0.2}_{-0.1}$ & 4.55 \\
53695.25--53707.58 ($\#4$) & 8.3 & 0.54 & $1.65^{+0.04}_{-0.05}$  &  $8.6\pm0.7$       & 3.68  & 0.61  &   $1.1\pm0.8$  & $10.0^{+0.4}_{-0.3}$  & 3.64 \\
53710.46--53722.19 ($\#5$) & 8.2 & 0.53 & $1.52^{+0.04}_{-0.05}$  &  $6.5^{+0.6}_{-0.5}$ & 3.45   & 0.73  &  $4.5\pm1.2$  & $10.4\pm0.3$   & 3.33 \\
53725.91--53736.34 ($\#6$) & 7.8 & 0.67 & $1.37\pm0.06$        &  $3.7\pm0.4$      & 2.51   & 0.80  & $8.8\pm1.7$  & $8.4^{+0.4}_{-0.3}$  & 2.37 \\
53795.10--53806.97 ($\#7$) & 7.9 & 0.68 & $1.70\pm0.06$        & $11.5^{+1.0}_{-0.8}$ & 4.56   & 0.65 &   $<0.9$   & $12.1\pm0.2$ & 4.58 \\ \hline
\end{tabular}\\
$A_1$ is the 1~keV normalization of the power law in units of
  $10^{-3}$ ph cm$^{-2}$ s$^{-1}$ keV$^{-1}$. $F_{2-10}$ is the
  observed/absorbed model flux in units of $10^{-11}$ erg cm$^{-2}$
  s$^{-1}$.  $I_{\rm Fe}$ is kept frozen at $7 \times 10^{-5}$ ph
  cm$^{-2}$ s$^{-1}$.  Each spectral fit is performed over the
  3--23~keV bandpass and has 45 $dof$.
\end{minipage}
\end{table*}

\clearpage

\clearpage

\subsection{MR~2251--178 (Sy~1.5/QSO)}

\begin{figure*}
\includegraphics[angle=-90,width=1.0\textwidth]{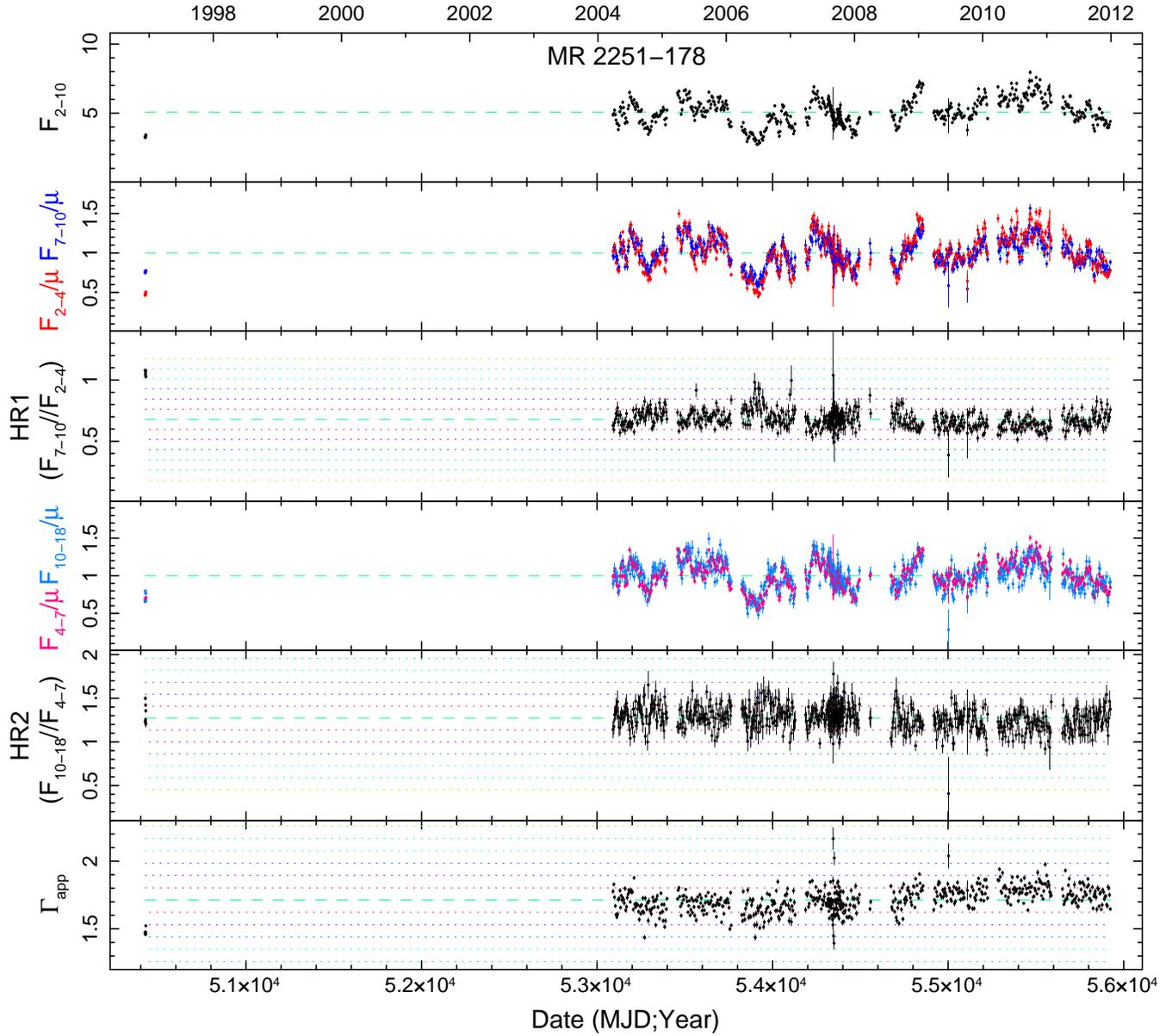}  
\caption{Same as Fig.~\ref{fig:mega3783}, but for the Sy~1.5/QSO MR~2251--178.}  
\label{fig:mega2251}  
\end{figure*}

$\bullet$ \textit{1996 (``Secure A'')} This source was observed by
\textit{RXTE} for 3~d in 1996 MJD 50426-9, then subjected to sustained
monitoring during 2004 March -- 2011 December, as seen in
Fig.~\ref{fig:mega2251}.  Values of $HR1$ in 1996 are $\sim1.0-1.1$,
a $\sim4\sigma$ deviation from the mean value of 0.68 from the
2004--2011 data.  $\Gamma_{\rm app}$ during 1996 is 1.5, a
$\sim2-3\sigma$ deviation.

The X-ray spectrum of this source is characterized by absorption
from gas spanning a wide range of ionization states; Gofford et
al.\ (2011) quantify five such zones using combined \textit{Suzaku} +
\textit{Swift}-Burst Alert Telescope spectra.  Furthermore, there is evidence for the
column densities of at least some of these zones to vary over
timescales of months--years (Kaspi et al.\ 2004).  \textit{RXTE} is
not strongly sensitive to the effect of the highest ionization zones
of absorption, log($\xi$) $\ga3$, e.g., "Zones 3--5" in Gofford et
al.\ (2011). Any absorbing column we detect in excess of $N_{\rm Gal}$
may potentially be analogous to (or even potentially identified as) an
overdense region of "Zone 1" from Gofford et al.\ (2011; with
log($\xi$) $\sim-0.23$ and $N_{\rm H} \sim 5\times 10^{20}$
cm$^{-2}$), or the absorber with log($\xi$) $\sim+0.02$ and $N_{\rm H}
\sim 2.4 \times 10^{21}$~cm$^{-2}$ measured by Gibson et al.\ (2005)
using a 2002 \textit{Chandra}-High-Energy Transmission Grating
Spectrometer (HETGS) observation.\footnote{"Zone 2"
  from Gofford et al.\ (2011), which has log($\xi$) $\sim2.21$, may
  also potentially have a modest impact on \textit{RXTE}
  spectra. However, at the column density measured by Gofford et
  al.\ (2011), $6\times 10^{21}$~cm$^{-2}$, the effect on the X-ray
  continuum near 3--5~keV is only at the $\sim4$ per cent level.}  Here, we
fit the \textit{RXTE} spectra assuming full-covering neutral
absorption only.



We sum all the data during the 1996 campaign; the good exposure for
the PCA was 90.1~ks.  The good exposure time per High-Energy X-ray
Timing Experiment (HEXTE) cluster is
20.1/20.5~ks for A/B, with the source detected only out to 25~keV, so
we use PCA data only.  We fit 3--23~keV data, and assume no
reflection component and no high-energy rollover (see Rivers
\etal\ 2011a).  We first apply Model 1, keeping the Fe K$\alpha$
line energy and width $\sigma$ frozen at 6.0~keV and 0.9~keV,
respectively. This fit yields $\Gamma=1.52\pm0.01$, 
but $\chi^2/dof$ is 175.2/42 = 4.2, with poor data/model residuals
showing strong curvature up to the $\pm$6 per cent level below 10~keV. We
apply Model 2, keeping $\Gamma$ free, and obtain a fit with
$\chi^2/dof$ = 95.8/41 and data/model residuals $\la2$ per cent,
consistent with PCA calibration. Best-fitting model parameters are
$N_{\rm H} = 6.6^{+0.8}_{-1.4} \times 10^{22}$~cm$^{-2}$,
$\Gamma=1.73^{+0.02}_{-0.06}$, and $I_{\rm Fe}$ fell to $3.1\pm1.3
\times 10^{-5}$ ph cm$^{-2}$ s$^{-2}$.

\begin{figure}
\includegraphics[angle=-90,width=0.48\textwidth]{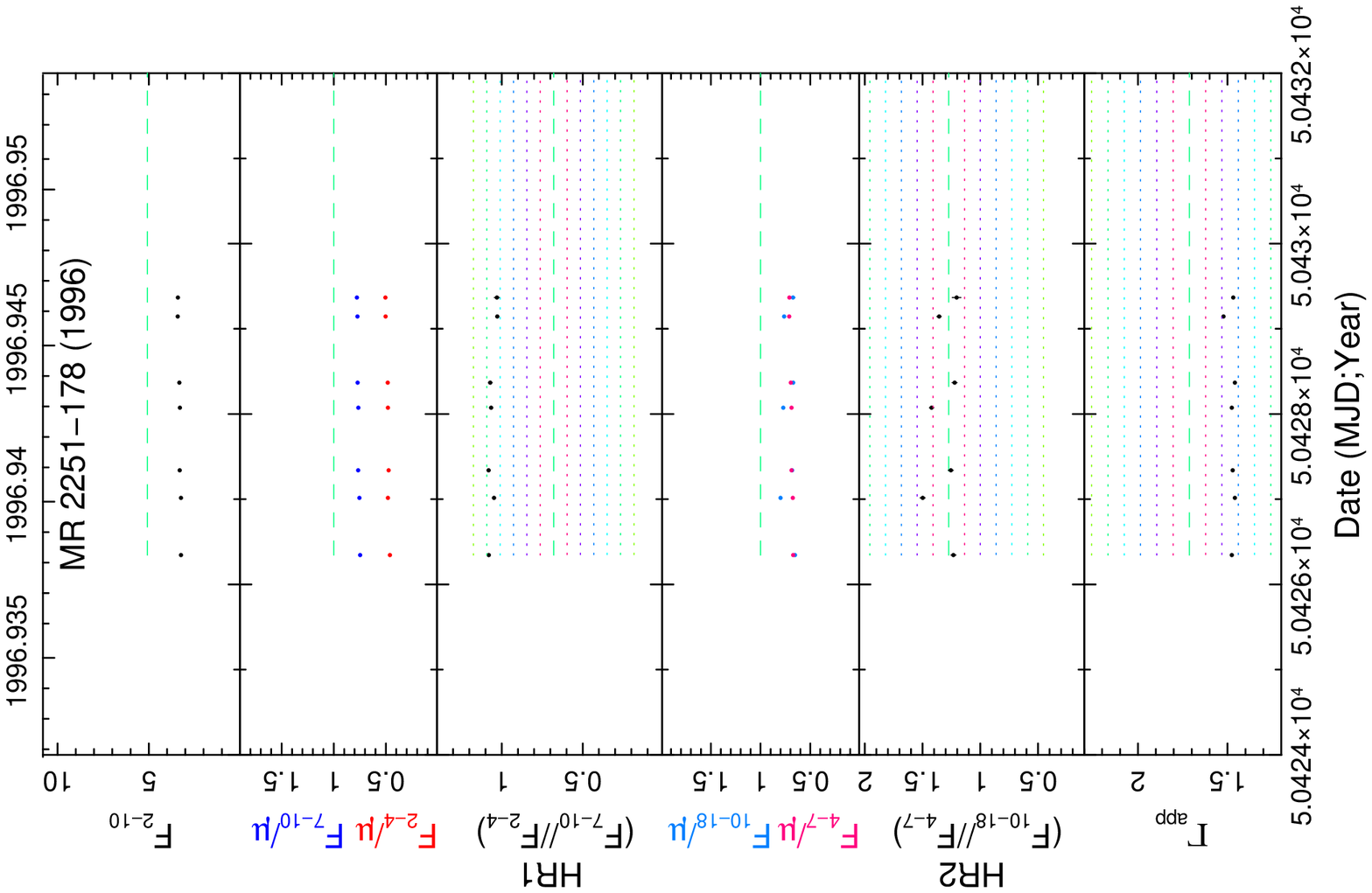} 
\caption{Same as Fig.~\ref{fig:mega2251}, but a zoom-in on the 1996 data.}  
\label{fig:mega2251zoom}  
\end{figure}

Fig.~\ref{fig:mega2251zoom} shows a zoom-in on the 1996 data;
we find no evidence then for strong variations in $HR1$ on timescales
$<$3~d.  Constraints on the duration of the event from \textit{RXTE} data alone are
poor; we can only set a lower limit of 3~d, the length of the 1996
campaign.  Peak $N_{\rm H}$ may of course be higher than that observed
in 1996.  Dadina (2007), fitting two \textit{BeppoSAX} observations in
1998 June and 1998 November, noted total columns consistent with
$\sim1-2\times$$N_{\rm H,Gal}$.  In the context of an absorption
event, this measurement suggests that the event had ended by 1998 June.
Reeves \& Turner (2000), fitting \textit{ASCA} spectra obtained in 1993
November--December, found $N_{\rm H} < 1 \times 10^{20}$~cm$^{-2}$,
implying an upper limit for the duration of the event of $\sim4.5$
yr.

\clearpage


\subsection{NGC~3227 (Sy 1.5)} 

\begin{figure*}
\includegraphics[angle=-90,width=1.0\textwidth]{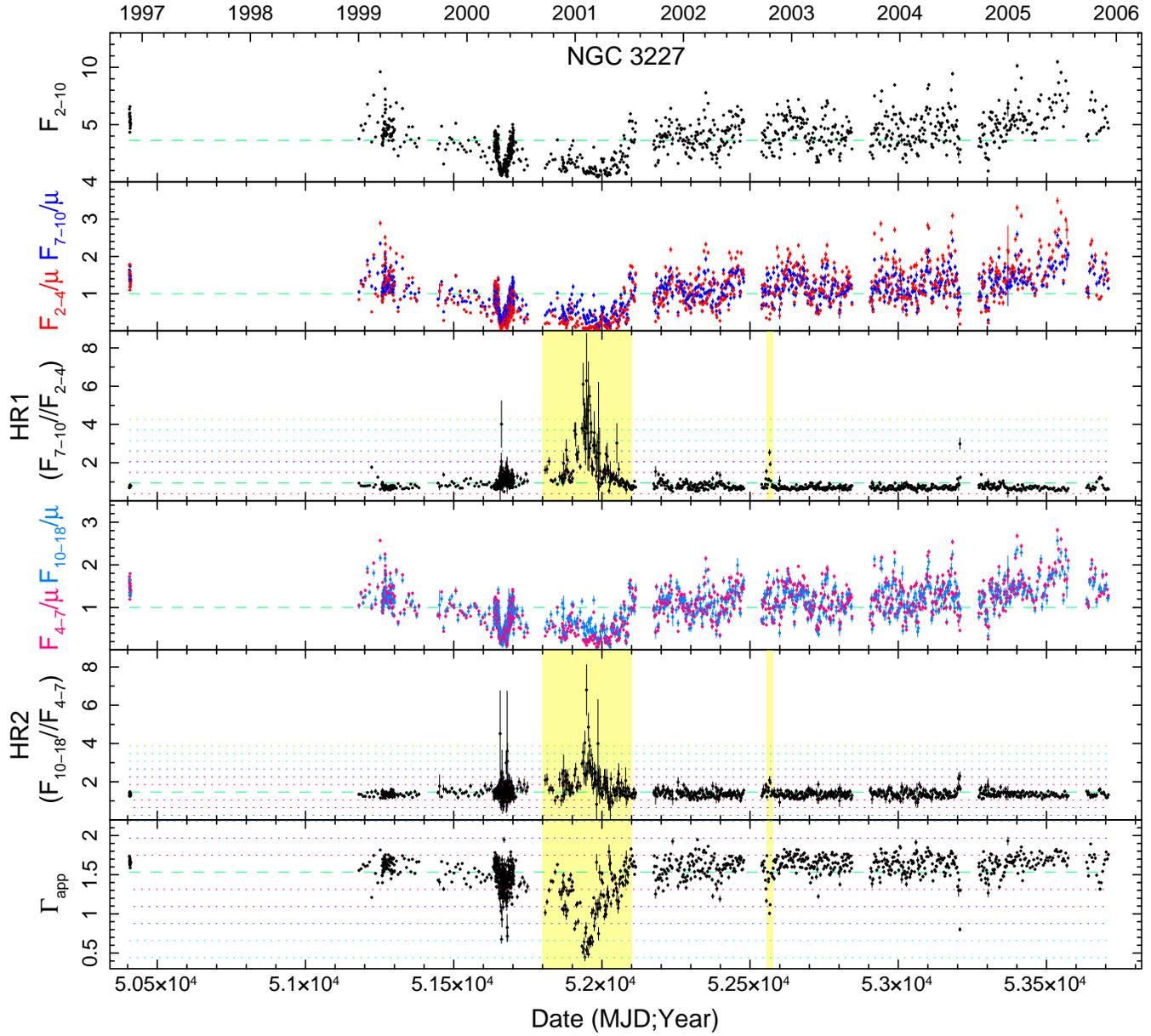} 
\caption{Same as Fig.~\ref{fig:mega3783}, but for the Sy 1.5
  NGC~3227. The yellow shaded areas indicate putative eclipse
  events. The reader is referred to Lamer \etal\ (2003) for details on
  the 2000--2001 event.  A zoom-in on late 2002 is shown in
  Fig.~\ref{fig:mega3227zoom02_TR}.  }
\label{fig:mega3227}  
\end{figure*}

The long-term light curves are plotted in Fig.~\ref{fig:mega3227}.

$\bullet$ \textit{2000.9--2001.2 (``Secure A''):} Lamer \etal\ (2003)
confirmed a complete eclipse event (ingress and egress) by a full-covering
cloud. They determined the cloud to be mildly ionized
via a contemporaneous \textit{XMM-Newton} observation, with 
log($\xi$, erg cm s$^{-1}$) $\sim -0.3 - 0$.
Our $HR1$, $HR2$, and $\Gamma_{\rm app}$ light curves show peak
deviations up to the $\ga6-8$, $\ga3-5$, and the 
$\sim4$$\sigma$ levels, respectively, and $\Gamma_{\rm app}$ gets as low as
$\sim0.5$.  We refer the reader to Lamer \etal\ (2003) for details of
time-resolved spectral fitting and fitting the $N_{\rm H}$($t$)
profile.  Lamer \etal\ (2003) found values of peak $N_{\rm H}$ of $19$
and $26 \times 10^{22}$~cm$^{-2}$ based on fits to the $N_{\rm
  H}$($t$) profile using a uniform-density sphere and a $\beta$-profile
fits, respectively; we adopt the latter value for the rest of this
paper.  Based on our $HR1$ light curve, we adopt a duration of 77--94
d, similar to values used by Lamer \etal\ (2003).

\begin{figure}
\includegraphics[angle=-90,width=0.48\textwidth]{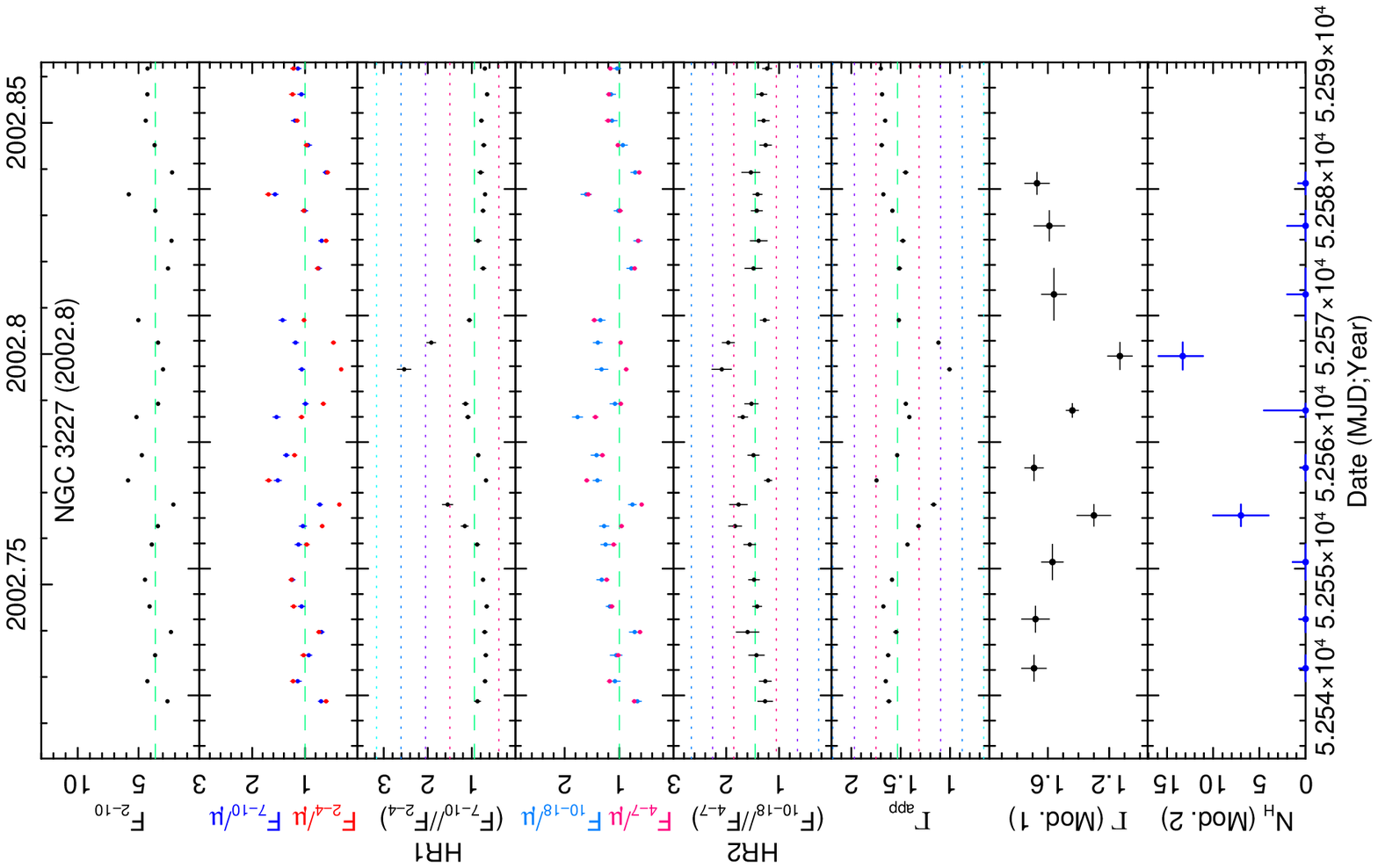} 
\caption{Same as Fig.~\ref{fig:mega3227}, but a zoom-in on the
  ``secure B'' occultation event in late 2002.  The bottom two panels
  show $\Gamma$ from Model 1 ($N_{\rm H}$ set to 0) and $N_{\rm H}$ in
  units of $10^{22}$~cm$^{-2}$ from Model 2 ($\Gamma$ frozen at
  1.61).}
\label{fig:mega3227zoom02_TR}  
\end{figure}

$\bullet$ \textit{2002.8 (``Secure B''):} As seen in
Fig.~\ref{fig:mega3227zoom02_TR}, two consecutive points at MJD
52565.7 and 52567.8 show anomalously high $HR1$ values: $2.54\pm0.15$
(2.9$\sigma$) and $1.92\pm0.10$ (1.8$\sigma$), respectively.
$\Gamma_{\rm app} = 1.01\pm0.02$ and $1.12\pm0.02$, $\sim2\sigma$
deviations.  $HR2$ shows only $\sim1.5\sigma$ deviations.

We sum spectra obtained during MJD 52541--52581 in groups of two to
match the time resolution of the putative event, and fit each 3--23
keV spectrum, keeping $I_{\rm Fe}$ frozen at $6\times10^{-5}$ ph
cm$^{-2}$ s$^{-1}$.  Fitting spectrum \#7 with Model 1 yields a fit
with $\chi^2_{\rm red}=1.34$ and with strong systematic curvature in
the data/model residuals $\sim3-7$~keV; $\Gamma$ is $1.13\pm0.08$.
Applying Model 1 to the other spectra yields an average value of
$\Gamma$ of 1.61.  Applying Model 2 to spectrum \#7 with $\Gamma$
frozen at 1.61 fixed data/model residuals; 
$\chi^2_{\rm red}=0.55$, and $N_{\rm H} = 13.3^{+2.6}_{-2.2} \times 10^{22}$
cm$^{-2}$. 
Results for all spectra are listed in 
Table~\ref{tab:3227TRtable02}.  From the start/stop times of the two
observations, limits on the duration of the event are 2.1--6.6~d.

Using \textit{XMM-Newton}, Markowitz et al.\ (2009) modeled two outflowing zones of ionized
absorption, both with $N_{\rm H} \sim 1-2 \times 10^{21}$~cm$^{-2}$,
and with log($\xi$) $\sim 1.2-1.4$ and $\sim$2.9. Neither are expected
to significantly impact the time-resolved PCA spectra.  Due to the
much higher column and lower ionization, it is likely that the eclipse
events detected with \textit{RXTE} are likely distinct from these outflowing
warm absorbers.

\begin{figure}
\includegraphics[angle=-90,width=0.48\textwidth]{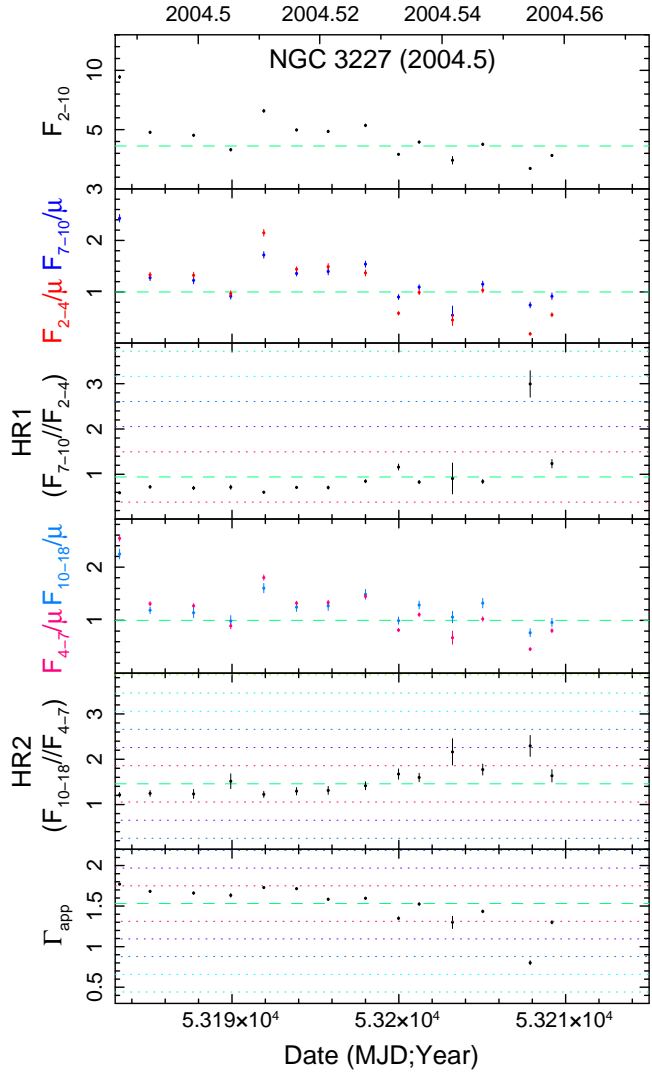}  
\caption{Same as Fig.~\ref{fig:mega3227}, but a zoom-in on the rejected event in 2004.}  
\label{fig:mega3227zoom2004}  
\end{figure}

$\bullet$ Additional events do not meet our selection criteria and
are rejected due to being single points only (failing criterion 2).
As seen in Fig.~\ref{fig:mega3227zoom2004}, a single point at MJD
53207 (2004 Jul 21) shows a very strong $HR1$ deviation up to
$3.0\pm0.29$, a $3.7\sigma$ deviation.  $\Gamma_{\rm app} =
0.80\pm0.03$, a 3.2$\sigma$ deviation; $HR2$ shows a $2\sigma$
deviation.

At 2000.3, $HR1$ goes to a 6$\sigma$ deviation.  There are four single
points in the period 2000.3--2000.4 where $HR2$ goes to a
$\geq$2$\sigma$ deviation, but with extremely large errors.  Again,
however, each of these events fails criterion 2 and is rejected.


The bottom two panels in Fig.~\ref{fig:mega3227zoom02_TR} also
suggest a significant increase in $N_{\rm H}$ for spectrum \#4, at MJD
52555.  Given the fact that the continuum flux was relatively low in
all wavebands, the near-identical values of $\chi^2_{\rm red}$ between
Models 1 and 2, and especially the fact that only 1 point shows a
$>$1$\sigma$ deviation in both the $HR1$ and $\Gamma_{\rm app}$ light
curves (thus failing criterion 2), we do not consider this as an
event.

\begin{table*}
\begin{minipage}{160mm}
\caption{Results of time-resolved spectral fits to NGC~3227, late 2002}
\label{tab:3227TRtable02} 
\begin{tabular}{lccccccccc} \hline
Start--stop            & Expo & \multicolumn{4}{c}{Model 1: $N_{\rm H}$ frozen at 0}        & \multicolumn{4}{c}{Model 2: $N_{\rm H}$ free; $\Gamma$ frozen} \\
   (MJD)               & (ks) & $\chi^2_{\rm red}$ & $\Gamma$      & $A_1$        & $F_{2-10}$ & $\chi^2_{\rm red}$ &  $N_{\rm H}$ ($10^{22}$~cm$^{-2}$) & $A_1$  & $F_{2-10}$ \\ \hline
52541.11--52543.16 (\#1) &1.7&0.58&  $1.69\pm0.08$ & $9.7^{+1.5}_{-1.3}$  & 4.12 & 0.66 & $<0.7$ & $8.3\pm0.3$ & 4.06 \\     
52544.99--52547.03 (\#2) &1.9&0.57&  $1.68\pm0.09$ & $7.9\pm1.3$       & 3.38 & 0.62 & $<0.7$ & $6.8\pm0.3$ & 3.34 \\     
52549.13--52551.92 (\#3) &2.0&0.62&  $1.57\pm0.07$ & $8.4\pm1.3$       & 4.33 & 0.64 & $<1.4$ & $9.1\pm0.3$ & 4.36 \\     
52553.36--52555.07 (\#4) &1.6&0.70&  $1.30\pm0.11$ & $3.3\pm0.7$       & 2.72 & 0.60 & $7\pm3$ & $7.1\pm0.6$ & 2.56\\    
52556.96--52558.98 (\#5) &1.9&0.33&  $1.69\pm0.06$ & $12.9^{+1.6}_{-1.4}$ & 5.42 & 0.46 & $<0.5$  & $11.1\pm0.3$ & 5.34\\    
52561.97--52563.04 (\#6) &1.7&1.03&  $1.44\pm0.04$ & $6.7^{+1.0}_{-0.9}$  & 4.23 & 0.72 & $<4.5$  & $10.0\pm0.5$ & 4.00\\   
52565.74--52567.87 (\#7) &1.8&1.34&  $1.13\pm0.08$ & $3.0^{+0.5}_{-0.4}$  & 3.18 & 0.55 & $13.3^{+2.6}_{-2.2}$ & $9.9\pm0.6$ & 2.97\\     
52569.63--52573.72 (\#8) &1.8&1.03&  $1.56\pm0.08$ & $7.4\pm1.1$       & 3.85 & 0.95 & $<2.0$ & $8.4\pm0.5$ & 3.71\\    
52575.90--52578.28 (\#9) &1.9&0.79&  $1.59\pm0.10$ & $6.2^{+1.2}_{-1.0}$  & 3.10 & 0.80 & $<2.0$ & $6.4\pm0.3$ & 3.11\\    
52579.57--52581.31 (\#10)&1.7&0.95&  $1.67\pm0.08$ & $9.7^{+1.5}_{-1.3}$  & 4.23 & 0.98 & $<0.8$ & $8.6\pm0.3$ & 4.19  \\ \hline   
\end{tabular}\\
$A_1$ is the 1~keV normalization of the power law in
  units of $10^{-3}$ ph cm$^{-2}$ s$^{-1}$ keV$^{-1}$. $F_{2-10}$ is
  the observed/absorbed model flux in units of $10^{-11}$ erg
  cm$^{-2}$ s$^{-1}$.  $I_{\rm Fe}$ is kept frozen $6 \times 10^{-5}$
  ph cm$^{-2}$ s$^{-1}$. For Model 2, $\Gamma$ is frozen at 1.61.
  Each spectral fit is performed over the 3--23~keV bandpass and
  has 45 $dof$.
\end{minipage}
\end{table*}

\FloatBarrier 
\clearpage


\subsection{Cen~A (NLRG)}

\begin{figure*}
\includegraphics[angle=-90,width=1.0\textwidth]{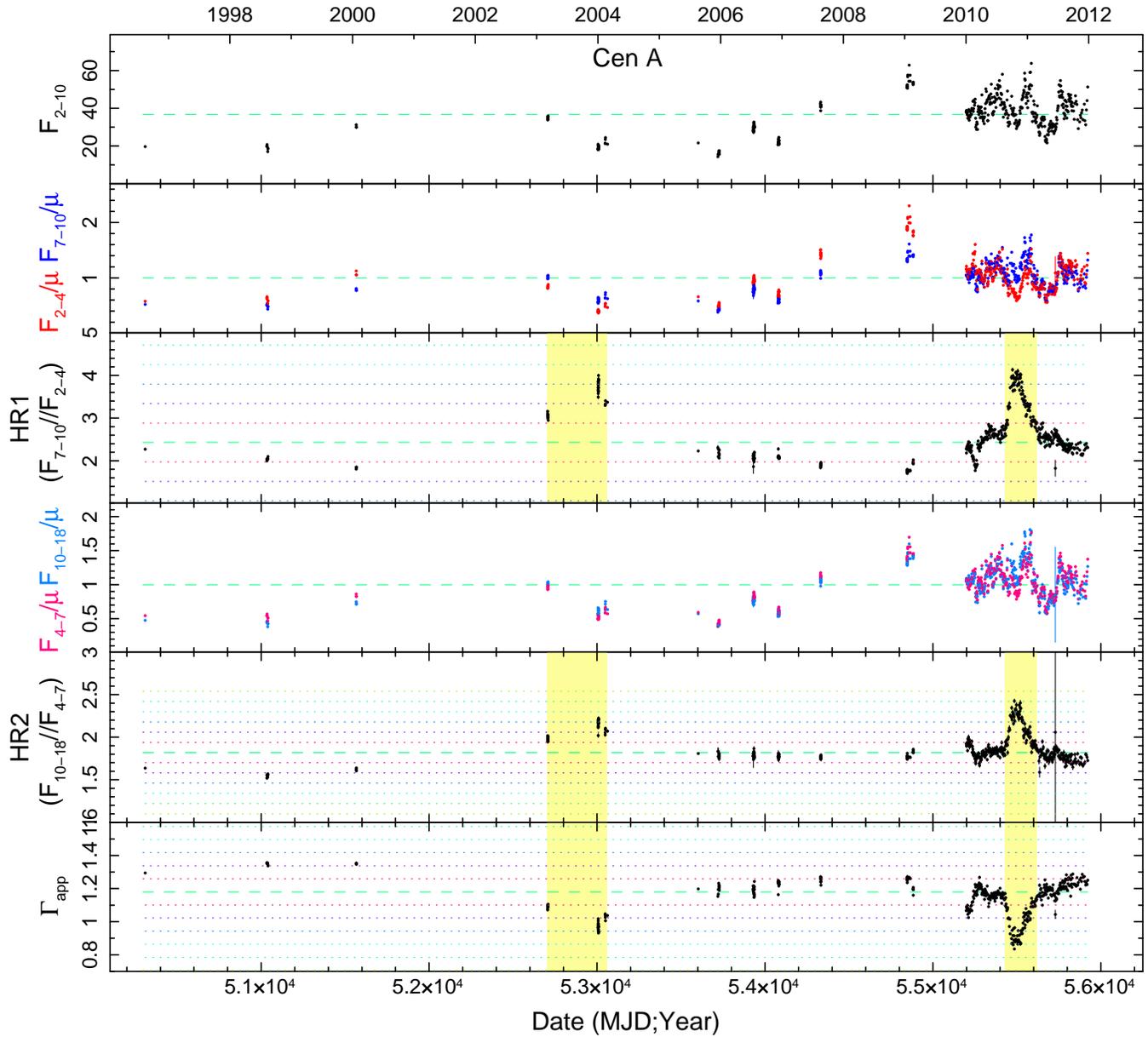} 
\caption{Same as Fig.~\ref{fig:mega3783}, but for the NLRG Cen
  A. The yellow shaded areas indicate eclipse events whose
  spectroscopic analyses can be found in by Rivers \etal\ (2011b) and
  Rothschild \etal\ (2011).}
\label{fig:megacena}  
\end{figure*}

\begin{figure}
\includegraphics[angle=-90,width=0.48\textwidth]{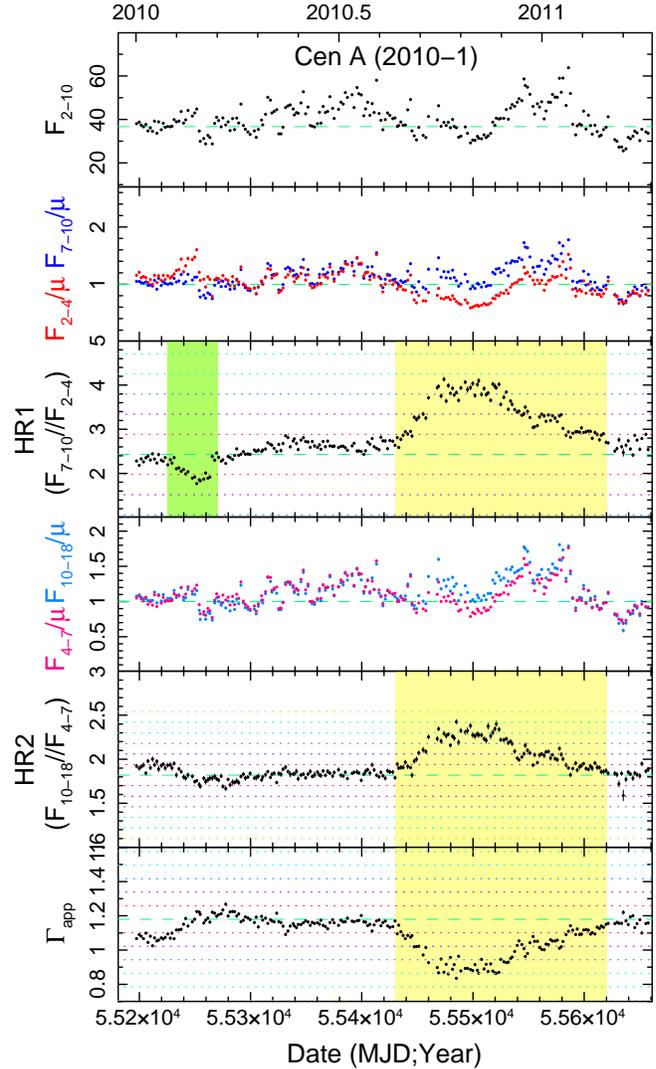}  
\caption{A zoom-in on the 2010--2011 occultation event in Cen~A; see
  Rivers \etal\ (2011b) for details of fits to the time-resolved
  spectra and to the resulting $N_{\rm H}$($t$) profile. The green
  shaded area denotes a dip in $HR1$ and thus $N_{\rm H}$; see
  Fig.~\ref{fig:cenasmalldip}.}
\label{fig:megacena1011zoom}  
\end{figure}

$\bullet$ \textit{2010--2011 (``Secure A'')}: Our $HR1$, $HR2$, and
$\Gamma_{\rm app}$ light curves (Fig.~\ref{fig:megacena}) each
suggest strong spectral flattening (at roughly 3--4$\sigma$ deviation)
over a $\sim$6 month period in 2010--2011 (MJD $\sim$ 55430--55620),
during a period of monitoring once every 2 d.  The means and
standard deviations of $HR1$, $HR2$, and $\Gamma_{\rm app}$ in
Fig.~\ref{fig:megacena} were calculated omitting data during this
period to avoid biases due to the large number of individual points
during the eclipse.  A zoom-in on this time period is shown in
Fig.~\ref{fig:megacena1011zoom}. Time-resolved spectroscopy by Rivers
\etal\ (2011b) successfully deconvolved $\Gamma$ and $N_{\rm H}$ and
confirmed a complete eclipse (ingress and egress) with peak
$\Delta$$N_{\rm H}$ of $8 \times 10^{22}$~cm$^{-2}$ above the
``baseline'' level of $\sim 20 \times 10^{22}$~cm$^{-2}$; the reader
is referred to that paper for details.  With spectra binned to once
every 10 d and an event duration of a little over 170 d, this
event is likely the best-observed eclipse so far in terms of resolving
the $N_{\rm H}$ profile.

The profile is consistent with a cloud whose density was symmetric
along the direction of transit (as opposed to a ``comet-shaped''
cloud, e.g., Maiolino \etal\ 2010); to the eye the $HR1$ and $HR2$
light curves may look slightly skewed, but fits to the $N_{\rm H}$
profile were consistent with a symmetric cloud.  A
centrally-concentrated sphere with a linear density profile was a
better fit than a constant-density sphere.  From our $HR1$ light
curve, we adopt a duration of 170.2--184.5~d.

$\bullet$ \textit{2003--2004 (``Secure A'')}: Similar changes in $HR1$
and $\Gamma_{\rm app}$ suggest an event with roughly similar
$\Delta$$N_{\rm H}$ peaking in 2003--2004. However, constraints on the
duration of the event and peak $N_{\rm H}$ are poor because 
\textit{RXTE} observed Cen~A during this time 
via three clusters of observations 
separated by months--years, as opposed to via sustained monitoring.
Spectral fits by Rothschild \etal\ (2011; see their fig.\ 4) reveal
$N_{\rm H} = 23 \times 10^{22}$~cm$^{-2}$ during the 2003 March
campaign and $24-26 \times 10^{22}$~cm$^{-2}$ during 2004 January and
February, compared to $15-18 \times 10^{22}$~cm$^{-2}$ measured
during the 2000 January and earlier campaigns and the 2005 August and
later campaigns.  The spectral fits of Rothschild \etal\ (2011) thus
indicate an observed $\Delta$$N_{\rm H}$ = $8 \pm 1 \times 10^{22}$
cm$^{-2}$ with a similar baseline $N_{\rm H}$ as the 2010--2011 event.
Constraints on the duration of the event from the $HR1$ light curve are
356--2036~d. 

\begin{figure}
\includegraphics[angle=-90,width=0.48\textwidth]{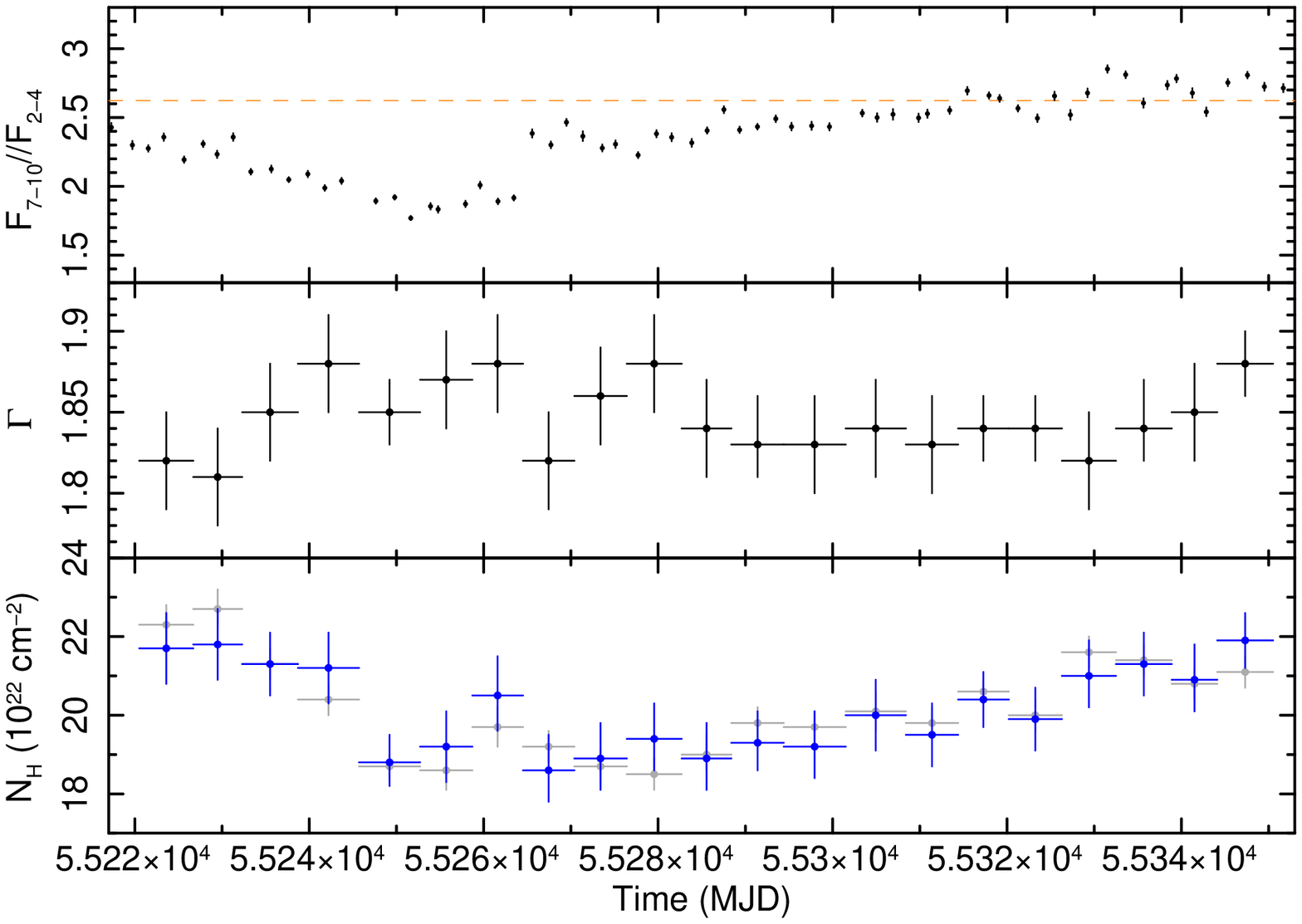}  
\caption{Time-resolved spectral fits for the first four months of
  monitoring of Cen~A in 2010, with a moderately higher time
  resolution compared to Rivers et al.\ (2011b).  The resulting
  $\Gamma$($t$) and $N_{\rm H}$($t$) light curves are depicted in the
  middle and bottom panels, respectively.  In the bottom panel, the
  blue and gray points denote $N_{\rm H}$($t$) with $\Gamma$ free and
  with $\Gamma$ frozen at 1.85, respectively.}
\label{fig:cenasmalldip}  
\end{figure}

We also present evidence that the ``baseline'' level of $N_{\rm H}$ is
not constant.  There is a small ``dip'' in the $HR1$ light curve in
2010, shortly after the 2-d monitoring commenced.  We sum up
individual spectra from the first four months of monitoring in groups
of three (6 d; Rivers et al.\ 2011b binned every five spectra / 10
d).  We leave $\Gamma$ free, but hold $I_{\rm Fe}$ frozen at
$4.2\times 10^{-4}$ ph cm$^{-2}$ s$^{-1}$.  The resulting values of
$\Gamma$($t$) and $N_{\rm H}$($t$) are displayed in
Fig.~\ref{fig:cenasmalldip} and listed in
Table~\ref{tab:cenasmalldiptable}.  The value of $N_{\rm H}$($t$)
reaches a maximum in bin \#2 (covering MJD 55226.7--55232.3), with
$N_{\rm H} = 21.7\pm0.9 \times 10^{22}$~cm$^{-2}$, reaches a minimum
in bin \#8 (MJD 55264.5--55270.4) with $N_{\rm H} = 18.6^{+0.9}_{-0.8}
\times 10^{22}$~cm$^{-2}$, and returns to a maximum value of
$21.9\pm0.7 \times 10^{22}$~cm$^{-2}$ in bin \#21 (MJD
55341.1--55350.5).  As the uncertainties on each $N_{\rm H}$ point are
90 per cent confidence, this difference is a $\sim2.2\sigma$ result.  
This result cannot be due to degeneracy between $\Gamma$ and
$N_{\rm H}$: an increase in $\Gamma$ is expected to be associated with
an \textit{increase} in $N_{\rm H}$, contrary to what is observed.
However, freezing $\Gamma$ at the average value of 1.85 does not
significantly change the $N_{\rm H}$($t$) light curve; those values
are plotted in gray in Fig.~\ref{fig:cenasmalldip}.  Summing over all
spectra, the total value of $\chi^2$/$dof$ is 1067.95/1113, compared
to 1037.92/1092 with $\Gamma$ free; this means that allowing $\Gamma$
to be free improves the fits only at 93.3 per cent confidence according to
an $F$-test.

\begin{table*}
\begin{minipage}{140mm}
\caption{Time-resolved spectral fits to Cen~A, early 2010}
\label{tab:cenasmalldiptable}
\begin{tabular}{ccccccc} \hline
Start--stop    & Expo &                    & $N_{\rm H}$          &                     &    $A_1$                           &   \\     
   (MJD)       & (ks) & $\chi^2_{\rm red}$ & ($10^{22}$~cm$^{-2}$) & $\Gamma$        & (ph cm$^{-2}$ s$^{-1}$ keV$^{-1}$) & $F_{2-10}$\\ \hline    \hline
55221.53--55225.65 (\#1) & 2.4  & 0.88 & $21.7\pm0.9$         & 1.82$\pm$0.03        &  $0.214^{+0.016}_{-0.014}$  &   34.62 \\ 
55227.83--55231.27 (\#2) & 2.4  & 0.97 & $21.8\pm0.9$         & 1.81$\pm$0.03        &  $0.214^{+0.016}_{-0.014}$  &   35.13 \\ 
55233.29--55237.65 (\#3) & 2.8  & 1.08 & $21.3\pm0.8$         & 1.85$\pm$0.03        &  $0.243^{+0.016}_{-0.014}$  &   37.66 \\ 
55239.84--55243.70 (\#4) & 2.2  & 1.02 & $21.2\pm0.9$         & 1.88$\pm$0.03        &  $0.254^{+0.020}_{-0.017}$  &   37.43 \\ 
55247.62--55251.64 (\#5) & 3.3  & 0.97 & $18.8^{+0.7}_{-0.6}$ & 1.85$\pm$0.02        &  $0.244^{+0.014}_{-0.012}$  &   39.85 \\ 
55253.90--55257.92 (\#6) & 2.5  & 0.84 & $19.2\pm0.9$         & 1.87$\pm$0.03        &  $0.187^{+0.016}_{-0.014}$  &   29.29 \\ 
55259.59--55263.45 (\#7) & 2.6  & 0.70 & $20.5^{+1.0}_{-0.9}$ & 1.88$\pm$0.03        &  $0.191^{+0.016}_{-0.014}$  &   28.64 \\ 
55265.56--55269.48 (\#8) & 2.6  & 0.97 & $18.6^{+0.9}_{-0.8}$ & 1.82$\pm$0.03        &  $0.190^{+0.014}_{-0.013}$  &   32.66 \\ 
55271.35--55275.12 (\#9) & 2.4  & 0.93 & $18.9^{+0.9}_{-0.8}$ & 1.86$\pm$0.03        &  $0.217^{+0.016}_{-0.014}$  &   35.00 \\ 
55277.71--55281.56 (\#10) & 2.4  & 0.75 & $19.4^{+0.9}_{-0.8}$ & 1.88$\pm$0.03        &  $0.220^{+0.017}_{-0.015}$  &   33.54 \\ 
55283.86--55287.54 (\#11) & 2.2  & 0.76 & $18.9^{+0.9}_{-0.8}$ & 1.84$\pm$0.03        &  $0.209^{+0.016}_{-0.014}$  &   34.64 \\ 
55289.33--55293.49 (\#12) & 2.8  & 1.08 & $19.3^{+0.8}_{-0.7}$ & $1.83^{+0.03}_{-0.02}$ &  $0.221^{+0.014}_{-0.013}$  &   37.10 \\ 
55295.29--55299.64 (\#13) & 2.8  & 1.42 & $19.2^{+0.9}_{-0.8}$ & 1.83$\pm$0.03        &  $0.184^{+0.014}_{-0.012}$  &   31.11 \\ 
55303.40--55306.94 (\#14) & 2.5  & 1.03 & $20.0\pm0.9$         & 1.84$\pm$0.03        &  $0.191^{+0.015}_{-0.013}$  &   30.76 \\ 
55309.87--55313.43 (\#15) & 2.6  & 0.66 & $19.5\pm0.8$         & 1.83$\pm$0.03        &  $0.219^{+0.015}_{-0.013}$  &   36.18 \\ 
55315.44--55319.17 (\#16) & 3.1  & 1.18 & $20.4\pm0.7$         & 1.84$\pm$0.02        &  $0.271^{+0.015}_{-0.013}$  &   42.99 \\ 
55321.25--55325.45 (\#17) & 2.9  & 0.59 & $19.9\pm0.8$         & 1.84$\pm$0.02        &  $0.225^{+0.015}_{-0.013}$  &   36.36 \\ 
55327.28--55331.54 (\#18) & 2.5  & 0.95 & $21.0^{+0.9}_{-0.8}$ & 1.82$\pm$0.03        &  $0.217^{+0.015}_{-0.014}$  &   35.43 \\ 
55333.61--55338.38 (\#19) & 2.4  & 1.03 & $21.3\pm0.8$         & $1.84^{+0.03}_{-0.02}$ &  $0.272^{+0.018}_{-0.016}$  &   42.31 \\ 
55339.42--55342.86 (\#20) & 2.1  & 1.24 & $20.9^{+0.9}_{-0.8}$ & 1.85$\pm$0.03        &  $0.250^{+0.018}_{-0.016}$  &   39.02 \\ 
55345.33--55349.51 (\#21) & 2.6  & 0.91 & $21.9\pm0.7$         & 1.88$\pm$0.02        &  $0.303^{+0.019}_{-0.017}$  &   43.94 \\ \hline 
\end{tabular}\\
Results of time-resolved spectroscopy for Cen~A to
  quantify the small "dip" in the baseline level of $N_{\rm H}$ in
  early 2010 (approx.\ MJD 55230--55330).  
  $A_1$ is the 1~keV normalization of the power
  law. $F_{2-10}$ is the observed/absorbed model flux in units of
  $10^{-11}$ erg cm$^{-2}$ s$^{-1}$. $I_{\rm Fe}$ is frozen at
  $4.2 \times 10^{-4}$ ph cm$^{-2}$ s$^{-1}$. Each spectral fit is
  performed using the 3--30~keV bandpass and has 52
  $dof$.
\end{minipage}
\end{table*}

\FloatBarrier
\clearpage
\subsection{NGC~5506 (Sy 1.9)}

\begin{figure*}
\includegraphics[angle=-90,width=1.0\textwidth]{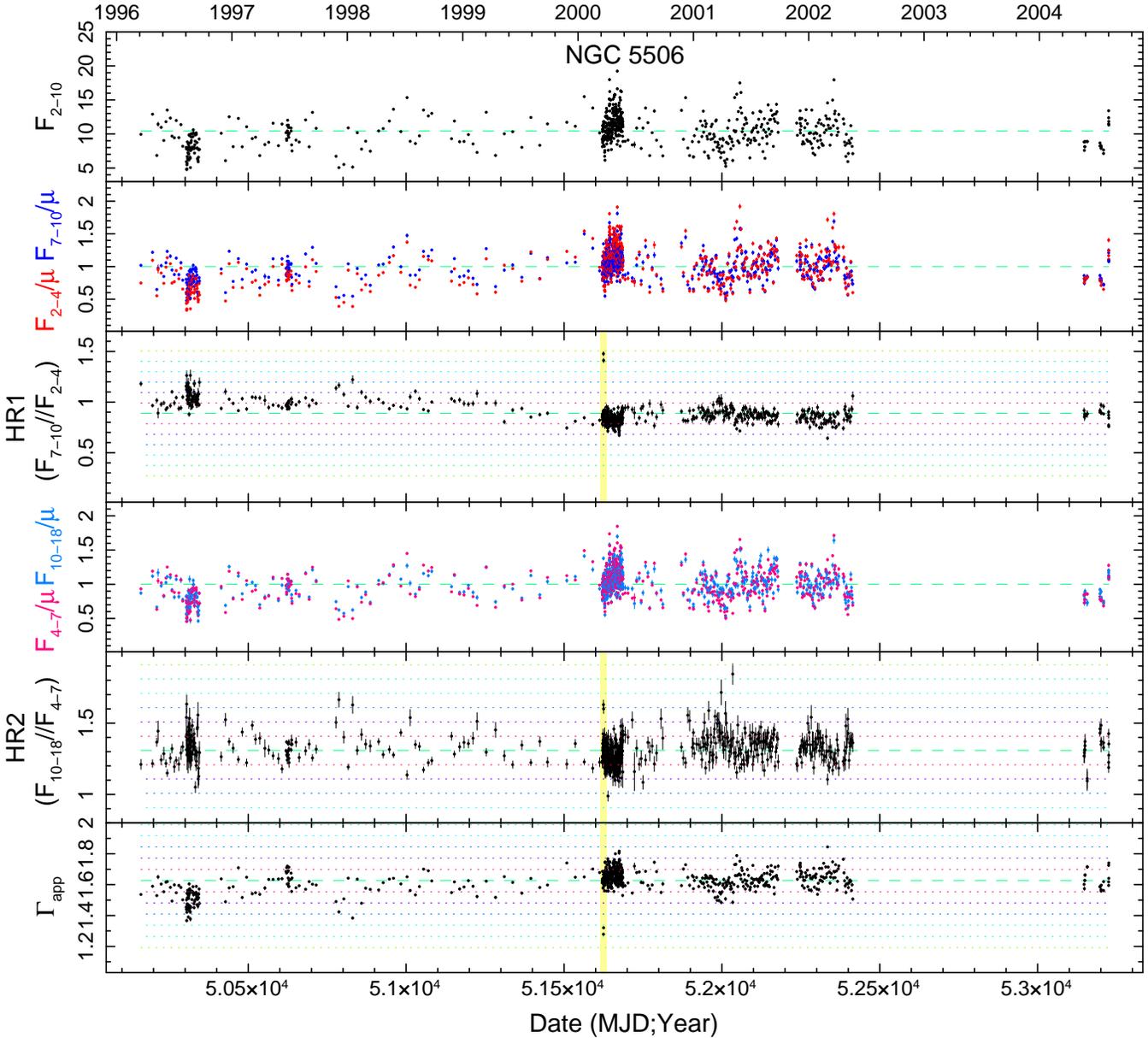} 
\caption{Same as Fig.~\ref{fig:mega3783}, but for the Sy~1.9
  NGC~5506. The yellow shaded area indicates a candidate eclipse event
  in 2000; see Fig.~\ref{fig:mega5506zoom_TR} for a zoom-in on that
  period.}
\label{fig:mega5506}  
\end{figure*}

\begin{figure}
\includegraphics[angle=-90,width=0.48\textwidth]{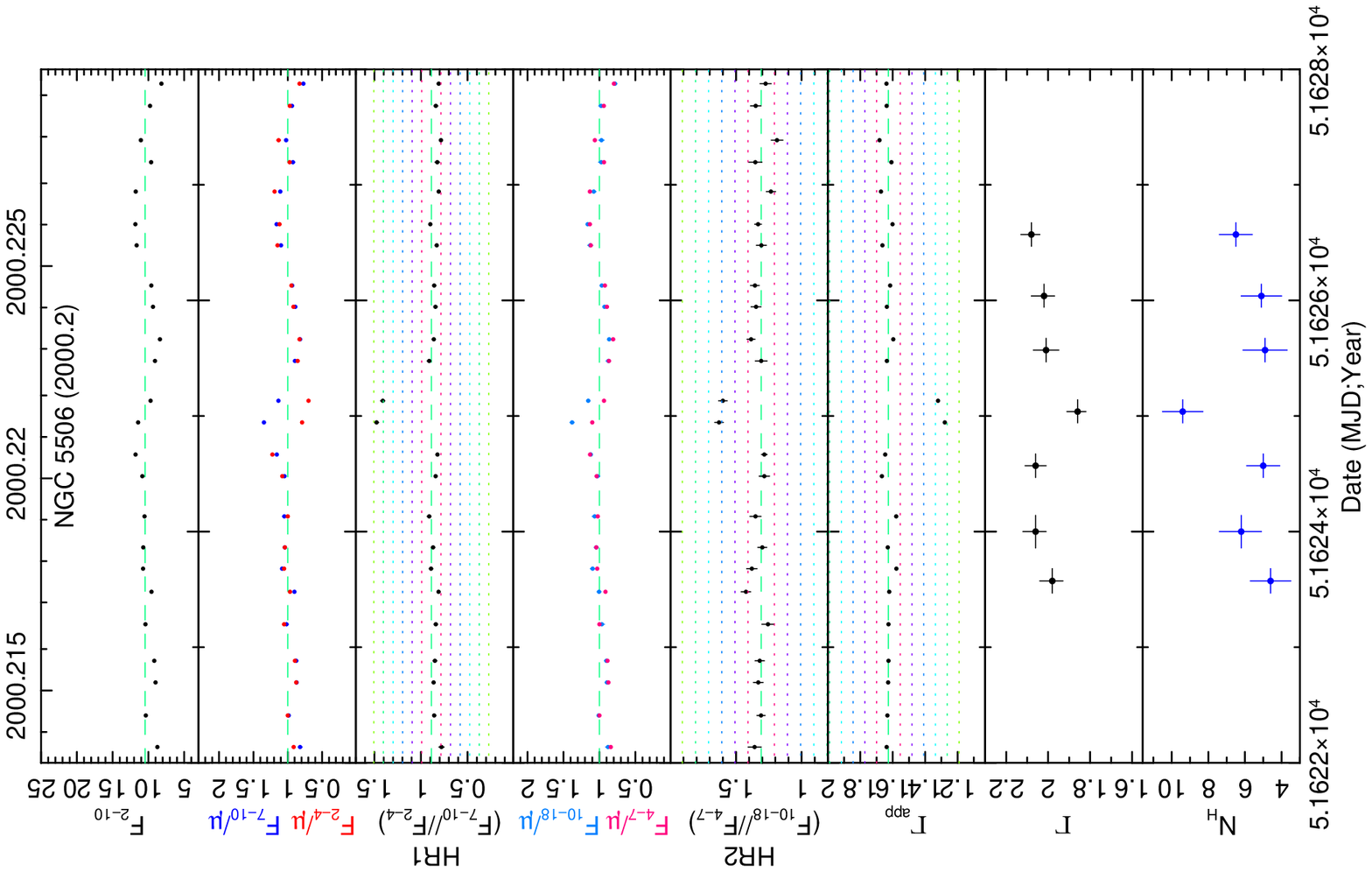}  
\caption{Same as Fig.~\ref{fig:mega5506}, but a zoom-in on the
  eclipse event in 2000.  The bottom two panels show $\Gamma$ and
  $N_{\rm H}$ in units of $10^{22}$~cm$^{-2}$.}
\label{fig:mega5506zoom_TR}  
\end{figure}

$\bullet$ \textit{2000.2 (``Secure A'')}: The long-term light curves
are plotted in Fig.~\ref{fig:mega5506}.  During a period of
intensive monitoring in early 2000 (Fig.~\ref{fig:mega5506zoom_TR},
with four observations per day, two consecutive observations (at MJD
51624.9 and 51625.1) show anomalous increases in $HR1$ from its mean
value of 0.88 with a standard deviation of 0.10 to values of
1.41$\pm$0.02 and 1.48$\pm$0.02, deviations of 5.2 and 6.0$\sigma$.
$\Gamma_{\rm app}$ drops from 1.6 to 1.3 ($>4\sigma$ deviation).

We perform time-resolved spectroscopy by summing spectra together in
groups of two.  We fit 3--23~keV data, allowing $N_{\rm, H}$,
$\Gamma$, $A_1$ and $I_{\rm Fe}$ to vary from their best-fitting
time-averaged values. Uncertainties on $I_{\rm Fe}$ within each time
bin are usually over 50 per cent, and so we re-fit with $I_{\rm Fe}$ held
at $2.5 \times 10^{-4}$ ph cm$^{-2}$ s$^{-1}$.  The results are listed
in Table~\ref{tab:5506TRtable}.  The resulting best-fitting parameters for
$\Gamma$ and $N_{\rm H}$ are plotted in
Fig.~\ref{fig:mega5506zoom_TR}.

$N_{\rm H}$ increases from an average of $5.4\pm1.0\times 10^{22}$
to $9.4\pm1.1\times 10^{22}$~cm$^{-2}$ during MJD
51624.9--51625.1.  For the MJD 51624.9--51625.1 spectrum, freezing
$\Gamma$ to 1.93 causes $\chi^2_{\rm red}$ to increase from 1.08 to
1.19; $N_{\rm H}$ goes to a value of $10.7\pm0.6\times 10^{22}$
cm$^{-2}$ (consistent with the value obtained with $\Gamma$ free), but
there are systematic data/model residuals around 5~keV, at the
$\sim8$ per cent level.

We conclude that a short-duration obscuration event with
$\Delta$$N_{\rm H} = 4.0\pm1.4 \times 10^{22}$~cm$^{-2}$ occurred.
Given the start and stop times of these observations and the
surrounding ones, the duration of the event must be in the range
$17.2-69.0$~ks.

\begin{table*}
\begin{minipage}{150mm}
\caption{Time-resolved spectral fits to NGC~5506, 2000.2}
\label{tab:5506TRtable} 
\begin{tabular}{ccccccc} \hline
Start--stop    & Expo &                   & $N_{\rm H}$          &                     &    $A_1$                      &   \\     
   (MJD)       & (ks) & $\chi^2_{\rm red}$   & ($10^{22}$~cm$^{-2}$) & $\Gamma$        & ($10^{-2}$ ph cm$^{-2}$ s$^{-1}$ keV$^{-1}$) & $F_{2-10}$\\ \hline    \hline
51623.47--51623.68 ($\#1$) & 1.7 & 0.61 & $4.6\pm1.1$         & $1.98^{+0.06}_{-0.05}$ & $4.1^{+0.6}_{-0.5}$ & 9.53  \\  
51623.86--51624.14 ($\#2$) & 1.6 & 0.71 & $6.2^{+1.2}_{-1.1}$    & $2.06^{+0.06}_{-0.05}$ & $5.1^{+0.7}_{-0.6}$ & 9.91  \\  
51624.47--51624.67 ($\#3$) & 2.3 & 0.66 & $5.0\pm0.9$         & $2.06\pm0.05$       & $5.3^{+0.6}_{-0.5}$ & 10.73  \\   
51624.94--51625.14 ($\#4$) & 2.0 & 1.08 & $9.4\pm1.1$         & $1.86^{+0.05}_{-0.04}$ & $4.0^{+0.5}_{-0.4}$ & 9.53  \\   
                           &     & 1.19 & $10.7\pm0.6$        & 1.93 (frozen)          & $4.7\pm0.1$         & 9.45 \\  
51625.47--51625.67 ($\#5$) & 2.3 & 0.79 & $4.9\pm1.2$         & $2.01\pm0.06$          & $3.7^{+0.5}_{-0.4}$ & 8.09  \\   
51625.94--51626.14 ($\#6$) & 2.0 & 0.71 & $5.1\pm1.1$         & $2.02^{+0.06}_{-0.05}$ & $4.2\pm0.6$         & 8.84  \\    
51626.47--51626.67 ($\#7$) & 2.3 & 0.91 & $6.5\pm0.9$         & $2.08^{+0.05}_{-0.04}$ & $5.9\pm0.6$         & 10.73   \\ \hline    
\end{tabular}\\
Results of time-resolved spectroscopy for NGC~5506 in
  early 2000.  $A_1$ is the 1~keV normalization of the power
  law. $F_{2-10}$ is the observed/absorbed model flux in units of
  $10^{-11}$ erg cm$^{-2}$ s$^{-1}$. $I_{\rm Fe}$ is frozen at
  $2.5 \times 10^{-4}$ ph cm$^{-2}$ s$^{-1}$.  Each spectral fit is
  performed over the 3--23~keV bandpass and has 44 $dof$.
\end{minipage}
\end{table*}

\clearpage
\subsection{Mkn~348 (Sy 2)}

\begin{figure*}
\includegraphics[angle=-90,width=1.0\textwidth]{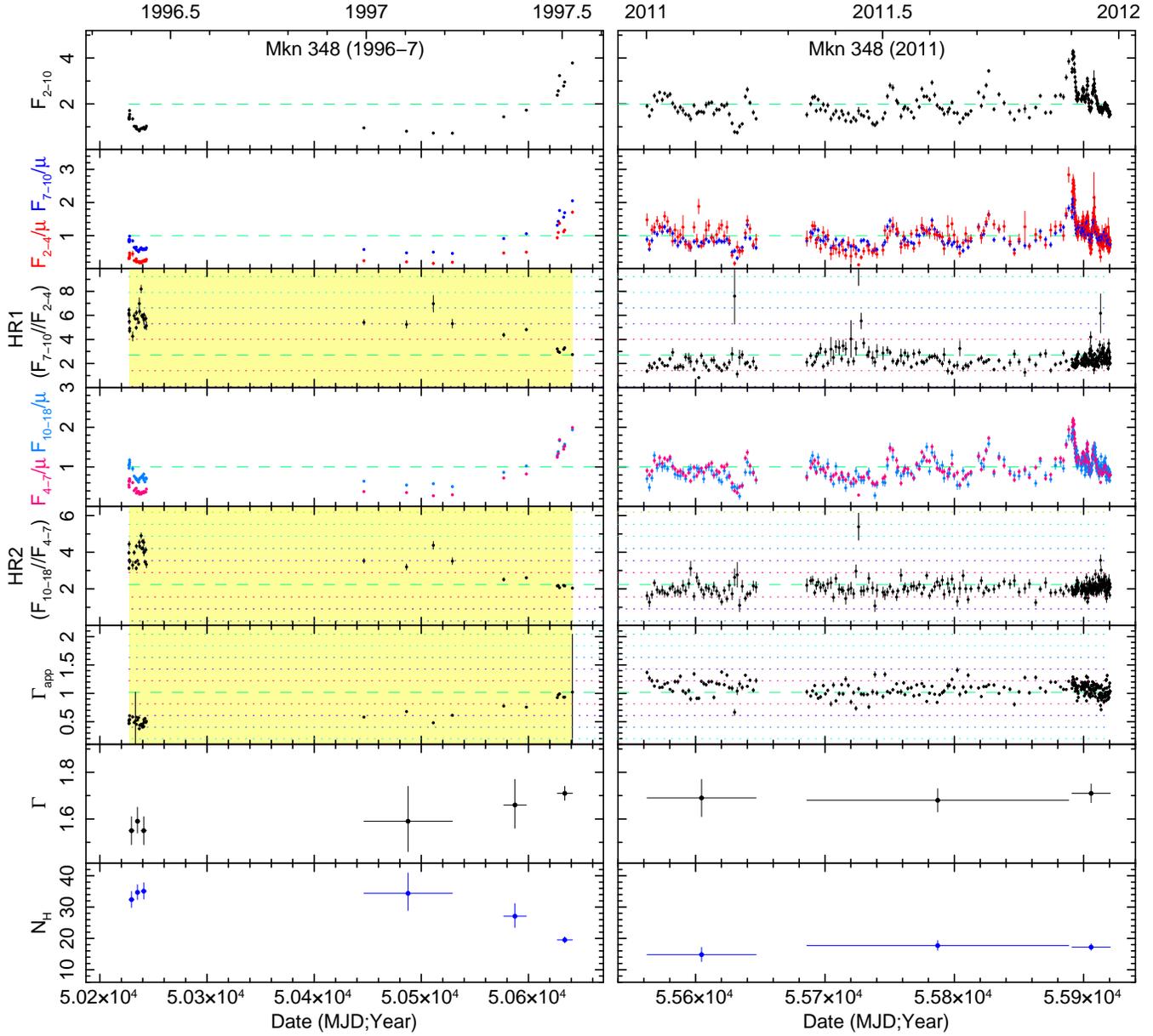} 
\caption{Same as Fig.~\ref{fig:mega3783}, but for the Sy~2
  Mkn~348. The yellow shaded area covers the 1996--1997 campaign and
  time-resolved spectroscopy indicating variations in $N_{\rm H}$ which
  have
  been published by Akylas \etal\ (2002).  The bottom two panels
  indicate $\Gamma$ and $N_{\rm H}$ in units of $10^{22}$~cm$^{-2}$
  from our time-resolved spectroscopic analysis, from fits where both
  parameters are free. }
\label{fig:mega348}  
\end{figure*}

$\bullet$ \textit{1996--1997 (``Secure A'')}: The long-term light
curves are plotted in Fig.~\ref{fig:mega348}. \textit{RXTE} observed
this source 25 times over a 16 d period in 1996 May--June, followed by
sporadic observations from 1996 December through 1997 July. The source
was monitored regularly once every two days during 2011 January --
November followed by four times daily monitoring for 30 d in 2011
November--December. Our $HR1$ light curve shows a decrease from 5--8
in 1996 ($\sim$2--3$\sigma$) to 3 by 1997, with values $\sim$1.5--3 in
2011.

Smith \etal\ (2001) and Akylas \etal\ (2002), fitting spectra from the 1996--1997 campaigns,
reported a steady decrease in $N_{\rm H}$ from $\sim$27 to 12 $\times
10^{22}$~cm$^{-2}$; they assumed $\Gamma$ frozen at 1.85 and
\textsc{angr} abundances.  We divide data from the 1996--1997
campaign into six coarse time slices, to confirm the decrease in
$N_{\rm H}$, albeit with a larger time resolution.  We assume a
simple full-covering neutral absorber and used the best-fitting model from
Rivers \etal\ (2013); we keep $R$ frozen at 0.3, and $I_{\rm Fe}$ frozen
at $3\times 10^{-5}$ ph cm$^{-2}$ s$^{-1}$ (when we allow $I_{\rm
  Fe}$ to vary, values are almost always consistent with this value
at the 90 per cent level). Assuming \textsc{angr} abundances, we obtain
best-fitting values of $N_{\rm H}$($t$) consistent with Akylas
\etal\ (2002; their fig.\ 2). Assuming \textsc{wilm} abundances instead, we
obtain values of $N_{\rm H}$ an average of 1.42 higher. That is, 
we measure a decrease from $\sim 32$ to $\sim20$ $\times 10^{22}$~cm$^{-2}$, as listed in
Table~\ref{tab:348TRtable}.

We also perform spectral fits to the 2011 data, dividing them into
three coarse time slices (MJD 55562--55647, 55684--55888,
and 55890--55921), with the divisions corresponding to a sun-angle gap and
the start of the intensive monitoring.  We apply the same model as
above, and we obtain best-fitting values of $N_{\rm H}$ of $\sim 15-17
\times 10^{22}$~cm$^{-2}$ (using \textsc{wilm} abundances),
with corresponding values of $\Gamma$ listed in Table~\ref{tab:348TRtable}.

As shown in Table~\ref{tab:348TRtable}, if we assume that the value of
$N_{\rm H}$ throughout 1996--1997 was $16.6 \times 10^{22}$~cm$^{-2}$
(the average of the three values obtained for 2011), we obtain fits
for spectra \# 1--5 with significantly worse values of $\chi^2_{\rm
  red}$, strong curvature in the data/model residuals, and values of
$\Gamma$ as low as 1.2.

Spectral fits to an \textit{XMM-Newton} EPIC observation in 2002 July
performed by Brightman \& Nandra (2011) yield $N_{\rm H} =
13.1^{+0.7}_{-1.3} \times 10^{22}$~cm$^{-2}$ (with
$\Gamma=1.68^{+0.07}_{-0.10}$, \textsc{angr} abundances, thus $\sim 18
\times 10^{22}$~cm$^{-2}$ with \textsc{wilm} abundances), consistent
with the idea that $N_{\rm H}$ has varied only weakly since 1997
July.  Assuming the ``baseline'' $N_{\rm H}$ level is $\sim
15-18 \times 10^{22}$~cm$^{-2}$ (\textsc{wilm} abundances), the
\textit{XMM-Newton} and 2011 \textit{RXTE} data thus support the
suggestion by Akylas \etal\ (2002) that \textit{RXTE} just caught the
tail end of the eclipse in 1997 July.

We assign a minimum duration of 399 d (from the start of the
\textit{RXTE} monitoring at MJD 50227 until the point when
\textit{RXTE} measured a value of $N_{\rm H}$ consistent with the
``baseline'' level, after MJD 50626. Using the peak value of $N_{\rm
  H}$ from spectrum \#3 and a baseline level of $16.6 \times 10^{22}$
cm$^{-2}$, we adopt $\Delta$$N_{\rm H} = 18\pm3 \times 10^{22}$
cm$^{-2}$.

The Tartarus database of AGN X-ray spectra observed with
\textit{ASCA}\footnote{http://heasarc.gsfc.nasa.gov/W3Browse/asca/tartarus.html}
indicates absorption by a column $\sim 10 \times 10^{22}$~cm$^{-2}$
for an observation on 1995 August 4 (MJD 49933), suggesting an upper
limit to the duration of 693 d.  If the event happened to start
immediately after the \textit{ASCA} observation, and if it had an
$N_{\rm H}$($t$) profile that was symmetric in time, then the event
peak would likely have occurred in 1996 July, just after the 1996
\textit{RXTE} observations started.  If this is the case then
\textit{RXTE} was lucky enough to observe the approximate peak $N_{\rm
  H}$ value.  We cannot rule out, however, that peak $N_{\rm H}$
occurred before the \textit{RXTE} observations started and thus might
be higher than that observed by \textit{RXTE}.

\begin{table*}
\begin{minipage}{160mm}
\caption{Results of time-resolved spectral fits to Mkn~348}
\label{tab:348TRtable} 
\begin{tabular}{lcccccccc}\hline
Start--stop            & Expo & \multicolumn{5}{c}{$N_{\rm H}$, $\Gamma$ free}              & \multicolumn{2}{c}{$N_{\rm H}$ frozen at $11.5\times 10^{23}$~cm$^{-2}$} \\
   (MJD)               & (ks) & $\chi^2_{\rm red}/dof$ & $N_{\rm H}$ & $\Gamma$      & $A_1$              & $F_{2-10}$ & $\chi^2_{\rm red}/dof$ &  $\Gamma$    \\ \hline
50227.05--50231.99 (\#1) & 28.8 & 1.48/46 & $32.4^{+2.6}_{-2.5}$ & $1.55\pm0.06$       &  $5.5^{+0.9}_{-0.7}$ & 1.27 & 4.25/47 & $1.20\pm0.03$ \\   
50233.16--50237.03 (\#2) & 79.0 & 2.62/46 & $34.7^{+2.6}_{-2.3}$ & $1.59^{+0.06}_{-0.05}$ &  $4.1^{+0.6}_{-0.5}$ & 0.85 & 6.96/47 & $1.19\pm0.02$\\
50238.44--50243.35 (\#3) & 65.8 & 2.36/46 & $35.1^{+2.7}_{-2.5}$ & $1.55\pm0.06$       &  $3.9\pm0.4$      & 0.87 & 6.17/47 & $1.15\pm0.02$\\
50446.44--50528.99 (\#4) & 18.2 & 0.50/46 & $34.4^{+6.5}_{-5.5}$ & $1.59^{+0.15}_{-0.13}$ &  $3.6^{+1.5}_{-1.0}$ & 0.74 & 1.05/47 & $1.19\pm0.06$\\
50576.94--50598.06 (\#5) &  7.5 & 0.50/46 & $27.1^{+4.0}_{-3.6}$ & $1.66^{+0.11}_{-0.10}$ &  $7.1^{+2.1}_{-1.6}$ & 1.46 & 0.98/47 & $1.40\pm0.05$ \\
50626.91--50641.07 (\#6) & 27.8 & 1.02/46 & $19.5^{+1.0}_{-0.9}$ & $1.71\pm0.03$       & $12.2\pm1.0$      & 2.67 & 1.51/47 & $1.63\pm0.02$ \\ \hline 
55562.45--55646.78 (\#7) & 38.4 & 1.31/44 & $14.8^{+2.3}_{-2.2}$ & $1.69\pm0.08$       &  $6.5^{+0.8}_{-0.7}$ & 1.64 & 1.32/45 & $1.74\pm0.04$\\   
55685.99--55888.42 (\#8) & 75.2 & 0.72/44 & $17.7^{+1.6}_{-1.5}$ & $1.68\pm0.05$       &  $7.1\pm0.9$      & 1.71 & 0.73/45 & $1.65\pm0.03$ \\  
55890.93--55920.54 (\#9) & 124.0& 1.16/44 & $17.2^{+1.1}_{-1.0}$ & $1.71\pm0.04$       &  $8.7^{+0.8}_{-0.7}$ & 2.01 & 1.15/45 & $1.69\pm0.02$  \\ \hline   
\end{tabular}\\
Results of time-resolved spectroscopy for Mkn~348; see also
  Akylas \etal\ (2002) for spectral fits to the 1996--1997 data.  
  Column densities, which are in units of $10^{22}$~cm$^{-2}$, 
  refer to total observed columns, i.e., baseline level plus eclipsing cloud.
  $A_1$ is the 1~keV normalization
  of the power law in units of $10^{-3}$ ph cm$^{-2}$ s$^{-1}$
  keV$^{-1}$. $F_{2-10}$ is the observed/absorbed model flux in units
  of $10^{-11}$ erg cm$^{-2}$ s$^{-1}$. $I_{\rm Fe}$ is frozen at
  $3\times 10^{-5}$ ph cm$^{-2}$ s$^{-1}$.  Each spectral fit is
  performed using the 3--23~keV bandpass.
\end{minipage}
\end{table*}

\clearpage

\subsection{Additional objects with ``candidate'' events:}
\begin{figure*}
\includegraphics[angle=-90,width=1.0\textwidth]{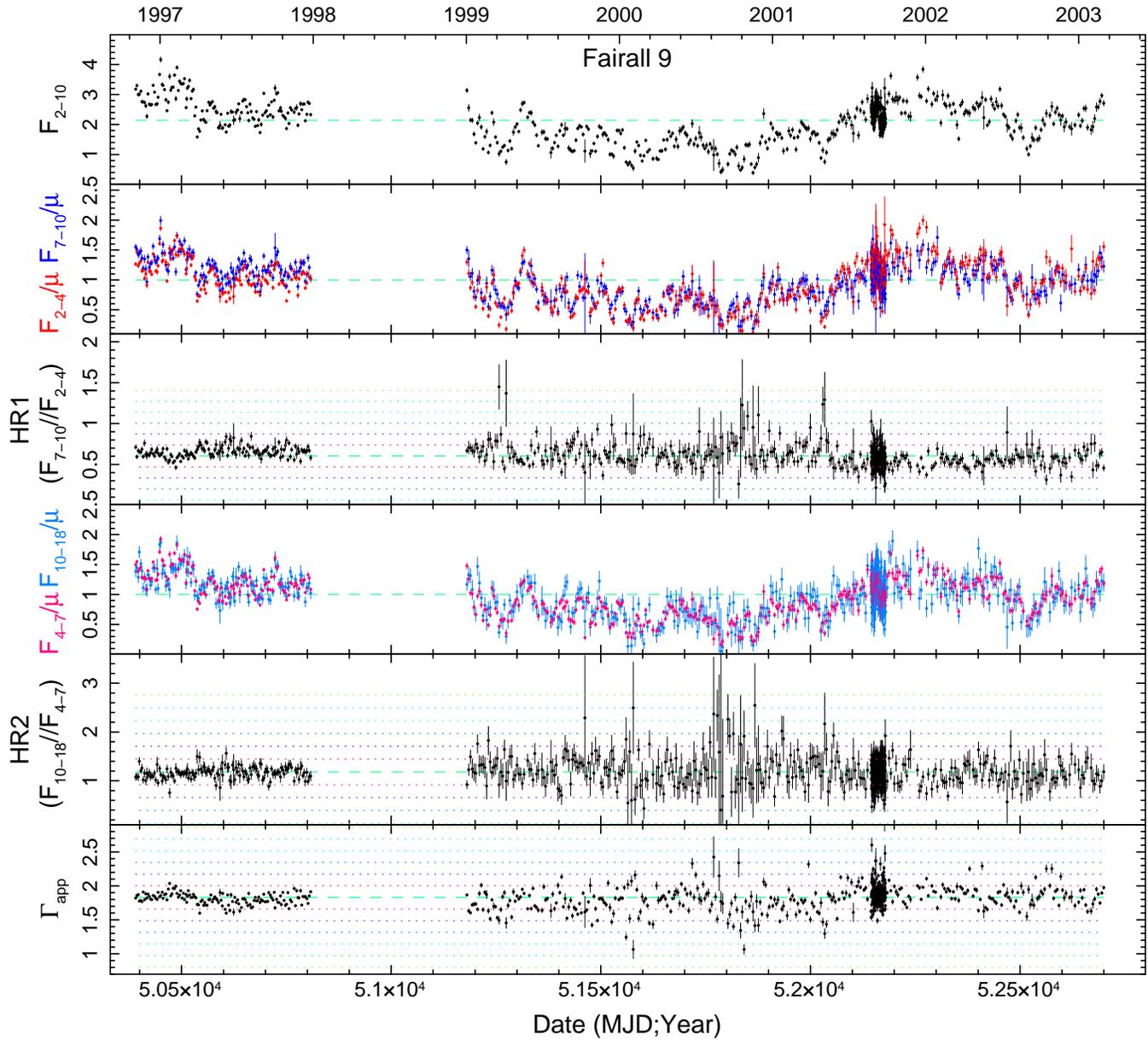}  
\caption{Same as Fig.~\ref{fig:mega3783}, but for the Sy~1 Fairall~9.}  
\label{fig:megafrl9}  
\end{figure*}

\textit{Fairall 9 (Sy 1), 2001.3 (``candidate'')}: The long-term light
curve is plotted in Fig.~\ref{fig:megafrl9}.

\begin{figure}
\includegraphics[angle=-90,width=0.48\textwidth]{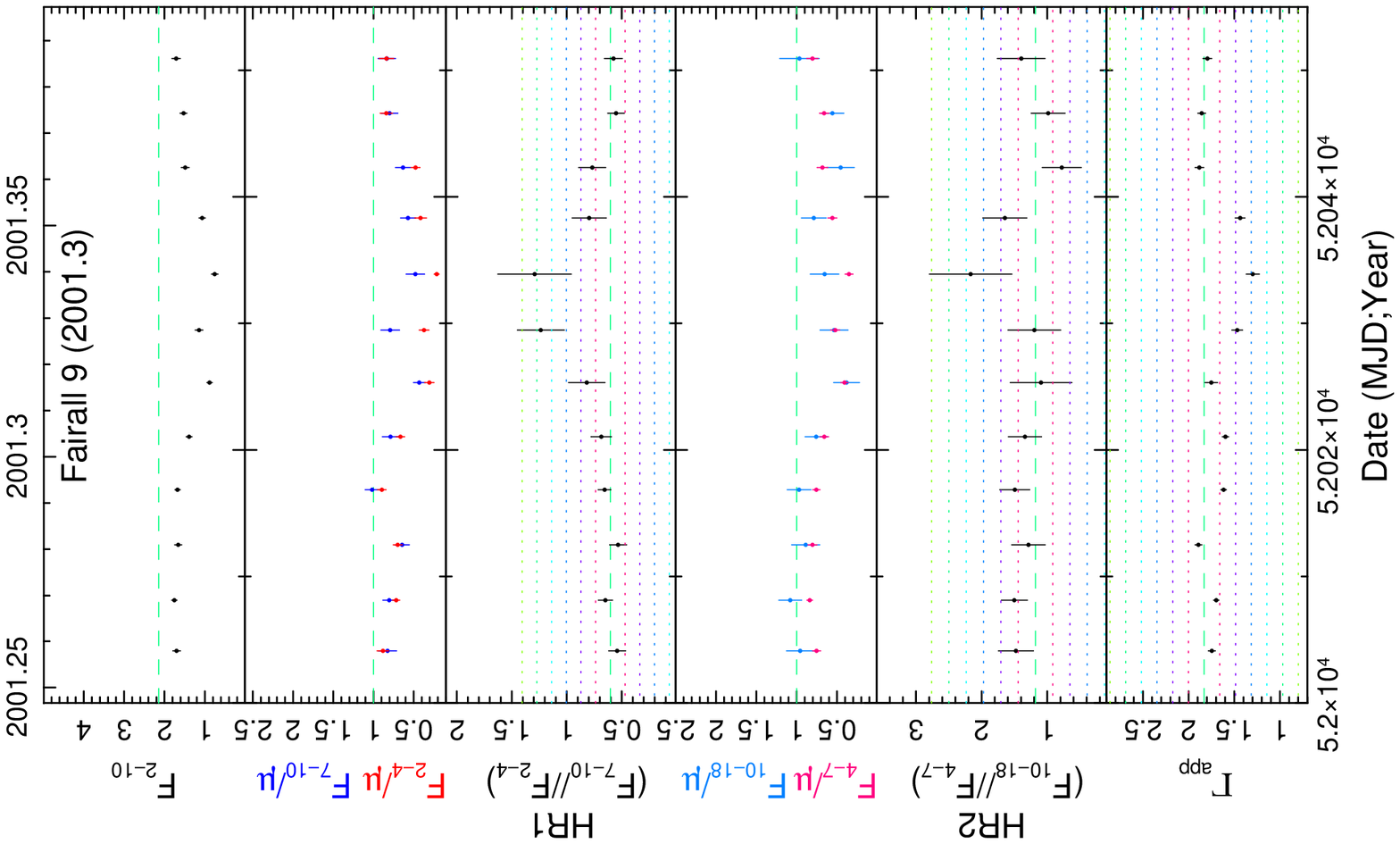}  
\caption{Same as Fig.~\ref{fig:megafrl9}, but a 
zoom-in on the "candidate" event in 2001.}  
\label{fig:megafrl9zoom}  
\end{figure}

There is a small spike in $HR1$ over two consecutive points, at MJD
52029 and 52033 ($HR1_{\rm peak} \sim 1.3$ ; $>3\sigma$), plotted in
Fig.~\ref{fig:megafrl9zoom}. However, the uncertainties are very
high in each case ($\sim$0.3-0.5).  $\Gamma_{\rm app}$ is roughly $1.2-1.4$
with uncertainties 0.06--0.08.  We sum the two spectra together for
spectral fitting, but the derived constraints are very poor: adopting
the parameter values from the time-averaged fit from Rivers \etal\
(2013), adding an extra column for neutral absorption using
\textsc{zphabs}, and freezing $\Gamma$ to 2.0 
yields $N_{\rm H} < 2.1 \times 10^{23}$~cm$^{-2}$. The event 
satisfies criteria 1--3, but with only an upper limit to $N_{\rm H}$
and with all continuum flux levels rather low, 
we classify this as a ``candidate'' event with a duration in the range 4--12 d .

$\bullet$ There are also two single point anomalously high values of
$HR1$ at MJD 51257 and 51274 with values of $HR1$ and $\Gamma_{\rm
  app}$ similar to the above ``candidate'' event, but they fail
criterion 2. Furthermore, spectral fitting to each spectrum yields $N_{\rm H} <
4 \times 10^{23}$~cm$^{-2}$, and so these events are rejected and not
discussed further.

Lohfink et al.\ (2012) note the trio of low-flux ``dips'' in the 2--10
and 10--20~keV continuum light curves of Fairall~9 in mid- to
late-2000 (MJD $\sim$ 51780, 51830, and 51860).  They explore if these
sudden decreases in the X-ray continuum flux could be associated with
structural changes in the inner accretion disk or with absorption by
Compton-thick clouds transiting the line of sight.  However, our $HR1$
and $HR2$ light curves do not reveal anything obvious during mid- to
late-2000: values are usually consistent with $\langle HR1 \rangle$ or
$\langle HR2 \rangle$ and/or have large error bars and large
point-to-point scatter.

\clearpage

\begin{figure*}
\includegraphics[angle=-90,width=1.0\textwidth]{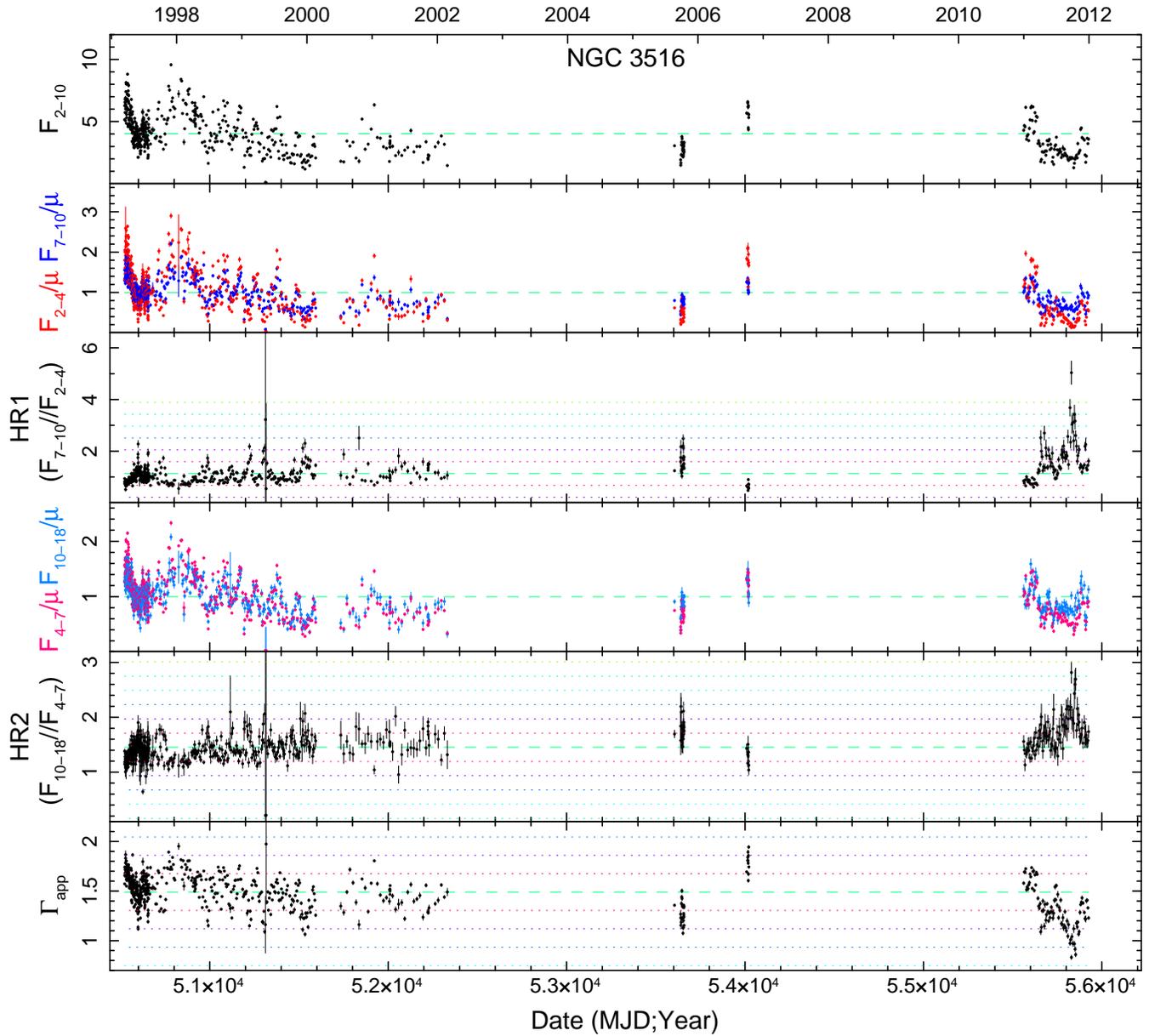}   
\caption{Same as Fig.~\ref{fig:mega3783}, but for the Sy~1.5 NGC~3516.}  
\label{fig:mega3516}  
\end{figure*}

\textit{NGC~3516 (Sy~1.5), 2011.7, ``candidate'' event:} This source's
$HR1$ light curve (Fig.~\ref{fig:mega3516}) shows systematic trends
that are visually not obviously correlated with 2--10 or 10--18~keV
continuum flux.  Of peculiar interest is a "spike" in $HR$ to $\sim$
2.4 ($\sim4\sigma$) during Fall 2011. $\Gamma_{\rm app}$ is usually
1.49$\pm$0.19 for NGC~3516 but reaches values as low as 0.8--0.9 in
Fall 2011.

The X-ray spectrum of this object is commonly characterized by complex
absorption, which can be variable on timescales of years, as
documented by Turner \etal\ (2008).  For example, Markowitz
\etal\ (2008) found the source to be in a Compton-thin absorbed state
in 2005.  Turner \etal\ (2008), fitting \textit{Chandra} HETGS and
\textit{XMM-Newton} EPIC and RGS spectra, list four "zones" of
absorption in addition to $N_{\rm H,Gal}$.  Zone 3 (using their
notation) is expected to impact modeling of PCA spectra of NGC~3516.
This absorber is a partial-covering, moderately-ionized (log$\xi$ =
2.19) absorber; Turner \etal\ (2008) argue that variations in the
covering fraction $CF$ on timescales of $\sim$ half a day can explain
observed spectral variability.  Rivers \etal\ (2011a), fitting the
time-averaged PCA + HEXTE spectrum, modeled Zone 3 with log($\xi$)
frozen at 2.19 and column density $N_{\rm H,WA}$ frozen at $2 \times
10^{23}$~cm$^{-2}$, the best-fitting parameters from Turner \etal\ (2008),
but leave the covering fraction $CF$ free, obtaining a best-fitting value
of $CF=0.55 \pm 0.10$.

For time-resolved spectroscopy, we sum the spectra from 2011 into
five bins spanning approximately 80 d each; these spectra typically
have exposure times of 11--22~ks and 25--50 $\times 10^3$ counts in the
3--23~keV band.  We fit 3--23~keV spectra using the best-fitting model
parameters from Rivers \etal\ (2011a), keeping log ($\xi$) and $N_{\rm
H,WA}$ frozen at 2.19 for simplicity.

We test three simple models for the spectral variability: a model
with $CF$ free, $\Gamma$ frozen at 1.85 (from Rivers \etal\ 2011a),
and $N_{\rm H,WA}$ frozen at $2 \times 10^{23}$~cm$^{-2}$ (which we call Model
PC1); $\Gamma$ frozen at 1.85, $CF$ frozen at 0.75, and $N_{\rm
  H,WA}$ free (Model PC2); and $\Gamma$ free, $CF$ frozen at the
arbitrary value of 0.75, and $N_{\rm H,WA}$ frozen at $2 \times
10^{23}$~cm$^{-2}$ (Model PC3).  $A_1$ and $I_{\rm Fe}$ are free
parameters in all fits. The results are summarized in Table~A8.

For Model PC1, best-fitting values of $CF$ span $0.34$ to $1.00$, with
the highest value occurring for the Fall 2011 spectrum ($\#4$).
However, uncertainties on $CF$ are extremely large; we also cannot
rule out full-covering for spectrum $\#2$.  Values of $\chi^2_{\rm
  red}$ between each of the three models were very similar and usually
$\la$1.  For model PC2, spectrum $\#4$ yields a value of
$N_{\rm H,WA}$ about twice that for spectra $\#2$, 3 and 5, but
$\chi^2_{\rm red}$ was 1.57, with strong data/model residuals (at the
$\sim10-20$ per cent level) below 4--5~keV.

For Model PC3, best-fitting values of $\Gamma$ are usually 1.7 and
greater, with $\Gamma=1.55\pm0.04$ for the Fall 2011 spectrum; such
values are not implausible based on empirical grounds.  Since we
cannot rule out flattening of the primary power law as the driver
behind the observed variations in $HR1$, we cannot confirm that
variations in properties of the ionized absorber occurred.  More
definitive results can be obtained if this source is subjected to
future long-term X-ray spectral monitoring using higher energy
resolution and a wider bandpass than the PCA, such as eROSITA or the
proposed \textit{Large Observatory For X-ray Timing}.


We thus classify the Fall 2011 event as a ``candidate'' full-covering
event, with a duration of $\sim57$ d based on the points in the $HR1$
light curve above $1\sigma$ deviation.  With best-fitting covering
fractions of 74, 100, and 79 per cent before, during, and after the event,
respectively, in Model PC1, we use a covering fraction of 23.5 per cent
and assign $\Delta$$N_{\rm H}$ =(0.235~$\times$~$2\times10^{23}$~cm$^{-2}$) = $4.7\times10^{22}$
cm$^{-2}$.

\begin{table*}
\begin{minipage}{160mm} 
\caption{Results of time-resolved spectral fits to NGC~3516, 2011.7}
\label{tab:3516TRtable} 
\begin{tabular}{lcccccccc} \hline

Start--stop                & Expo & \multicolumn{4}{c}{Model PC1}                            & \multicolumn{3}{c}{Model PC2}              \\
   (MJD)                   & (ks) & $\chi^2_{\rm red}$ & $CF$     & $A_1$           & $F_{2-10}$ & $\chi^2_{\rm red}$ & $N_{\rm H,WA}$ & $A_1$   \\ 
                           &      & (44 $dof$)       &          &                 &          &  (45 $dof$)      &              &                 \\ \hline
55561.55--55637.70 ($\#1$) & 17.4 & 0.67 & $0.34^{+0.10}_{-0.09}$ & $ 6.3\pm1.1$         & 4.60  & 0.76 & $8.7^{+1.3}_{-2.0}$ & $13.5\pm0.4$ \\ 
55641.58--55721.90 ($\#2$) & 17.7 & 0.84 & $0.84^{+0.16}_{-0.14}$ & $12.8\pm1.0$         & 2.55  & 1.02 & $35\pm5$            & $11.4\pm0.7$     \\ 
55725.88--55769.57 ($\#3$) & 10.8 & 0.50 & $0.74^{+0.24}_{-0.18}$ & $10.1\pm1.3$         & 2.44  & 0.49 & $28^{+7}_{-5}$      & $10.0\pm0.8$ \\ 
55773.55--55873.27 ($\#4$) & 22.2 & 0.98 & $1.00^{+0.00}_{-0.08}$ & $14.5^{+0.3}_{-0.4}$ & 2.01  & 1.57 & $65^{+2}_{-4}$      & $12.3\pm0.3$         \\ 
55877.52--55925.39 ($\#5$) & 11.4 & 1.08 & $0.79^{+0.19}_{-0.15}$ & $13.4^{+1.2}_{-1.3}$ & 2.92  & 1.13 & $30\pm5$            & $12.3\pm0.9$  \\ \hline            
\end{tabular}
\begin{tabular}{lccc}\hline
Start--stop                & \multicolumn{3}{c}{Model PC3} \\
   (MJD)                   & $\chi^2_{\rm red}$ &  $\Gamma$ &   $A_1$    \\ 
                           & (45 $dof$)      &           & \\ \hline 
55561.55--55637.70 ($\#1$) & 1.53 & $2.12\pm0.03$ & $30.3^{+0.9}_{-1.2}$ \\
55641.58--55721.90 ($\#2$) & 0.90 & $1.78\pm0.04$ & $9.5\pm0.7$      \\ 
55725.88--55769.57 ($\#3$) & 0.49 & $1.86\pm0.05$ & $10.4^{+1.1}_{-1.0}$ \\ 
55773.55--55873.27 ($\#4$) & 1.03 & $1.55\pm0.04$ & $ 5.0\pm0.4$ \\
55877.52--55925.39 ($\#5$) & 1.11 & $1.83\pm0.04$ & $11.8\pm1.0$ \\ \hline
\end{tabular} \\
Results of time-resolved spectroscopy for the 2011 campaign of 
NGC~3516.  Each model incorporates a partial-covering absorber with
log($\xi$)=2.19. Covering fraction $CF$ is free in Model PC1, and
frozen at 0.75 in Models PC2 and PC3. $N_{\rm H,WA}$ is free in
Model PC2 and frozen at $20 \times 10^{22}$~cm$^{-2}$ in Models
PC1 and PC3. $\Gamma$ is free in Model PC3, and frozen at
1.85 in Models PC1 and PC2.  $N_{\rm H}$ is listed in units of
$10^{22}$~cm$^{-2}$.  $A_1$ is the 1~keV normalization of the
\textit{covered} power law only in units of $10^{-3}$ ph cm$^{-2}$
s$^{-1}$ keV$^{-1}$. $F_{2-10}$ is the observed/absorbed model flux in
units of $10^{-11}$ erg cm$^{-2}$ s$^{-1}$; values for Models PC2
and PC3 were virtually identical to those for model PC1. $I_{\rm
  Fe}$ is left free, with best-fitting values and uncertainties typically
$\sim6\pm3\times 10^{-5}$ ph cm$^{-2}$ s$^{-1}$ for Models PC1 and
PC2 and typically $\sim8\pm4\times 10^{-5}$ ph cm$^{-2}$ s$^{-1}$
for Model PC3.
\end{minipage}
\end{table*}

\clearpage

\subsection{Summary of eclipse events manifested via flux-flux plots}

\begin{figure*}
\includegraphics[angle=-90,width=1.0\textwidth]{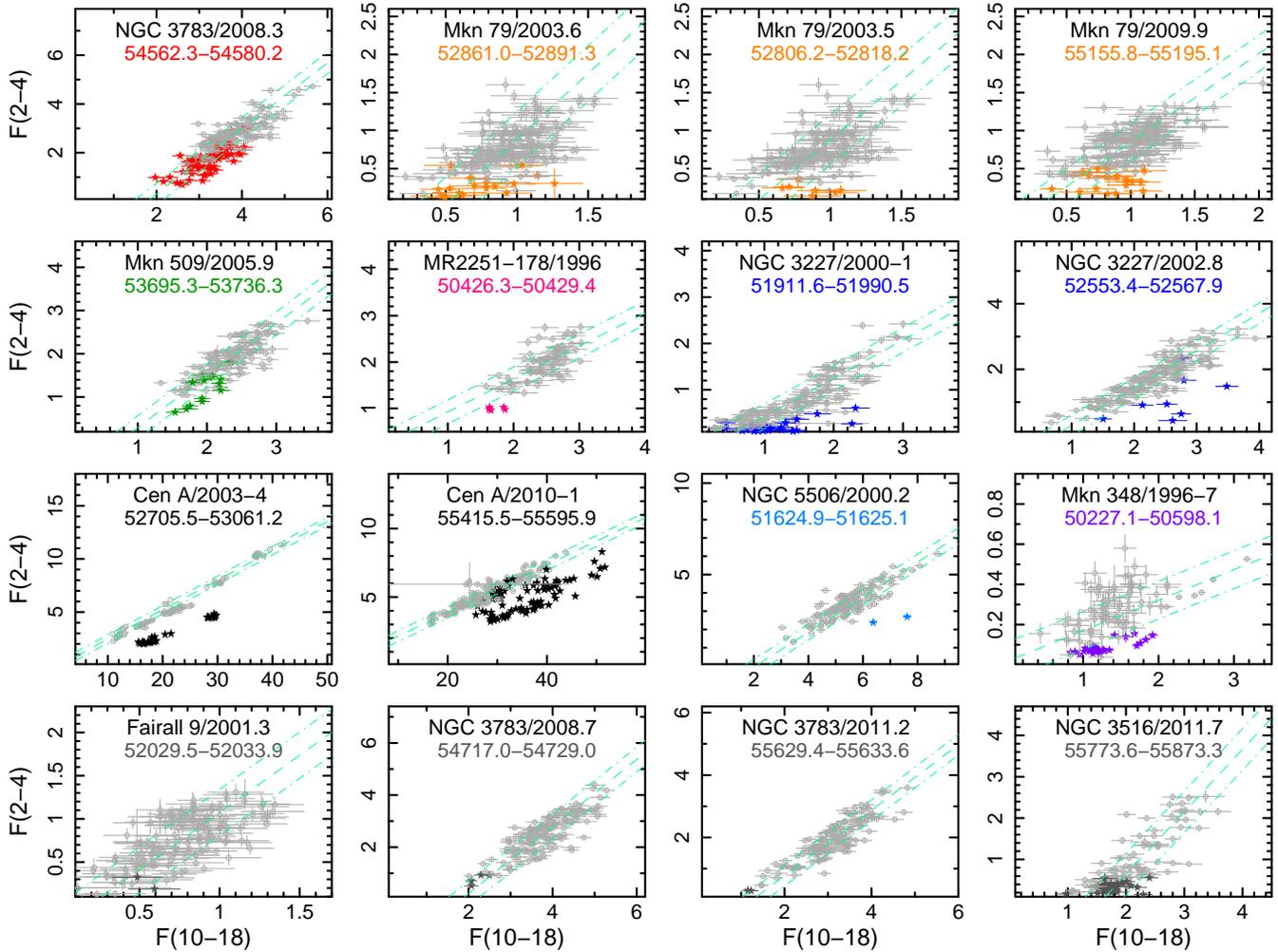}   
\caption{
2--4 keV flux plotted against 10--18 keV flux for each observation
during the secure/candidate eclipse events (colored stars) and for the
75 observations before and after each putative eclipse event
(excluding other candidate/secure events; gray open circles). The 
12 secure events are in the top three rows; the candidate events
are in the bottom row. The green dashed line is a best fit to the 
non-eclipsed points, with the green dot-dashed lines indicating 
their $\pm 1\sigma$ distribution. For each of the secure events, 
the 2--4 keV band is affected independently of the 10--18 keV flux, 
indicating an eclipse by material with $N_{\rm H} \sim 10^{22-23}$ 
cm$^{-2}$.  However, for each candidate event, there exists some 
ambiguity in modeling the spectral behavior, usually due to low 
flux in the power-law component (as probed by the 10--18 keV flux).   
}\label{fig:fluxflux}
\end{figure*}

If there is an eclipse by a discrete cloud with $N_{\rm H} \sim
10^{22-23}$ cm$^{-2}$, as implied by the above spectral fits, then the
2--4 keV band should be impacted much more strongly than the 10--18
keV band. The 10--18 keV band probes the uneclipsed portion of the
continuum power-law component in this case. In
Figure~\ref{fig:fluxflux}, we plot 2--4 keV flux as a function of
10--18 keV flux for each observation during each eclipse event, along
with flux points representing non-eclipsed periods (the gray points).
We use only the 75 data points before and after the putative eclipse
(excluding other putative eclipse events) for clarity, to avoid
overcrowding the figure. The green dashed line is a best fit to the
gray points, using a linear regression algorithm that accounts for
uncertainties in both quantities (Fasano \& Vio 1988). The green
dot-dashed lines indicate the $\pm 1\sigma$ distribution of the gray
points. When there is no eclipse, the 2--4 and 10--18 keV bands are
both probing the continuum and track variations in the power-law
continuum. During each of the secure events, the flux-flux points
track the spectral variability away from the main distribution,
especially near the peak of each event. These plots do not directly
yield quantitative information about $N_{\rm H}$ or $\Gamma$; those
quantities are best obtained via time-resolved spectroscopy. However,
these plots confirm that the 2--4 keV band is impacted independently
of the behavior of the primary power law. For each of the candidate
events, however, the low levels of 10--18 kev continuum flux (as
indicated in bottom row of panels) introduce ambiguity in terms of our
ability to use time-resolved spectroscopy to rule out flattening of
the primary power law as the cause of observed spectral flattening.

\FloatBarrier

\subsection{Type IIs with $\geq$0.6 years of monitoring}

\begin{figure*}
\includegraphics[angle=-90,width=1.00\textwidth]{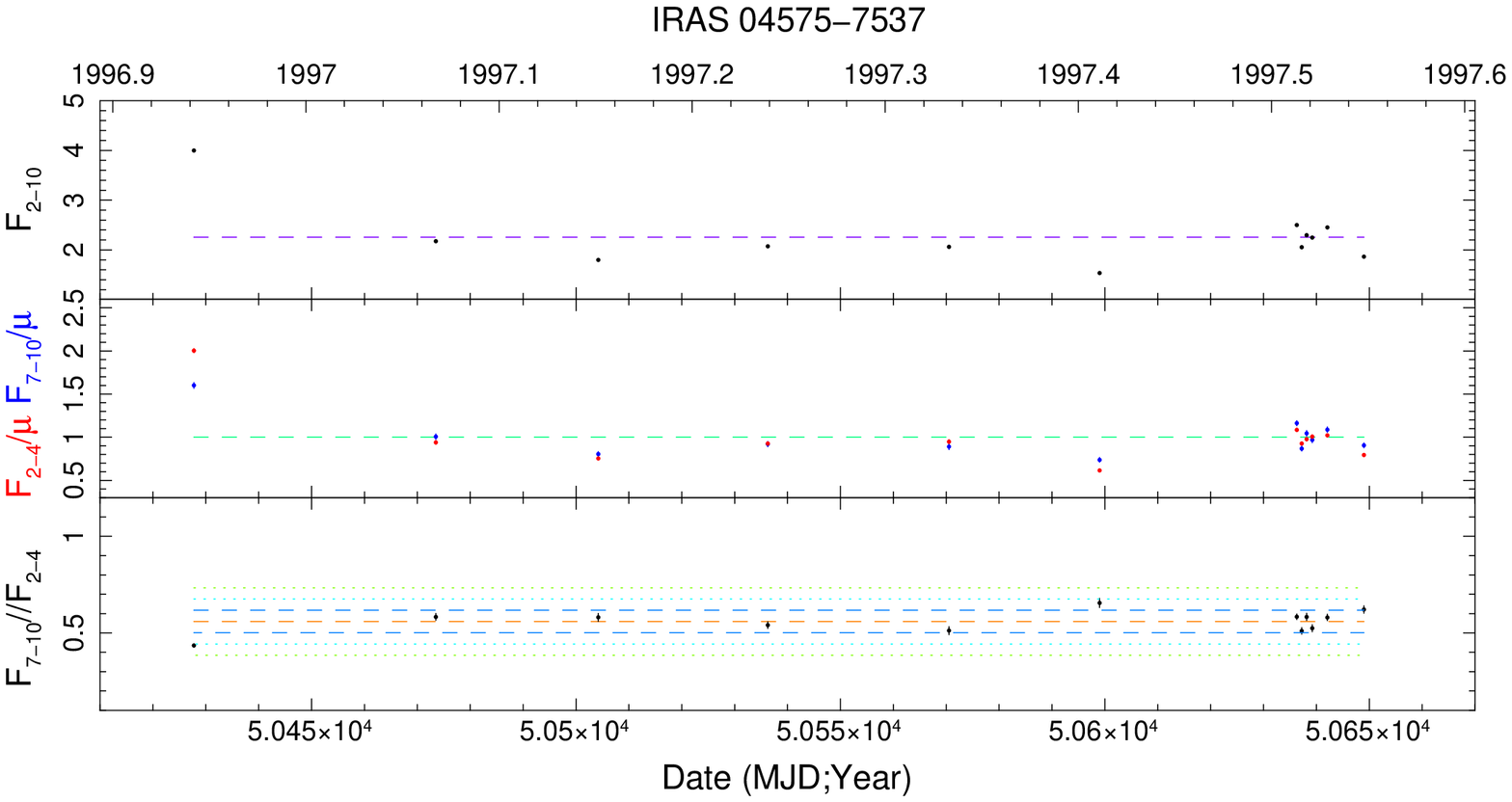}  
\caption{Long-term continuum flux [$F_{\rm 2-10}$, top panel; $F_{\rm
    2-4}$ (red) and $F_{\rm 7-10}$ (blue), middle panel] and $HR1$
  light curves (bottom panel) for the Sy~2 IRAS~04575--7537.}
\label{fig:sy2iras}  
\end{figure*}

\begin{figure*}
\includegraphics[angle=-90,width=1.00\textwidth]{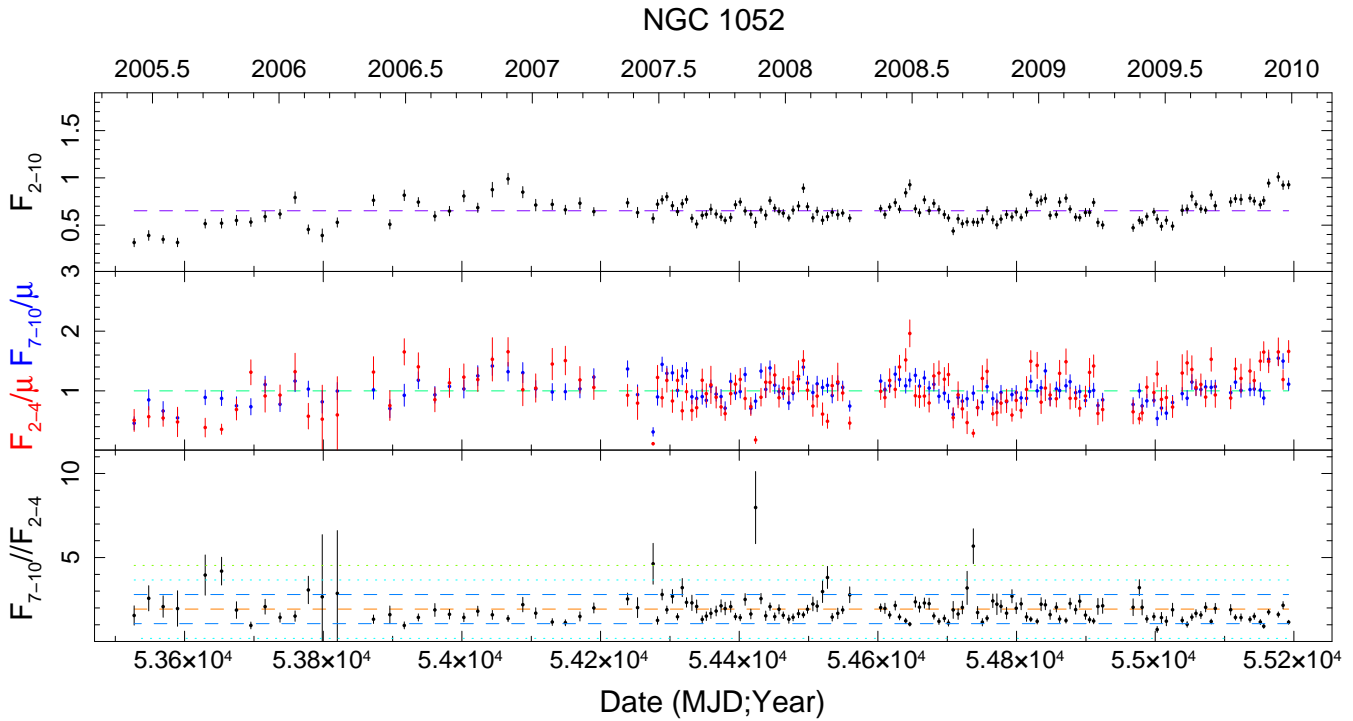}       
\caption{Same as Fig.~\ref{fig:sy2iras}, but for the radio-loud Sy~2 NGC~1052.}
\label{fig:sy21052}  
\end{figure*}

\begin{figure*}
\includegraphics[angle=-90,width=1.00\textwidth]{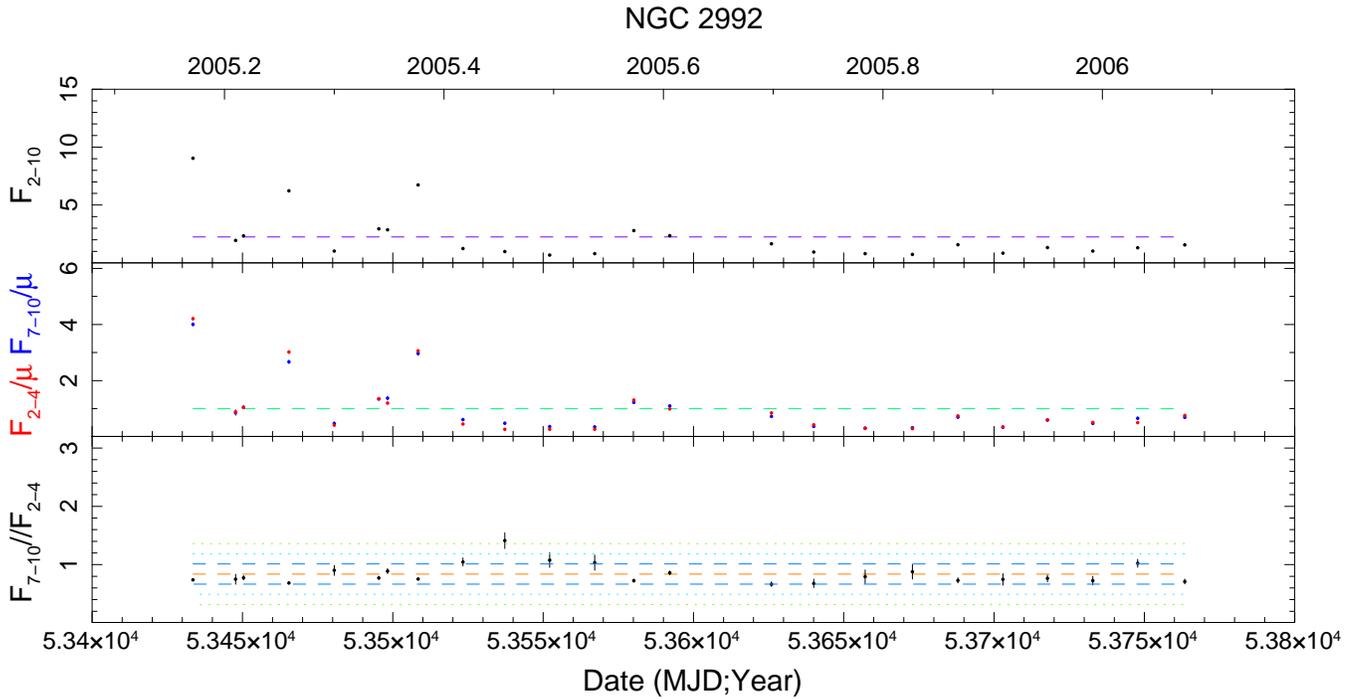}       
\caption{Same as Fig.~\ref{fig:sy2iras}, but for the Sy~2 NGC~2992.}
\label{fig:sy22992}  
\end{figure*}

\begin{figure*}
\includegraphics[angle=-90,width=1.00\textwidth]{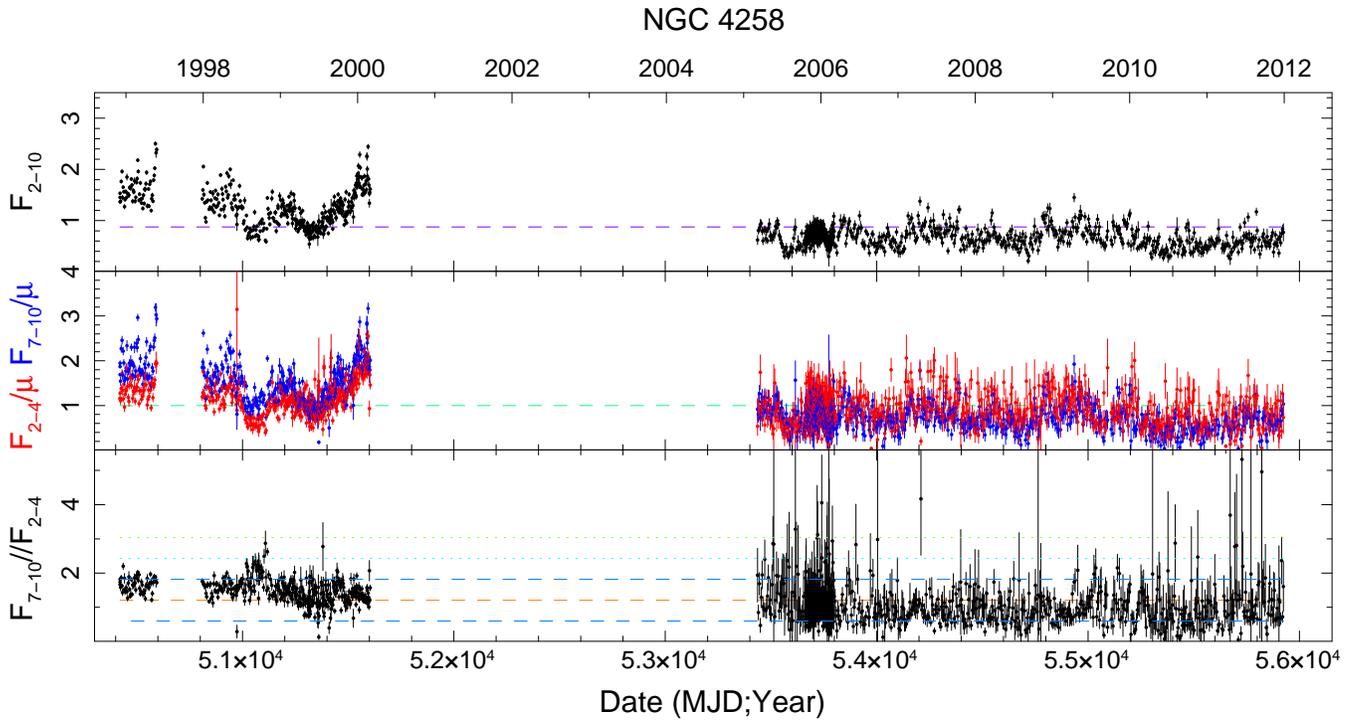}       
\caption{Same as Fig.~\ref{fig:sy2iras}, but for the Sy~2/LLAGN NGC~4258.}
\label{fig:sy24258}  
\end{figure*}

\begin{figure*}
\includegraphics[angle=-90,width=1.00\textwidth]{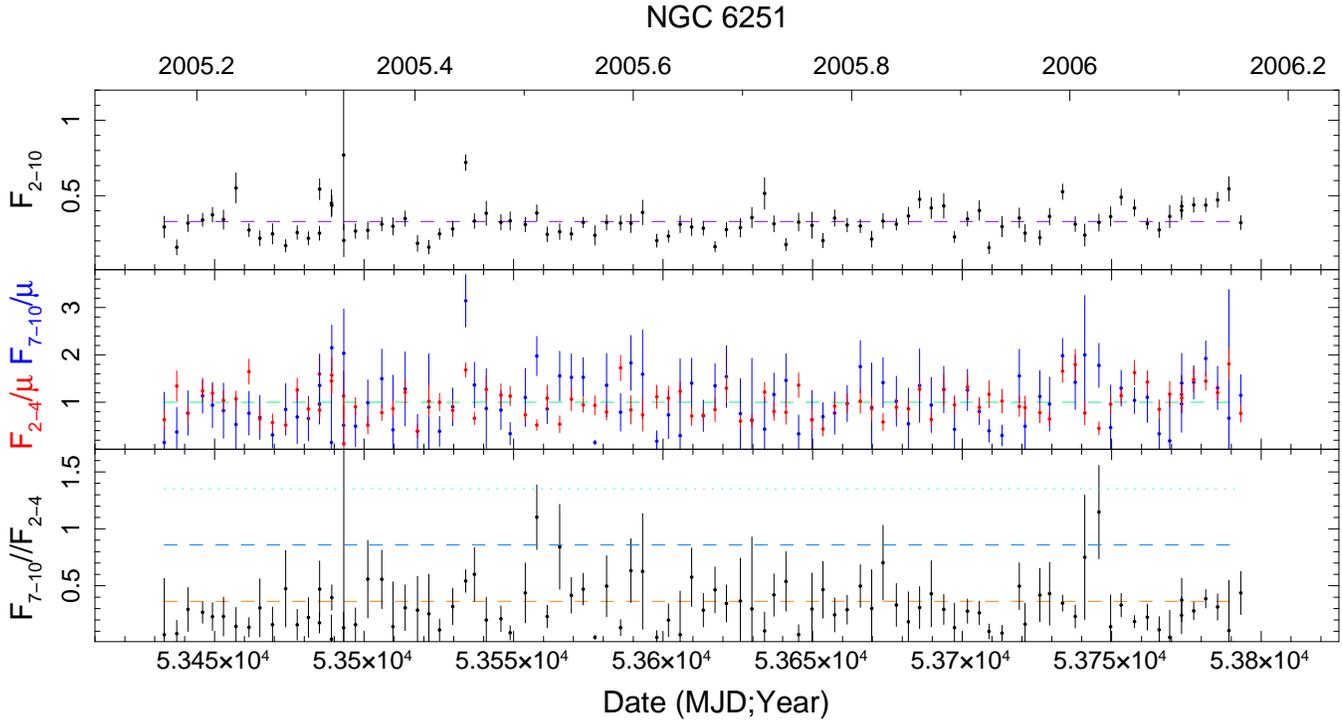}       
\caption{Same as Fig.~\ref{fig:sy2iras}, but for the Sy~2/LERG NGC~6251.}
\label{fig:sy26251}  
\end{figure*}

\begin{figure*}
\includegraphics[angle=-90,width=1.00\textwidth]{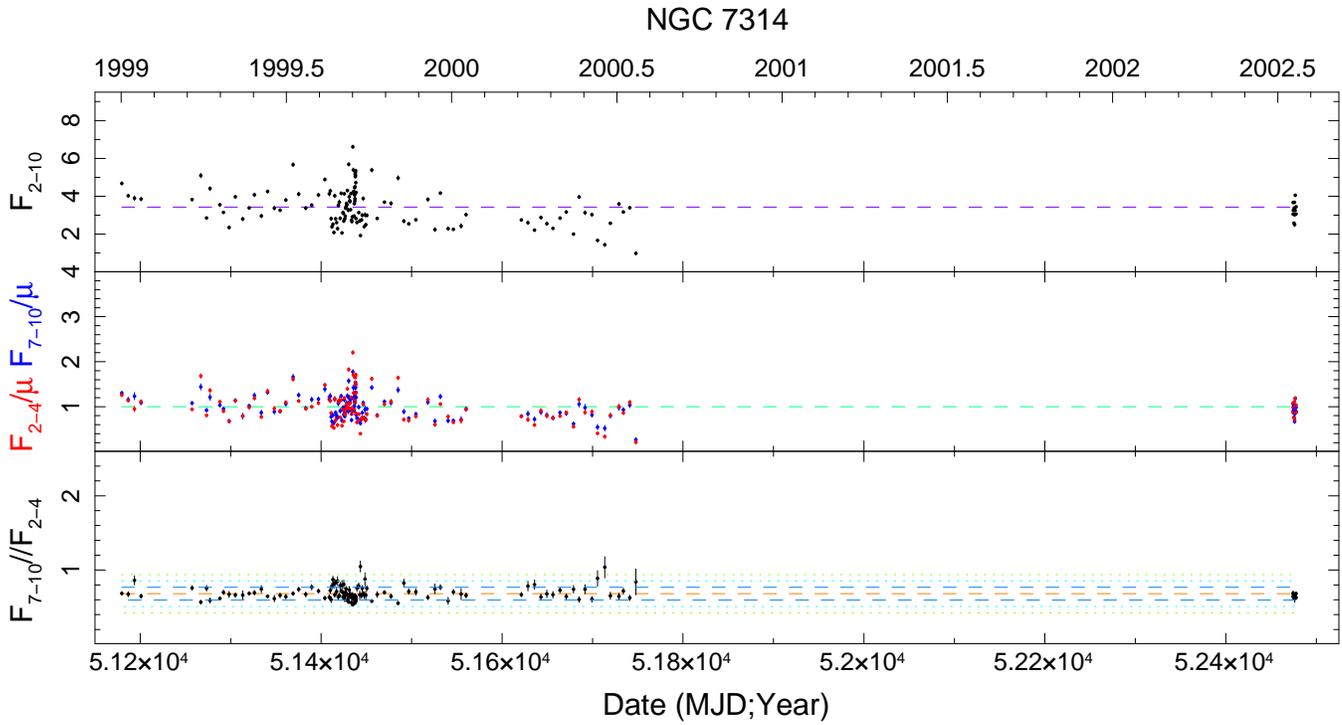}       
\caption{Same as Fig.~\ref{fig:sy2iras}, but for the Sy~1.9 NGC~7314.}
\label{fig:sy27314}  
\end{figure*}

In Figs.~\ref{fig:sy2iras}--\ref{fig:sy27314} we plot the long-term
continuum and $HR1$ light curves for six type II objects with at least
0.6 years of sustained monitoring.  X-ray spectral observations
typically confirm the presence of an X-ray column $\ga10^{22}$
cm$^{-2}$ in each of these objects (e.g., REN02).  As hardness ratios
are sensitive to changes in either $\Gamma$ or $N_{\rm H}$, the lack
of any obvious systematic trends (above 2.0$\sigma$ from $\langle HR1
\rangle$, and/or trends $\sim50$ per cent greater than $\langle HR1 \rangle$)
suggests the lack of strong variations in either of these parameters.
If we assume that $\Gamma$ is intrinsically constant, then we can use
the average and 1$\sigma$ standard deviation of $HR1$ and the best-fitting
time-averaged parameters from Rivers et al.\ (2013) to estimate the
corresponding maximum change in $N_{\rm H}$; these estimates of
$\Delta$$N_{\rm H}$ are listed in Table~\ref{tab:boringsy2s}.

\begin{table*}
\begin{minipage}{160mm}
\caption{Estimated sensitivity to $\Delta$$N_{\rm H}$ for type~IIs based on $HR1$ light curves}
\label{tab:boringsy2s} 
\begin{tabular}{lccllc}  \hline
                    &                       &          &  Typical                                       & Typical                                      &  $\Delta$$N_{\rm H}$   \\
Source              & $\langle HR1 \rangle$ & $\sigma$ & $N_{\rm H}$ ($10^{22}$~cm$^{-2}$)                 & $\Gamma$                                     &  ($10^{22}$~cm$^{-2}$)  \\ \hline
IRAS~04575--7537    & 0.56                  & 0.06     & $3.6\pm2.6$ (R13)                              & $2.48\pm0.22$ (R13)                          & $\sim0.6$     \\
Mkn~348 (2011 only) & 2.35                  & 0.84     & $16.6\pm1.6$ (TW); $\sim11-16$ (R02)           & $1.69\pm0.06$ (TW)                           & $\sim5$       \\
NGC~1052            & 1.93                  & 0.87     & $13.6\pm5.2$ (R13); $31^{+11}_{-26}$ (B11)       & $1.71\pm0.29$ (R13); $1.7^{+0.1}_{-0.2}$ (B11)  & $\sim6$       \\
NGC~2992            & 0.84                  & 0.17     & 0.4--1.6 (R02); $0.41\pm0.17$ (B11)            & $1.78\pm0.18$ (R13); $1.59\pm0.03$ (B11)     & $\la4$   \\
NGC~4258            & 1.21                  & 0.61     & $8.4\pm3.9$ (R13); $\sim9-14$ (R02)            & $1.80\pm0.17$ (R13)                          & $\sim5$       \\
NGC~5506$^{\dagger}$  & 0.89                  & 0.10     & $1.9\pm0.5$ (R11); $\sim2-4$ (R02)             & $1.93\pm0.03$ (R11); $1.82^{+0.05}_{-0.04}$ (B11)& $\sim1$      \\
NGC~6251            & 0.36                  & 0.49     & $\la0.4$ (G09)                            & $2.38\pm0.23$ (R13); $1.67\pm0.06$ (E06)      & $\la9$  \\
NGC~7314            & 0.68                  & 0.09     & $\sim0.8-1.3$ (R02); $0.60^{+0.01}_{-0.03}$ (B11) & $1.99\pm0.10$ (R13); $1.95^{+0.02}_{-0.01}$ (B11)& $\sim2$      \\ \hline 
\end{tabular} \\
Estimates of $\Delta$$N_{\rm H}$ corresponding to the standard deviations $\sigma$ of $HR1$.
$^{\dagger}$For NGC~5506, the two observations at MJD 51624.94 and 51625.14 during an eclipse event are excluded.
References for typical values of $\Gamma$ and $N_{\rm H}$ are:
B11 =  Brightman \& Nandra (2011);
E06 = Evans et al.\ (2006);
G09 = Gonz\'{a}lez-Mart\'{i}n et al.\ (2009);
R11 = Rivers et al.\ (2011a);
R13 = Rivers et al.\ (2013);
R02 = Risaliti, Elvis \& Nicastro (2002) and references therein;
TW = this work (Mkn~348: average of the three values derived from time-resolved fitting).
\end{minipage}
\end{table*}

For NGC~4258, the average value of $HR1$ drops from 1.46
(stand.\ dev.\ 0.35) for the 1997--2000 data to 1.07
(stand.\ dev.\ 0.68) for the 2005--2011 data. While such a drop is
consistent with the average values of $\Gamma$ or $N_{\rm H}$
systematically varying, spectral fitting on data summed over
1997--2000 and over 2005--2011 failed to provide evidence for changes
in $\Gamma$ or $N_{\rm H}$, as uncertainties on these parameters were
extremely large.  Similarly, the average value of $HR1$ for NGC~5506
(Fig.~\ref{fig:mega5506}) also seems to fall very slightly towards the
end of 1999: $\langle HR1 \rangle$ = 1.03 (stand.\ dev.\ 0.08) for
data before MJD~51300, and 0.85 (stand.\ dev.\ 0.06) for data after
MJD~51500 (excluding the eclipse event covering
the two observations at MJD 51624.94 and 51625.14).  
Spectral fitting on these two data sets,
however, yields values of $N_{\rm H}$ consistent with the average at
90 per cent confidence.

\FloatBarrier \clearpage

\section{Inferred radial distances of various emission/absorption components in our eight primary sources}

In this appendix, we provide the inferred distances from the central
black hole of various emission/absorption components in our eight
sources with ``secure'' eclipse events; Fig.~\ref{fig:plotrs} is
based on these values.  In Table~B1, 
we list the X-ray clumps' minimum and maximum
distances as calculated in $\S$5.2: assuming a range of log($\xi$)
from --1 to +1 for all events except for the two events in 
Cen~A (log($\xi$) = --1
to 0) and NGC~3227/2000--1 (--0.3 to 0).  We also list inferred
locations of IR-emitting tori from either interferometry or
reverberation mapping, BLR emission lines, and Fe K$\alpha$
line-emitting gas.  The reader is reminded that even if distances from
the black hole are commensurate, structures may not physically overlap
(e.g., emission originating from out of the line of sight versus X-ray
absorption along the line of sight).

\begin{table*}
\begin{minipage}{160mm}
\caption{Inferred radial distances of various structures in objects with ``secure'' eclipses} 
\label{tab:BLRdust}
\begin{tabular}{llccl}\hline
Source        & Component  & Radius (light-days) & Radius ($R_{\rm g}$) & Ref. \\  \hline
NGC 3783      & He~\textsc{ii}              &  $1.4^{+0.8}_{-0.5}$  &  $850^{+480}_{-300}$            & P04   \\ 
              & Si~\textsc{iv}              &  $2.0^{+0.9}_{-1.1}$  &  $1210^{+550}_{-670}$           &  P04 \\
              & C~\textsc{iv}               &  $3.8^{+1.0}_{-0.9 }$ &  $2300^{+610}_{-550}$           & P04 \\
              & H$\beta$                    &  $10.2^{+3.3}_{-2.3}$ &  $6180^{+2000}_{-1390}$          &  P04    \\
              & He~\textsc{i}$\lambda$2.058$\mu$ & $14.7^{+4.7}_{-3.2}$ & $8570^{+2740}_{-1870}$        &  R02 \\
              & Br$\gamma$ (broad)          &  $21.4^{+1.5}_{-1.3}$ & $1.253^{+0.083}_{-0.075}\times 10^4$   &  R02 \\
              & Fe K$\alpha$ width, HETGS   &  $66^{+39}_{-20}$    &  $3.9^{+2.2}_{-1.2} \times 10^4$ &    S10 \\ 
              & IR torus (B08, Model A)     & 48--76             &   $2.8-4.4  \times 10^4$      & B08 \\
              & $H$-band-emitting dust        &  $66^{+6}_{-7}$  &      $3.9\pm0.4\times 10^4$     & L11 \\
              & $K$-band-emitting dust        &  $76^{+11}_{-17}$ &      $4.4^{+0.6}_{-1.0} \times 10^4$     & L11 \\
              & X-ray clump                 &  55--400           &  3.2--23 $\times 10^4$        & This work       \\
              & $R_{\rm d}$                  &  240                & $13.9   \times 10^4  $     & This work \\
              & IR torus (B08, Model B)     & 250--357           & $14.6-20.9  \times 10^4$      & B08 \\     \hline
Mkn~79        & Fe K$\alpha$ width, EPIC                    &  $3.0^{+0.7}_{-0.8}$  &  $990^{+210}_{-240}$       &  Ga11 \\
              & H$\beta$                                    &  $9.0^{+8.3}_{-7.8}$  & $ 3100^{+2800}_{-2600}$    &  P04\\
              & H$\beta$                                    &  $16.0^{+6.4}_{-5.8}$ &  $ 5400^{+2200}_{-2000}$   &  P04\\
              & H$\beta$                                    &  $16.1\pm6.6$      &  $ 5500\pm 2200$        &  P04\\ 
              & C~\textsc{iv}                               &  16.9              &   5590                  &  K07 \\
              & Br$\gamma$                                  &  20.5              &  6750                   & L08 \\ 
              & H$\alpha$                                   &  23.3              &  7660                   & L08 \\ 
              & H$\beta$                                    &  26.4              &  8690                   & L08 \\ 
              & He~\textsc{i}$\lambda$5876                   &  27.9              &  9170                   & L08 \\ 
              & O~\textsc{i}$\lambda$1.1287$\mu$              & 30.7              &  10090                 & L08 \\ 
              & Pa$\alpha$, Pa$\beta$, Pa$\delta$, Pa$\epsilon$ & 29.0--31.5        & 9530-- 10360           & L08 \\ 
              & O~\textsc{i}$\lambda$8446                    &  32.6             &  10720                 & L08 \\ 
              & He~\textsc{i}$\lambda$1.0830$\mu$             &  57.1             & 18800                  & L08  \\      
              & $R_{\rm d}$                                   &  340               & $11.1   \times 10^4  $     & This work \\ 
              & X-ray clump                                  &  60--1020          &  $2.0-34 \times 10^4$  & This work     \\ \hline 
Mkn~509      & He~\textsc{ii}                        &  $33.5^{+8.2}_{-7.1}$    &   $3870^{+950}_{-820}$  &  P04 \\ 
             & He~\textsc{i}$\lambda$5876            &  59                    &  6550    & L08 \\
             & H$\beta$                             &  68                    &  7570    & L08 \\
             &  H$\beta$                           &  $79.6^{+6.1}_{-5.4}$    &  $9190^{+700}_{-620}$   &  P04 \\ 
           & Pa$\alpha$, Pa$\beta$, Pa$\gamma$, Pa$\delta$, Pa$\epsilon$ &  93--120 & 10330--13340   & L08 \\
             & O~\textsc{i}$\lambda$8446             &  99                    & 11020    & L08 \\
             & H$\alpha$                            & 100                    & 11180    & L08 \\
             & H$\gamma$                            & 103                    & 11450    & L08 \\
             & Br$\gamma$                            & 111                    & 12390    & L08 \\
             & He~\textsc{i}$\lambda$1.0830$\mu$     & 119                    & 13320    & L08 \\
             & Fe K$\alpha$ line width, HETGS      & $123^{+254}_{-89}$         & $1.4^{+2.8}_{-1.0} \times 10^4$   &  S10 \\  
             & O~\textsc{i}$\lambda$1.1287$\mu$      & 126                    & 14010    & L08 \\
             & $R_{\rm d}$                            &  610                   &  $6.7   \times 10^4  $     & This work \\
             & X-ray clump                         &  230--2800              &  2.6--31  $\times 10^4$  & This work \\   \hline
MR~2251--178  & Fe K$\alpha$ width, \textit{Suzaku}  & $>$9                &  $>$750                   & Go11  \\    
              & H$\beta$                             &    27               & 2300                & S07 \\
              & C~\textsc{iv}                        & $85^{+15}_{-13}$      & $7230^{+1310}_{-1040}$ & S07 \\  
              & $R_{\rm d}$                 &     910             &  $8.1   \times 10^4  $       & This work \\
              &  X-ray clump               &    200--2500        & 1.7--120 $\times 10^4$        & This work \\ \hline
\end{tabular}
\end{minipage}
\end{table*}

\setcounter{table}{0}

\begin{table*}
\begin{minipage}{160mm}
\caption{Inferred radial distances of various structures in objects with ``secure'' eclipses, continued} 
\begin{tabular}{llccl}\hline
Source        & Component  & Radius (light-days) & Radius ($R_{\rm g}$) & Ref. \\  \hline
NGC~3227      & He~\textsc{i}$\lambda$5876         & 2.0              &   4580                  & L08 \\
              & H$\beta$                           & 3.4              &   7640                  & L08 \\
              & Pa$\epsilon$                       & 3.4              &   7690                  & L08 \\
              & H$\gamma$                          & 3.7              &   8330                  & L08 \\
              & O~\textsc{i}$\lambda$8446          & 4.0              &   8980                  & L08 \\
              & H$\beta$                           & $3.8\pm0.8$      &  $9000\pm1890$  &  D10  \\
              & Br$\gamma$                         & 4.0              &   9010                  & L08 \\
              & H$\alpha$                          & 4.4              &  10010                  & L08 \\
              & O~\textsc{i}$\lambda$1.1287$\mu$   & 4.9              &  11220                  & L08 \\
              & He~\textsc{i}$\lambda$1.0830$\mu$  & 5.7              &  12860                  & L08 \\
              & Pa$\beta$, Pa$\delta$              & 6.0              &  13600                  & L08 \\
              & Fe K$\alpha$ width, EPIC           & $7.2^{+12.7}_{-4.9}$  &  $1.7^{+3.0}_{-1.2} \times 10^4$  & M09 \\ 
              & H$\alpha$                          & $18.9^{+8.7}_{-11.3}$ &  $4.48^{+2.06}_{-2.68} \times 10^4$ &  P04 \\
              & $K$-band-emitting dust               & $\sim$20           &  $4.74 \times 10^4$ &  S06  \\
              & X-ray clump                        & 6.5--91            &  $1.5 - 21 \times 10^4$ &  This work \\  
              & $R_{\rm d}$                          & 85                 &  $19.2   \times 10^4  $     & This work \\
              & Fe K$\alpha$ variability           & $<$ 700            &  $< 1.7 \times 10^6$  & M09 \\ \hline
Cen~A         & $R_{\rm d}$                            & 42                &  $1.22   \times 10^4  $     & This work \\ 
              & Fe K$\alpha$ width, \textit{Suzaku}  &  $>$ 66           &   $>1.9 \times 10^4$        & Ma07\\ 
              & IR Torus                             &  120--360         &   3.5--11 $\times 10^4$    &  Me07  \\   
              & X-ray clump                          &   94--710         &   2.7--20 $\times 10^4$    & This work \\  \hline
NGC~5506      & X-ray clump                          &  15--220          &  $0.3-4.4 \times 10^4 $          & This work  \\ 
              & Pa$\beta$                            &  $\sim$190        &  $3.7 \times 10^4 $               & N02 \\
              & Fe K$\alpha$ width, HETGS            &  $220^{+764}_{-127}$ &  $4.4^{+15.2}_{-2.5} \times 10^4 $  &  S10 (68$\%$ err.) \\  
              & $R_{\rm d}$                            & 240               &  $4.8   \times 10^4  $     & This work \\ 
              & Br$\gamma$                           &  $250^{+40}_{-30}$  &  $5.0^{+0.7}_{-0.6} \times 10^4 $       & N02 \\
              & Br$\alpha$                           &  $420^{+70}_{-60}$  &  $8.3^{+1.6}_{-1.2} \times 10^4$       & L02  \\ \hline
Mkn~348       & H$\beta$                             & $1.19\pm0.03$        &  1340$\pm$30              & T95\\
              & H$\alpha$                            & $1.47^{+0.10}_{-0.11}$  &  $1650^{+130}_{-110} $       & T95\\
              & Br$\gamma$                           & $\sim$ 16.6          &  $1.87 \times 10^4  $     & V97\\
              & Pa$\beta$                            & $\sim$ 30.4          &  $3.42 \times 10^4  $     & V97 \\
              & $R_{\rm d}$                            & 235                  &  $26.8   \times 10^4  $     & This work  \\ 
              & X-ray Clump                          & 140--1290            &  $16-150 \times 10^4$      & This work \\   \hline 
\end{tabular}\\
For BLR line lags from reverberation mapping (P04, D10), $\tau_{\rm
  cent}$ is used if available. For Mkn~79, we use ``unflagged''
H$\beta$ values only. $R_{\rm d}$ denotes the approximate outer
boundary of the DSZ, i.e., dust residing at
distances greater than $R_{\rm d}$ likely does not sublimate, while
distances smaller than $\sim\frac{1}{2}-\frac{1}{3}R_{\rm d}$ are
likely to be dust-free.
References for Column (5) are: B08 = Beckert \etal\ (2008), D10 =
Denney \etal\ (2010), Ga11 = Gallo \etal\ (2011), Go11 = Gofford et
al.\ (2011), 
K07 = Kelly \& Bechtold (2007), L02 = Lutz \etal\ (2002), L08 = Landt
\etal\ (2008), L11 = Lira \etal\ (2011), Ma07 = Markowitz \etal\
(2007), M09 = Markowitz \etal\ (2009), Me07 = Meisenheimer \etal\
(2007), N02 = Nagar \etal\ (2002), P04 = Peterson \etal\ (2004), R03 =
Reunanen \etal\ (2003), S06 = Suganuma \etal\ (2006), S07 = Sulentic
\etal\ (2007), S10 = Shu \etal\ (2010; 68 per cent uncertainties
used), T95 = Tran (1995), V97 = Veilleux \etal\ (1997).
\end{minipage}
\end{table*}

\end{document}